\documentclass{ucthesis}
\usepackage{graphicx}
\usepackage{amsmath}
\usepackage{amstext}
\usepackage{mapleenv}
\usepackage{maplestyle}
\usepackage{array,tabularx,hhline}
\begin{document} 

\def\km    {~km}
\def\cm    {~cm}
\def\fm    {~fm}
\def\mm    {~mm}
\def\um    {~$\mu$m}
\def\m     {~m}

\def\deg   {$^\circ$}
\def\T     {~T}
\def\V     {~V}
\def\MV    {~MV}
\def\kV    {~kV}
\def\Vcm   {~V/cm}
\def\kVcm  {~kV/cm}

\def\evcc  {~eV/c$^2$}
\def\mev   {~MeV}
\def\mevc  {~MeV/c}
\def\mevcc {~MeV/c$^2$}
\def\gev   {~GeV}
\def\gevc  {~GeV/c}
\def\gevcc {~GeV/c$^2$}
\def\gevfmfmfm  {~GeV/fm$^3$}
\def\tev   {~TeV}

\def\piz   {$\pi^0$}
\def\gzepz {${\gamma}Z{\rightarrow}e^-e^+Z$}
\def\pizgg {$\pi^0\rightarrow\gamma\gamma$}
\def\ele   {$e^-$}
\def\pos   {$e^+$}
\def\ee    {$e^+e^-$ \ }
\def\AuAuee {$AuAu \rightarrow Au^*Au^* + e^+e^-$}
\def\sqnn    {$\sqrt{s_{_\mathrm{NN}}}$}

\def\au      {$^{^{197}}$Au}
\def\auau    {\au+\au}
\def\pb      {$^{^{208}}$Pb}
\def\pbpb    {\pb+\pb}

\def\armGas  {10\% CH$_4$ and 90\% Ar}

\def\ecm     {E$_\mathrm{CM}$}

\def\dedx {$dE/dx$}
\def\y    {y}
\def\pt   {p$_t$}
\def\pz   {p$_z$}
\def\px   {p$_x$}
\def\py   {p$_y$}
\def\xypl {$xy$-plane}
\def\rzpl {$rz$-plane}
\def\szpl {$sz$-plane}

\def\etal {{\it. et al}}

\title {Electron-Positron Production in Ultra-Peripheral Heavy-Ion \\ 
Collisions with the STAR Experiment}
\author{Vladimir Borisovitch Morozov}

\degreeyear{Fall 2003}
\degree{Doctor of Philosophy}
\chair{Professor Marjorie D. Shapiro}
\othermembers{Doctor Spencer R. Klein \\
Professor Richard Marrus \\
Professor Steven N. Evans}
\numberofmembers{4}
\prevdegrees{B.\ S.\ (Moscow Institute of Physics and Technology) 1997\\
M.\ A.\ (University of California at Berkeley) 1999}
\field{Physics}
\campus{Berkeley}
 
\maketitle
\approvalpage
\copyrightpage

\begin{abstract}
This thesis presents a measurement of the cross-section of the purely
electromagnetic production of \ee pairs accompanied by mutual nuclear
Coulomb excitation \AuAuee, in ultra-peripheral gold-gold collisions
at RHIC at the center-of-mass collision energy of $\sqrt{s_{NN}} =
200$ GeV per nucleon. These reactions were selected by detecting
neutron emission by the excited gold ions in the Zero Degree
Calorimeters. The charged tracks in the \ee events were reconstructed
with the STAR Time Projection Chamber.

The detector acceptance limits the kinematical range of the observed
\ee pairs; therefore the measured cross-section is extrapolated to
$4\pi $ with the use of Monte Carlo simulations. We have developed a
Monte Carlo simulation for ultra-peripheral \ee production at RHIC
based on the Equivalent Photon Approximation, the lowest-order QED \ee
production cross-section by two real photons and the assumption that
the mutual nuclear excitations and the \ee production are independent
(EPA model).  
various kinematic regions.

We compare our experimental results to two models: the EPA model and a
model based on full QED calculation of the \ee production, taking the
photon virtuality into account. The measured differential
cross-section $d\sigma /dM_{inv}$ ($M_{inv}$ -- \ee invariant mass)
agrees well with both theoretical models. The measured differential
cross-section $d\sigma /dp_{\perp}^{tot}$ ($p_{\perp}^{tot}$ -- \ee
total transverse momentum) favors the full QED calculation over the
EPA model.

\vspace{-0.5in}
\abstractsignature
\end{abstract}
 
\begin{frontmatter}
 
\begin{dedication}
\null\vfil
{\large
"The {\it central} and {\it peripheral} collisions of relativistic heavy ions may be compared to the case of two potential lovers walking on the same side of the street, but in opposite directions. If they do not care, they can collide frontally... It could be a good opportunity for the beginning of {\it strong interactions} between them. ... On the other hand, if they pass far from each other, they can still exchange glances (just {\it electromagnetic} interaction!), which can even lead to their {\it excitation}. ...the effects of these {\it peripheral} collisions are sometimes more interesting than the violent frontal ones."\\
}
\vspace{12pt}
\begin{flushright}
G. Baur and C. A. Bertulani \\
Physics Reports {\bf 163}, 299, (1989) \\
\end{flushright}
\vfil\null
\end{dedication}
 
\tableofcontents
\listoffigures
\listoftables
\begin{acknowledgements}
A large number of people have contributed directly or indirectly to the presented results and this thesis, many more than could possibly be mentioned in this brief account. I would like to thank each and every one of the people who collaborated with me on this analysis and helped me in the creation of this thesis. I thank my adviser Dr. Spencer Klein for introducing me to the world of ultra-peripheral physics and for guiding me from start to finish of my dissertation. I would also like to thank my UC Berkeley sponsor and co-adviser Professor Marjorie Shapiro for supervising my progress over the past 5 years.

A special note of appreciation goes to my ultra-peripheral working group colleague Dr. Falk Meissner. It has been truly a privilege and a great pleasure to work with somebody as knowledgeable, helpful and friendly as Falk. All of the particle physics analysis methods, tricks and hacks I know, I owe to Falk. This thesis has been infinitely improved from its original version through Falk's very careful and critical review. Thank you Falk, I wouldn't have made it this far without your help!

I would also like to thank the stuff of the LBNL Relativistic Nuclear Science group headed by Dr. Hans Georg Ritter for support and help in my graduate studies. I thank Dr. Joakim Nystrand for helping me get started in the ultra-peripheral collisions working group. I also thank Dr. Ian Johnson for help on the electron-positron track studies in the STAR TPC, and Dr. Kai Schweda for reading my thesis and helping me clean up all the bugs and typos! And I thank Dr. Kai Hencken of the University of Basel for the theoretical results.

Finally, I thank all my family members and friends for their support and trust, which have always been such an inspiration!

\end{acknowledgements}
 
\end{frontmatter}

\chapter{Introduction}
\label{ch:Intro} 

This thesis presents a study of the production of \ee   pairs in ultra-peripheral AuAu collisions at the Relativistic Heavy Ion Collider (RHIC) observed with the STAR detector (Solenoidal Tracker at RHIC). 
While most of the RHIC physics program is concerned with the hadronic AuAu interactions, 
we focus on the ultra-peripheral interactions, where the Au ions interact only via the long-range forces. The \ee   pairs are a product of purely electromagnetic interaction of the virtual photon fields emitted by the Au ions. The process of two photons converting into an \ee   pair has been studied experimentally with a good precision in quantum electrodynamics (QED) for pair energies up to 100 GeV.

The study of this process at RHIC, however, opens a door for several new interesting features. Since the fields from the proton constituents of Au ions (with charge $Z=79$) add coherently, the electromagnetic field strength scales as $Z^2$ and the interaction rate as $Z^4$. Therefore, RHIC will produce a very large two-photon interaction rate. Additionally, the field coupling constant is scaled up by the charge of Au ions, $Z\alpha \sim 0.6$, and the photon-photon interactions enter a strong interaction regime. We wish to measure the cross-section of the \ee   production in this regime.

The STAR detector is optimized for detecting charged particles at mid-rapidity and in the transverse momentum range from about 100 MeV/c up to several GeV/c. The fraction of \ee   pairs with both electron and positron tracks reaching the STAR Central Trigger Barrel (located at mid-rapidity) is close to zero; therefore triggering on the exclusive reaction $AuAu \rightarrow AuAu + e^{+}e^{-}$ in STAR is very difficult. We chose to focus our attention on a closely related reaction -- electromagnetic production of an \ee   pair with mutual nuclear Coulomb excitation of the colliding ions: \AuAuee. The neutrons emitted when the nuclear excitations decay provide a tag on this reaction in the forward neutron calorimeters. 

\begin{figure}
\centering
\includegraphics[width=200pt]{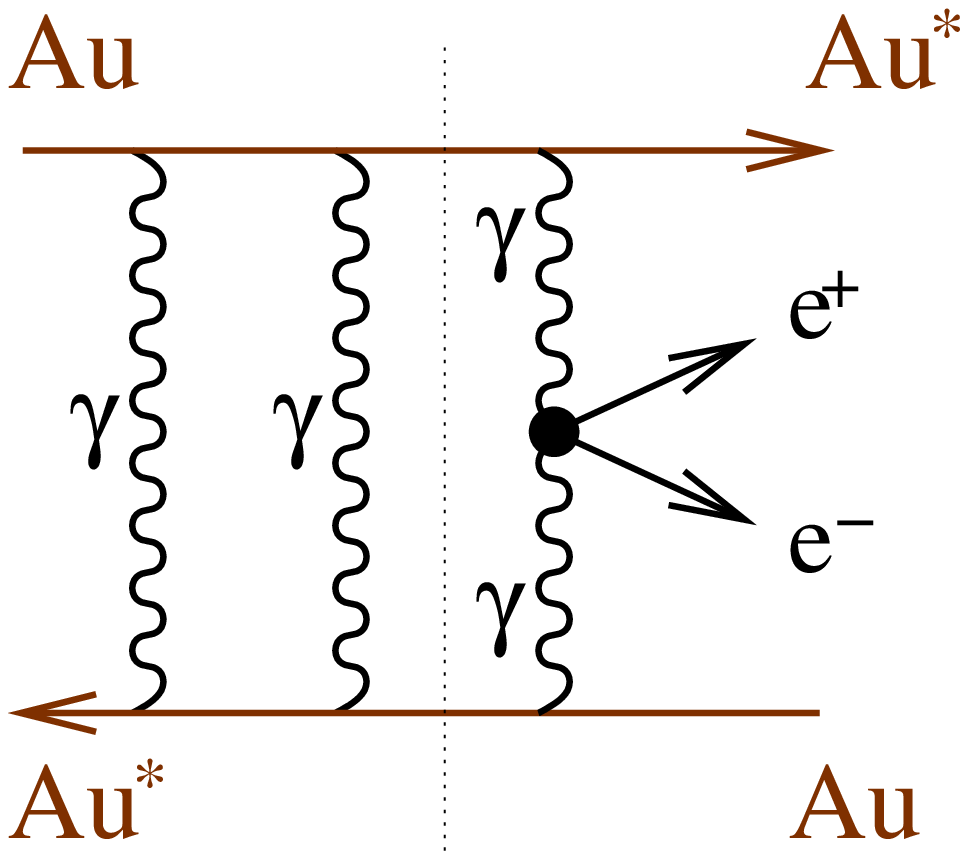}
\caption[Electron-positron production with a mutual Coulomb excitation]{\label{fig:eewithCoulomb}Electron-positron production with a mutual Coulomb excitation. The \ee   pair production (right of the dashed line) is independent of the simultaneous nuclear excitation (left of the dashed line).}
\end{figure}

Figure \ref{fig:eewithCoulomb} gives a schematic view of this process. Each Au ion emits a field of virtual photons, but this process doesn't disrupt the emitting ion. The virtual photons can produce an \ee   pair in collisions with the photons from the ion in the opposite beam. Additionally, photons can cause an excitation of the Au ions in the opposite beam. We trigger on the AuAu collisions  in which both Au ions are excited via a photon exchange.

Recent theoretical and experimental studies of mutual Coulomb excitation in heavy-ion collisions suggest that mutual excitation is independent of other reactions in the collision. However, Coulomb excitation has not been observed in coincidence with electromagnetic \ee   pair production in any previous heavy-ion collision experiments.

The study of the reaction $AuAu\rightarrow Au^{*}Au^{*}+e^{+}e^{-}$ in STAR thus gives us a chance to examine two important physics questions. First is the test of QED at high fields -- are there indications that at the available energies more than two photons are involved in an \ee pair creation? The second question is whether pair production is independent of photonuclear excitation. 

\chapter{Physics of Electron-Positron Pair Production}
\label{ch:Physics} 

Electron-positron pair production is a purely electromagnetic process; therefore we start with a brief overview of quantum electrodynamics (QED). We then discuss the special features of \ee   pair production in relativistic heavy-ion collisions. We present the lowest-order QED framework for computing the \ee   production cross-section, and discuss its higher-order extensions. Finally we present a model for the computation of the \AuAuee \ cross-section. 

\section{Overview of Quantum Electrodynamics}
\label{sec:QEDIntro}
Quantum electrodynamics is one of the simplest Abelian gauge quantum field theories of nature. 
The theory incorporates the ideas of Maxwell's electromagnetism with quantum mechanics, such as quantization and the spin-$1/2$ nature of electrons and positrons. The key results of the theory can be summarized as follows:
\begin{list}{-  }
\item The electromagnetic field consists of photons -- the massless spin-1 corpuscles of energy of the field.
\item
\item The interaction of electrons and positrons with the field (and each other) can be represented as the interaction between the electrons/positrons and the photons.
\item The electrons/positrons and photons can be born or annihilated in interactions with each other.
\end{list}
\addvspace{10pt}
\newpage

In accordance with this idea, the QED Lagrangian consists of three terms: a term from relativistic quantum mechanics describing free electrons/positrons, a term describing the free Maxwell field, and a term responsible for the interaction between electrons/positrons and the field:
\begin{equation}
\label{eqn:Lagrangian}
\QTR{cal}{L}_{\text{QED}}=\QTR{cal}{L}_{\text{electron}}+\QTR{cal}{L}_{\text{Maxwell}}+\QTR{cal}{L}_{\text{Interact}}=\overline{\psi }\left( i\widehat{\partial }-m_e\right) \psi -\frac 14\left( F_{\mu \nu }\right) ^2-e\overline{\psi }\gamma ^\mu \psi A_\mu 
\end{equation}
where $\psi$ is the Dirac wave-function of electrons/positrons, $A_{\mu}$ is the electromagnetic vector potential, $F_{\mu \nu }=\partial _\mu A_\nu -\partial _\nu A_\mu $ is the electromagnetic field tensor, and $e$ and $m_{e}$ are the charge and the mass of the electron. We use the natural system of units where $c=1$ and \ $\hbar = 1$. 

The Lagrangian (\ref{eqn:Lagrangian}) considers only one fermion family -- electrons and positrons. Amazingly, this very simple Lagrangian can account for nearly all observed phenomena from macroscopic scales down to $10^{-13}$ cm. 
At distance scales below $10^{-13}$ cm (or, equivalently, for interaction energies above 100 GeV) phenomena explained by other theories, such as electro-weak theory or quantum chromodynamics become significant.

One of the central applications of the QED Lagrangian is to find the transition amplitude for some initial state which, generally speaking, consists of particles 1, 2, etc. with momenta $p_{1}$, $p_{2}$, etc. at time $t_{0}$ (denoted $| \text{Initial}\rangle$), to make a transition into a final state consisting of particles A, B, etc.  with momenta $p_{A}$, $p_{B}$, etc. at time $t_{1}$ (denoted $\langle\text{Final}|$). According to the rules of quantum mechanics the transition amplitude is:
\begin{equation}
\label{eqn:TimeEvolution}
\langle\text{Final}| \text{Initial}\rangle  = \langle p_{1}, p_{2} ...| \exp \left\{ -i\int_{t_0}^{t_1}dt\int d^3x\cdot \QTR{cal}{L}_{\text{Interact}}(t,x)\right\}  | p_{A}, p_{B} ...\rangle
\end{equation}
where the expression sandwiched between the initial and final momenta is the usual time-evolution operator, based on the interaction part of the QED Lagrangian (\ref{eqn:Lagrangian}). The computation of cross-sections for all QED processes is based on variations of formula (\ref{eqn:TimeEvolution}).\footnote{The formula given here is for conceptual demonstration only, for details see the standard texts\cite{Landafshitz, Peskin}.}

\subsection{Perturbative Expansions in Quantum Electrodynamics}
\label{sec:Feynman}
The analytical computation of the transition amplitude (\ref{eqn:TimeEvolution}) is impossible for most non-trivial initial/final state combinations. Fortunately, in 1949 Feynman\cite{Feynman} proposed that the exponential in the time-evolution operator can be expanded as:
\[\exp \left\{ -i\int dtd^3x\cdot L_{\text{Interact}}(t,x)\right\} = \]
\[1\! +\! (-i)\int dtd^3x\cdot L_{\text{Interact}}(t,x)\! +\! (-i)^2\iint dt_1d^3x_1dt_2d^3x_2\cdot L_{\text{Interact}}(t_1,x_1)\cdot L_{\text{Interact}}(t_2,x_2)\! +\! ..\! = \]
\begin{equation}
\label{eqn:Perturbative}
\! = \! 1 \! +\! (-i)\sqrt{4\pi \alpha }\int dtd^3x\overline{\psi }\gamma ^\mu \psi A_\mu \! +\! 4\pi \alpha (-i)^2\iint dt_1d^3x_1dt_2d^3x_2\overline{\psi }_1\gamma ^\mu \psi _1A_{1\mu }\cdot \overline{\psi }_2\gamma ^\mu \psi _2A_{2\mu }\! +\! ...\ 
\end{equation}
where $\alpha = e^2/4\pi \approx 1/137 $ is the fine structure constant. The fine structure constant, often referred to as the coupling constant, represents the coupling strength of the electromagnetic field to the electron charge. Effectively Equation (\ref{eqn:Perturbative}) is a Taylor expansion in the orders of $\alpha ^{1/2}$.

\begin{figure}
\centering 
\includegraphics[bbllx=0pt,bblly=50pt,bburx=1050pt,bbury=720pt,width=270pt]{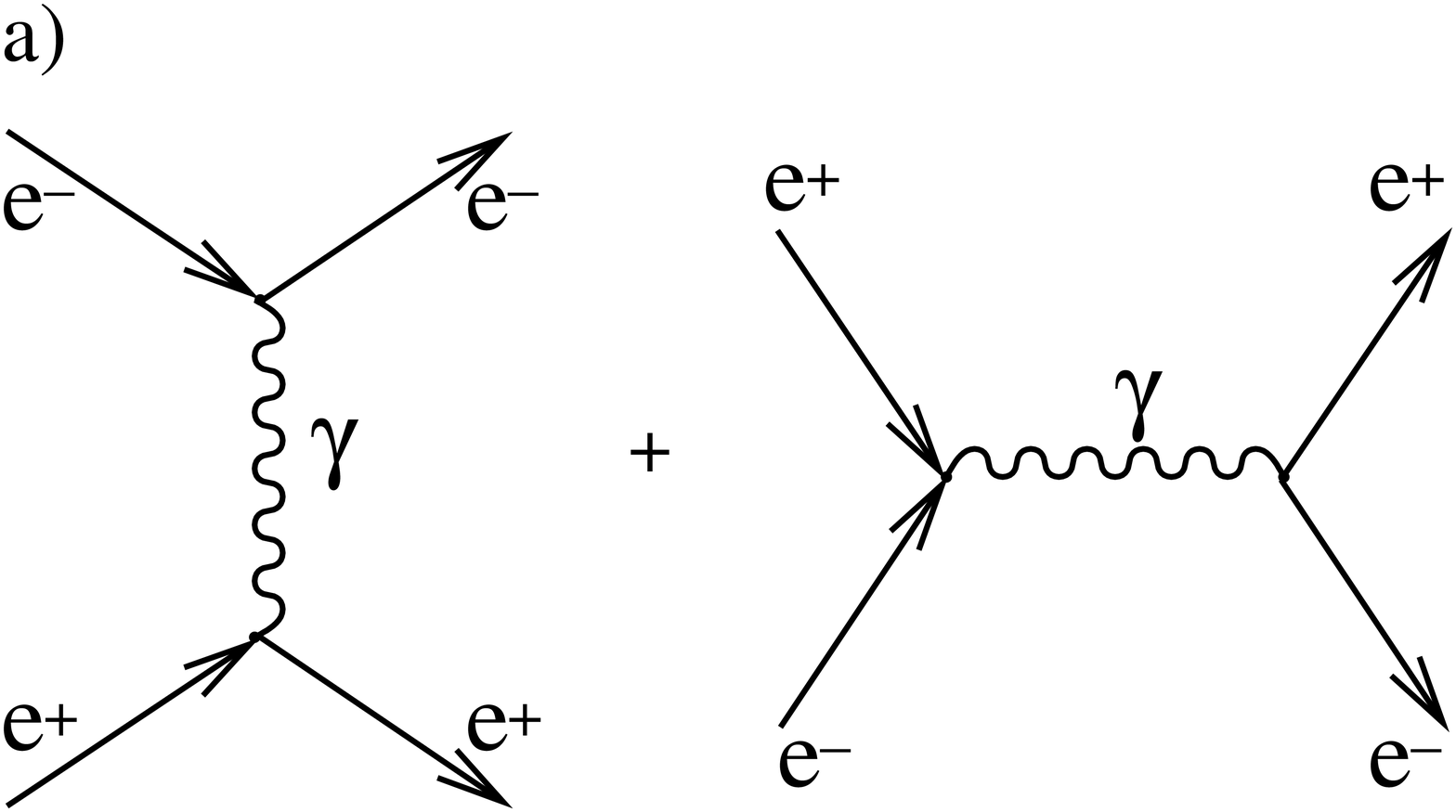}
\includegraphics[bbllx=0pt,bblly=5pt,bburx=430pt,bbury=675pt,width=110pt]{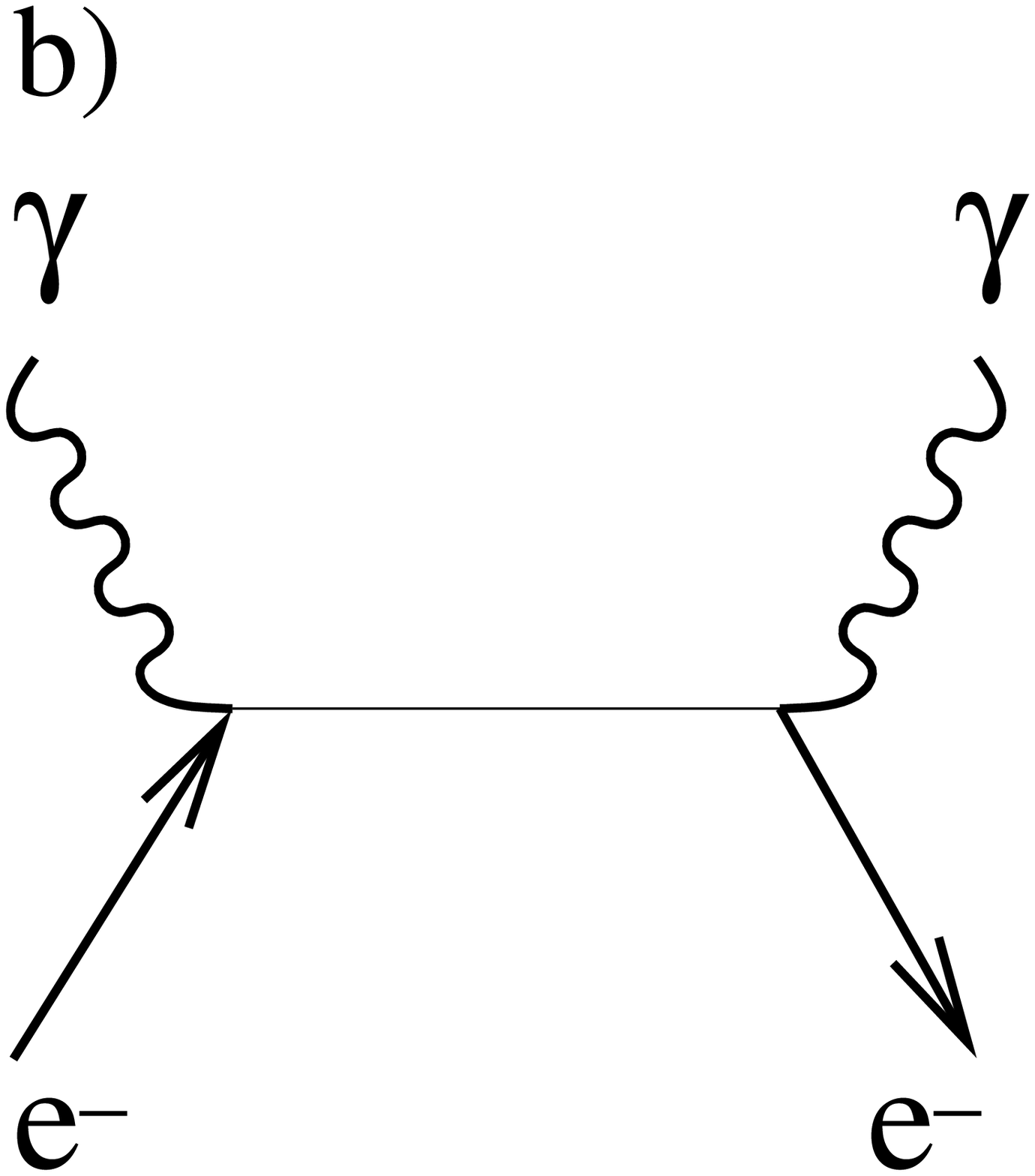}
\includegraphics[bbllx=0pt,bblly=50pt,bburx=450pt,bbury=720pt,width=115pt]{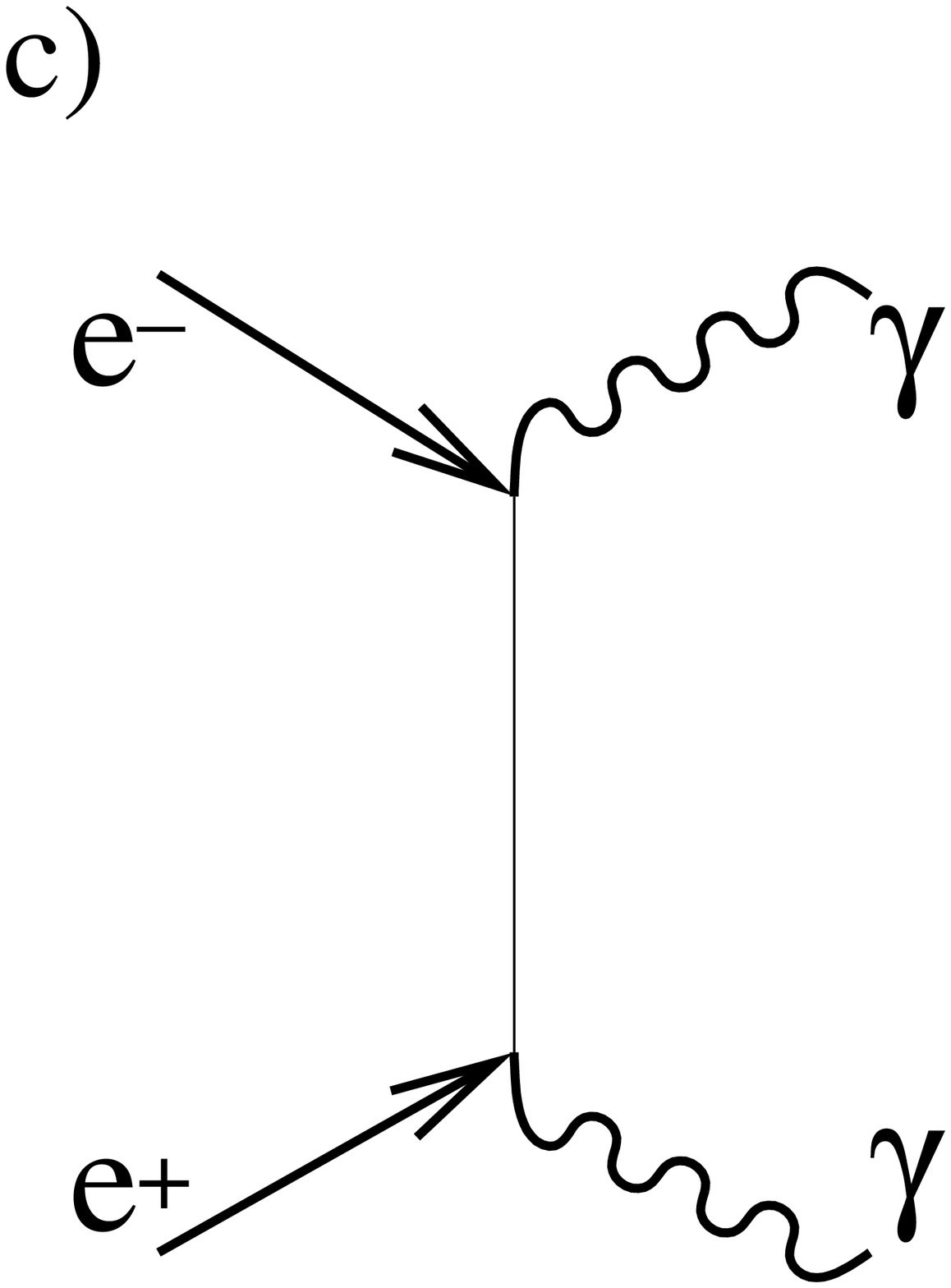}
\includegraphics[bbllx=-20pt,bblly=25pt,bburx=430pt,bbury=695pt,width=115pt]{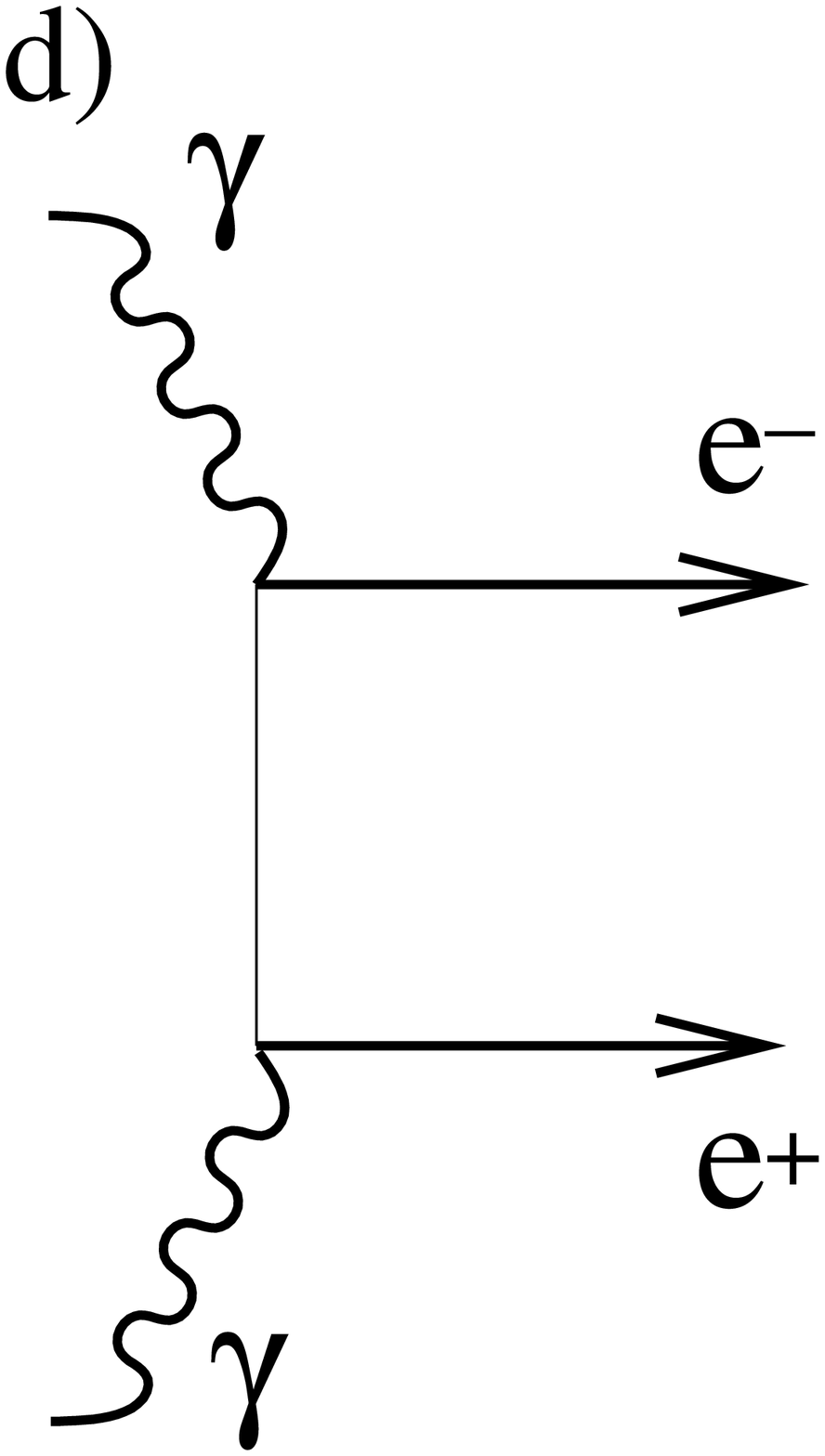}
\caption[Lowest order QED Feynman diagrams]{\label{fig:twophoton}Lowest order Feynman diagrams for: a) Bhabha scattering; b) Compton scattering; c) pair annihilation; d) pair creation. Topologically identical diagrams not presented in b), c) and d).}
\end{figure}

For the non-identical initial and final states, the lowest-order non-zero matrix element in this perturbative expansion is of the order $\alpha$\cite{Landafshitz}. Figure \ref{fig:twophoton} shows the lowest-order QED Feynman diagrams for:  $e^{+}e^{-} \rightarrow  e^{+}e^{-}$ (Bhabha scattering), $e^{-}\gamma \rightarrow e^{-}\gamma $ (Compton scattering), $e^{+}e^{-}\rightarrow \gamma \gamma $ (pair annihilation) and the reverse process $\gamma \gamma \rightarrow e^{+}e^{-}$ (pair creation).\footnote{The processes involving only photons in the initial state (e.g. Figure \ref{fig:twophoton} (d)) are referred to as 'two-photon  physics'.}

The lowest order 
is a good approximation to the total amplitude if the energies of the particles involved in the reaction are small, so that each n-th term of the power expansion is dominated by the small coefficient $\alpha^{n/2}$. However, the higher order corrections can be measured in high-precision experiments, and provide further confirmation to the QED\cite{Kinoshita}. In fact, to this day QED is the most stringently tested and the most successful of all physical theories, with agreement between the theory and the data up to eight significant digits.

\subsection{Kinematical Properties of $\gamma \gamma \rightarrow $\ee   Process}
\label{sub:GammaGammaEE}

\begin{figure}
\includegraphics[height=1.9in]{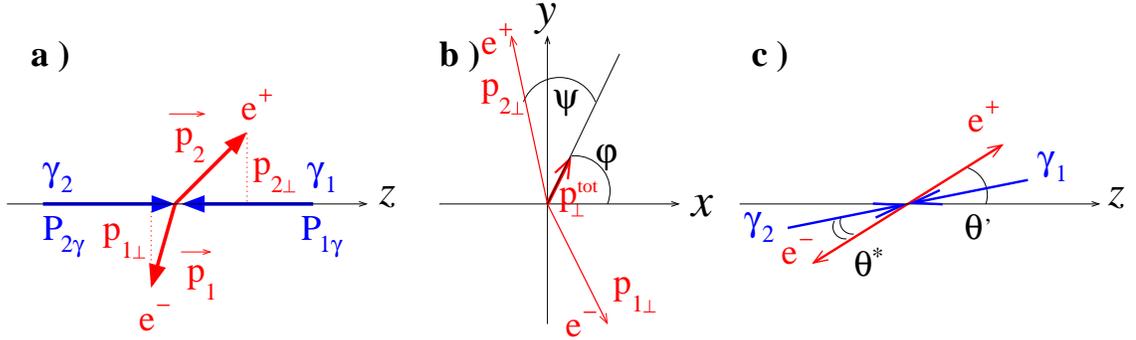}
\caption[Pair kinematics]{\label{fig:kinematics} Pair kinematics: a) longitudinal projection in the lab frame, b) transverse projection in the lab frame, c) longitudinal projection in the center of mass frame of the $\gamma \gamma $ pair.}
\end{figure}

Let's consider the properties of two-photon annihilation into an \ee   pair closely. We assume that the two colliding photons have very large momenta $p_{1\gamma}$ and $p_{2\gamma}$ of opposite signs along the axis $z$ and very small momenta in the plane transverse to the $z$ axis.\footnote{This will always be the case for the photons produced in the heavy-ion collisions, as we explain later.} Figure \ref{fig:kinematics} shows the kinematical variables used in the description of this reaction.

The six variables ($p_{x,y,z}^1$ and $p_{x,y,z}^2$) describing the individual track kinematics uniquely define the six variables describing the pair kinematics: pair invariant mass ($M_{inv}$), pair rapidity ($Y$), absolute value of the pair total transverse momentum ($p_\bot ^{tot}$), the azimuthal angle of the total transverse momentum ($\varphi$),\footnotemark[4] the angle between the total momentum and the momentum of the negative track  in transverse plane ($\psi$),\footnotemark[4] and the polar angle of the negative track in the center-of-mass frame ($\theta '$). The formulae for computing these variables from the two tracks' momenta are provided in Appendix \ref{app:variables}.

\footnotetext[4]{For non-degenerate $p_\bot ^{tot} \ne 0$ cases.}
In the center-of-mass frame, the kinematics of the reaction is defined by the angle between the track momenta and the direction of the photons $\theta^*$ (Figure \ref{fig:kinematics}, c). This angle is the same for the electron and positron tracks. Determination of the $\theta^*$ generally requires the knowledge of photon momenta $(p_{\gamma x},p_{\gamma y}, p_{\gamma z})$; however, if the photon transverse momenta are approximately zero, $\theta^*$ can be approximated by $\theta '$. We use a Monte Carlo simulation in Chapter \ref{ch:MonteCarlo} to examine the accuracy of this approximation. In the lab frame the angles between the track momenta and the $z$ axis are different ($\theta _{e^+}$ and $\theta_{e^-}$). Traditionally,  $\theta _{e^+}$ (or $\theta_{e^-}$) is expressed via pseudorapidity, defined as 
\begin{equation}
\label{pseudorapidity}
\eta ^{track}  =  - 0.5\ln \left( {\tan \left( {\frac{1}{2} \theta _{track} } \right)} \right)
\end{equation}

The computation of the cross-section for the \ee   pair production (in the pair's center-of-mass frame) in the lowest QED order is performed according to the Feynman rules for the diagram in Figure \ref{fig:twophoton} (d)\cite{somebodyCrossSection}. For two colliding circularly polarized non-virtual photons the angular distribution of the cross-section in the center-of-mass frame is:

\begin{equation}
\label{eqn:AngularCrossSection}
\frac{{d\sigma _{\gamma \gamma  \to e^ +  e^ -  } }}{{d\cos \theta ^ 
 *  }} = 16\pi ^2 \alpha ^2 \sqrt {\frac{{\left( {s - 4m^2 } \right)}}
 {{s^3 }}} \left[ {\frac{{\sqrt s  - \sqrt {s - 4m^2 } \cos \theta ^ 
 *  }}{{\sqrt s  + \sqrt {s - 4m^2 } \cos \theta ^ *  }} + \frac{{\sqrt 
 s  + \sqrt {s - 4m^2 } \cos \theta ^ *  }}{{\sqrt s  - \sqrt {s - 
 4m^2 } \cos \theta ^ *  }}} \right]
\end{equation}
where $s=M_{inv}^2$ is the squared invariant mass of the system, $m$ is the mass of the electron, and $\alpha$ is the fine structure constant. Appendix \ref{app:variables} discusses the effects of photon polarization, which are negligible in the rangle of energies available at RHIC. 

For the large $\gamma \gamma $ system invariant masses ($m^2 \ll s$) the right side of Equation ($\ref{eqn:AngularCrossSection}$) can be approximated as $ d\sigma / d\cos \theta ^ * \sim (1 + \cos ^2\theta ^ * )/ \sin ^2\theta ^ * $. Thus the electron/positron production is peaked  in the direction of the incoming photons. 

Integrating the angular cross-section (\ref{eqn:AngularCrossSection}) one obtains the full cross-section to produce an \ee   pair:

\begin{equation}
\label{eqn:CrossSectionGammaGammaEE}
\sigma _{\gamma \gamma  \to e^ +  e^ -  } \left(\! s \right) \! = \! 8\pi 
 \alpha ^2 \frac{1}{s}\left( {1\! + \! \frac{{4m^2 }}{s}\! - \! \frac{{8m^4 }}
 {{s^2 }}} \right)\ln \left( {\frac{{\sqrt s }}{{2m}}\! + \! \sqrt {\frac{s}
 {{4m^2 }}\! - 1} } \right)\! + \! 4\pi \alpha ^2 \left( {\frac{1}{s}\! +\! \frac
 {{4m^2 }}{{s^2 }}} \right)\sqrt {1\! - \! \frac{{4m^2 }}{s}}
\end{equation}
As a general property of all $\gamma \gamma $ processes, the cross-section depends only on $\sqrt{s}$. In their center-of-mass frame the two photons have equal energies and therefore only one variable defines the reaction.

The cross-section of the \ee   pair  production has a peak at $s \sim  (2.5m)^2$, and drops asymptotically as $ln(s)/s$ for $\sqrt{s} \gg m$. The differential cross-section for \ee   pair production is maximal at the invariant masses close to the mass of the electron and in the forward region.

\section{QED in Relativistic Heavy-Ion Collisions}
\label{sec:QEDatRHIC}

To understand $e^{+}e^{-}$ with nuclear excitation, it is useful to first consider exclusive $e^{+}e^{-}$ production. We will first describe the fundamental approach to the two-photon processes in the ultra-peripheral relativistic heavy-ion collisions -- the Equivalent Photon Approximation. This approximation uses the lowest-order term of perturbative QED. Next we will discuss the corrections to the cross-section from the higher-order terms and from non-perturbative calculations. Finally, we will discuss the previous experimental measurements of \ee   pair production at \ee colliders and heavy-ion colliders.
  
\subsection{Pair Production in Ultra-Peripheral Heavy-Ion Collisions in the Lowest-order QED Approximation}
The basic framework for studying the two-photon physics in the high-energy collisions was first introduced by Fermi \cite{Fermi} and then developed in detail by Landau and Lifshitz\cite{Landafshitz}. Their idea was that high-energy projectiles emitting the electromagnetic fields retain almost all of their initial momenta in the ultra-peripheral interaction, and thus can be assumed to follow classical straight-line trajectories with constant velocities. Therefore, the ultra-peripheral process can be thought of as a convolution of three processes: an emission of a photon by one of the colliding particles, photon emission by the other particle, and annihilation of two photons into a final state (labeled $X$)\cite{GammaGammaEPA}. The two photons' densities can be combined into a new quantity: the two-photon luminosity $L_{\gamma \gamma}$ per ion pair. Using the two-photon luminosity, the cross-section of the two-photon production of the state $X$ in the ultra-peripheral collision of two heavy ions (AA) becomes:  

\begin{figure}
\centering 
\resizebox*{0.41\textwidth}{!}{\includegraphics{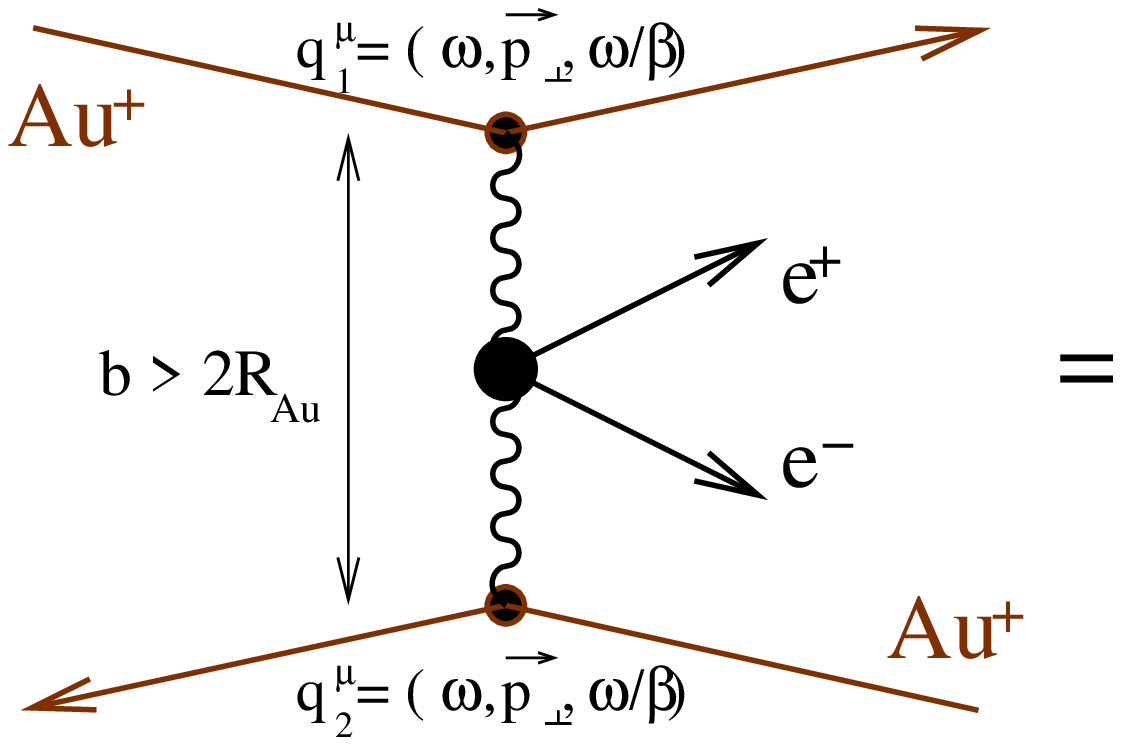}}
\resizebox*{0.35\textwidth}{!}{\includegraphics{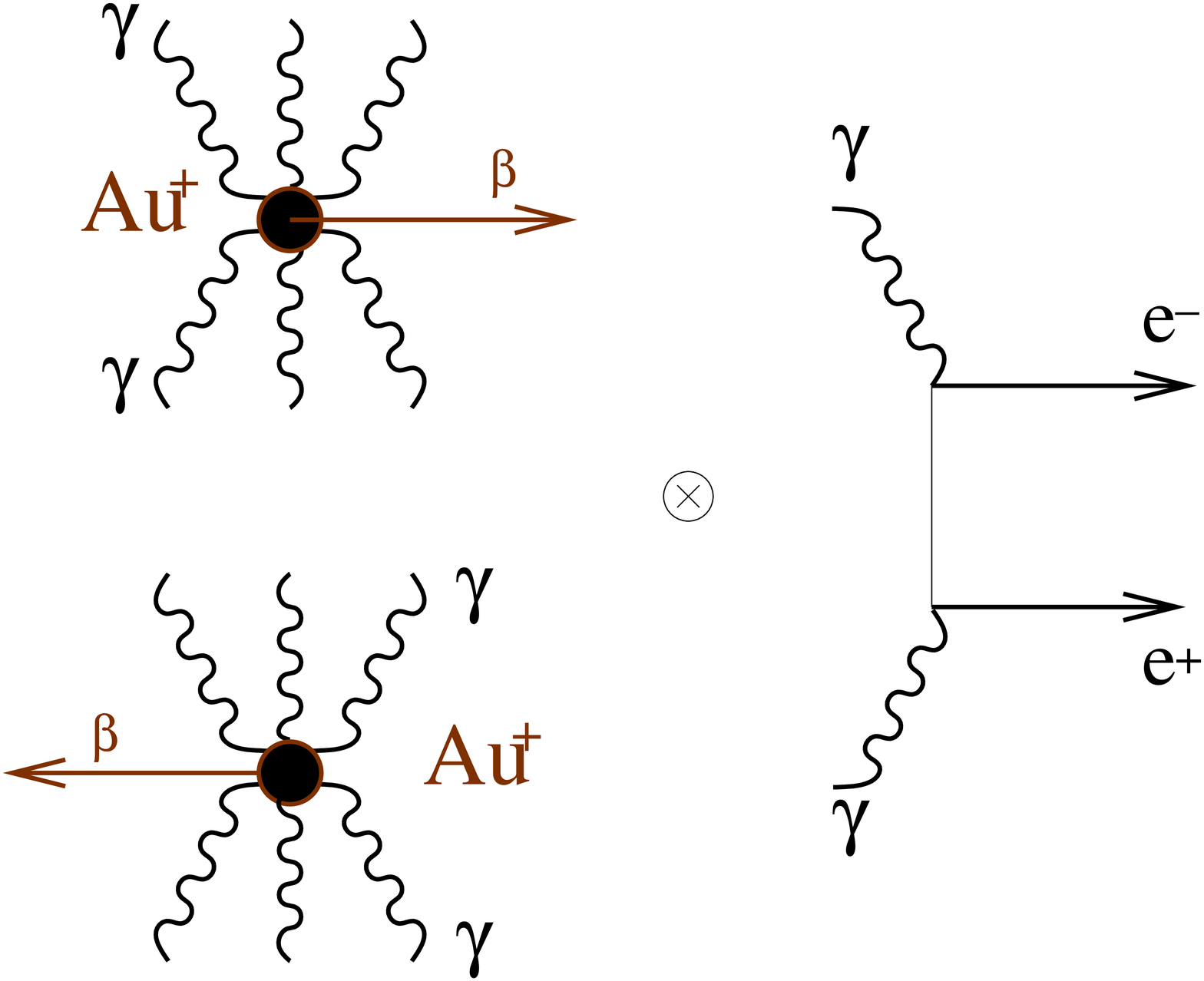}}
\caption[Equivalent Photon Approximation]{\label{fig:EPA}Equivalent Photon Approximation. Coupling of the fields to the Au ions is assumed to be classical. $b$ is the impact parameter -- the distance between the colliding ions. The exchange photons have four-momenta $q_{1}^{\mu }$ and $q_{2}^{\mu }$.}
\end{figure}

\begin{equation}
\label{eqn:GammaGammaLuminosity}
\frac{d\sigma _{AA \to AA+X}}{dM_{inv}dY}=\frac{dL_{\gamma \gamma }}{dM_{inv}dY}\cdot \sigma _{\gamma \gamma \to X}\left( M_{inv},Y\right) 
\end{equation}
where $M_{inv}$ and $Y$ are the invariant mass and rapidity of the system.

Figure \ref{fig:EPA} demonstrates the idea  graphically for the case of \ee pair production. The exchange photons have four-momenta $
q^\mu   = \left( {\omega ,\overrightarrow {p_ \bot  } ,p_z  = \omega /\beta } \right)$, 
where $\omega $ is the energy of the exchange photon and $\beta $ is the velocity of the nucleus. The virtuality of such photons is $
Q^2 = -q^2  = \omega ^2 /\gamma ^2  + p_ \bot ^2 $, and $\gamma  = 1/\sqrt {\left( {1 - \beta ^2 } \right)} $ is the Lorentz boost of the nucleus in the lab frame. 

\subsubsection{Equivalent Photon Approximation}
Ignoring the photon virtuality, the photon flux accompanying a heavy nucleus is given by the Weizs\"acker Williams approximation\cite{WeizsackerWilliams}. This approximation considers the ions as point-like sources of electric field,\footnote[5]{Since we consider only the fields outside of the ions, the exact details of the charge distribution inside the ions are unimportant for the calculation of the photon energy spectrum.} and the field distribution in the lab frame is taken as a field of the charge $Ze$ moving with the speed $\overrightarrow{\beta}$. For $|\overrightarrow{\beta}| \approx 1 $ the $\overrightarrow E$ and $\overrightarrow B$ fields are mostly in the plane transverse to the ion's velocity, with vector $\overrightarrow E$ pointing radially away from the ion and vector $\overrightarrow B$ perpendicular to $\overrightarrow E$. Thus at each point in the transverse plane, the $(\overrightarrow E, \overrightarrow B)$ field combination can be approximated by a plane electromagnetic wave, propagating in the direction of $\overrightarrow{\beta}$. The equivalent photon spectrum (energy per unit frequency interval per unit area) as a function of the distance to the ion $b$ is\cite{WeizsackerWilliams}:
\begin{equation}
\label{eqn:WeizsackerWilliams}
\frac{d^2n}{d^2\overrightarrow{b}}\left( \overrightarrow{b},\omega \right) =\frac{Z^2\alpha }{\pi ^2}\left( \frac \omega \gamma \right) ^2\frac 1{\beta ^4}K_1^2\left( x\right) ,\ \ x=\frac{\omega |\overrightarrow b |}{\gamma \beta }
\end{equation}
where $K_1$ is the modified Bessel function of order one. The Bessel function becomes very small for $x > 1$. Since we are considering fields outside of the nuclei ($b > R$), this means that the photon energies are limited to $\omega  < \gamma /R $, where $R$ is the nuclear radius.
The physical interpretation of this cutoff is that in order for the electromagnetic field to couple coherently to the ions, the photon wavelength in the rest frame of the ions should be greater than the size of the ion: $\lambda > R$. Thus, in the nucleus rest frame coherence limits the transverse and longitudinal momenta of exchange photons to $p_ \bot   < 1/R$ ( $\sim 30$ MeV/c for Au), and in the lab frame (with the Lorentz boost in the longitudinal direction) the photon energy is limited to $\omega  < \gamma /R \approx 3 $ GeV $\gg p_{\bot }$.

To calculate an overall two-photon density $F_{\gamma \gamma}$ we need to integrate expression (\ref{eqn:WeizsackerWilliams}) over all space and over all impact parameters. This introduces one important complication. When the nuclei physically collide $(b < 2R)$, then the hadronic interactions will completely overshadow the electromagnetic ones. Therefore, in calculating the usable two-photon density, this overlap region must be excluded\cite{BaurFerreira}: 

\begin{equation}
\label{eqn:computeGammaGammaLumi}
F_{\gamma \gamma }  = \int\limits_{R_{Au}}^{\infty} {d^2 \overrightarrow {b_1 } } \int\limits_{R_{Au}}^{\infty} {d^2 \overrightarrow {b_2 } } \frac{{d^2 n}}{{d^2 \overrightarrow b_1 }}\left( {\overrightarrow b_1 ,\omega _1 } \right)\frac{{d^2 n}}{{ d^2\overrightarrow b_2 }}\left( {\overrightarrow b_2 ,\omega _2 } \right) \cdot \Theta \left( {\left| {\overrightarrow {b_1 }  - \overrightarrow {b_2 } } \right| - 2R } \right)
\end{equation}
the $\theta$-function above reflects an exclusion of the nuclear overlap region from the density calculation.

We can convert the two photon energies $\omega _1 , \omega _2$ into the invariant mass and the rapidity of the $\gamma \gamma $ pair in the lab frame: $M_{inv} = 2\sqrt{ \omega _1 \omega _2 }$ and $ Y = 1/2 \log(\omega _1 / \omega _2 ) $, and the two-photon differential luminosity can be expressed as a function  of these two variables:\footnote[6]{For the derivation, see Appendix \ref{app:variables}.}

\begin{equation}
\label{eqn:GammaGammaDiffLumi}
\frac{{d^2 L_{\gamma \gamma } }}{{dM_{inv} dY}}\left( {M_{inv},Y} \right) = \frac{2}{{M_{inv} }}F{}_{\gamma \gamma }\left( {\frac{{M_{inv} }}{2}\exp \left( Y \right),\frac{{M_{inv} }}{2}\exp \left( { - Y} \right)} \right) 
\end{equation}

\begin{figure}
\centering
\includegraphics[width=280pt]{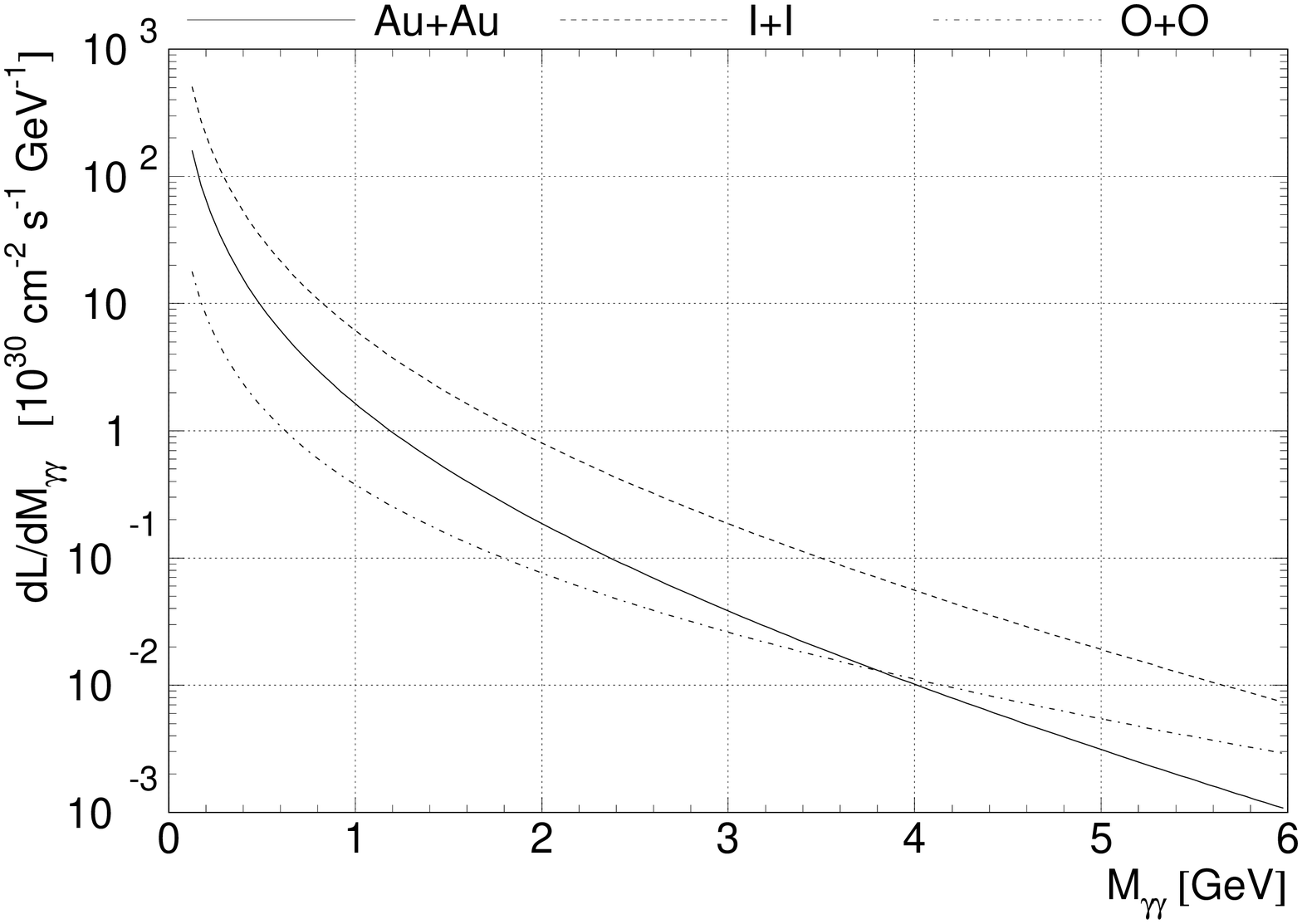}
\caption[Two photon differential luminosities vs. $\gamma \gamma $ pair invariant mass at RHIC for gold, iodine and oxygen beams.]{\label{fig:GammaGammaLumi}Two photon differential luminosities vs. $\gamma \gamma $ pair invariant mass $M_{\gamma \gamma}$ at RHIC for gold (solid), iodine (dashed) and oxygen (dot-dashed) beams \ \cite{SpencerNystrand}.}
\end{figure}

Klein and Nystrand in \cite{SpencerNystrand} used the above method to calculate total two-photon differential luminosity 
$dL_{\gamma \gamma } ^{tot} /dM_{inv}= L_{hadronic} \cdot \int\limits_{-\infty} ^\infty \frac{d^2L_{\gamma \gamma }}{dM_{inv}dY}dY$ at RHIC for several ion species for the photon center-of-mass energies up to 6 GeV. For gold beams $\gamma _{Au} = 108$ and $L_{AuAu}=2.0 \cdot 10^{26} \text{ cm}^{-2} \text{s}^{-1}$ design RHIC Lorentz boost and luminosity are assumed ($\gamma _{I}= 104$ and $L_{II}=2.7 \cdot 10^{27} \text{ cm}^{-2} \text{s}^{-1}$ for iodine, $\gamma _{O} = 135$ and $L_{OO}=9.8 \cdot 10^{28} \text{ cm}^{-2} \text{s}^{-1}$ for oxygen). Figure \ref{fig:GammaGammaLumi} shows the results of the calculation. The highest two-photon luminosity is obtained in I+I interactions. The lower $Z$ is compensated by higher nuclear luminosity and smaller nuclear radius. The two-photon luminosity drops off quickly with increasing two-photon center-of-mass energy, which contributes to the rapidly decreasing total cross-section $\sigma _{AuAu\to AuAu+X}$ as a function of the state $X$ invariant mass.

\subsubsection{Applicability of the Lowest-order Approximation for \ee   Production at RHIC}
The use of the lowest-order QED approximation and the EPA explicitly assumes that the \ee   pair is produced by exactly two real photons. Photon virtuality at RHIC is typically on the order of $Q^2 \sim 1/R^2 _{Au} = (30)^2 \text{ MeV}^2$.  This is much greater than the electron mass squared ($m_e ^2 = (0.51)^2 \text{ MeV}^2$), and the photon virtuality cannot be neglected for some regions of electron-positron phase space.  Additionally, the field coupling to the Au ions is very strong -- $Z\alpha \sim 0.6$ at RHIC -- and taking only the lowest-order term in the perturbative expansion (\ref{eqn:Perturbative}) might not be accurate. 
Lastly, the Compton wavelength of the electron $\lambda^e _{C} = 1/m_e = 386$ fm is much larger than the typical impact parameter at RHIC for exclusive pair production ($\sim$ 100 fm), and the \ee   pair production is poorly localized. 
The following subsection gives an overview of the theoretical papers dealing with these complications, and presents a prediction of the \ee production cross-section in the lowest QED approximation and corrections from higher-order terms.

\subsection{Electron-positron Production Cross--section at RHIC}
\label{sub:TheoryPapers}

The total \ee   pair production cross-section in ultra-peripheral heavy-ion collisions can be computed as a convolution of the two-photon luminosity and the cross-section for $\gamma \gamma \rightarrow e^+e^-$. 
To avoid problems with photon virtuality, it is customary to set the lower integration limits in Equation (\ref{eqn:computeGammaGammaLumi}) equal to  $\lambda^e _{C}$ instead of $R_{Au}$. It has been shown that the total \ee pair production cross-section is dominated by large impact parameters, therefore cutting out the region $b<\lambda^e _{C}$ does not affect the cross-section significantly. Alscher {\it et al.} in \cite{BaurHencken} compute the resulting total cross-section for the exclusive \ee   pair production at RHIC at $\sqrt{s_{NN}} = 200$ GeV to be 33 kb. 
However, most of the particles are produced at low invariant masses (below 10 MeV) and into the very forward direction, so the fraction of the cross-section visible to the detectors is very small. We will estimate the observable cross-section in the next chapter.

At impact parameters below $\lambda^e _C$ this approach breaks down for the \ee pairs. Formally, we can write the probability to produce an $e^{+}e^{-}$ pair if colliding ions are at impact parameter $b$ as:
\begin{equation}
\label{eqn:EPAPofB}
P(b)=\int\limits_{\omega _1^{\min }\ }^\infty \int\limits_{\omega _2^{\min }\ }^\infty \frac{dn(\omega _1,b)}{db}\frac{dn(\omega _2,b)}{db}\frac{d\omega _1}{\omega _1}\frac{d\omega _2}{\omega _2}\sigma _{\gamma \gamma }\left( \omega _1,\omega _2\right) 
\end{equation}
where the photon densities $n(\omega, b)$ are from (\ref{eqn:WeizsackerWilliams}), and the cross-section $\sigma _{\gamma \gamma }$ is from (\ref{eqn:CrossSectionGammaGammaEE}). Since the two-photon cross-section is highly peaked at $\sim 2m_e$, we can approximate $\sigma _{\gamma \gamma }(\omega _1 , \omega _2 ) \approx \sigma _{\gamma \gamma }(m_e , m_e ) \sim  \alpha ^2 / m^2 _e$ and assume $\omega^{\min } = m_e$. With this modification, (\ref{eqn:EPAPofB}) becomes:
\begin{equation}
\label{eqn:EPAPofB1}
P(b) \propto \int\limits_{m_e }^\infty  {Z^2 \alpha b\left( {\frac{{\omega _1 }}{\gamma }} \right)^2 } K_1^2 \left( {\frac{{b\omega _1 }}{\gamma }} \right)\frac{{d\omega _1 }}{{\omega _1 }}\int\limits_{m_e }^\infty  {Z^2 \alpha b\left( {\frac{{\omega _1 }}{\gamma }} \right)} ^2 K_1^2 \left( {\frac{{b\omega _1 }}{\gamma }} \right)\frac{{d\omega _2 }}{{\omega _2 }} \cdot \frac{{\alpha ^2 }}{{m_e^2 }}
\end{equation}

Using the identity   $\int\limits_c^\infty  x K_1^2 \left( {xa} \right)dx \! = \! \frac{1}{{a^2 }}\ln \left( {\frac{\delta }{{ca}}} \right)$, where $ \delta \! \approx \! 0.681 $ is a constant, related to Euler constant, $P(b)$ is proportional to:
\begin{equation}
\label{eqn:MorePofB}
P(b) \propto \left( {Z\alpha } \right)^4 \frac{1}{{m_e^2 b^2 }}\ln ^2 \left( {\frac{{\delta \gamma }}{{m_e b}}} \right)
\end{equation}

A more careful calculation presented, for example, in \cite{Bertulani} yields:
\begin{equation}
\label{eqn:PofB}
P^{(1)}\left( b\right) = \frac{14}{9\pi ^2}\left( Z\alpha \right) ^4\frac 1{m_e^2b^2}\ln ^2\left( \frac{\gamma \delta }{2m_eb}\right) \
\end{equation}
where  $\gamma =2\gamma _{\text{Lab}}^2-1$ is the Lorentz factor in the target frame. Equation (\ref{eqn:PofB}) shows that at RHIC energies ($\gamma_{\text{Lab}} = 108$) and for impact parameters less than the $\lambda^e _C$, this probability exceeds 1 (unitarity violation), which clearly demonstrates a breakdown of the lowest order QED approach for \ee production at RHIC.

\begin{figure}
\centering 
\includegraphics[width=440pt]{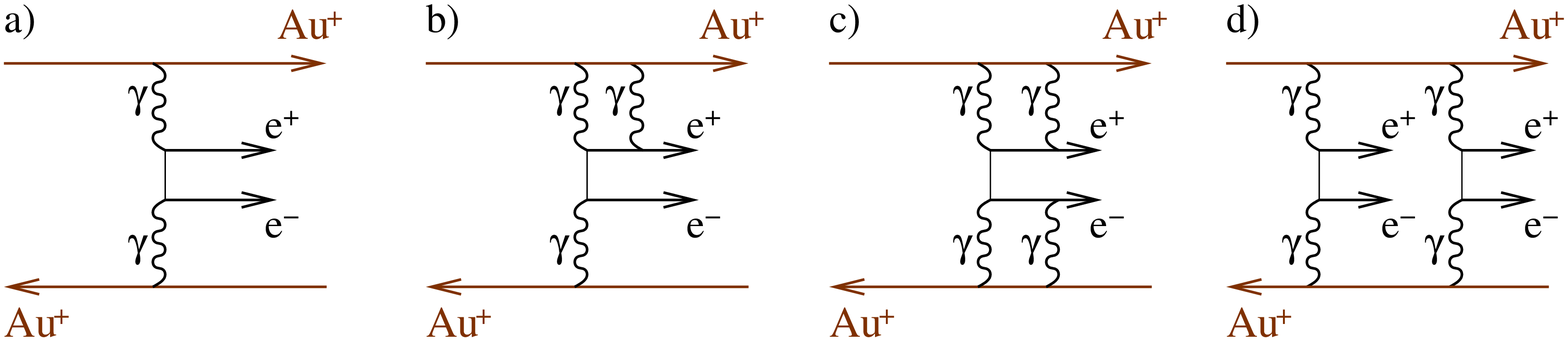}
\caption[Feynman diagrams for electron/positron production at RHIC]{\label{fig:RHICFeynman}a) Lowest order Feynman diagram for the \ee   production b) next to lowest order three-photon exchange diagram c) four-photon exchange diagram, d) double \ee pair production dominant contribution.}
\end{figure}

Further analysis\cite{BaurPhysRev} showed that production of multiple \ee   pairs (e.g. Figure \ref{fig:RHICFeynman} d)) is significant in the energy range where EPA prediction violates unitarity. Including multiple pairs in the cross-section computation restores unitarity. The pair production probability is then distributed according to a Poisson distribution:

\begin{equation}
P_N (b) = \frac{{\left( {P^{\left( 1 \right)} \left( b \right)} \right)^N \exp \left( { - P^{\left( 1 \right)} \left( b \right)} \right)}}{{N!}}
\label{eqn:Poisson}
\end{equation}
where lowest order contribution $P^{(1)}\left (b\right)$ is the mean of the Poisson distribution. 
Multiple pair production is dominant over the single pair production for $b < \lambda^e _C$ at RHIC energies. For AuAu collisions at $\sqrt{s_{NN}}=200$ GeV at RHIC, each ultra-peripheral \ee pair in the invariant mass range $M_{inv} \sim 10 \div 20 $ MeV is expected to be accompanied on the average by one more \ee pair. However, the second pair should be typically low-energy, and not visible to the detectors\cite{BaurHencken}. 

If one wishes to compute higher-order QED contributions to the $e^{+}e^{-}$ cross-section, a number of complications arise. There are many higher-order Feynman diagrams for this reaction. Two of such possible diagrams including an exchange of three and four photons are presented in Figure \ref{fig:RHICFeynman} (b) and (c). Adding an extra photon to the higher-order corrections diagram suppresses the contribution by $Z\alpha \sim 0.6$ which is not significantly smaller than the previous lowest-order contribution.

Diagrams like the one in Figure \ref{fig:RHICFeynman} (b), where either the electron or the positron couples to the Au ion by more than one photon, make an especially large contribution. This correction is historically called 'Coulomb correction', since it represents a re-scattering of the electron or positron in the field of the nucleus. If observable, the Coulomb corrections should make the distributions of $d\sigma_{e^+e^-} / dp_{e^+}$ and $d\sigma_{e^+e^-} / dp_{e^-}$ ($p$ is the momentum of the electron or positron) to be different, since the electron and positron will scatter differently in the field of the positively charged Au ions.

Authors of \cite{Bertulani} approached the computation of the Coulomb corrections in the collisions of charges $Z_1e$ and $Z_2e$ using the results of the Bethe-Heitler process $\gamma +Z\rightarrow e^{+}e^{-}+Z$ where higher-order effects are well-known. Such a treatment requires an ad hoc symmetrization with respect to $Z_1$ and $Z_2$ and the results of this analysis are being disputed by several authors \cite{Ivanov}.

Some authors abandoned the Feynman perturbative expansion in Equation (\ref{eqn:Perturbative}) and solved the Dirac's equation for the $e^{+}e^{-}$ in the electromagnetic field of Au ions non-perturbatively. In this approach the authors work in the light-cone coordinates where the effect of electromagnetic fields has a form of a delta-function. Thus the Green's function for the exact wave function of $e^{+}e^{-}$ at the interaction point is found. The transition amplitude is constructed from the Green's functions. The results were found to match the perturbative calculations with Coulomb correction\cite{BaltzMcLerran, LeeMilstein, BaurPhysReports}.

The agreement of the non-perturbative calculations for a single pair production with the perturbative calculations is somewhat surprising, given the large coupling constant $Z\alpha$. However, it was also observed that for multiple pair production, the non-perturbative result was smaller than the perturbative result\cite{KleinCapri}. Since multiple pair production is naturally a higher order process, it's not surprising that a difference appears. 

\subsubsection{Comparing Lowest-order and Non-perturbative Calculations for RHIC}

\begin{figure}
\centering
\includegraphics[width=280pt]{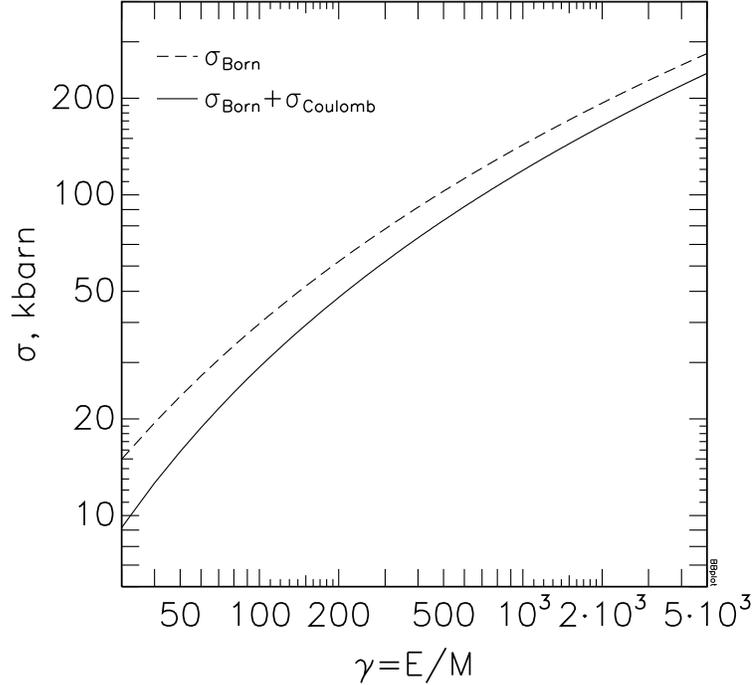}
\caption[Comparison of the lowest-order QED cross-section and Coulomb corrections]{\label{fig:Coulomb}Comparison of \ee   production cross-section in the lowest QED order ('Born cross-section') and cross-section with Coulomb corrections as a function of beam Lorentz factor \ \cite{Ivanov}.}
\end{figure}

Using the perturbation theory to all orders the authors of \cite{Ivanov} find that Coulomb corrections to the lowest-order total \ee cross-section are negative and equal to $ \sim -25\%$ for Au ions at $\sqrt{s_{NN}}=200$ GeV. Figure \ref{fig:Coulomb} compares the lowest-order QED cross-section and the cross-section with Coulomb correction as a function of beam energy.

The total cross-section of the \ee   pair production is dominated by the low invariant mass pairs ($M_{inv} \approx 2 m_e$). However, STAR detector can only observe \ee pairs above a certain minimal invariant mass cutoff ($M^{\min}_{e^+e^-} \gg 2 m_e$), and the magnitude of Coulomb corrections in the STAR observable range is unclear. Qualitatively, we expect that the Coulomb corrections become less significant for such pairs. Compared to the lowest-order Feynman diagram, the Coulomb correction diagrams include at least one extra photon propagator. The photon propagator suppreses the diagram by the factor $\sim 1/q^2$\cite{Landafshitz}. Since high invariant mass pairs are created by high energy exchange photons, the virtuality of the exchange photons is typically higher for the high-$M_{inv}$ pairs than for the low-$M_{inv}$ pairs (virtuality $ Q^2  = \omega ^2 /\gamma ^2  + p_ \bot ^2 $). Therefore, adding an extra photon to the Feynman diagram suppresses higher-order contributions more strongly for the high-$M_{inv}$ pairs than for the low-$M_{inv}$ pairs. However, the exact computation of Coulomb corrections for high-$M_{inv}$ pairs is not yet available\cite{PrivateTonyBaltz}.

\subsection{Transverse Momenta in Equivalent Photon Approximation}
The equivalent photon approximation assumes zero transverse momenta for the equivalent photons.
To estimate the photon transverse momentum spectra authors of \cite{Vidovic} consider a coupling of a photon with 4-momentum $q^\mu $ to the ion charge with an extended form factor $F(t)$. The number of photons with energy $\omega$ and transverse momentum $p_ \bot$ is given by:
\begin{equation}
\label{eqn:Hencken}
\frac{{dN\left( {\omega ,p_ \bot  } \right)}}{{d\omega dp_ \bot  }} \propto p_ \bot  ^3 \left( {\frac{{F(t)}}{t}} \right)^2 
\end{equation}
where $t =  - q^2  = \omega ^2 /\gamma ^2  + p_ \bot ^2 $ and $F(t)$ is a Fourier transform of the nuclear charge density $\rho (r)$. The nuclear charge density is assumed to have a Woods-Saxon distribution: 
\begin{equation}
\label{eqn:WoodsSaxon}
\rho (r)\propto \frac{1}{{1 + \exp \left( {\frac{{r - R_{Nuc} }}{a}} \right)}}
\end{equation}
where nuclear radius is calculated from $R_{Nuc}=r_o A^{1/3}$ with $r_o = 1.16(1.-1.16A^{-2/3})$ fm, and a constant skin-thickness of $a =$ 0.535 fm is used. These parametrizations are obtained from electron-nucleus scattering data\cite{WoodsSaxon}.

For a heavy nucleus $\rho (r)$ can be well approximated by a convolution of a hard-sphere and a Yukawa potential. The Fourier transform of $F(t)$ is then a product of separate Fourier-transforms\cite{SpencerJoakimExclusiveVM}:  
\begin{equation}
\label{eqn:FormFactor}
F\left( t \right)\propto t^{-3/2} \left( {\sin \left( {t^{1/2} R_{Nuc} } \right) - t^{1/2} R_{Nuc} \cos \left( {t^{1/2} R_{Nuc} } \right)} \right) \cdot \left( {\frac{1}{{1 + tc^2 }}} \right) \text{ where } c=0.7
\end{equation}

The distribution of the photons as a function of the photon energy $\omega $ and transverse momentum $p_\perp$ is then:
\begin{equation}
\label{eqn:pperp}
\frac{{dN(\omega ,p_ \bot  )}}{{d\omega dp_ \bot  }} \! \propto \! \frac{{\sin (\sqrt t R_{Nuc} ) \! - \! \sqrt t R_{Nuc} \cos (\sqrt t R_{Nuc} )}}{{\left( {(\omega /\gamma )^2  + p_ \bot ^2 } \right)^{3/2} }} \cdot \frac{{p_ \bot ^3 }}{{\left( {(\omega /\gamma )^2  + p_ \bot ^2 } \right)^4 }} \cdot \frac{1}{{\left( {1 + \left( {(\omega /\gamma )^2  + p_ \bot ^2 } \right)^2 c^2 } \right)}}
\end{equation}
where $\gamma = 108$ is the Lorentz factor in the lab frame. Equation (\ref{eqn:pperp}) shows that for a photon of energy $\omega $ the perpendicular momentum cannot be much higher than $\omega /\gamma$, or $\omega /108$. The transverse momenta of the photons are negligible compared to the photon energies, and the use of $p_\bot = 0$ approximation in the calculation of the \ee   production cross-section is justified.

In general, the low transverse momentum of the equivalent photons and the low total transverse momentum of the produced final state is one of the defining characteristics of ultra-peripheral interactions at heavy-ion accelerators. We will use this property for separating the true signal from the background in our analysis. 

\subsection{Previous Measurements}

The subject of two-photon physics first received wide attention at the Kiev Conference in 1970, with the reports of Brodsky\cite{somebodyCrossSection} and Balakin\cite{Balakin}. In 1971 the first observation of 'electroproduction' reaction $e^+ e^-   \to e^+  e^-   + e^+  e^-$ was made at the colliding beam machine VEPP-2 in Novosibirsk\cite{Balakin1}. This experiment measured about 100 of \ee pairs produced at the colliding beams energies of $\sim 510$ MeV, and the  cross-section in the acceptance region was found to be 20.0 mb.

\begin{figure}
\centering
\includegraphics[bbllx=0pt,bblly=250pt,bburx=434pt,bbury=454pt,width=200pt]{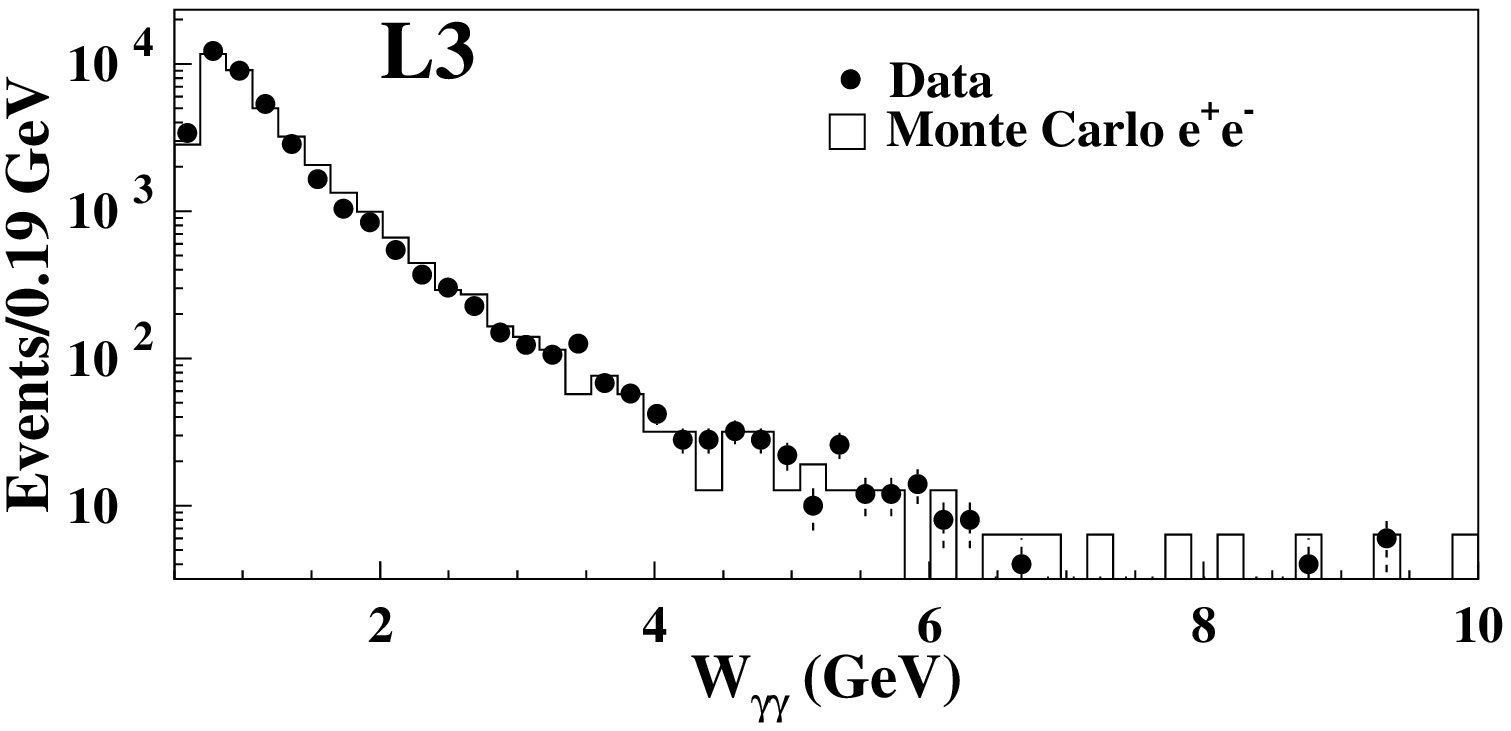}
\includegraphics[bbllx=0pt,bblly=215pt,bburx=434pt,bbury=419pt,width=200pt]{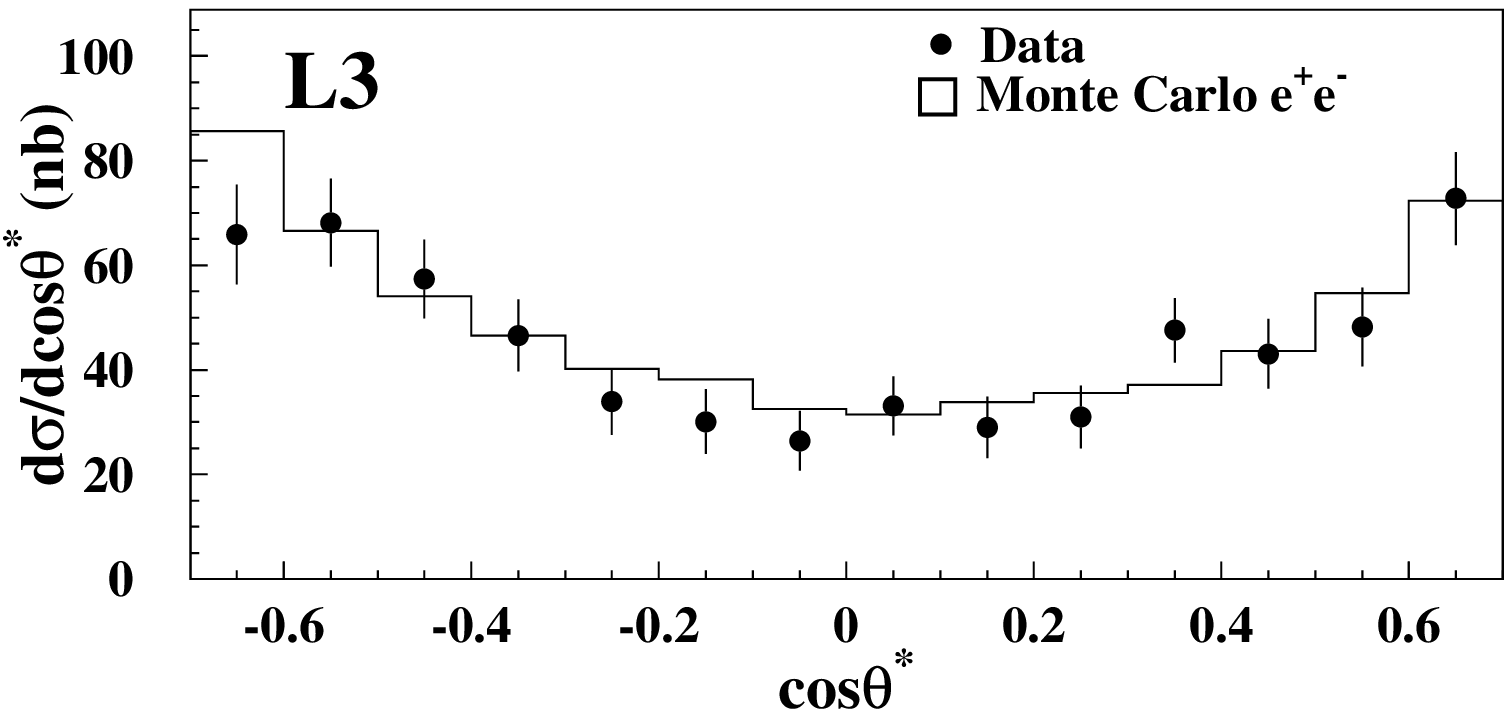}
\caption[L3 collaboration observed \ee pairs]{Left: L3 observed invariant mass spectrum of \ee   pairs. Right: L3 measured differential cross-section $d\sigma /d\cos \theta ^* $ \ \cite{L3}.}
\label{fig:L3}
\end{figure}

Since then a large number of experiments measured two-photon \ee production at electron-positron colliders. L3 experiment at LEP collider observed the untagged reactions $e^+e^-  \to  e^+e^- + e^+e^-$ at $\sqrt{s}=91$ GeV. The total available integrated luminosity was $\sim 52 \text{ pb} ^{-1}$\cite{L3}, which yielded good event statistics. Figure \ref{fig:L3} compares Monte Carlo predictions to the observed \ee pair invariant mass ($W_{\gamma \gamma}$) spectrum and the differential cross-section $d\sigma /d\cos \theta ^* $.\footnote[7]{$\theta ^*$ is defined as lepton polar angle in the center-of-mass frame (equivalent to angle $\theta '$ in Figure \ref{fig:kinematics})} Due to detector acceptance the observable kinematical range was limited to (44\deg $<  \theta ^* <  $ 136\deg ) and $W_{\gamma \gamma} \ge 500$ MeV. The total observable cross-section was measured to be $\sigma _{data}  = 2.56 \pm 0.01({\rm stat}{\rm .}) \pm 0.05({\rm syst}{\rm .}){\rm  \ nb}$, which is in excellent agreement with the lowest-order QED Monte Carlo.

\begin{figure}
\centering
\includegraphics[bbllx=155pt,bblly=535pt,bburx=435pt,bbury=795pt,width=133pt,clip=true]{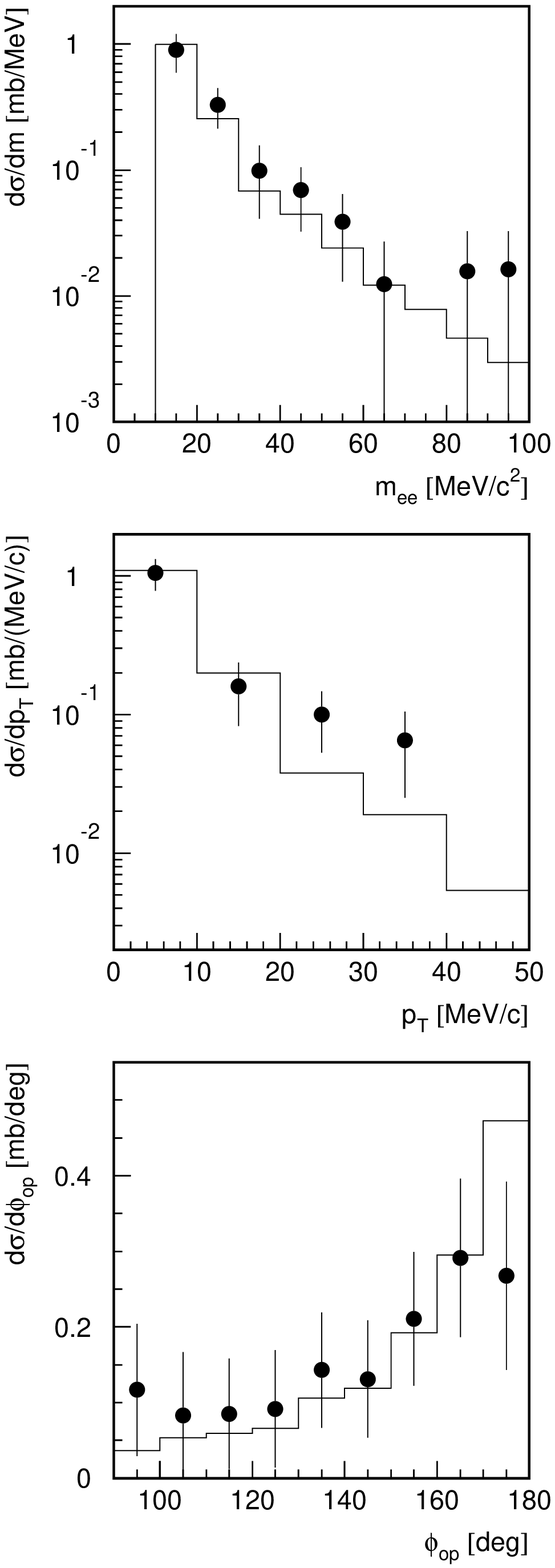}
\includegraphics[bbllx=155pt,bblly=0pt,bburx=435pt,bbury=260pt,width=133pt,clip=true]{CERES}
\includegraphics[bbllx=155pt,bblly=270pt,bburx=435pt,bbury=530pt,width=133pt,clip=true]{CERES}
\caption[CERES collaboration observed \ee pairs]{CERES measured differential cross-sections for \ee pairs. Left: $d\sigma /dm_{ee}$, center: $ d\sigma /d\phi_{op} $, right: $ d\sigma /dp_ \bot $. Dots - data, solid line - Monte Carlo \ \cite{CERES}.}
\label{fig:CERES}
\end{figure}

The first measurement of the electromagnetic \ee production in heavy-ion collisions was made by the CERES/NA45 Collaboration\cite{CERES}. CERES/NA45 is a fixed-target experiment dedicated to the measurement of \ee production in proton-proton, proton-nucleus and nucleus-nucleus collisions. For the ultra-peripheral \ee production studies sulphur beams incident on lead target were used (${}^{{\rm 16}}{\rm S}_{32}$ and  ${}^{{\rm 82}}{\rm Pb}_{207} $), with the beam energy of 200 GeV per nucleon. Only events with no hadronic interactions taking place were accepted for the study. Figure \ref{fig:CERES} shows the differential cross-section as a function of the pair invariant mass $m_{ee}$, pair transverse momentum $p_ \bot$ and the pair azimuthal opening angle $\phi_{op}$. The detector acceptance limited the kinematic variables to $10\text{ MeV} \le m_{ee}  \le 100\text{ MeV}$, $E_{e^ \pm  }  \ge 25$ MeV, $141{\rm \ mrad} \le \theta _{e^ \pm  }  \le 260{\rm \  mrad}$ and 90\deg $\le \phi_{op} \le$ 180\deg.\footnote[8]{$E_{track}$ is track energy, and $\theta_{track}$ is the track polar angle in the lab frame.} The data agreement with the lowest-order QED Monte Carlo is good for the invariant mass distribution. The $p_\bot$ distribution shows a slight enhancement for $p_\bot>20$ MeV/c, which correlates with a slight excess at azimuthal opening angles $\phi_{op}<120$\deg. This disagreement is attributed to statistical uncertainties and possibly imperfect background subtraction. The total cross-section was found to be ${{\sigma  = 13.9 \pm 3.1({\rm stat}{\rm .}) + 3.5} \mathord{\left/
 {\vphantom {{\sigma  = 13.9 \pm 3.1({\rm stat}{\rm .}) + 3.5}  - }} \right.
 \kern-\nulldelimiterspace}  - }1.9({\rm syst}{\rm .})$ mb, which is in very good agreement with the lowest-order QED prediction.

\section{Electron-Positron Pairs with Mutual Coulomb Excitation in Relativistic Heavy-Ion Collisions}
\begin{figure}
\centering
\includegraphics[width=200pt]{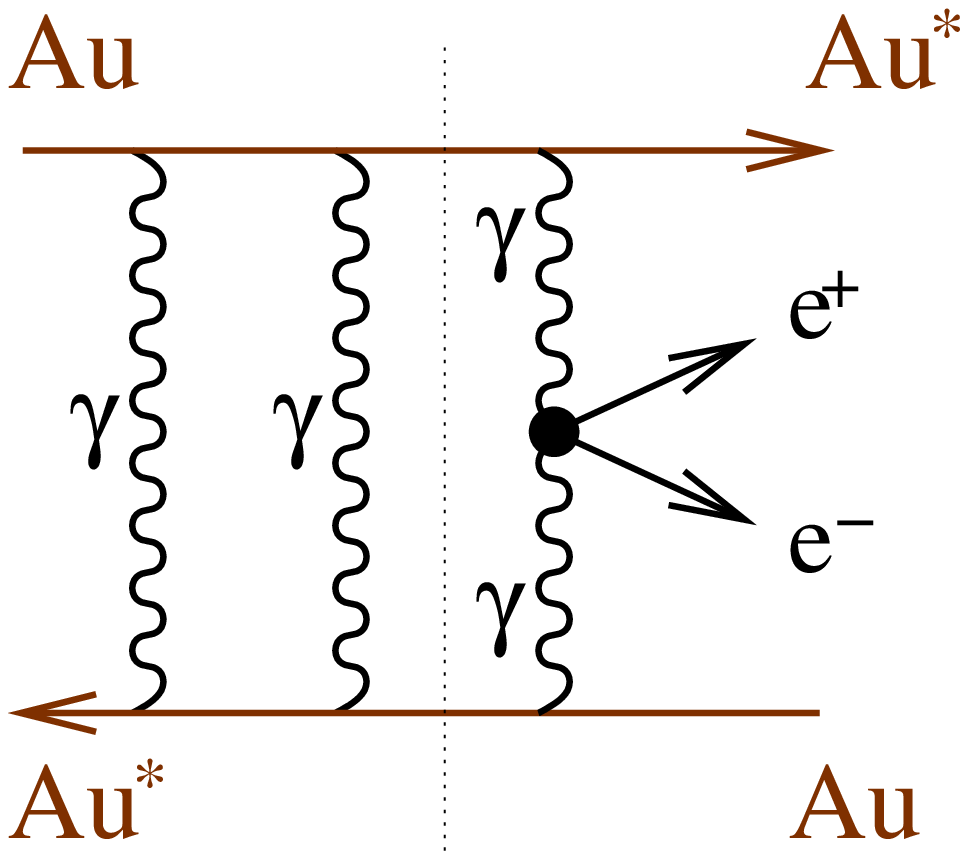}
\caption[Electron-positron pair production with mutual nuclear excitation]{\label{fig:EEwithCoulomb}\ee   production with mutual nuclear excitation. We assume that the pair production (right of the dashed line) is independent of the nuclear excitation (left of the dashed line).}
\end{figure}

The gold ions moving in the accelerator beams are surrounded by the flux of virtual photons. 
In addition to producing an \ee   pair these photons are capable of exciting gold ions from the opposite beam. The diagram in Figure \ref{fig:EEwithCoulomb} shows a production of an \ee   pair accompanied by an exchange of two photons between the Au ions, leading to the mutual Coulomb excitation of the ions. The leading mode of the excitation is excitation into the state of the Giant Dipole Resonance (GDR) \cite{GDR}. This collective excitation usually decays by a single neutron emission. Upon emission the neutrons move in the longitudinal direction with approximately the same momentum as the beam.

The diagram in Figure \ref{fig:EEwithCoulomb}, corresponding to the \ee   pair production by two photons plus an exchange of two photons, is the dominant lowest-order diagram for \ee   pair production with Coulomb breakup. In  1955 S. N. Gupta demonstrated that emission of photons involved in \ee   pair production is independent of the emission of both photons which are involved in the excitation of the ions\cite{Gupta}. This means that the probability of the \ee   production at a given impact parameter $b$ in the lowest order is independent of the probability of the excitation and breakup of the ions. This property is called factorization\cite{Hencken,GammaGammaFactorization}. If factorization holds, the total cross-section for the ultra-peripheral \ee   production with simultaneous nuclear breakup is:

\begin{equation}
\label{eqn:factorization}
\sigma _{e^{+}e^{-}\text{ with mutual breakup}}=\int d^2\overrightarrow{b}P_{e^{+}e^{-}}(\left| \overrightarrow{b}\right| )P_{XnXn}(\left| \overrightarrow{b}\right| )P_{\text{no hadronic}}(\left| \overrightarrow{b}\right| )
\end{equation}
where $P_{e^+e^-}$ is the probability for an \ee pair production,  $P_{\text{no hadronic}}$ is a probability of no hadronic interaction happening between the nuclei, and $P_{XnXn}$ is a probability of a simultaneous nuclear excitation with breakup.\footnote[9]{We distinguish between the case where both excited nuclei emit exactly one neutron $(1n,1n)$ and the case where both excited nuclei emit one or more neutrons $(Xn,Xn)$.}

\begin{figure}
\centering
\includegraphics[width=280pt]{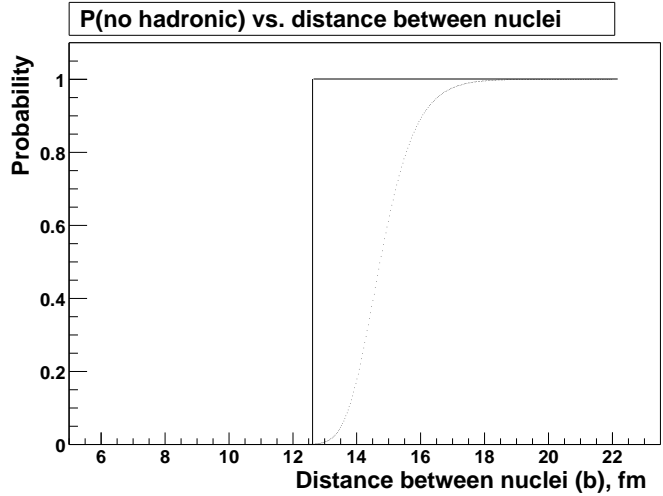}
\caption[Probability of no hadronic interaction taking place between the ions as a function of nuclear impact parameter]{\label{fig:PnoHadr}Probability of no hadronic interaction taking place between the ions as a function of nuclear impact parameter. Solid line -- theta-function cut-off, dotted line -- $P_{\text{no hadronic}}(b)$ computed with Woods-Saxon nuclear form-factor.}
\end{figure}

The quantity $P_{\text{no hadronic}}(b)$ modifies the $\theta$-function cut-off previously used in Equation (\ref{eqn:computeGammaGammaLumi}). While the theta-function represents a hard-sphere nuclear model, the computation of $P_{\text{no hadronic}}$ assumes a nuclear form-factor with smooth edges.
Using a Glauber Model we get:
\begin{equation}
\label{eqn:Glauber}
P_{\text{no hadronic}}(b) = 
 {\rm exp}\left( {{\rm  - }\sigma _{{\rm nn}} \int {T_A (\overrightarrow r )T_B (\overrightarrow b  - \overrightarrow r )d^2 \overrightarrow r } } \right)
\end{equation}
where $\sigma _{{\rm nn}} = $ 52 mb is the hadronic nucleon-nucleon cross-section (at $\sqrt{s_{NN}}=$ 200 GeV) and $T(r)$ is a nuclear thickness function. The nuclear thickness function is calculated from the  nucleon density $\rho (b)$ distributed according  to the Woods-Saxon formula (Equation (\ref{eqn:WoodsSaxon})):
\begin{equation}
\label{eqn:T}
 T(\overrightarrow b ) = \int {\rho (} \overrightarrow b ,z)dz 
\end{equation}
Figure \ref{fig:PnoHadr} compares the hard-sphere cut-off (at $b=2R_{Au}$) and $P_{\text{no hadronic}}$. 

The probability of mutual nuclear excitation with breakup $P_{XnXn}$ was first computed by Baltz, Klein and Nystrand in \cite{SpencerWithBaltz} for the case of ultra-peripheral photo-nuclear $\rho ^0 $ production accompanied by a mutual nuclear dissociation. 
The computation assumes the  flux of virtual photons to be distributed according the Weizs\"acker-Williams approximation (\ref{eqn:WeizsackerWilliams}). The lowest-order probability for an excitation of a colliding beam ion to any state which emits one or more neutrons $(Xn)$ is:
\begin{equation}
\label{eqn:P1Xn}
P_{C(Xn)}^1(b)=\int \frac{d^2n\left( \omega , \overrightarrow b \right) }{d^2 \overrightarrow b}\sigma _{\gamma A\to A^{*}}\left( \omega \right) \frac{d\omega }\omega   
\end{equation}
where $\sigma _{\gamma A \! \to \!  A^ *  } \left( \omega \right)$ is cross-section of the excitation with breakup of nucleus $A$ by a single photon of energy $\omega$. This quantity has been determined by measurement at a wide range of energies \cite{nuclearExcitation}.

\begin{figure}[t]
\centering
\includegraphics[width=300pt]{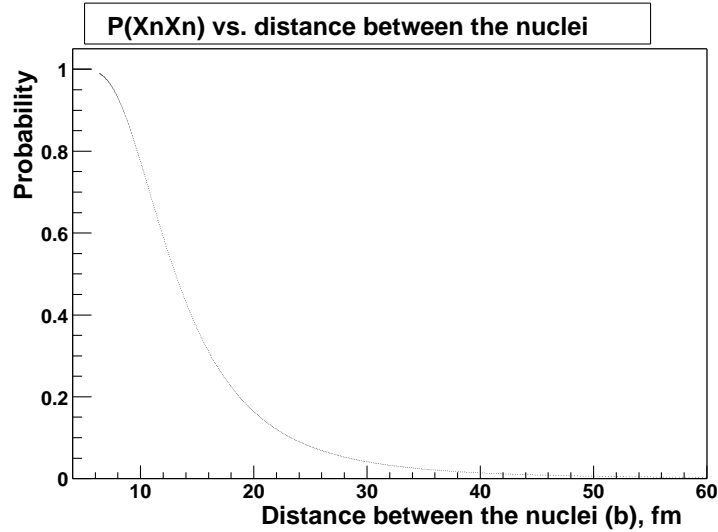}
\caption[Probability of mutual nuclear excitation with breakup as a function of the nuclear impact parameter]{\label{fig:Pxnxn}Probability of mutual Coulomb nuclear excitation with breakup as a function of the nuclear impact parameter.}
\end{figure}

At small impact parameters, $P_{C(Xn)}^1 (b)$ can exceed 1 and can be interpreted as a mean number of excitations. The unitarization procedure, similar to the unitarization for \ee   pair production probability, needs to be utilized. The probability of having at least one Coulomb excitation is then $P_{C(Xn)} (b) = 1 - \exp \left( - {P_{C(Xn)}^1 (b)} \right)$.  In the reaction involving the dissociation of both ions, each individual breakup occurs independently \cite{factorization}. The probability is thus the square of the individual breakup probabilities, i.e. $P_{XnXn} (b) = \left( {P_{C(Xn)} (b)} \right)^2 $.

Figure \ref{fig:Pxnxn} shows the probability of mutual Coulomb nuclear excitation with breakup computed for 120 values of $b$ between 5 fm and 60 fm for Au ions at $\sqrt{s_{NN}}=200$ GeV. For the excitation and breakup of Au ions to be likely, the impact parameter must be small ($b < 30$ fm).

\section{Other Related Processes at RHIC}
\label{sec:otherprocesses}
\subsection{Ultra-Peripheral Electromagnetic Processes}
The bound-free pair production is another pure QED process in ultra-peripheral relativistic heavy-ion collisions which is of practical importance in the collider. It is the process, where a pair is produced but with the electron not as a free particle, but into an atomic bound state of one of the nuclei. Using the Equivalent Photon Approximation and the approximate wave functions for bound state and continuum, the cross-section for this process for Au beams at RHIC was found to be about 90 b\cite{BaurPhysRev}. This changes the charge of the ions in the beam, causing beam loss. 

Electron and positron can also form a bound state, positronium. For the two-photon processes, the positronium can only be in the para-positronium state $^1S_0$. An interesting state with a negative charge parity -- orthopositronium ($^3S_1$ state of positronium) -- can be created in three-photon interactions\cite{orthopositronium}, like the one in Figure \ref{fig:RHICFeynman}. Cross-section for this process is suppressed with respect to para-positronium production only by $(Z\alpha)^2$ and in the case of Au ions at RHIC the production of orthopositronium should be comparable to the production of para-positronium.

A wide range of final states not involving \ee   pairs is possible in the two-photon interactions at RHIC. The photons couple to any state with internal charge constituents (i.e. quarks) and spin/parity $J^{PC} = 0^{-+}, 0^{++}, 2^{-+}$ or $2^{++}$ (for $J \le 2$). The spin-1 final state in a two-photon reaction is impossible, because spins of massless photons can only be aligned or antialigned. We discuss some of the final states below, grouping them by the area of physics interest. These topics represent excellent future two-photon physics research opportunities at RHIC\cite{Krauss, SpencerNystrand, Bertulani}.
\begin{list}{$\bullet $ }
\item  High-field QED: $\tau ^+ \tau ^-$ and $\mu ^+ \mu ^-$ pairs

These events are created by the same mechanism as \ee   pairs. The cross-sections are low, due to the very high masses of these leptons. 

\item

\item  Meson Spectroscopy: Search for glueballs and other exotica

These events consist of a single meson or an exotic particle (e.g. glueball) in the final state. Because two photons couple to $(\text{charge})^4$, the cross-section $\sigma ( \gamma \gamma \rightarrow X )$ is a direct measurement of the quark and isospin content of the mesonic final state. Consequently, final states consisting of charged particles ( $q \bar{q}$ ) are possible, but pure gluon final states are not possible.

\item Meson form factors (pairs)

Two-photon interactions also produce meson pairs via $\gamma \gamma \rightarrow X \bar{X}$. At the hadron level, photons couple only to charged mesons, so $\pi ^+ \pi^-$ should be produced, but not $\pi ^0 \pi^0$.  At high enough energies photons couple directly to the quark content of the mesons, and both $\pi ^+ \pi^-$ and $\pi ^0 \pi^0$ final states are produced in comparable numbers. By comparing the rates of the two final states, the transition can be studied, and the size of the mesons determined.

\end{list}
\subsection{Ultra-Peripheral Photo-nuclear Processes}
\label{sub:rho0}
An ultra-peripheral collision of heavy ions might involve only one photon, interacting with the hadronic field of another ion, in which case a reaction is called 'photo-nuclear' interaction. One of the possible descriptions of this process is that a photon from one ion can fluctuate into a virtual $q \bar{q}$ pair, which scatters on the other ion and emerges as a real vector meson.\footnote[10]{This interaction mechanism is called 'vector dominance model', and believed to be dominant at RHIC energies. There are a number of other approaches to photo-nuclear processes, including photon-Pomeron scattering.} For Au ions at RHIC design luminosity this reaction produces several vector meson species: $\rho ^0$ at the rate of $\sim 100$ Hz, $\omega $ at the rate of 12 Hz and $\phi $ at the rate of $7.9$ Hz\cite{SpencerNystrandVectorMeson}.

The  $\rho ^0 $ is a short-lived resonance (width $\Gamma = 150$ MeV) which quickly decays into a  $\pi ^+ \pi ^-$ pair. Additionally, $\pi ^+ \pi ^-$ can be created in RHIC collisions via the non-resonant channel $\gamma \rightarrow \pi ^+ \pi ^-$. These processes have been recently observed at RHIC by the STAR collaboration at $\sqrt{s_{NN}}=130$ GeV and $\sqrt{s_{NN}}=200$ GeV both as an exclusive channel and in coincidence with mutual nuclear excitations\cite{SpencerRhoPaper,Falk200GeV}. This is a potential background for \ee production, as discussed in Chapter \ref{ch:Simu}.

\begin{figure}[t]
\centering
\includegraphics[height=150pt,clip=t]{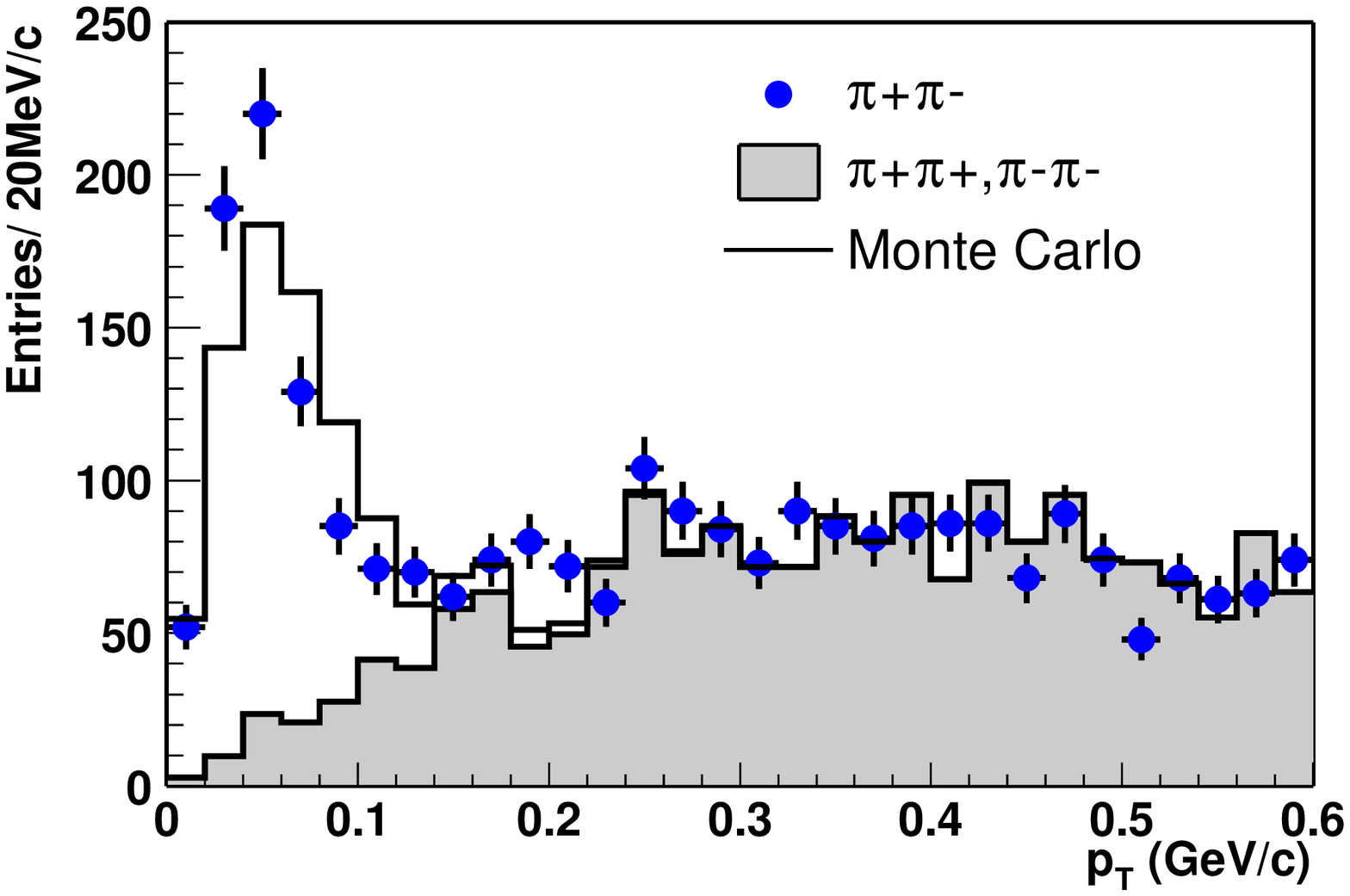}
\includegraphics[height=150pt,clip=t]{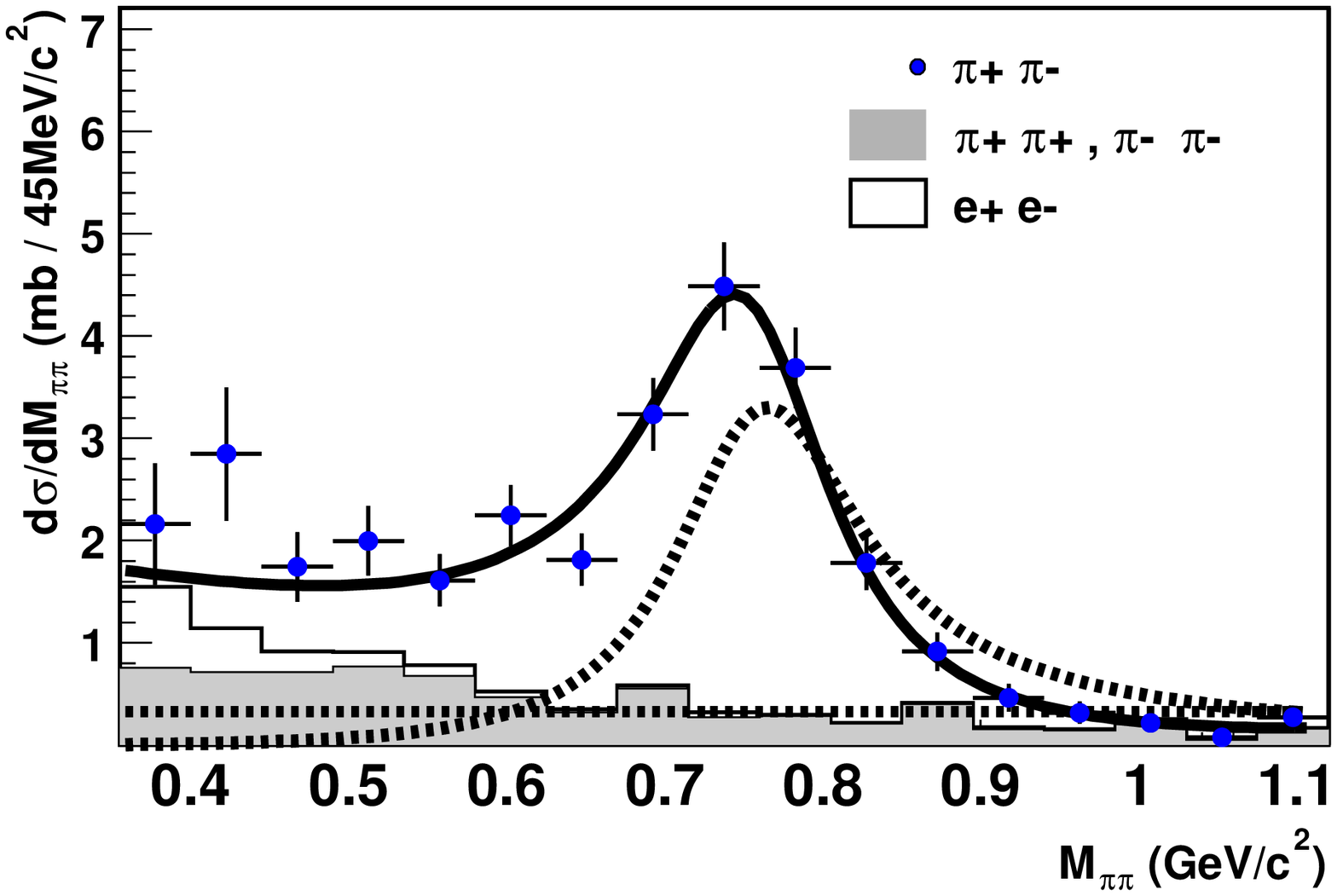}
\caption[STAR collaboration observed ultra-peripheral $\pi ^+ \pi ^-$ pairs]{\label{fig:Rho}STAR collaboration observed ultra-peripheral $\pi ^+ \pi ^-$ pairs. Left: pair transverse momentum distribution. Right: $d\sigma / dM_{\pi \pi}$ distribution (dashed lines represent direct $\pi ^+ \pi ^-$ and $\rho ^0$ contributions) \ \cite{SpencerRhoPaper}. }
\end{figure}

Figure \ref{fig:Rho} shows the pair transverse momentum and invariant mass distribution of identified ultra-peripheral coherent events $Au+Au \to Au^* + Au^* + \pi ^+ \pi^-$ at $\sqrt{s_{NN}}=130$ GeV. The transverse momentum distribution shows a prominent peak at low momenta. The incoherent $\pi ^+ \pi ^-$ pair spectrum is modelled with the same-sign pair combinations ($\pi ^+ \pi ^+$, $\pi ^- \pi ^-$, shaded histograms in Figure \ref{fig:Rho}), and doesn't show a peak at low transverse momenta. The invariant mass distribution is split into contributions from the coherently produced $\rho ^0$ mesons (Breit-Wigner distribution), direct $\pi ^+ \pi ^-$ (flat in $M_{inv}$) plus the interference between these two sources and background from un-identified \ee pairs at very low invariant masses.  

\chapter{Monte Carlo Event Generator}
\label{ch:MonteCarlo}

\begin{figure}
\centering
\resizebox*{0.45\textwidth}{!}{\includegraphics{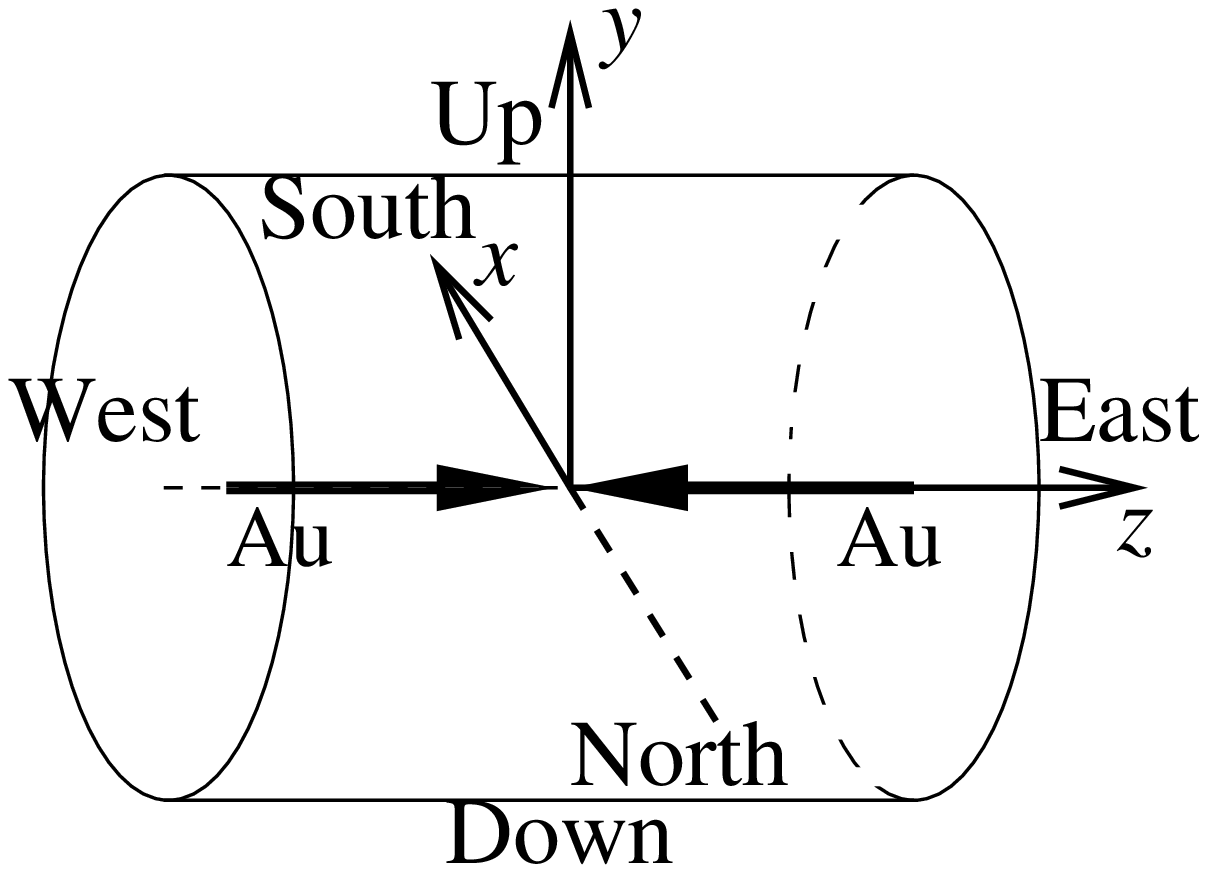}}
\resizebox*{0.45\textwidth}{!}{\includegraphics[clip=true]{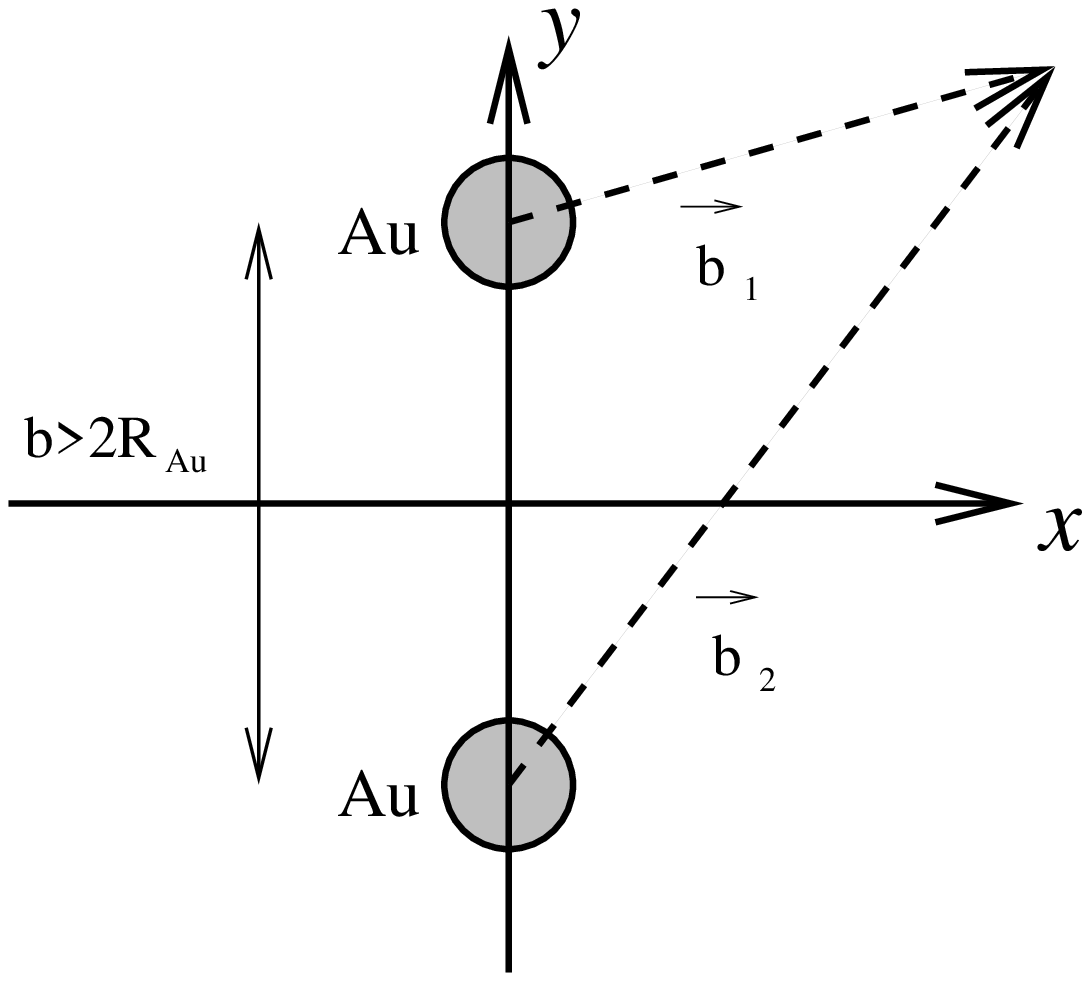}}
\caption[STAR coordinate system]{Left: STAR coordinate system with a schematic view of the STAR TPC and the RHIC beams. Right: two-dimensional photon density integration in STAR coordinate system.}
\label{fig:TPCgeometry}
\end{figure}

The following chapter describes a method we used for generating \(e^{+}e^{-}\) with the nuclear excitation. The method is based on the Equivalent Photon Approximation. The output of the event generator will be used in Monte Carlo studies of detector acceptance, efficiencies and systematic effects (Chapter \ref{ch:Simu}). The events are generated in the STAR laboratory frame of reference.
Figure \ref{fig:TPCgeometry} shows the STAR coordinate system definition, used throughout the rest of this thesis.

\section{Differential Cross-section Computation}

To generate \ee   events, we need first to compute the differential cross-section \newline ${{d\sigma _{e^ +  e^ -  } } \mathord{\left/ {\vphantom {{d\sigma _{e^ +  e^ -  } } {dM_{inv} dY}}} \right. \kern-\nulldelimiterspace} {dM_{inv} dY}} $. 
As we see from the equation (\ref{eqn:GammaGammaLuminosity}) this quantity depends on the two-photon luminosity with nuclear excitation. We perform numerical integration over two impact parameters $\overrightarrow {b_1 } $ and $\overrightarrow {b_2 }$ in $(x,y)$ plane to find the two-photon density (see Figure \ref{fig:TPCgeometry}):
\begin{equation}
\label{eqn:LumiWithExcitation}
F_{\gamma \gamma }=\int\limits_{R_{Au}}^{R_{\max }}d^2\overrightarrow{b_1}\int\limits_{R_{Au}}^{R_{\max }}d^2\overrightarrow{b_2}\frac{d^2n}{d^2\overrightarrow{b}_1}\left(\overrightarrow  b_1,\omega _1\right) \frac{d^2n}{d^2\overrightarrow{b}_2}\left(\overrightarrow  b_2,\omega _2\right) P_{\text{no hadr}}\left( \left| \overrightarrow{b_1}-\overrightarrow{b_2}\right| \right) P_{XnXn}\left( \left| \overrightarrow{b_1}-\overrightarrow{b_2}\right| \right) 
\end{equation}

Equation (\ref{eqn:LumiWithExcitation}) is derived from the original Equation (\ref{eqn:computeGammaGammaLumi}) with the addition of the requirement of the simultaneous nuclear excitation ($P_{XnXn}$) and replacement of a $\theta$-function cut-off with a probability of no hadronic interaction between the ions ($P_{\text{no hadr}}$). The upper limits of integration are taken as the distance from the nuclei where the field becomes very small: $R_{\max} \sim 4\gamma _{Au} /\omega $. 

The value of the two-photon density (\ref{eqn:LumiWithExcitation}) is computed for 100 values of the pair invariant mass ($M_{inv} = 2\sqrt{ \omega _1 \omega _2 }$) and pair rapidity ($ Y = 1/2 \log(\omega _1 / \omega _2 ) $) in the range of 100 MeV $< M_{inv}  <$ 300 MeV and $ |Y| < 1.3$. We chose these values because they represent the kinematical range in which the STAR detector has non-zero acceptance to the \ee pairs. All mass and rapidity values are equidistant, with the distances $\Delta M_{inv}  = 2.0 $ MeV and  $\Delta Y = 0.026$. Each value in this two-dimensional table is then converted into the two-photon luminosity (Equation (\ref{eqn:GammaGammaDiffLumi})), and multiplied by the lowest-order QED differential cross-section for the two-photon annihilation into a \ee   pair, taken at the corresponding values of the pair mass and rapidity (Equation (\ref{eqn:CrossSectionGammaGammaEE})). As the result we get a tabulated differential distribution $
{{d\sigma _{e^ +  e^ -  } } \mathord{\left/
 {\vphantom {{d\sigma _{e^ +  e^ -  } } {dM_{inv} dY}}} \right.
 \kern-\nulldelimiterspace} {dM_{inv} dY}} $.

We expect that the use of EPA for cross-section studies of the \ee  pairs with high invariant masses is justified, since the photon virtuality is significantly smaller than the \ee pair invariant mass ($Q^2 \sim 1/R^2_{Au} \sim (30)^2 \text{ MeV}^2 \ll M^2 _{inv}$). The use of $R_{Au}$ as a lower intergration limit in (\ref{eqn:LumiWithExcitation}) is also justified by this circumstance. 

\section{Event Generator}
We generate electron/positron tracks (track momenta $\overrightarrow{p}_{e^+}$ and $\overrightarrow{p}_{e^-}$) in the detector lab frame. The origin of the tracks is taken to be at $(0,0,z)$, where $z$ is generated as a Gaussian random variable with mean zero and $\sigma = 30$ cm to approximate the distribution of vertex position in the data.\footnote{The smearing of the $z_{vert}$ in the data is significant ($\sigma_z \approx 30$ cm, see Chapter \ref{ch:Experiment}). Therefore the simulations include $z_{vert}$ smearing to study its possible effects on the track reconstruction. The smearing of the vertex radial position is small ($\sigma_R \approx 0.25$ mm) and is expected to have negligible effect on the reconstruction. We do not include transverse vertex smearing in the simulations.}  The following describes how we draw $\overrightarrow{p}_{e^+}$ and $\overrightarrow{p}_{e^-}$.

\subsection{Drawing $Y$ and $M_{inv}$}
Once we have obtained a tabulated differential cross-section distribution we can draw two correlated random variables $M_{inv}$ and $Y$ according to this distribution. A two-step procedure needs to be used, whereby the value of $M_{inv}$ in the range $(M_{inv}^{\min}, M_{inv}^{\max})=(100 \text{ MeV}, 300 \text{ MeV})$ is drawn according to its marginal distribution, and then the value of $Y$ in the range $(Y^{\min},Y^{\max})=(0, 1.3)$ is determined according to the conditional distribution $f_{Y} \left(Y' | M_{inv} \right)$. 

We draw a random number from a uniform $(0,1)$ distribution and obtain a value of $M_{inv}^0$ comparing this random number to a table of values of 
\begin{equation}
\label{eqn:drawM}
\frac{{\int\limits_{M_{inv}^{\min } }^{M_{inv}} {\int\limits_{Y^{\min } }^{Y^{\max } } {dYdM'_{inv} } \frac{{d\sigma }}{{dM_{inv} dY}}} }}{{\int\limits_{M_{inv}^{\min } }^{M_{inv}^{\max } } {\int\limits_{Y^{\min } }^{Y^{\max } } {dYdM'_{inv} } \frac{{d\sigma }}{{dM_{inv} dY}}} }}
\end{equation}

Then this value $M_{inv}^0$ is used to determine $Y^0$ by comparing another draw from uniform  $(0,1)$ distribution to a table of values of 
\begin{equation}
\label{eqn:drawY}
\frac{{\int\limits_{Y_{\min } }^Y {dY'\frac{{d\sigma }}{{dM_{inv} dY}}} }}{{\int\limits_{Y_{\min } }^{Y_{\max } } {dY'\frac{{d\sigma }}{{dM_{inv} dY}}} }}
\end{equation}
and randomly assigning the sign of $Y^0$.\footnote{We use the property that the distribution of $Y$ is symmetric around zero.} Using these values of $Y^0$ and $M_{inv}^0$ the momenta of the interacting photons can be calculated.

\subsection{Drawing Transverse Momentum of the Photons}
\label{sub:TransverseMomentum}

We draw a value of the photon transverse momentum after we determine the photon energy (for EPA photons $ \left| {p_z } \right| = E $). Since the transverse momentum is much smaller than the energy of the photon, this is a reasonable approach. Adding the transverse momentum after the photon energy has been determined makes photons slightly virtual, since $
\left( {q^\mu  } \right)^2  = \left( {E,p_x ,p_y ,p_z } \right)^2  = E^2  - p_ \bot ^2  - E^2  < 0$. For a given value of photon energy $\omega $ the photon transverse momenta $p_{\perp}$ are distributed according to the distribution (\ref{eqn:pperp}). It is impractical to use the same method as we used for $M_{inv}$ and $Y$ to draw a value of $p_{\perp}$ since a new table of the distribution function $F_{p_{\perp}} (\omega )$ would need to be generated for each new value of $\omega $. Instead, we use a Von Neumann accept/reject method \cite{pdg} to make a single random draw from a distribution $F_{p_{\perp}}(\omega)$ for a photon with energy $\omega $. 

The $x$ and $y$ projections of the photon momentum are obtained using a randomly chosen (uniformly distributed) azimuthal angle. The transverse momentum of the two-photon system $p_{\perp} ^ {tot}$ is the sum in quadratures of the transverse momenta of the two individual photons. 

At this stage we can find the momenta of the photons in their center of mass system, by boosting the lab photon momenta by a Lorentz-factor 
\begin{equation}
\label{eqn:beta}
\overrightarrow \beta   = {{\left( {p_x^{tot} ,p_y^{tot} ,p_z^{tot} } \right)} \mathord{\left/
 {\vphantom {{\left( {p_x^{tot} ,p_y^{tot} ,p_z^{tot} } \right)} {\left( {E_1  + E_2 } \right)}}} \right.
 \kern-\nulldelimiterspace} {\left( {E_1  + E_2 } \right)}}
\end{equation}

Due to non-zero transverse momenta, the two-photon rest frame is not collinear with the $z_{\text{cm}}$-axis. The Euclidean rotation which yields the frame collinear with photon momenta is specified by two parameters: the axis of rotation $\overrightarrow \nu$ and the angle of rotation $\phi$:
\[
\overrightarrow \nu   = \overrightarrow {p^{tot}_{\text{cm}} }  \times \overrightarrow z_{\text{cm}} , \ \footnote{$
\overrightarrow z  = \left( {0,0,1} \right)$ in the CM frame.}
\]
\begin{equation}
\label{eqn:NuPhi}
\phi  = \cos ^{ - 1} \left( {{{\overrightarrow {p_{cm}^{tot} }  \cdot \overrightarrow {z_{cm} } } \mathord{\left/
 {\vphantom {{\overrightarrow {p_{cm}^{tot} }  \cdot \overrightarrow {z_{cm} } } {\left| {\overrightarrow {p_{cm}^{tot} } } \right|}}} \right.
 \kern-\nulldelimiterspace} {\left| {\overrightarrow {p_{cm}^{tot} } } \right|}}} \right)
\end{equation}

\subsection{Generating Angular Distribution of the \ee   Pairs}

We start by generating electron and positron track momenta in the \ee   center of mass frame. The angular distribution for the pair (in the frame of reference collinear with photon momenta)  is given by (\ref{eqn:AngularCrossSection}). We use accept/reject method to draw the value of $\theta ^*$, and we draw azimuthal angle of the pair $\varphi$ from  a uniform $(0,2\pi )$ distribution. To transform the momenta into the frame of reference  which is collinear with the lab frame, we perform a Euclidean rotation of the generated momenta around axis $\overrightarrow \nu$ by the angle $-\phi$, defined in Equation (\ref{eqn:NuPhi}).

The final step is to transform the electron and positron track momenta from the center of mass frame into the lab frame. This requires a Lorentz boost by $-\overrightarrow{\beta}$ (defined in Equation (\ref{eqn:beta})). Appendix \ref{app:variables} provides formulae for the boost and rotation transformations.

\section{Monte Carlo Results for  \AuAuee}
\label{sec:EvgenResults}
\subsubsection{Kinematical Distributions}
\begin{figure}
\centering
\includegraphics[width=280pt]{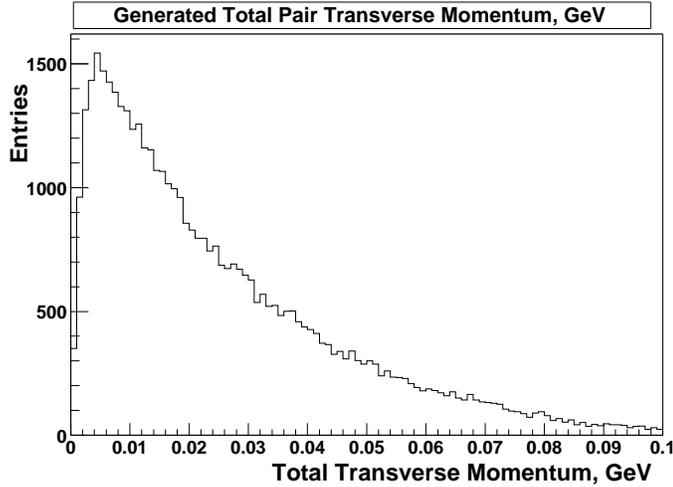}
\caption[Total pair $p_{\perp}^{tot}$ distribution for pairs with $100 \text{ MeV }<M_{inv}<300 \text{ MeV}$ and $|Y|<1.3$]{Total pair $p_{\perp}^{tot}$ distribution for pairs with $100 \text{ MeV }<M_{inv}<300 \text{ MeV}$ and $|Y|<1.3$.}
\label{fig:pperp}
\end{figure}

A few interesting observations can be made at this point. First, we now can plot the distribution of the total transverse momenta of all ultra-peripheral \ee   pairs. This spectrum is a convolution of the photon energy spectrum and the energy-dependent single photon transverse momentum distribution. Figure \ref{fig:pperp} shows this distribution for AuAu collisions at $\sqrt{s_{NN}} = 200$ GeV. The distribution displays a prominent peak at $p_{\perp}^{tot} \sim 5$ MeV/c, which is a defining signature of ultra-peripheral reactions at heavy-ion colliders.

\begin{figure}
\centering
\includegraphics[width=190pt]{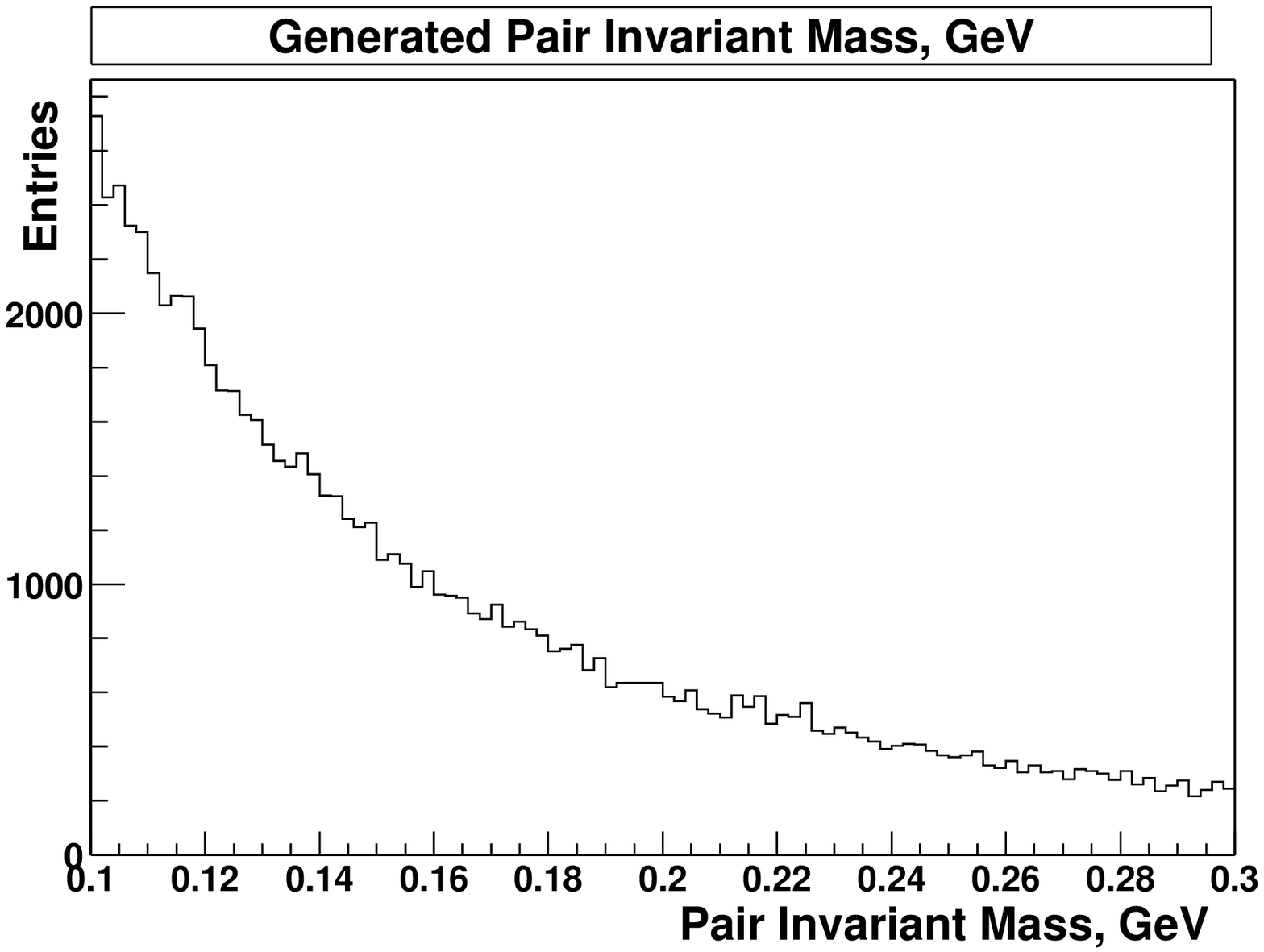}
\includegraphics[width=190pt]{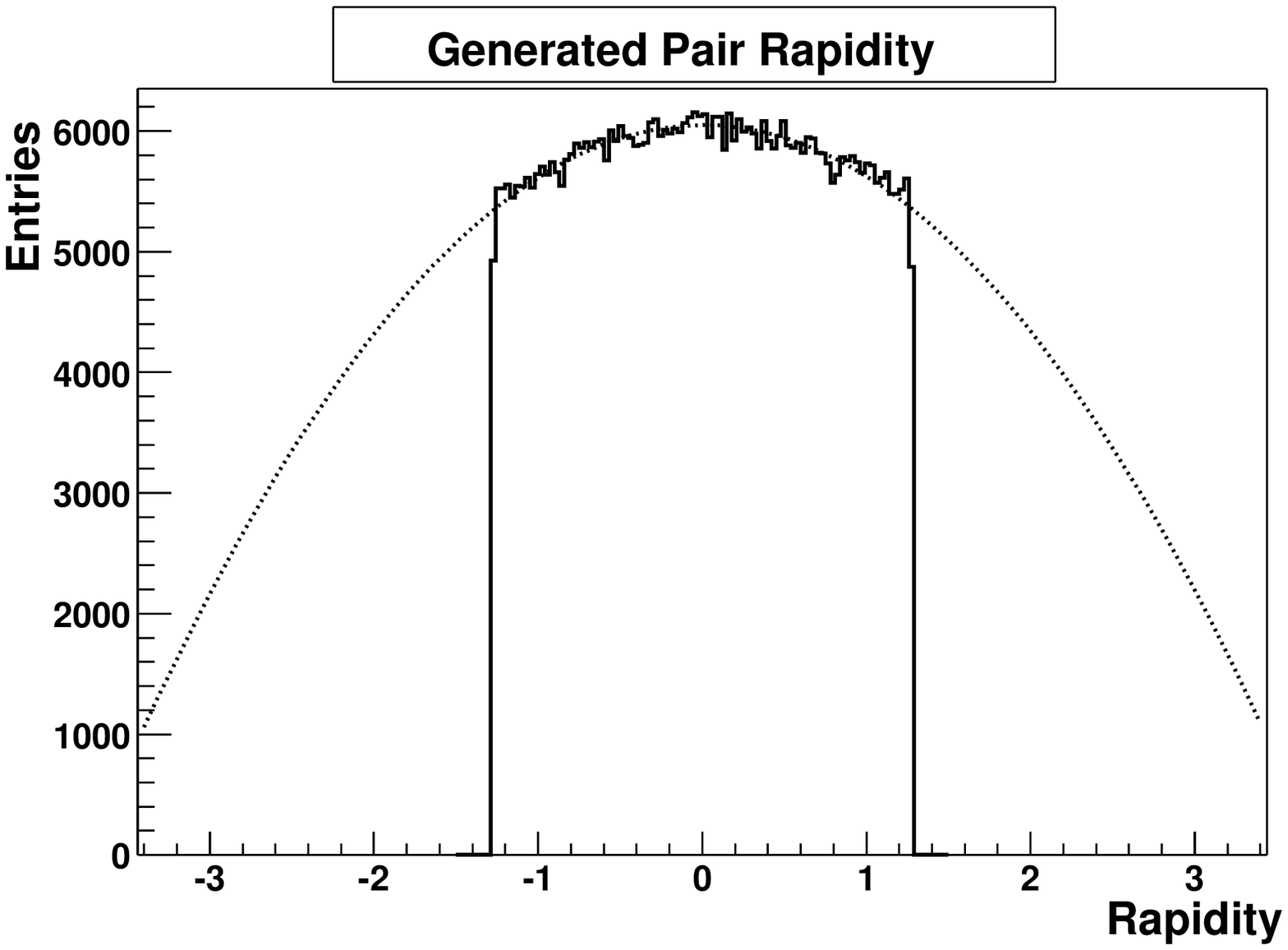}
\caption[Generated invariant mass and pair rapidity spectra]{Left: generated invariant mass spectrum. Left: pair rapidity spectrum, dotted - theoretical prediction, solid - generated.}
\label{fig:EvgenMinvAndY}
\end{figure}

We also present the distribution of the generated pair invariant mass and the pair rapidity (Figure \ref{fig:EvgenMinvAndY}). The invariant mass spectrum falls off very quickly with increasing $M_{inv}$. Total pair rapidity distribution is shown with a theoretical prediction (dashed line) between -3.2 and 3.2. The distribution reaches maximum at zero, and drops off for high values of $|Y|$.

\begin{figure}
\centering
\includegraphics[width=210pt]{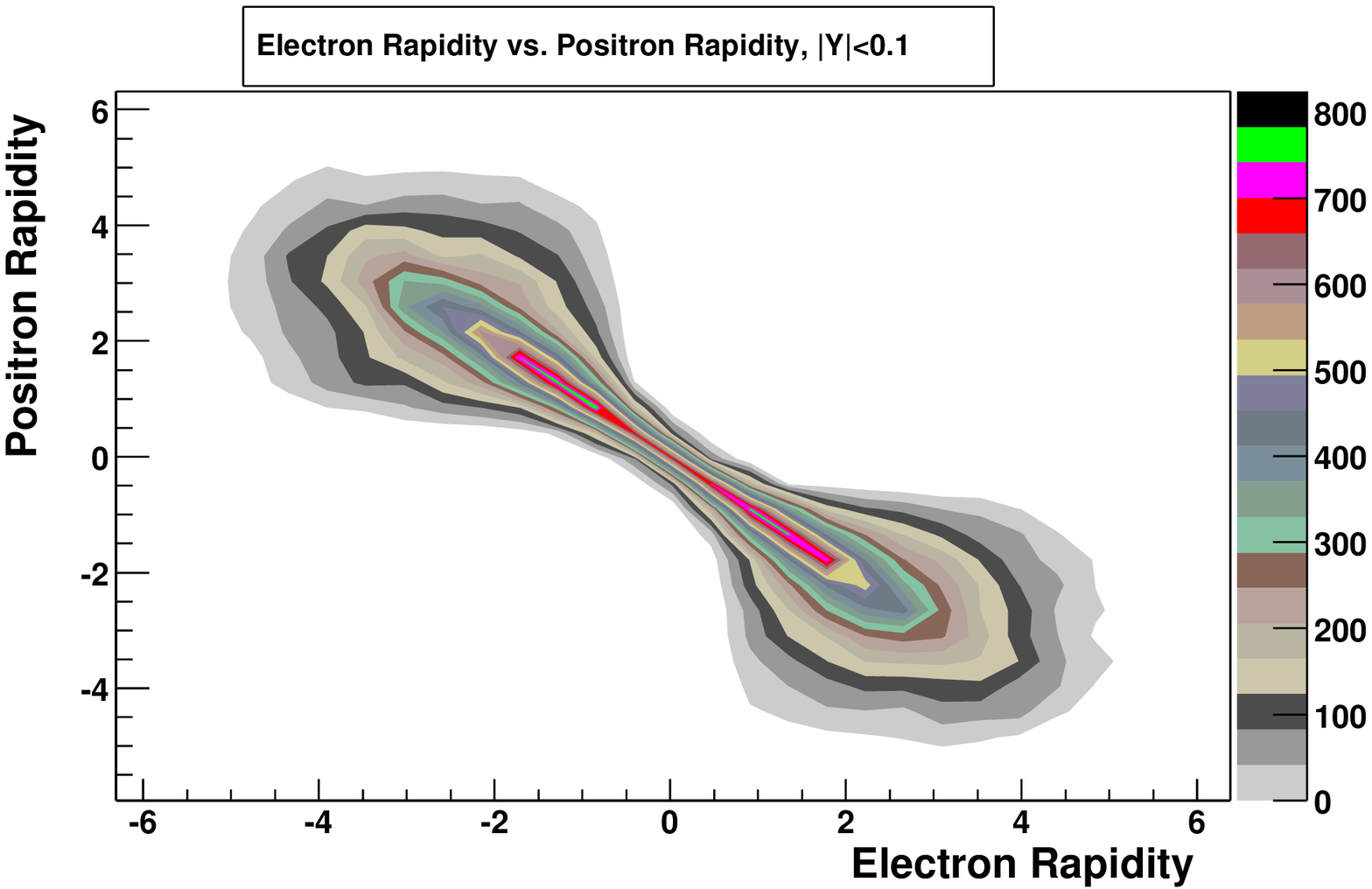}
\includegraphics[width=210pt]{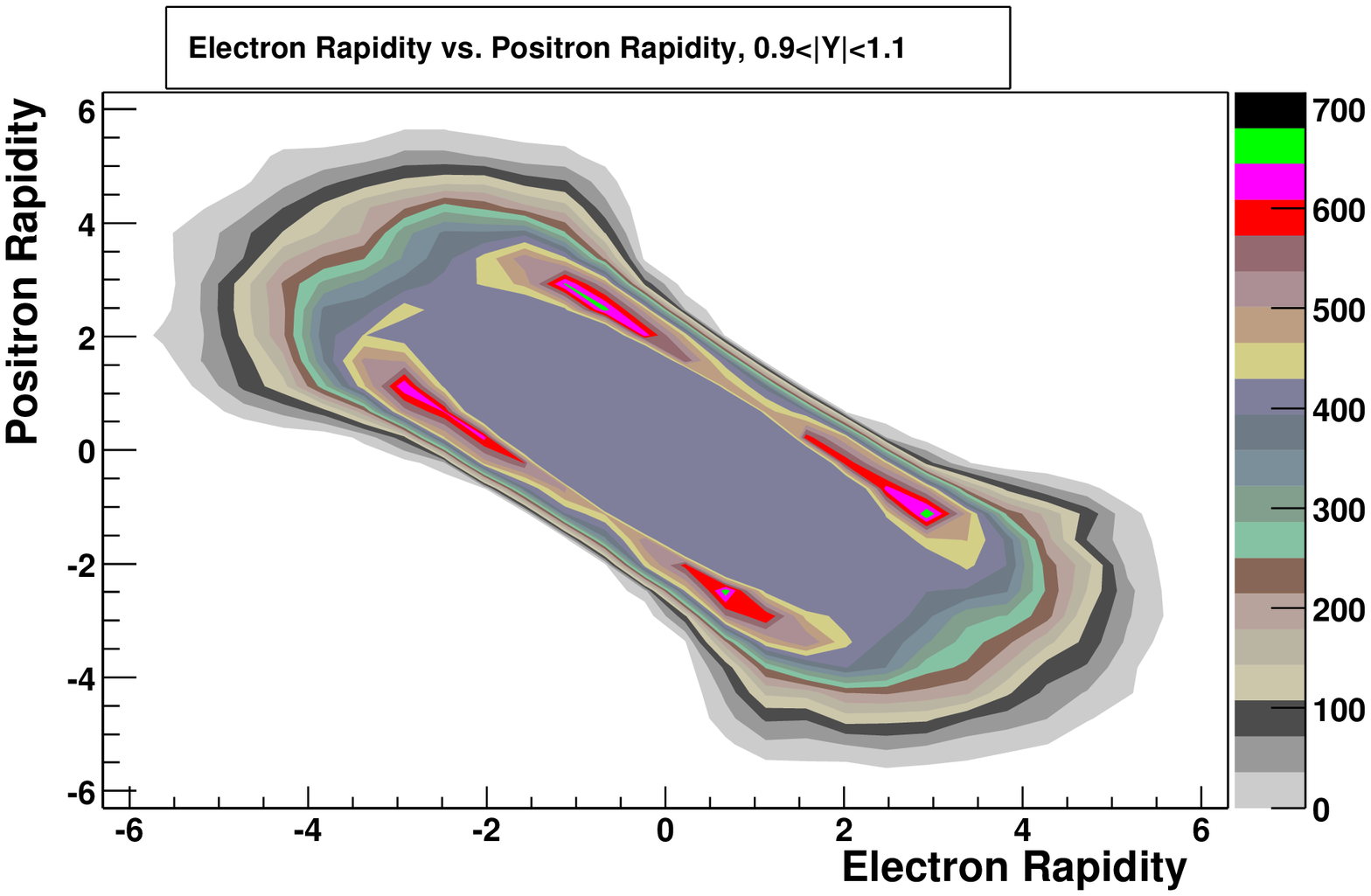}
\caption[Monte Carlo individual track $\eta ^{track} $ distributions]{Contour plots of electron vs. positron rapidity for pairs with $|Y|<0.1$ (left) and $0.9<|Y|<1.1$ (right).}
\label{fig:Eta1VsEta2}
\end{figure}

Unlike the total pair rapidity, the individual track rapidities are centered away from zero rapidity, as shown in Figure \ref{fig:Eta1VsEta2}. For a pair with a very central rapidity ($|Y|<0.1$) the individual track rapidity reaches a maximum at $\sim 1.5$ and  reaches up to 5.0. Typically, electron and positron tracks have individual rapidities of similar absolute value, but of opposite signs. For the pairs with non-central rapidity ($0.9<|Y|<1.1$) the track rapidity peaks at $\sim 2.0$ and reaches up to 6.0. Since both the electron and positron tracks are ultra-relativistic, the individual track rapidities ($Y_{e^+}$ and $Y_{e^-}$) are extremely close to the individual track pseudorapidities ($\eta_{e^+}$ and $\eta_{e^-}$). 

\vspace{10pt}

\begin{figure}[h]
\centering
\includegraphics[width=210pt]{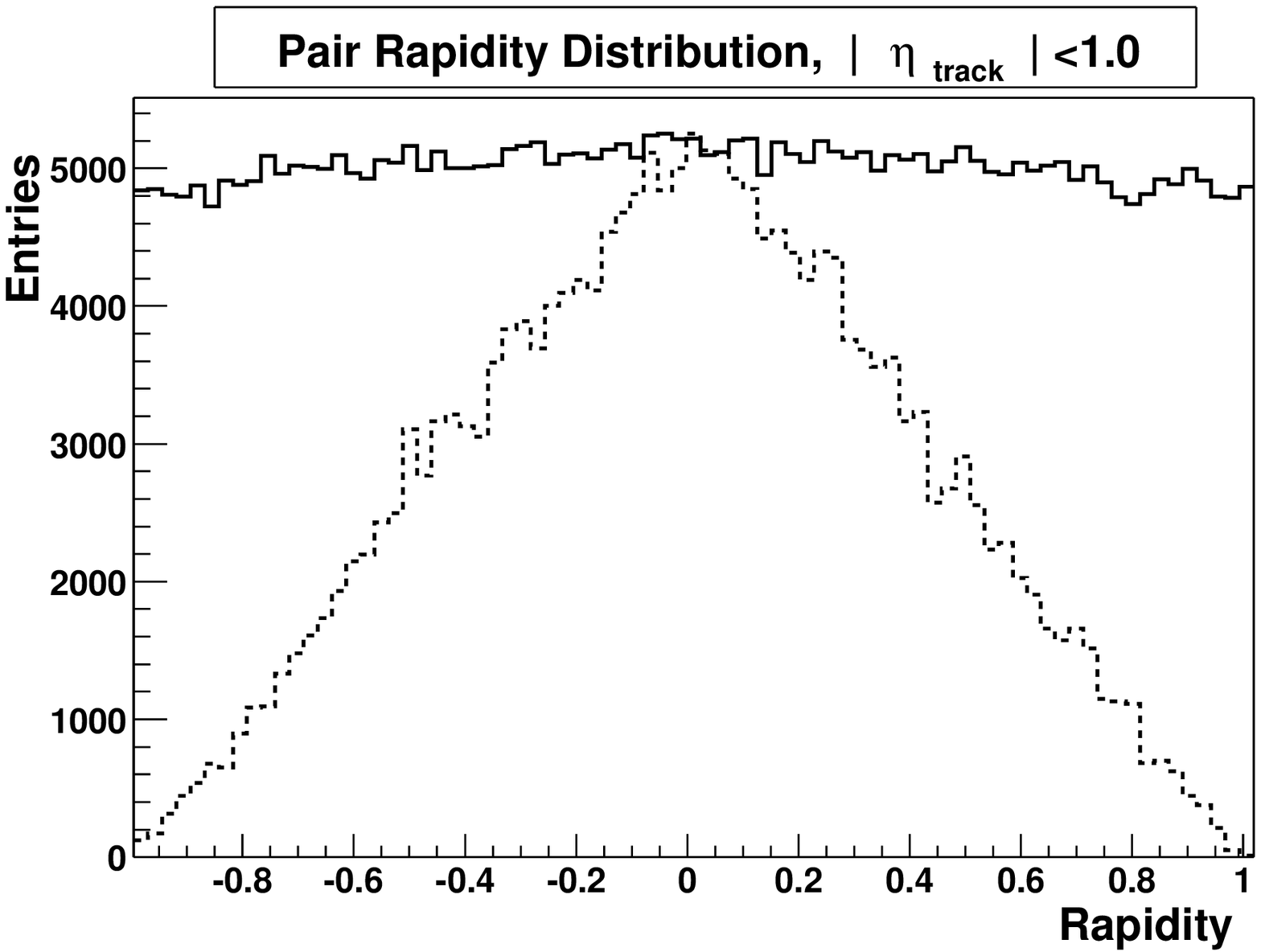}
\includegraphics[width=210pt]{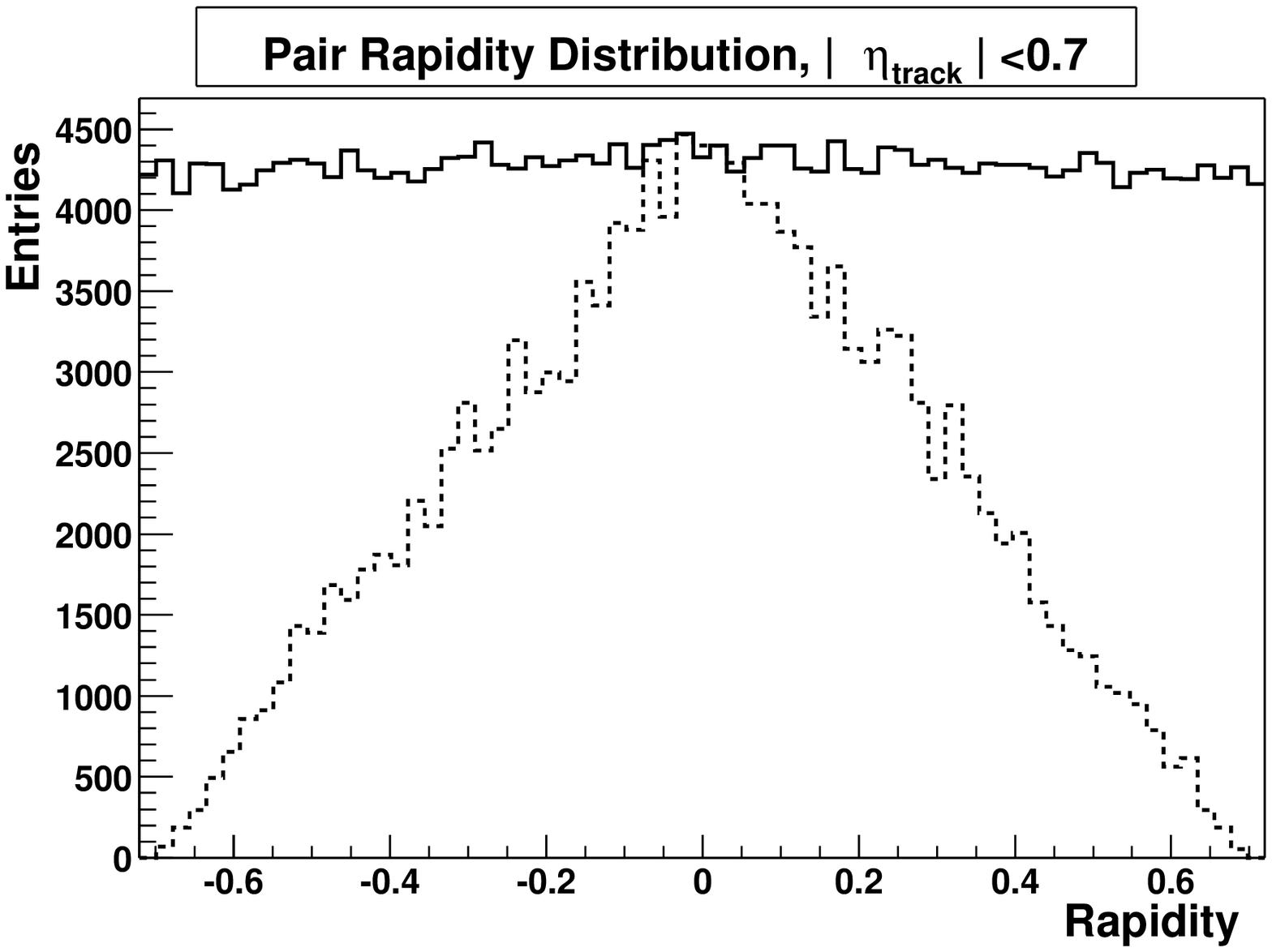}
\caption[Effect of the $\eta ^{\max}$ cut on the pair rapidity distribution]{Effect of the $\eta ^{\max}$ cut on the pair rapidity $Y$ distribution. Solid: generated $Y$ distribution (scaled), dashed: $Y$ distribution after $\eta ^{\max}$ cut}
\label{fig:YvsEta}
\end{figure}

Figure \ref{fig:YvsEta} shows a dependence of the total pair rapidity $Y$ distribution on the cut on the maximal absolute value of individual track pseudorapidity. The cuts $\left [ |\eta _{e^+}| < \eta ^{\max} \text{, } |\eta _{e^-}| < \eta ^{\max} \right ]$ effectively limit the total pair radity: $|Y|<\eta ^{\max}$. The suppression is strongest for the larger values of $|Y|$.

\vspace{10pt}

\newpage

\begin{figure}
\centering
\includegraphics[width=200pt]{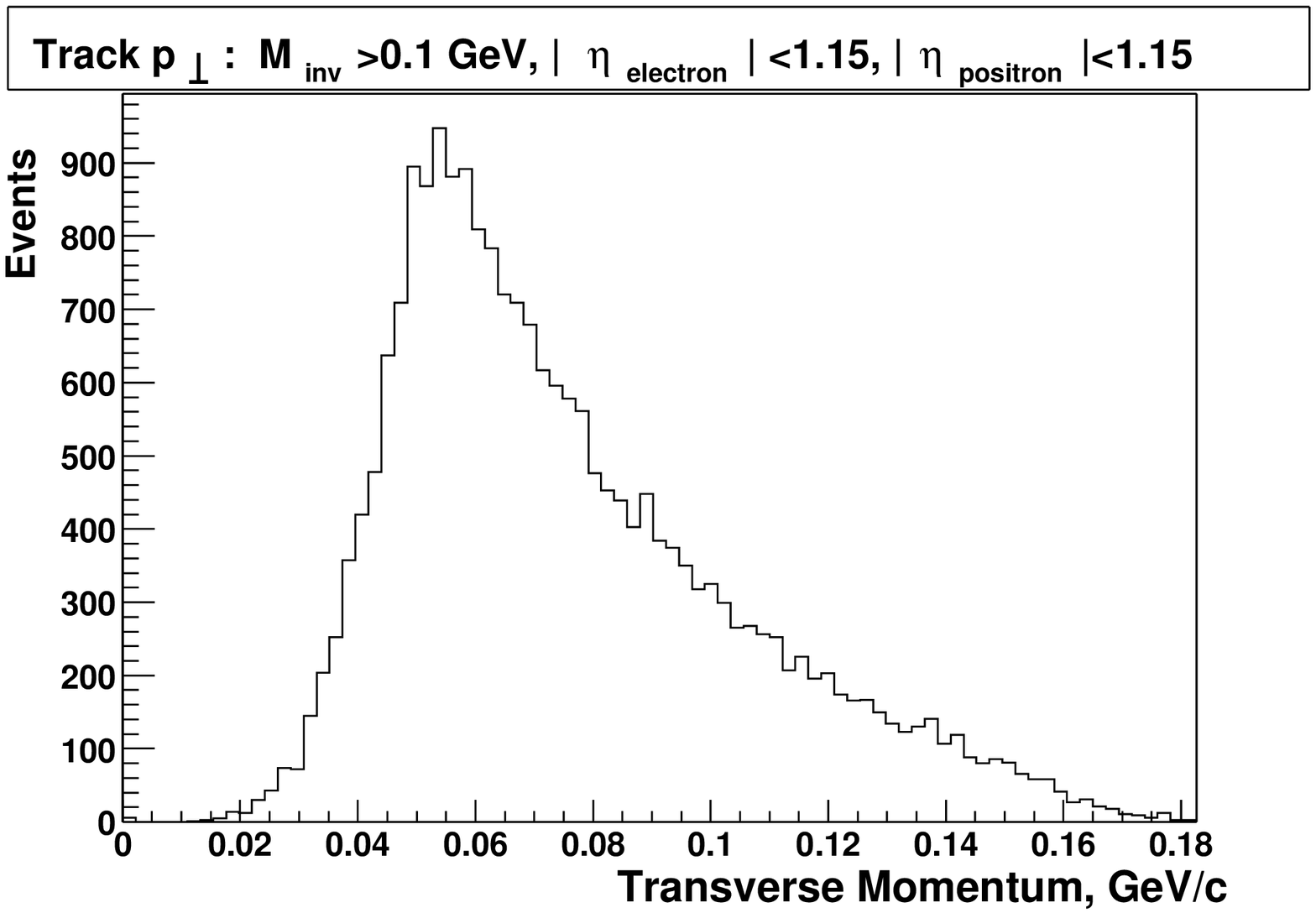}
\includegraphics[width=200pt]{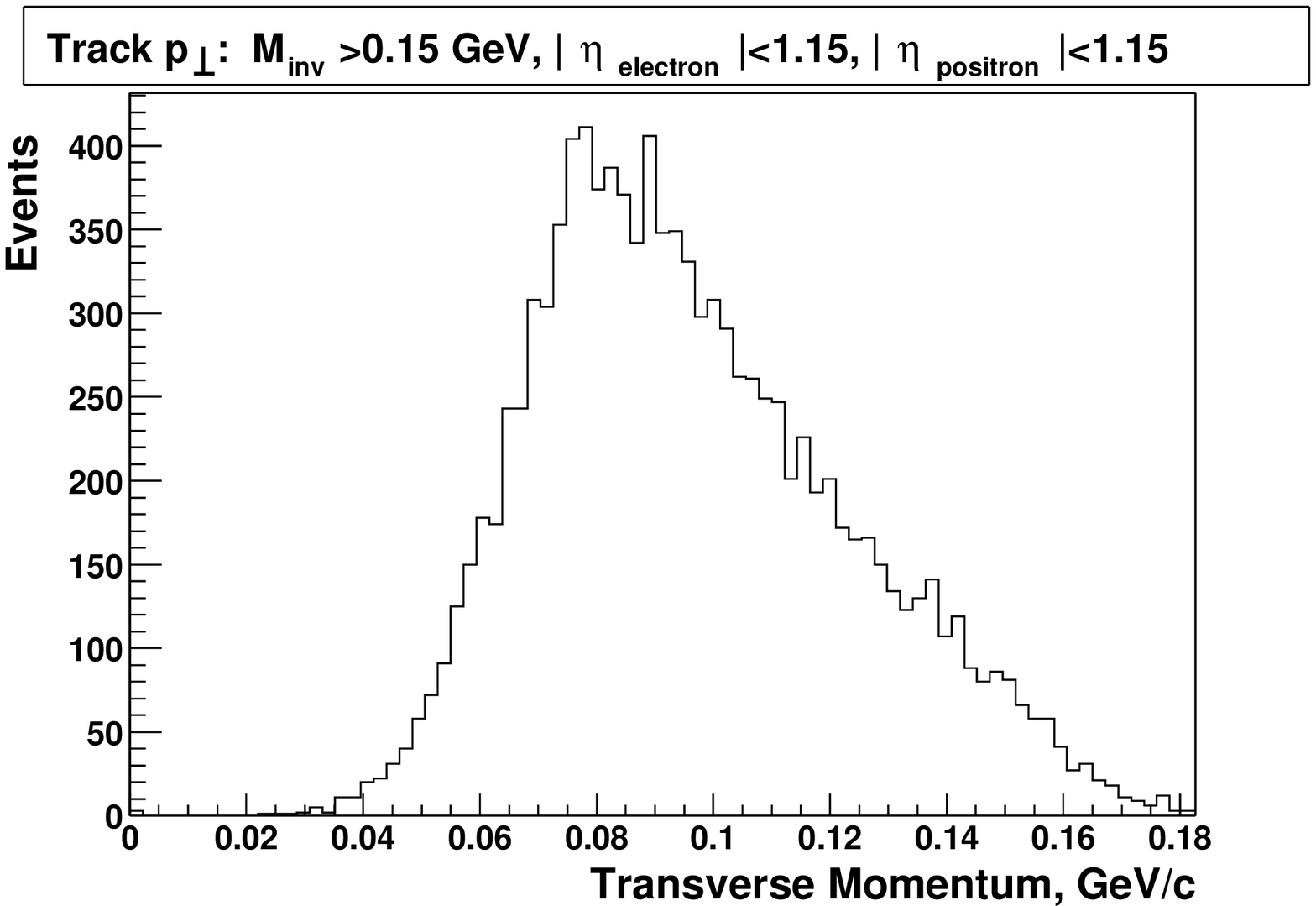}
\caption[Monte Carlo individual track $p_{\perp}^{track}$ distributions]{Single track transverse momenta for pairs with $M_{inv}>100$ MeV (left) and for pairs with $M_{inv}>150$ MeV (right). Track pseudorapidities are limited to $|\eta ^{track} |<1.15$.}
\label{fig:PtIndividual}
\end{figure}

It is also important to examine the individual track transverse momentum spectra for tracks at mid-rapidity ($|\eta_{e}|<1.15$). Figure \ref{fig:PtIndividual} shows the spectra for such tracks with a cut applied to the lowest invariant mass of the \ee   pair. We see that applying a cut on the invariant mass limits from below the individual track transverse momentum.

\vspace{-5pt}

\subsubsection{Approximating $\theta ^*$ by $\theta '$}
\begin{figure}[h]
\centering
\includegraphics[width=190pt]{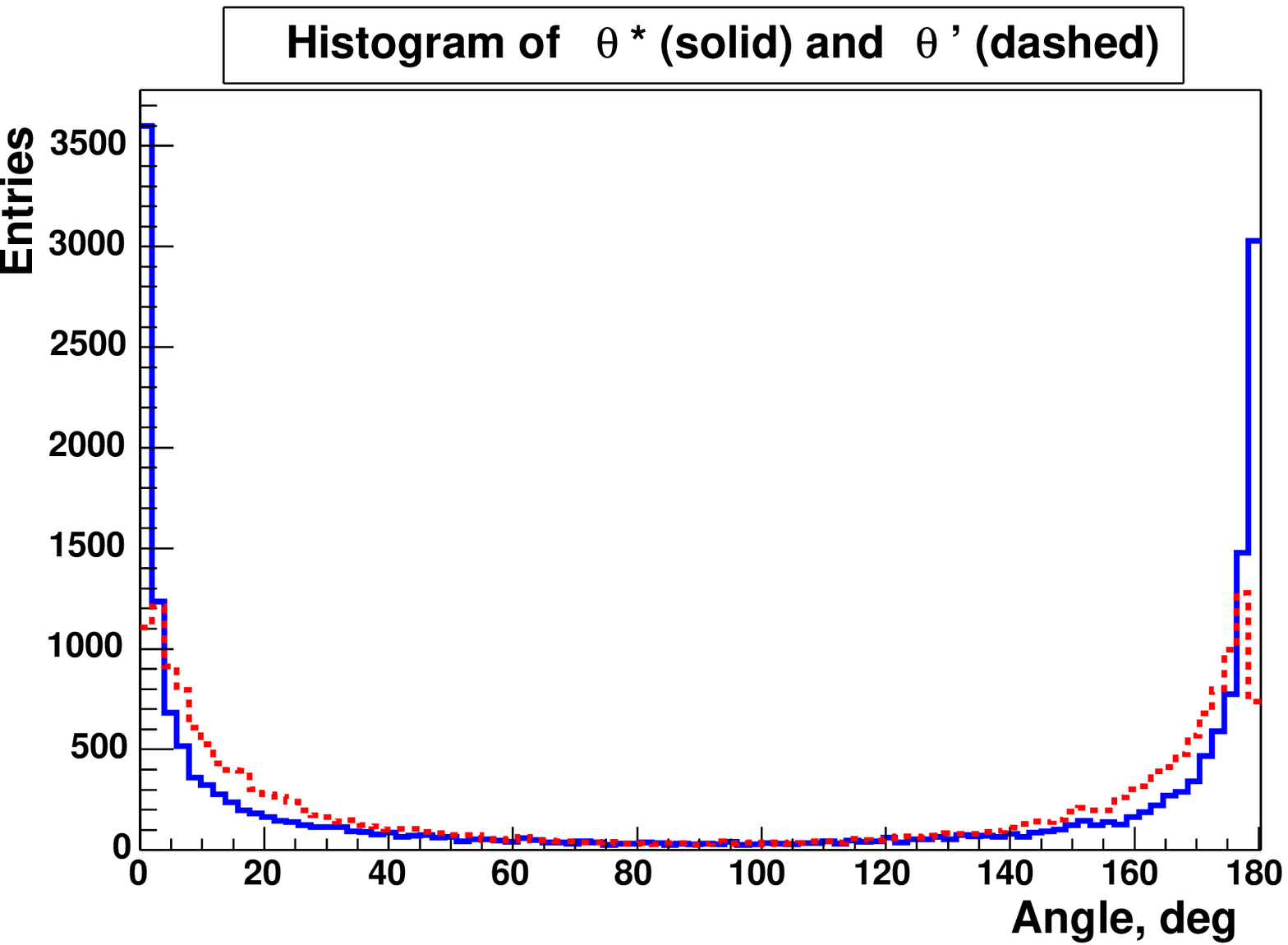}
\includegraphics[width=190pt]{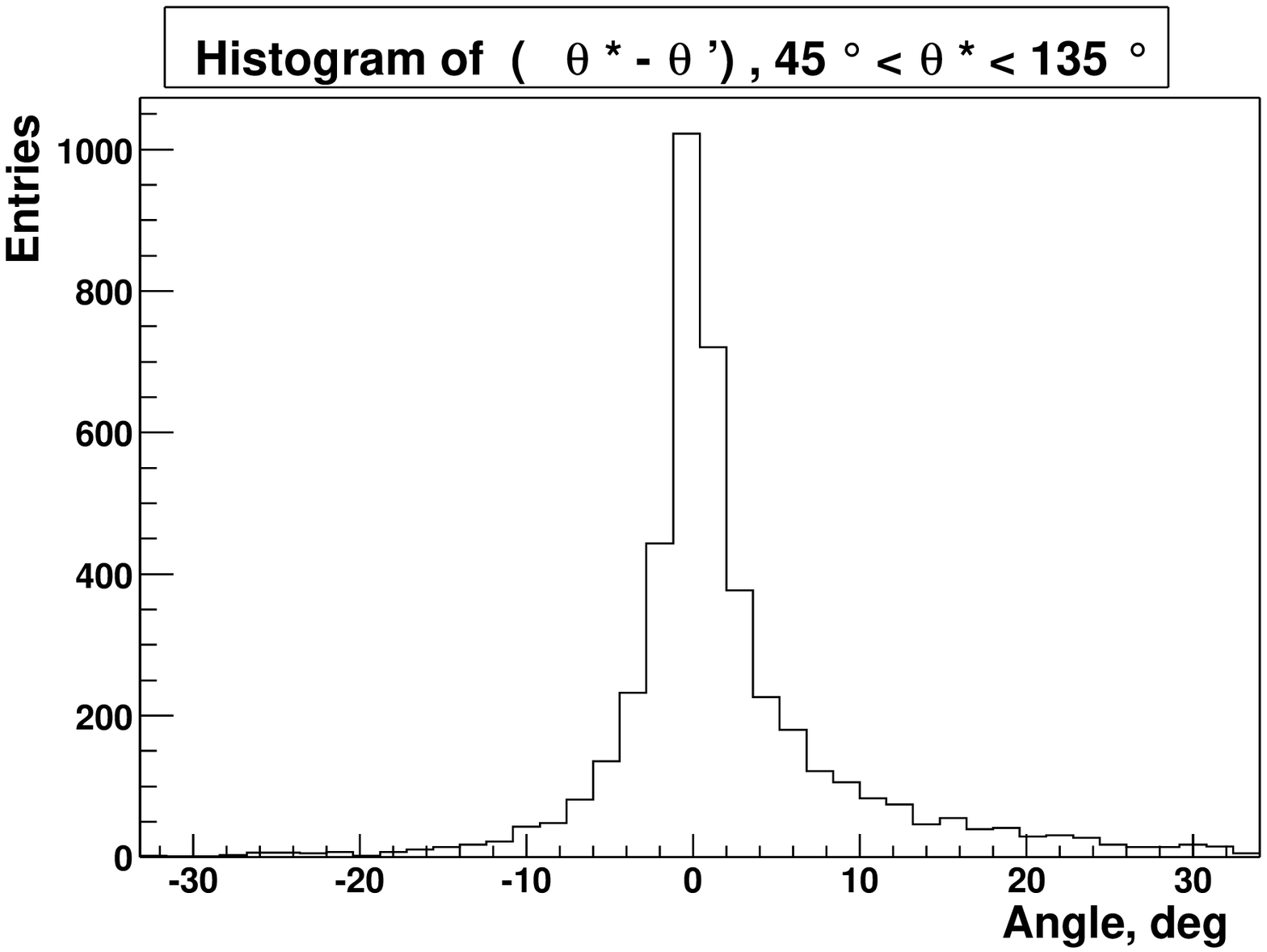}
\caption[Comparison of $\theta ^*$ and $\theta '$ distributions]{Comparing $\theta ^*$ vs. $\theta '$ for generated \ee pairs. Left: 1D histograms of $\theta ^*$ (solid) and $\theta '$ (dashed). Right: $\theta ^* - \theta '$ for $45 ^\circ < \theta ^* < 135 ^\circ$.}
\label{fig:ThetaStarVsThetaPrime}
\end{figure}

Since the event generator retains the information about the generated angle $\theta ^*$ for each event, we can compare these values to the values of angle $\theta '$ computed solely from the observed momenta of \ee tracks. Figure \ref{fig:ThetaStarVsThetaPrime} compares the two distributions. We see that the variable $\theta '$ approximates $\theta ^*$ very well in the range of $ 45 ^\circ < \theta ^* < 135 ^\circ$, and there is a significant difference between the two variables for $\theta ^* \approx 0 ^\circ$ and $\theta ^* \approx 180 ^\circ$. This is due to the fact that the photon polar angle ($\theta _{photon} \approx 1/\gamma _{Au}$) cannot be neglected in comparison with $\theta ^*$ if $\theta ^* \approx 0 ^\circ$  or $\theta ^* \approx 180 ^\circ$.

\vspace{-5pt}

\newpage

\subsubsection{Cross-section Predictions in the Limited Kinematical Range}
The Monte Carlo routine we described can provide a prediction of the \ee   pair production cross-section within a limited kinematical range. This is important since the STAR detector has a limited acceptance for individual charged tracks in $p_{\perp}^{track}$ and $\eta ^{track}$. 
Table \ref{tab:MCsigma} shows several Monte Carlo cross-section predictions for \ee   production at RHIC with nuclear excitation in a limited kinematical range.

\begin{table}[h]
\centering
\begin{tabular}{|l|c|}  
\hline \multicolumn{1}{|c|}{\bf Cuts}& \multicolumn{1}{|c|}{\bf $\sigma$ (mb) } \\ 
\hline $\ \ 10 \text{ MeV }\! <\! M_{inv}$ & $5\cdot 10^4$  \\ 
\hline $100 \text{ MeV }\! <\! M_{inv}$ & 195  \\ 
\hline $100 \text{ MeV }\! <\! M_{inv}$ and $|Y|\! <\! 3.0$ & 142  \\ 
\hline $140 \text{ MeV }\! <\! M_{inv}\! <\! 265$ MeV and $|Y|\! <\! 1.15$ & 46.40  \\
\hline $140 \text{ MeV }\! <\! M_{inv}\! <\! 265$ MeV, $|Y|\! <\! 1.15$ and $\cos(\theta ') \! < \! 0.7$& 4.95  \\
\hline $140 \text{ MeV }\! <\! M_{inv}\! <\! 265$ MeV, $|Y|\! <\! 1.15$ and $|\eta ^{track}|\! <\! 1.15$ & 2.58  \\
\hline $140 \text{ MeV }\! <\! M_{inv}\! <\! 265$ MeV, $|Y|\! <\! 1.15$ , $|\eta ^{track}|\! <\! 1.15$ and $p_{\perp} ^{track}\! >\! 65$ MeV/c & 2.08  \\
\hline
\end{tabular}
\caption[Electron-positron pair production with nuclear excitation cross-section at RHIC in STAR detector acceptance]{$e^{+}e^{-}$ production with mutual nuclear excitation cross-sections for various acceptances.}
\label{tab:MCsigma}
\end{table}

\chapter{Experimental Apparatus}
\label{ch:Experiment}
RHIC is a new accelerator facility in Brookhaven National Laboratory which started functioning in the summer of 2000. It can collide head on protons and a wide range of heavy nuclei at very high energies. 

There are four heavy-ion experiments at RHIC named BRAHMS, PHOBOS, PHENIX and STAR placed at four beam intersection regions at the collider. Each of these detectors is optimized for measuring different final states, but there are overlaps in their capabilities so that consistency checks can be made between them. STAR detector has the advantage of the complete $2\pi$ azimuthal angle coverage over the central rapidity region. Each of the RHIC intersection regions is also equipped with two identical Zero Degree Calorimeters for monitoring the collisions.

\section{RHIC System}
\label{sec:RHIC}
\begin{figure}
\centering %
\includegraphics[width=350pt]{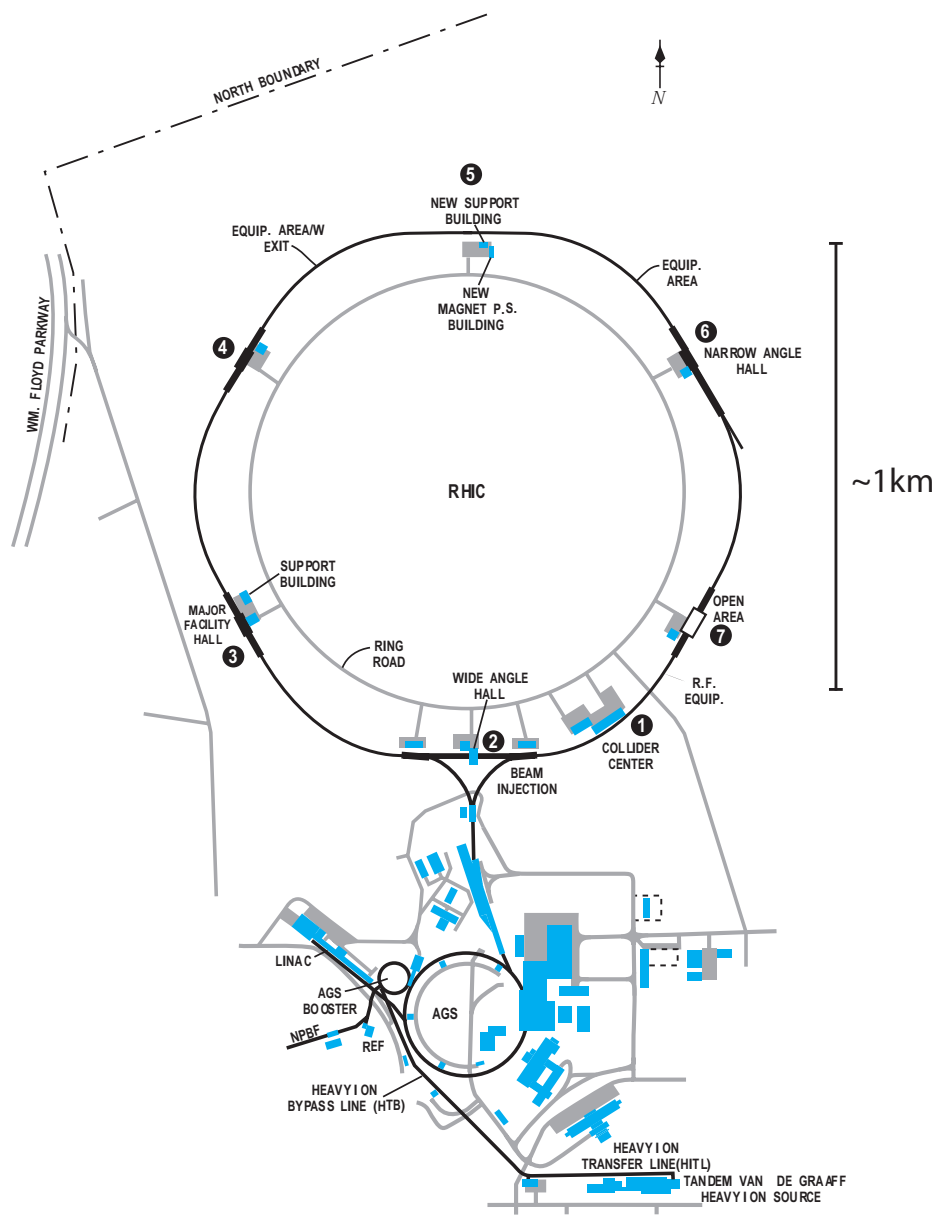}
\caption[RHIC system]{RHIC system. STAR is at the point labeled 2 \ \cite{RHICpaper}.}
\label{fig:RHIC}
\end{figure}

The RHIC accelerator complex reflects a long history of collider/accelerator development at Brookhaven National Laboratory. Figure \ref{fig:RHIC} shows the layout of the accelerator facility, starting from the production of a gold beam at the Tandem Van de Graaf to acceleration to full energy at the main RHIC ring.

At the beginning, partially ionized gold atoms are emitted from a source, such as a high temperature gold filament. The positively charged ions are accelerated through the Tandem Van de Graaff's two 15 million volt electrostatic accelerators, and are passed through thin sheets of gold foil, which further ionize the gold atoms. Ions exiting the tandem enter the Alternating Gradient Synchrotron (AGS), where they are accelerated in a 257-meter diameter radio-frequency synchrotron to a total energy of 11 GeV/nucleon. The AGS employs focusing technique with its 240 magnets situated along the acceleration ring. Each magnet successively alternates its magnetic field gradient inward and outward from the ring, allowing the beam to be focussed in both horizontally and vertically. The final complete ionization of the gold ions also happens at the AGS. From the AGS the ions are diverted to a transfer line to the main RHIC ring, where a switching magnet injects ions into a counterclockwise and clockwise rings. 

The main accelerating rings at RHIC known as a yellow (clockwise) and blue (counterclockwise) rings, are each about $\sim$ 610 m in radius\cite{RHICpaper}. The rings are filled by the ions in a boxcar fashion, resulting in 57 distinct bunches of ions (each containing $\sim 10^9$ $^{197}Au^{+79}$ ions) in each ring. The acceleration of bunches in these synchrotron rings is achieved by radio-frequency cavities and bending the beam into a circular shape by super-conducting magnets, positioned around the ring. Using the super-conducting technology RHIC is able to collide the heavy ions at the center-of-mass energies higher than any other machine in world (table \ref{tab:compareaccelerators}). The top center-of-mass energy achievable at RHIC for heavy ions is $\sqrt{s_{NN}} \sim$ 200 GeV per nucleon in the lab frame. 

Once the desired collision energy is reached the beams are synchronized to cross in six interaction regions. The design frequency at which RHIC bunches crossed, which depends on the energy of the beams, was 9.37 MHz in 2001. In the interaction regions, the beams are focussed and steered by quadrupole magnets for collisions at approximately 180 degree angle (head-on collisions). The design length of the region of space where collisions take place ('interaction diamond') is about 20 cm. In the 2001 run the interaction diamond had a (almost) Gaussian profile, with a sigma of $\sim$ 60cm. At top energy, 200 GeV, in 2001 RHIC has achieved AuAu luminosities of $\sim 10^{26} {\rm cm}^{-2}{\rm s}^{-1}$ \cite{Zhangbu}. 

The length of time that RHIC can continuously provide collisions at a single time (called 'fill') is limited by a few factors. First is the beam loss. This is mostly attributed to the collisions of the beam ions with the residual atoms inside the vacuum beampipe, called the beam-gas collisions. The next most significant source of beam loss was the capture by the positive Au ions of an electron from an $e^{+}e^{-}$ pair, discussed in Section \ref{sec:otherprocesses}. Another process which contributed to the beam loss, though to the lesser extent, was single photonuclear excitation of the beam ions followed by nuclear breakup\cite{ZDCpaper}. The synchrotron radiation was not a significant source of beam loss, due to the very high masses of the accelerated ions. The second reason for interrupting the fill is the slow dispersion of the bunches in the longitudinal to the beam direction. This causes the interaction diamond to become very long, and a large fraction of collisions unusable. The typical fill length was four to six hours in the 2001 run. 

\begin{table}
\centering
\begin{tabular}{lllcrrr@{.}l}
\hline
\multicolumn{1}{c}{Year}
&\multicolumn{1}{c}{Facility} &\multicolumn{1}{c}{Location}
&\multicolumn{1}{c}{Species}
&\multicolumn{1}{c}{E$_{beam}$}  &\multicolumn{1}{c}{\ecm} \\
\hline
1974~--~1991    & BEVALAC & LBNL & \auau &   2\gev &   2\gev \\ 
1994~--~present &     AGS & BNL  & \auau &  11\gev &   5\gev \\ 
1994~--~present &     SPS & CERN & \pbpb & 158\gev &  17\gev \\ 
2000~--~present &    RHIC & BNL  & \auau & 100\gev & 200\gev \\ 
\hline
\end{tabular}
\caption[Comparisons of heavy-ion facilities]
{Some beam species and collision energies for various heavy-ion accelerators. The facilities are given by a facility acronym and location. Beam energies correspond to the top per nucleon beam energy achievable by the facility. The {\ecm} column shows the per nucleon energy of the system in the center-of-mass frame \ \cite{IanThesis}.}
\label{tab:compareaccelerators}
\end{table}

\subsubsection{Relativistic Nuclear Physics Program at RHIC}
The central focus of the RHIC physics program is study of nuclear matter at high temperatures and densities. In the central collision of the highly relativistic nuclei enormous energy densities ($ >1{\text{GeV/fm}}^{3}$) are created. These conditions may create a system of the theoretically proposed state of deconfined quarks and gluons - a Quark Gluon Plasma \cite{QGP}.\footnote{Our Universe may have been in the state of the Quark Gluon Plasma a few microseconds after the Big Bang.} This system expands rapidly and the temperature and density will drop below critical values causing the formation of the gas of interacting hadrons. As the hadronic system expands and dilutes, inelastic interactions between hadrons become more and more scarce. At the point these inelastic interactions stop, the system reaches a so-called chemical freeze-out\cite{freezout}. Eventually, the system expands so much that even the elastic interactions between the hadrons cease, and the system is said to have reached a thermal freeze-out. RHIC was specifically designed with the goal to observe the novel state of matter QGP and to study the exact details of freeze-out of this state into a normal hadronic matter. 

Another fundamental question addressed by RHIC is a spin structure of nucleon. The total spin-1/2 of a proton is the  sum of the spin contributions of the quark constituents of the hadron, their angular moment and the gluons. In the present understanding of the nuclear spin, quarks contribute only $1/3$ of the total nuclear spin, and the gluon spin contribution is non-negligible\cite{GluonSpin}, but no measurement exists yet. RHIC is planning to collide polarized beams of protons to study the details of the gluon spin term.
 
\newpage

\subsubsection{Zero Degree Calorimeters}
\label{sub:zdc}

\begin{figure}
\centering %
\includegraphics[height=200pt,width=250pt,clip=true]{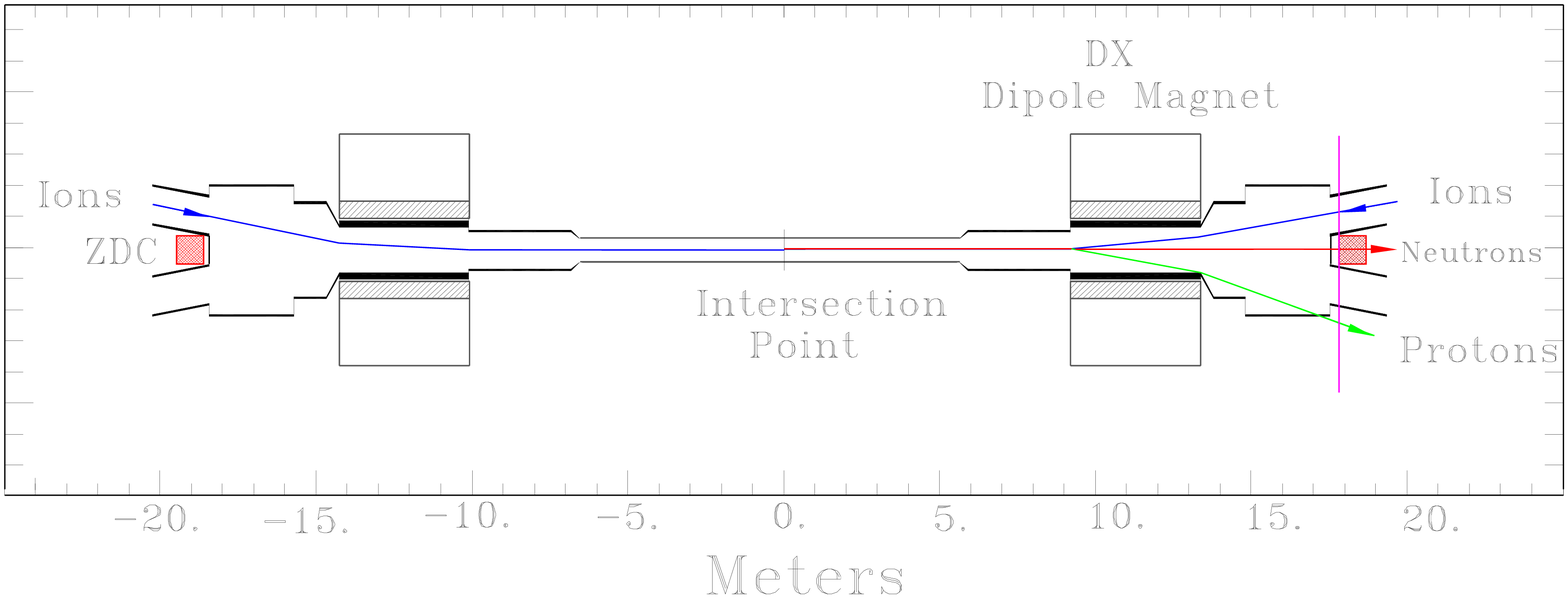}
\includegraphics[width=150pt,height=180pt,clip=true]{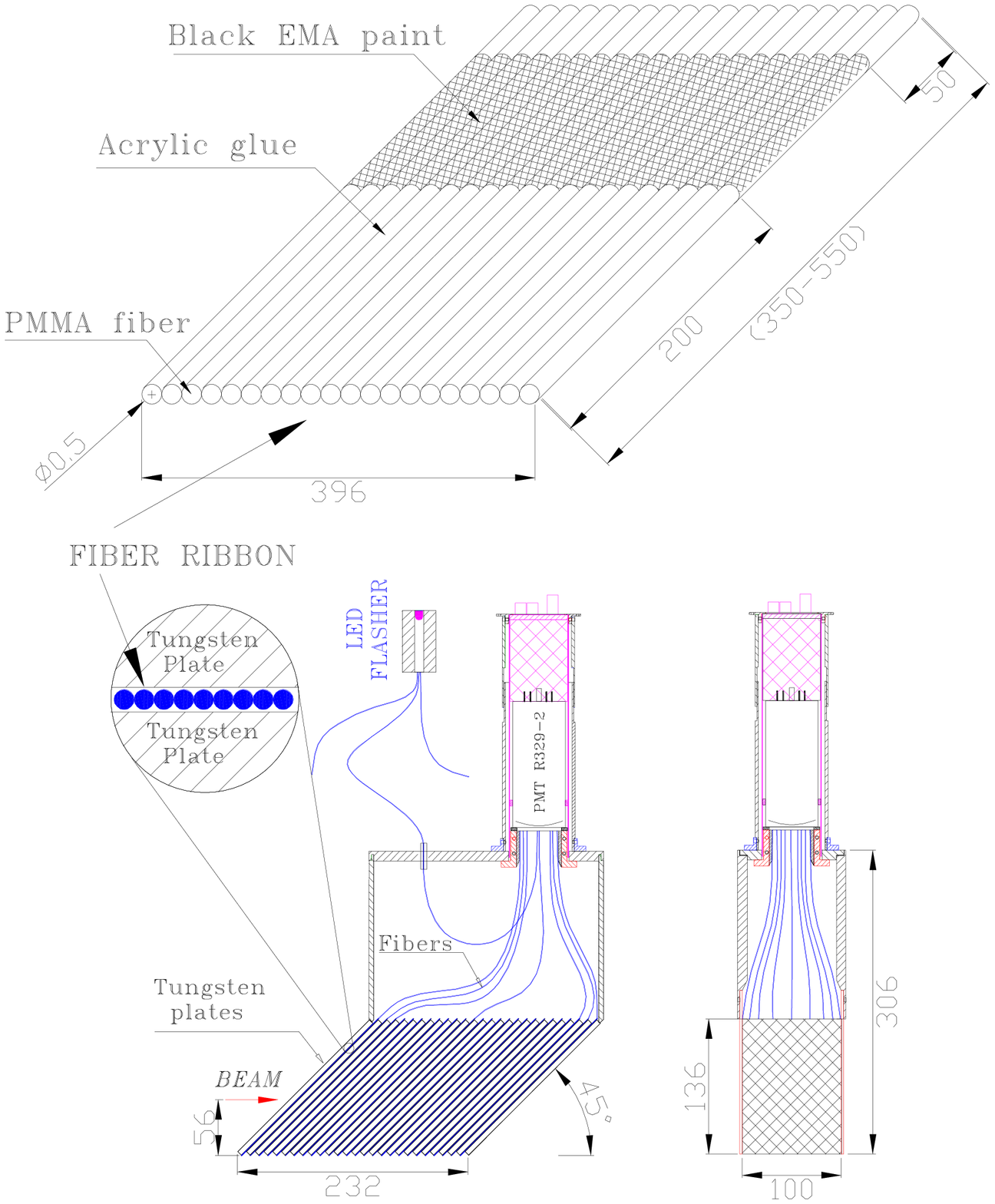}
\caption[Zero Degree Calorimeters]{Left: Location of the ZDCs around the interaction point. Right: hardware components of the ZDCs \ \cite{ZDCpaper}.}
\label{fig:ZDC}
\end{figure}

The main devices for monitoring the rate of RHIC collisions are Zero Degree Calorimeters. These detectors are located on either side of each of the six interaction regions at RHIC, 18 meters from the interaction regions along the beam pipe. In STAR, the two ZDCs are labeled 'East ZDC' and 'West ZDC'. The ZDCs are capable of detecting and measuring the energy of uncharged particles emitted from the interaction region in the direction of the beam\cite{ZDCpaper}.
The detectors consist of layers of Cherenkov fibers sandwiched between the tungsten absorber plates (10 $ \times$ 10 cm in size in the direction transverse to the beam, tilted at 45{\deg} angle with respect to the beam), as shown in Figure \ref{fig:ZDC}. When hadrons hit the detector hadronic showers are developed in tungsten absorbers, and generate a signal in the Cherenkov fibers.\footnote{This is a slight modification of the traditional sampling hadronic calorimeter design, which uses scintillators for hadronic shower sampling.} The signal from the fibers is then sent to the photomultiplying tubes (PMTs), with the summed analog output of the PMTs forming the ZDC signal (ZDC ADC signal).

The detectors have a $98\% \pm 2\%$ efficiency (flat over all detector area) for detecting uncharged particles within a 2 mrad cone around the beam direction\cite{ZDCpaper}. This acceptance region is sufficient to detect all of the spectator neutrons in AuAu collisions. Such neutrons are typically moving at the speed of the beam in the direction of the beam, and their transverse momentum $p_ \bot$ is mostly due to the Fermi motion ($\sim 5 $ MeV/c)\cite{nuclearExcitation}.\footnote{For Mutual Coulomb Dissociation neutrons $p_ \bot$ distribution is even more narrow than Fermi distribution\cite{ZDCpaper}.} Thus the angle between the direction of the beam and the momentum of the neutrons is of the order of 1.4 mrad, which is within the acceptance. 

The ZDCs are read out every RHIC bunch crossing and provide information whether the crossing resulted in a collision or not. The detectors also provide the timing information, measuring the arrival time of the neutral products of the heavy-ion collision to the ZDC on either side of the interaction point. In effect, the arrival time difference in two ZDCs provides a measurement of the primary interaction location in the direction of the beam. The resolution of the method can reach $\sigma = 3.2$ cm\cite{ZDCpaper}.

\section{STAR Detector}
\begin{figure}
\centering %
\includegraphics[width=400pt]{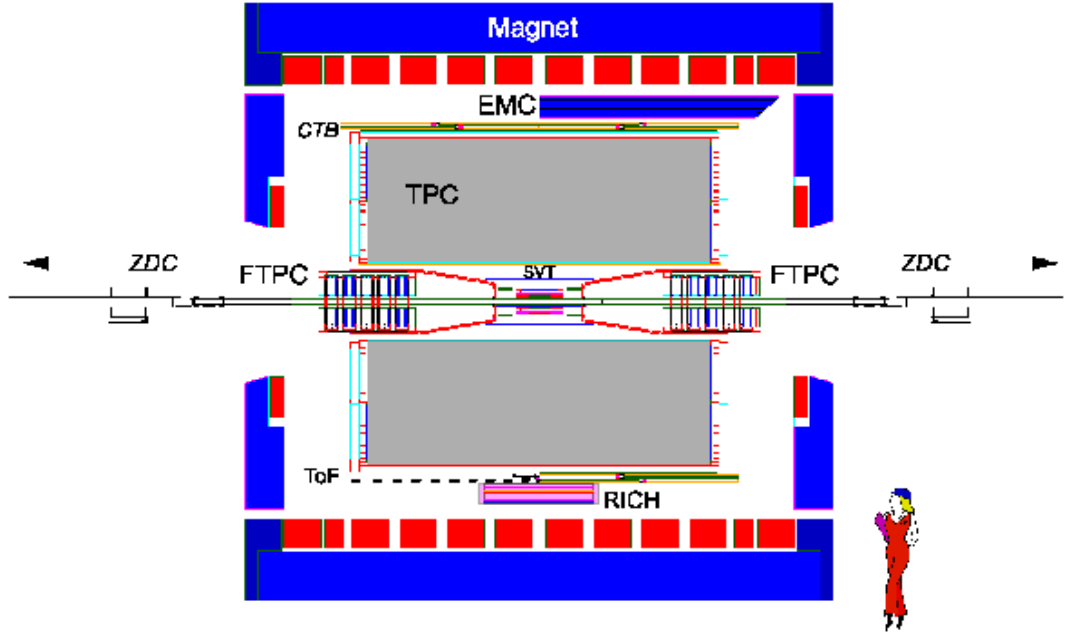}
\caption[Schematic view of STAR detector]{Schematic view of STAR detector \ \cite{STARoverview}.}
\label{fig:star}
\end{figure}

The Solenoidal Tracker at RHIC (STAR) shown in Figure \ref{fig:star} is a detector designed primarily for measurement of hadron production over a large solid angle. STAR features a detector system for high precision tracking, momentum analysis, and particle identification at the mid-rapidity. The large acceptance of STAR makes it particularly well suited for event-by-event characterization of heavy-ion collisions and for reconstruction of particle decays, such as $K_S^0\rightarrow \pi \pi $ and $\Lambda \rightarrow p\pi $\cite{STARoverview}. 

The detectors in STAR can be roughly split into two groups - fast detectors, which can read out data at the frequencies close to the frequency of RHIC bunch crossings, and slow detectors, which operate significantly under the RHIC frequency, but can provide a much more detailed information. A discussion of each of the detectors used in the 2001 run follows.

\subsection{STAR Magnet}
\label{sub:magnet}
The main STAR detector magnet is a 1100 ton solenoidal structure that consists of a 130-turn aluminum solenoid conductor. The magnet structure encloses most of the other STAR detectors, as shown in Figure \ref{fig:star}. It can provide 0.5 T of magnetic field when energized to a current of 4500 A, and 0.25 T with the current of 2250 A\cite{Magnet}. The field flux is returned through the poletips at either end of the solenoid, and then through a set of flux return bars, which are situated on the cylindrical surface of the solenoid. The purpose of the flux return is to keep the field uniform within the volume of the solenoid. 

The STAR magnet serves two purposes for the detection of charged products of the heavy-ion collisions at RHIC. First, it allows the momenta and sign of charged particles to be calculated by measuring the curvature of the particle track as it passes through the field volume. Second, the B-field produced by the STAR magnet is oriented in the beam direction, as is the drift field used in the TPC. As we explain in the following section, this reduces the dispersion in ionization produced in the TPC, increasing the position resolution of the TPC.

\subsection{STAR Time Projection Chamber}
\label{sec:TPC}

STAR Time Projection Chamber (TPC) is a main tracking detector in STAR. It is a large gas filled chamber capable of measuring three-dimensional space-points along charged particle trajectories. In comparison to silicon detectors, the TPC has coarser position resolution, but can make multiple measurements over a large volume. TPC is an intrinsically slow detector, since its readout speed is always limited by the drift speed of the ionized gas atoms ( $\sim40\mu s$)\cite{TPC}.

\begin{figure}
\centering %
\includegraphics[width=330pt]{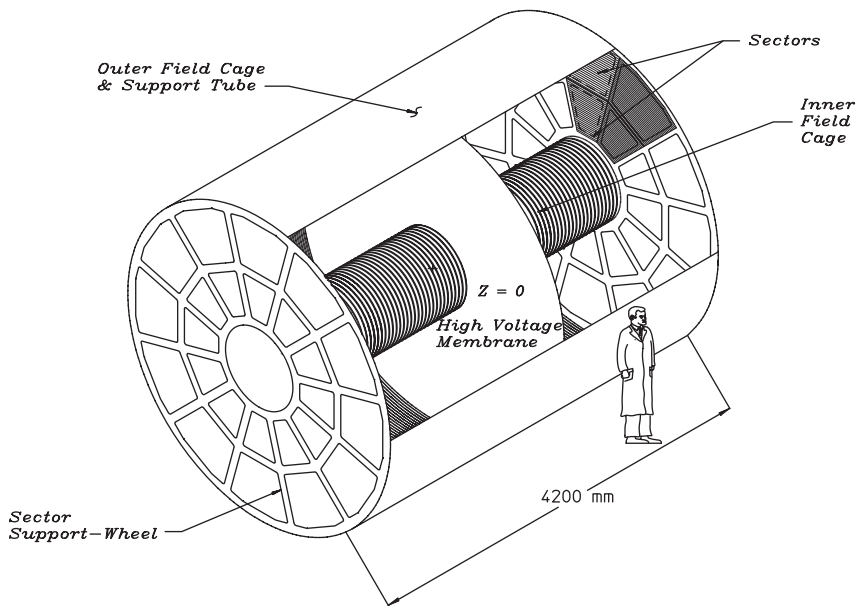}
\caption[STAR Time Projection Chamber]{Cylindrical geometry of the STAR TPC's gas volume and the sector layout on the endcaps. The boundaries of the TPC gas volume are the Inner Field Cage (IFC), Outer Field Cage (OFC) and the TPC endcaps \ \cite{TPC}.}
\label{fig:tpc}
\end{figure}

\subsubsection{TPC Main Gas Volume}

The STAR TPC gas chamber has a cylindrical geometry which extends 4.2{\m} in length and 2{\m} in radius, as shown in Figure \ref{fig:tpc}. The ionization region or active volume of the TPC is more than 45\m$^3$. This volume is kept slightly above atmospheric pressure (2 mbar) and filled with {\armGas} gas (P10). Signals originate from electrons that are freed when moving charged particles ionize the gas. The positive ions and free electrons move apart under the influence of a 147{\Vcm} electric field between the central membrane and end caps of the TPC. The positive ions are carried to the cathode at the central membrane and the electron clouds drift towards the ends of the detector. Positive ions are neutralized when they reach the cathode plane and the electron clouds are amplified in a Multi-wire Proportional Chambers (MWPC) close to the end caps. Since the drift velocity of the electrons is known, one coordinate ($z$) of the starting point can be deduced from the time taken for the electrons to drift to the MWPC. The other two coordinates are found through the projection of the signal onto a pad plane mounted below the MWPC. The pad plane lies perpendicular to the beam axis and is segmented into 136,608 pads. The electronics are capable of recording 512 time bins from each pad, of these, about 348 are read out between the central membrane and the MWPC. In total, the volume is effectively divided up into more than 47 million space-points.

\begin{figure}
\centering
\includegraphics[height=130pt]{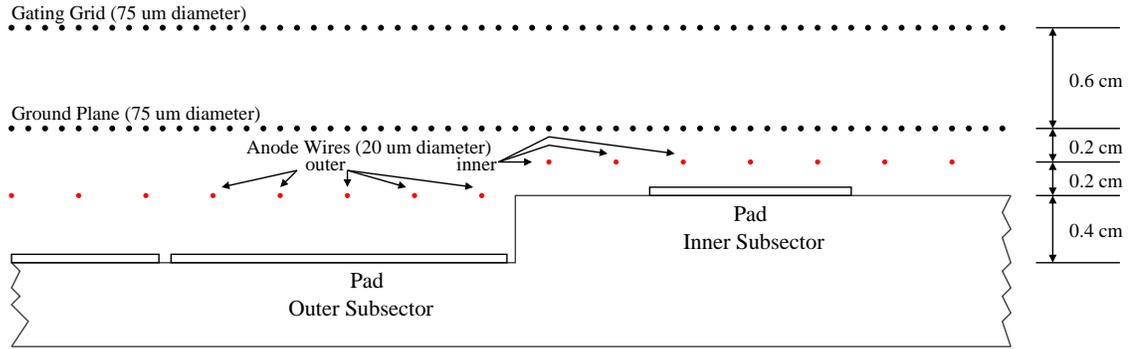}
\caption[Multi-wire Proportional Chambers of the STAR TPC]
{The three wire planes of the Multi-wire Proportional Chambers of the STAR TPC. The outer and inner sector geometries are shown on the left and right, respectively \cite{TPC}.}
\label{fig:mwpc}
\end{figure}

The Multi-wire Chambers shown in Figure \ref{fig:mwpc} have three planes of wires. The anode wires are closest to the pad plane and are 20{\um} in width. The inner sector anode wires are set to 1170{\V} and the outer to 1390{\V}. The combination of the fine width and high voltage on the anode wires produce a strong radial field near the surface of the wires. Drift electrons create ionization avalanches as they accelerate towards the positive anode wires. Positive ions created in these avalanches produce image charges on the pad plane.

The ground plane grid is the middle plane of wires that separates the drift volume from the amplification region. The ground plane grid has three main functions: to provide a ground plane for the drift field, to shield the pad plane from the gated grid and to capture some of the positive ions created near the anode wires. The drift field is established between the -31{\kV} central membrane and ground grid plane. The ground plane grid significantly reduces the signals induced on the pads when the gated grid opens. This prevents these induced signals from compromising the resolution of ionization signals at the beginning of the drift period. A large fraction of the slowly drifting positive ions created near the anode wires are neutralized on the wires of the ground plane grid. Positive ions that drift into the active volume, leakage current, cause distortions in the drift field.

The gated grid is furthest from the pad plane. The main purpose of the gated grid is to stop non-triggered ionization from reaching the amplification region and stop positive ions created in the amplification region that leak past the ground plane grid from reaching the active volume. 
In the closed state, adjacent gated grid wires alternate from positive to negative potentials. These potential differences set up electric fields between the wires that are perpendicular to the drift direction. The fields capture the non-triggered electrons and positive ions. In the opened state, the voltage on the gated grid wires is set to the corresponding equipotential surface of the drift field. In this state the gated grid is transparent to the drift electrons.

\begin{figure}
\centering
\resizebox*{0.75\textwidth}{!}{\includegraphics{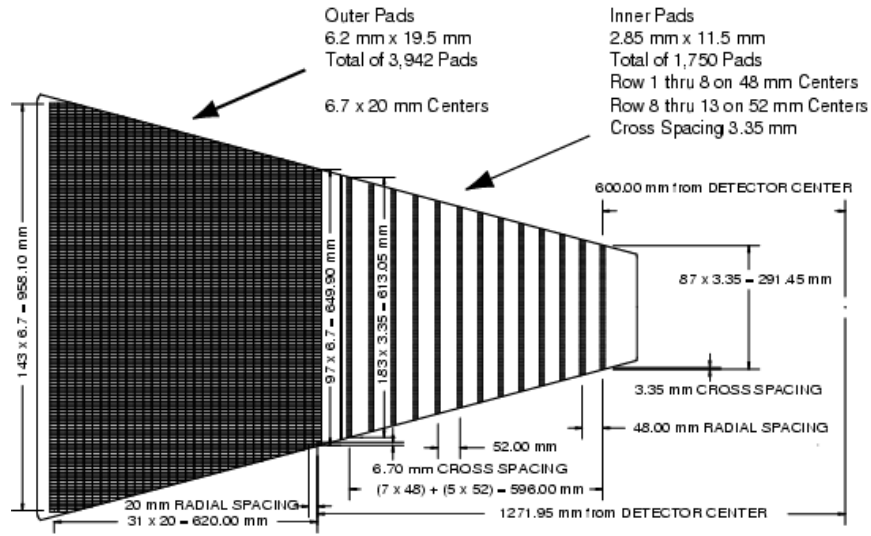}}
\caption[Pad plane of one TPC sector]
{Pad plane of one TPC sector. Each sector contains 5692 pads. The inner and outer pad geometries differ to compensate for the radially decreasing hit density \ \cite{TPC}.}
\label{fig:STARPadPlane}
\end{figure}

The TPC pads are laid out in sectors that cover 30{\deg} in azimuth, as shown in \ref{fig:tpc}. There are 24 identical sectors mounted on the east and west ends of the TPC. Each sector has 13 inner and 32 outer pad rows, as shown in Figure \ref{fig:STARPadPlane}. Effectively, the pads are plate capacitors. Local electric field changes are created on the surface of the pads by the slowly drifting positive ions created in avalanches near the anode wires. These local field changes induce currents on the pads and subsequently in the TPC electronics.

\subsubsection{TPC Electronics and Data Acquisition}
The TPC readout electronics boards are mounted on the back of each sector. Each sector has 181 analog Front End Electronics boards (FEE) and six digital readout (RDO) boards\cite{FEEs}. The circuitry on each FEE is separated into two parallel 16 channel circuits and is capable of covering up to 32 pads. The analog signals on the TPC pads are amplified, shaped, stored and digitized in two chips on the FEE. The Pre-Amp/Shaper-Amp (SAS) amplifies and shapes the signal. The SAS feeds the 512 slot switched capacitor array (SCA). This chip is an analog storage unit that also contains an analog to digital converter. The chip allows fast, low-noise sampling of the signal with minimal power consumption. It also permits digitization and readout of data at a reduced rate. Upon request the SCA chip digitizes the stored voltages on the capacitors and passes them onto a multiplexer on the RDO board. The multi-plexer communicates via fiber optic links with the data acquisition (DAQ) crates\cite{DAQ}. DAQ Receiver Boards (RB) receive the data from RDO boards and perform pedestal subtraction, zero-suppression and charge cluster finding along the TPC time dimension. The data is then sent into the DAQ Event Builder (EVB), which is responsible for gathering the event fragments from all of the sectors of all detectors, packing the data into DAQ files, buffering the data on disk and then sending the data to the High Performance Storage System (HPSS). 

\subsection{Additional Detectors}
A number of other detectors were present in STAR in the 2001 run, but not used in this analysis. Here is a brief description of them.

Forward Time Projection Chambers (FTPC)\cite{FTPC}, located on both sides of the TPC, provide track momentum reconstruction in the forward region (pseudorapidity interval of $2.5 \! < \! \mid \eta \mid \! < \! 4.0$). These detectors are especially useful for asymmetric collision studies, such as p+A collisions, and any other events where tracking in the forward region is desired. The 2001 run was a commissioning run for these detectors.

The Central Trigger Barrel (CTB)\cite{CTB} is a cylindrical detector positioned just outside of the outer diameter of the TPC, as shown in Figure \ref{fig:star}. The slats of the barrel are connected to photomultiplier tubes which give a response proportional to the number of charged particles which interact with the scintillator medium inside the slats. Thus CTB is a tool for measuring the charged multiplicity in the central rapidity region of $\mid \eta \mid < 1.0 $. The CTB can be read out every RHIC bunch crossing (every 104 ns). 

The Barrel Electromagnetic Calorimeter (EMC)\cite{EMC} is also located just outside of the outer field cage of the TPC. The detector is a barrel made up of 120 modules of a lead scintillator sampling calorimeter. This detector is intended to study rare, high-$p_{\perp}$ processes and photons, electrons, $\pi^0$ and $\eta$ mesons in the same rapidity region as the CTB. The detector can be read out on the nanosecond scale. In the 2001 run the EMC was commissioning the first set of modules.

The Multi-wire Proportional Chambers (MWPC)\cite{MySTARNote} is a detector which is a part of the STAR TPC, that can also be used as an independent charged multiplicity counter in the region of $1.0 < \mid \eta \mid < 2.0$. This detector is located on the endcaps of the STAR TPC, and can provide a pixelized count of the charged particles traversing its active volume for each RHIC strobe. The MWPC was expected to be of great use for studying ultra-peripheral events with forward-peaked tracks, and a STAR Note 434\cite{MySTARNote} describes a simulations package to estimate its efficiency for detecting such tracks. Due to noise, the MWPC could not be used as an independent detector in 2001. 

The Silicon Vertex Tracker (SVT)\cite{SVT} is a detector positioned right around the interaction region, outside of the beam pipe and inside the Inner Field Cage of the TPC. Its main goal is to provide additional tracking in the area directly adjacent to the interaction region, thus greatly increasing STAR tracking resolution and vertex finding accuracy.

The Ring Imaging Cherenkov Detector (RICH)\cite{RICH} was placed just outside the CTB. The RICH covers a small area, 1 m $\times $ 1 m and is designed to provide high precision velocity measurements for the high momentum particles that pass through it. This enhances the particle identification capabilities of STAR for high momentum tracks.

\subsection{Material Table in STAR}
\label{sub:MaterialTable}
The amount of material in the path of a charge particle can be expressed in terms of a radiation length\cite{pdg}.\footnote{Radiation length is a mean distance over which a high-energy electron loses all but $1/e$ fraction of its energy.} Table \ref{tab:material} summarizes the thickness of material (expressed in radiation lengths) traversed by the track between the interaction region and the active volume of the main tracking detector TPC. 

\footnotetext[5]{For events with vertex position within $|z_{vert}|<75$ cm.}
\footnotetext[6]{For a particle traversing a straight line between 50cm and 100cm in radial direction.}
\footnotetext[7]{Does not affect the scattering of detected particles inside the TPC.}
\begin{table}
\centering
\begin{tabular}{|l|c|}  
\hline {\bf Structure } & {\bf Radiation Length $(\%)$ } \\
\hline RHIC Beampipe\footnotemark[5]   & 0.28 \\
\hline SVT &  6.00 \\ 
\hline IFC & 0.62 \\
\hline TPC gas\footnotemark[6] & 0.39\\
\hline OFC\footnotemark[7] & 1.26 \\
\hline {\bf Total without OFC} & {\bf 7.29 } \\
\hline
\end{tabular}
\caption[Material traversed by a particle in STAR]{Material traversed by a particle in STAR. Table compiled from \cite{TPC} and \cite{SVT}.}
\label{tab:material}
\end{table}

The traversed radiation length defines the amount of secondary scattering suffered by a particle. The total angle of the secondary scattering over the length of the track can be approximated as a Gaussian random variable with zero mean and the width:

\begin{equation}
\label{eqn:secondaryScattering}
\theta_{0} = \frac{{13.6\text{ MeV/c }}}{{\beta p}}Z\sqrt {X_0 } \left[ {1 + 0.038\ln \left( {X_0 } \right)} \right]
\end{equation}
where $p$ is a particle momentum, $\beta$ - velocity, $Z$ is the charge of the particle and $X_0$ is the traversed radiation length. Deviations from the particle's trajectory as it travels through the material in STAR can be calculated using Equation (\ref{eqn:secondaryScattering}) and for a 100 MeV/c particle will be on the order of $\theta _0 \sim 0.37$ rad. This will cause mis-reconstruction of the scattered track momentum and limit the precision with which the track momenta can be reconstructed by STAR.

\section{Triggering in STAR}
\label{sec:trigger}
As we mentioned before, STAR can read out and store on tape the data at the rate of ~100 Hz, where as the RHIC bunch crossing frequency (maximal possible collision rate) was 9.37 MHz. This is a frequency at which the RHIC RF cavities operate, and this frequency is supplied to all experiments at RHIC, called a 'RHIC strobe'. We need a set of detectors which are capable of reading out information at the frequency of the RHIC strobe and making a rough decision if a particular bunch crossing might contain an interesting collision. Such detectors are the CTB, the EMC, the MWPC and the ZDCs. In the 2001 run only the CTB and the ZDCs were used for triggering.

\subsection{STAR Trigger System Design}

The on-line analysis of trigger detector information (or 'triggering') in STAR is done in several steps, called Levels 0 through 3 of the trigger. Each increasing level of trigger has access to a more detailed information from the trigger detectors and correspondingly takes more time. The whole trigger system is synchronized with the RHIC strobe, so that the trigger detectors are read out and the data is passed between the different levels of the trigger system only when the beams cross in the interaction region. Below we discuss the STAR trigger design in the 2001 run.

\subsubsection{Hardware Triggers (L0 - L2)}

Level 0 is the basic hardware trigger layer. This layer looks at every RHIC crossing, deciding whether to accept the event or not. Level 0 consists of a multi-layer system of data storage and manipulation boards (DSMs)\cite{trigger}. Each DSM consists mainly of a field programmable gate array (FPGA) which can be configured by using the VHD-Language. For each RHIC strobe, trigger detector data is passed into the first layer of the DSMs, from where the data is pipelined into the following layers of DSMs. The physical algorithms implemented in the FPGAs reduce the data along the DSM layer structure, yielding a final trigger decision, which combines the data from the trigger detectors (e.g. charged particle multiplicity count in CTB). In the final DSM layer this information is combined with detector LIVE/BUSY signal from the slow detectors and a trigger is issued if the slow detectors are live.\footnote[8]{'Live' is a term which means that the slow detectors are finished processing the previous event and are ready to accept a new event.} The time allowed between the RHIC crossing and Level 0 decision is 1.5 microseconds.

Trigger Level 1 uses CPUs to analyze the output of the first layer of the DSM tree during the TPC drift time. Level 1 has 100 microseconds to either abort the event pass it to Level 2. Level 2 also consists of CPUs which have access to the full trigger information (first-layer DSM inputs, most finely grained information). The Level 2 analysis takes place during the TPC digitization time ($\sim$ 5 milliseconds). If Level 2 doesn't abort the event, the event is sent to the data acquisition system.

\subsubsection{On-line Event Reconstruction (Level 3 Trigger)}
The Level 3 trigger is a processor farm which can use the information from the STAR TPC to perform an on-line reconstruction of events at the rate of $\sim$ 100Hz\cite{L3trigger}. Given a reconstructed event, Level 3 is able to analyze the properties of this event (such as total charge in the event, number of tracks in the STAR TPC, track dE/dx, etc) and make a decision if this event should be written out on tape or not. Thus Level 3 trigger can select events based on physics observables, such as rare particles like $J\left/ \Psi \right.$ or antinuclei. This trigger is particularly useful for selecting events with low multiplicity and specific event topological signature, such as ultra-peripheral collisions. 

\subsection{Types of Triggers Available in STAR 2001 Run}
\label{sub:TriggerTypes}
Given the trigger capabilities described above, a few different trigger conditions were programmed into the STAR trigger logic, called 'trigger types'. We discuss the trigger types that were used for studying ultra-peripheral collisions.

The Minimum Bias trigger was programmed at Level 0 to require a coincidence signal in the East and West ZDCs (more than $\sim 40 \%$ of a single neutron energy deposition in both ZDCs). The dominant fraction of events collected with this type of trigger was hadronic collisions with charged multiplicities in STAR TPC between $\sim 0 \text{ and } 3000$, since nearly all hadronic interactions of the colliding beams result in the emission of spectator neutrons in both directions of the beam. Additionally, a fraction of ultra-peripheral collisions can occur simultaneously with the photonuclear excitation of both Au ions, resulting in a Minimum Bias trigger.

The next modification of the trigger was to accept at Level 0 events which satisfy the Minimum Bias trigger conditions and show a collision that has a vertex position in the direction of the beam (determined by the ZDCs) within 30 cm from the center of the interaction region. This trigger type was called a Minimum Bias Vertex trigger. The reason for this type of trigger is that we want the events that happen in the center of the STAR TPC, where the geometrical acceptance is symmetrical in the direction of the beam.

Another trigger type was called a 'topology trigger' and was designed specifically to trigger on 2-track ultra-peripheral events. This trigger was programmed in the Level 0 of the STAR trigger system to accept events which had charged multiplicity 1 in both north and south quadrants of the STAR CTB, but nothing in the top and bottom quadrants (to ensure cosmic ray events would not be accepted). Optionally, this trigger could also utilize the capacities of Level 3, requiring that events have zero total charge and have a vertex position in the beam direction within 100 cm of the center of the interaction region.

An important feature of STAR trigger system is that it can simultaneously look for events satisfying either one of several different trigger types. This is called parallel triggering. This allows a more complete use of the RHIC run time, since several different kinds of physics events can be collected in the same run. For instance, for a significant portion of the 2001 run the Minimum Bias and topology triggers were run in parallel, allowing the STAR physicists to collect both ultra-peripheral data and hadronic collisions\cite{trigger}.

\chapter{Event Reconstruction and Simulations}
\label{ch:Reco}
The following Sections describe the off-line event reconstruction in the STAR TPC, with a special emphasis on low-momentum electron tracks. With a few exceptions, this reconstruction chain also applies to the simulated events, which we discuss in the Section \ref{sec:SimuChain}. 

\section{Reconstruction}

In order to extract meaningful physics from the raw data, the pixel information which is stored from each tracking detector must be 'converted' into reconstructed tracks. This is implemented with event reconstruction software, organized into a chain of reconstruction procedures (reconstruction 'makers'). In this analysis the only tracking detector used is the TPC, and the main purpose of the reconstruction  software is to determine charge clusters in pad time-space and convert them into position coordinates in the TPC. Using the cluster information, pattern recognition software finds tracks which are in turn used to accurately determine the position of the beam interaction point (primary vertex). The tracks that are found at this stage are either 'secondary' (not coming from the primary vertex, such as cosmic muon tracks, for instance) or 'primary' (produced at the interaction vertex), so a re-fit of the tracks is undertaken, including the interaction point in the fit. Given the curvature of the track and the magnetic field, we can measure the rigidity of the particle -- momentum divided by the charge. To get the momentum, tracking software assumes an absolute value of the charge equal to the charge of the electron.

\subsection{TPC Hit and Global Track Finding}
\label{sub:hitfinder}

Electron-cluster finding along the TPC time dimension ($z$) is performed in DAQ crates, before the data is shipped to HPSS. The remaining hit reconstruction is accomplished off-line by scanning along rows for adjacent pad signals. Each pad row is treated separately and hits are never reconstructed across different pad rows. Thus, a track crossing the entire active gas volume of the TPC from the inner field cage to the outer field cage can have up to 45 hits. A cluster is then a set of signal sequences on adjacent pads on a pad-row which have some overlap in time. The distributions are fit with a two dimensional Gaussian function to determine the centroid of the cluster along the pad-row and in the time direction. This is assumed to be a hit position. Once the 3-dimensional positions of all hits are determined, the hits are used by a TPC tracking algorithm, where a hit represents a possible point on a track. 

\begin{figure}
\centering

\resizebox*{0.35\textwidth}{!}{\includegraphics{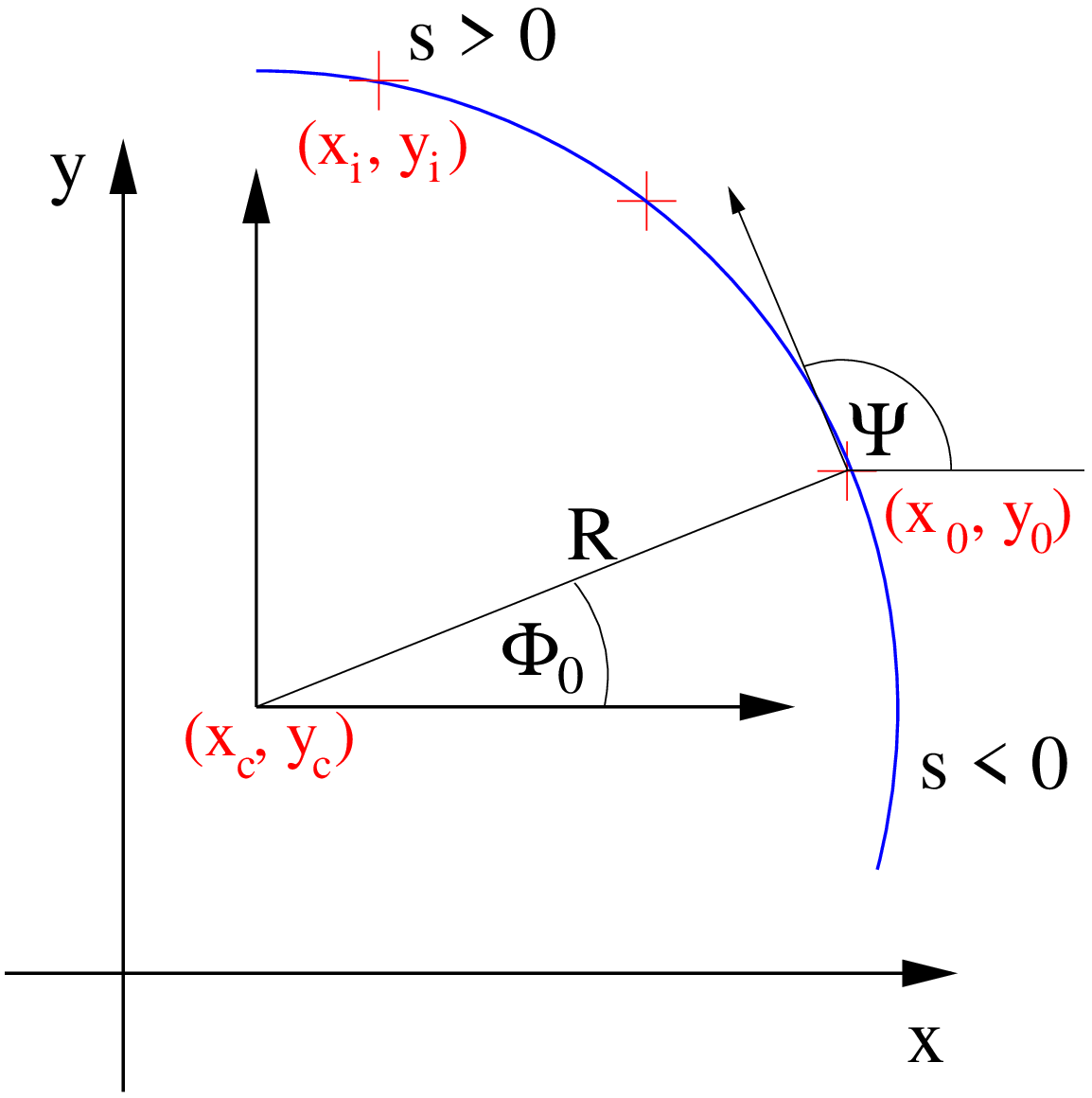}}
\resizebox*{0.35\textwidth}{!}{\includegraphics{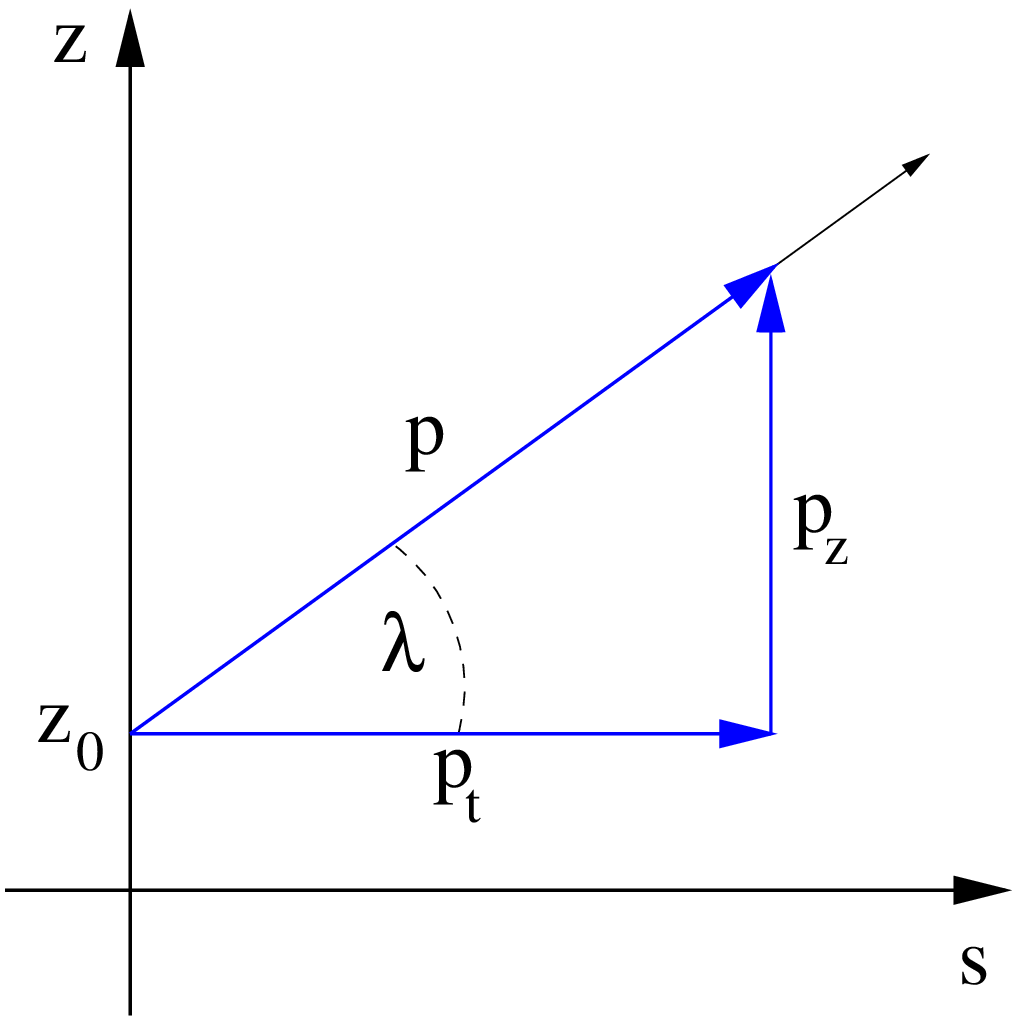}}
\caption[Geometric definitions of some commonly used track parameters]
{Geometric definitions of some commonly used track parameters. The crosses mark possible data points. $\Psi $ is a azimuthal angle of the track momentum at the point $(x_0,y_0)$. $p_t$, $p_z$ and $p$ are the track transverse, longitudinal and full momenta. $\lambda$, $R$ and  $\Phi _0 $ are the helix dip angle, radius of curvature and phase at the point $(x_0,y_0)$ \ \cite{tracking_starnote}.}
\label{fig:TrackParameters}
\end{figure}

Tracking starts at the outermost pad-row of the TPC, where the hit density is lowest. It searches for groups of three hits that lie close in physical space. These groups are the initial ``seeds'' of tracks. Track segments are formed from seeds through linear extrapolations to adjacent pad rows and clusters in $z$. Since the STAR magnetic field guides charged tracks into helical trajectories, segments are extended inward and outward with helical extrapolations. After these extensions, a minimum number of 5 hits is required on each track so that the five parameters of the track's helix model are uniquely defined. If this requirement is not met, then the track is flagged as a bad fit. Tracks are fit in two independent projections of the helix. A circular projection is fit in the plane perpendicular to the magnetic field, the \xypl. A line is fit in the \szpl, where $s$ is the length along a track's circumference \cite{tracking_starnote}. Tracks found with such a procedure are called 'global tracks'. 
Kinematic variables of a track are calculated from the parameterization of the helix; found via the circular and linear fits, and the direction and magnitude of the magnetic field. The definitions of some track parameters are given in Figure \ref{fig:TrackParameters}, and a more complete list of commonly used track parameters and their definitions can be found in Appendix \ref{app:variables}. Each global track is saved in the computer memory along with its kinematical and geometrical parameters and the basic information about the track fit, such as the number of hits used in the track fit and the $\chi ^2$ of the helix fit divided by the number of hits used.

The tracks which make a full $360^{\circ }$ rotation (in transverse projection) in STAR TPC will be reconstructed as two separate tracks, each making only a half-turn. A side-on view of such tracks is given in Figure \ref{fig:spiraling}, each of the two real tracks in the event is reconstructed as two half-spirals. One of the tracks will point towards the true interaction vertex in coordinate $z$ (tracks 1 and 3), while the other will not (tracks 2 and 4).  Tracks like the tracks 2 and 4 are called spiraling tracks. Spiraling tracks cause the contamination of clean 2-particle events with low-momentum tracks, as we will discuss in Chapter \ref{ch:Analysis}.

\subsection{Low Multiplicity Vertex Finder}
\label{sub:LMV}
An accurate knowledge of the interaction position (primary vertex) for each collision is required as this helps distinguish between track which originate from the primary vertex (primary tracks) and those from backgrounds (cosmics, beam-gas interactions, etc) or from weak decays. The primary vertex coordinates can also be included in a track re-fit for primary tracks which leads to a better determination of their momentum. In this section we describe a low multiplicity vertexer (LMV) -- a routine which is used to find primary vertices for events with low multiplicity.

The vertex finder procedure starts with an assumption that the true vertex is located at $(0,0)$ in \xypl.\footnote{This is a well justified assumption, since typically the vertex position is only within $\sigma = 0.25$ mm, from $(0,0)$ in the \xypl, and the $z$-position of the vertex is within $\sigma = 60$ cm from the center of the interaction diamond.} Any track which does not come to this point closer than a certain minimal value $R_{max}$ is removed from the vertex finding routine. The remaining tracks are modeled as helices and extrapolated inside the IFC, and an analysis is run to determine the amount of multiple scattering suffered by these tracks in the TPC gas, the TPC IFC, the SVT and the beampipe. The more secondary scattering the track experienced, the less weight this track will be given in the vertex fitting. Finally, a primary vertex is determined as a point $(x_0,y_0,z_0)$ which minimizes the weighted sum of squared distances from each helix to this point. A check procedure is performed to find the helix which is the farthest from the found vertex, and to remove the track if that distance is greater than a pre-set maximum $DCA_{max}$. If a track was removed from an analysis at this stage, a new vertex fit is performed, if all tracks were retained, the vertex finding routine is finished. The error squared of the vertex position is set equal to the weighted sum of squared distanced from the tracks to the primary vertex. 

There are a number of reasons why a vertex finding routine can fail for an event consisting of several low-momentum tracks which come from a true primary vertex within an interaction diamond. Fist of all, a track's kinematic parameters may be sightly mis-reconstructed, and the track may be found to pass within more than $R_{max}$ from the beam in the \xypl. Secondly, after the vertex fit is performed, some tracks may be found to pass within more than $DCA_{max}$ from the primary vertex, this happens again due to mis-reconstruction of tracks' kinematic parameters. Since each of these occurrences results in a removal of an outlier track from the analysis, the number of tracks remaining for the determination of the vertex may be less than two, which automatically causes the vertex finding routine to fail. Thus, there is a certain probability that for the low multiplicity events consisting of low-momentum tracks a vertex finder will find no vertex, even when in fact there is a vertex. We will estimate this probability in Section \ref{sub:VertexingEfficiency}.

\subsection{Primary Track Fitting and Kalman Filter}
\label{sub:Kalman}
Including the primary vertex in the track fits can significantly improve the accuracy of the track reconstruction. A global track pointing back to the primary vertex and having an extrapolated distance of closest approach to the vertex less than 3 cm is considered to have originated from the vertex. Global tracks which satisfy this criteria are termed primary tracks and are subjected to a re-fit of the track. The primary vertex is included in the fit.
Since the initial estimate of the tracks' kinematic parameters are know from the helix fit of the global tracks, a more sophisticated procedure, called Kalman filter can be utilized (see \cite{Kalman} for for a general description of Kalman filter). This is more realistic than a simple helix fit as it takes into account multiple scattering and energy loss in the inner field cage and the gas inside the TPC in each step of the fitting routine, in order to obtain the best estimate of the track momentum. This approach requires three passes through the data. The first pass, known as filtering, starts at the outermost pad-row and works its way in, removing hits from the track that fail very crude cuts. The second pass, referred to as smoothing, starts at the innermost radius and works its way outwards. Again, outliers are removed from the track but this time the rejection cuts are much harsher. The third pass is for the evaluation of the track parameters and the total $\chi^2$.

In order to estimate the degree of multiple scattering and energy loss, an estimate of the mass of the particle must be used. This is accomplished by assuming that all particles are pions. This is true for approximately 70-80\% of tracks in a typical STAR high-multiplicity event, but obviously not correct for the current analysis. Although it is desirable to fit every track with a correct mass hypothesis, this was not performed in the 2001 run reconstruction, as the process is extremely CPU intensive. We will study the effect of the incorrect mass hypothesis on the reconstructed primary track momenta in Section \ref{sec:smearing}. 

\subsection{Particle Identification with the TPC by Specific Energy Loss}
\label{sec:dedx}
\begin{figure}
\centering 
\includegraphics[width=350pt]{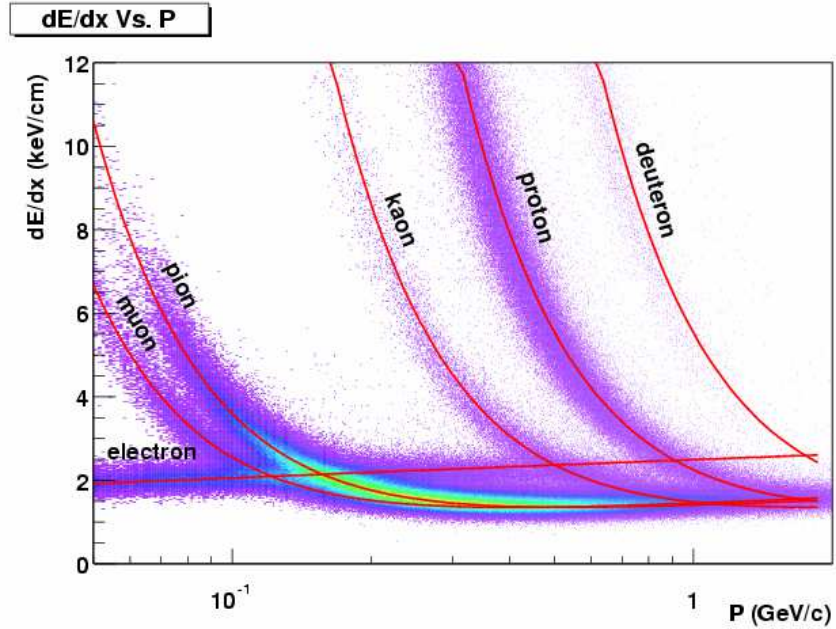}
\caption[Scatter plot of track \dedx \ measured by the STAR TPC vs. track momentum]{Scatter plot of track \dedx \ measured by the STAR TPC vs. track momentum \cite{TPC}.}
\label{fig:dedx}
\end{figure}

Charged particles lose energy while traversing matter. The main process contributing to this energy loss for heavy particles is Coulomb scattering of the traveling particles with electrons in atomic orbits, causing ionization of the traversed medium. The signal heights of the pads in a hit cluster give a measure of the energy loss in the TPC gas. These signals are summed over all pads in a cluster. The energy loss per unit length for each hit is then equal to this sum divided by the length of the track over the pad row. The distribution of the energy loss samples for hits along a track has a Landau-like distribution\cite{pdg}. An estimate of the mean of this distribution is calculated by computing the average of the lower 70\% of the distribution -- this procedure, called truncation, is necessary since it ensures that the truncated mean is not sensitive to the extremely large values of $dE/dx$ in a small fraction of clusters. The result is a measure of the energy loss per unit distance along a track, $dE/dx$ (KeV/cm) in the TPC gas. Figure \ref{fig:dedx} shows the scatter plots of the $dE/dx$ distribution vs. the momentum of the tracks, measured by STAR TPC.

For moderately relativistic charged particles with $m\gg m_e$ the ionization energy loss is a dominant process, and can be described by the Bethe-Bloch function\cite{pdg}:

\begin{equation}
\label{eqn:BetheBloch}
-\left\langle dE/dx\right\rangle _{truncated}^{mean}
= Kq^2\frac{Z}{A}\frac{1}{\beta^2}\left(\frac{1}{2}ln\frac{2m_ec^2\beta^2\gamma^2T_{max}}{I^2} - \beta^2 - \frac{\delta}{2}\right)
\end{equation}

\begin{table}
\centering
\begin{tabular}{lll}
\hline
Symbols &Definition                   &Units or Value \\
\hline
$m_e$   &electron mass                &0.51\mevcc \\
$r_e$   &classical electron radius    &2.81 fm \\
$N_A$   &Avogadro's number            &6.022$\times10^{-23}$ mol$^{-1}$ \\
$K$     &$4{\pi}N_A{r_e}^2c^2$        &0.307{\mev}g$^{-1}$cm$^2$ \\
$q$     &charge of the particle       & \\
$Z$     &Atomic number of the medium  & \\
$A$     &Atomic mass of the medium    & \\
$I$     &Mean excitation energy       & MeV \\
$T_{max}$ &\multicolumn{2}{l}{maximum kinetic energy that can be imparted to a free electron in a collision} \\
$\delta$  &\multicolumn{2}{l}{Bethe-Bloch density effect correction factor} \\
\hline
\end{tabular}
\caption[Physical constants used in the Bethe-Bloch function]
{Physical constants used in the Bethe-Bloch function, Equation (\ref{eqn:BetheBloch}).}
\label{tab:dedxVarTable}
\end{table}

The variables used in the Equation (\ref{eqn:BetheBloch}) are defined in table \ref{tab:dedxVarTable}. According to  Equation (\ref{eqn:BetheBloch}), ionization loss only depends on a particle's charge $(q)$, velocity $(\beta)$ and the properties of the medium that is traversed.  To get the expected mean $dE/dx$ as a function of the particle momentum in STAR TPC we need to assume a mass and a charge for a particle. Figure \ref{fig:dedx} shows in solid lines the resulting functions for several particle mass and charge assumptions.

For lighter particles, such as electrons, several other processes are important - Bremsstrahlung and Moller scattering (Bhabha scattering for positrons) are the two most significant. Bremsstrahlung is particularly important, since electrons are very light, and in the region of momentum of the order of dozens of MeV/c they are ultra-relativistic. The total length of the STAR TPC in the transverse direction (150 cm between the inner field cage and the outer field cage) is equal approximately to 1.17\% of the radiation length. Thus, an electron traversing a TPC with a momentum of $\sim$100 MeV/c will lose $\sim 0.0117\cdot \left( \frac{e-1}e\right) \cdot 100$ MeV$\approx 0.7$ MeV due to Bremstrahlung, which is comparable to the ionization energy loss over 150 cm of traversed P10 gas. The energy loss by Bremsstrahlung will (mostly) not be detected by the STAR TPC, therefore the Bethe-Bloch formula is not adequate for the description of energy loss by electrons in the STAR TPC. In general, we observe that the description of the energy loss by electrons in the STAR TPC is not well described in the GEANT detector simulation package (next section).

Despite the aforementioned difficulties, for the electrons in the region of momenta from $\sim$10 MeV/c to $\sim$100 MeV/c, star TPC detects $dE/dx$ energy loss which is clearly different from other species of particles (the lowest band on the left in Figure \ref{fig:dedx}). The average $dE/dx$ for the electrons shown as a solid line was not computed with the Bethe-Bloch formula (\ref{eqn:BetheBloch}), but rather determined experimentally from a clean sample of electrons and positrons from photon conversions \cite{IanThesis}. The electron band is sharply crossed by other particle bands as they approach their minimum ionizing values, as illustrated in Figure \ref{fig:dedx}.

With a high level of confidence particles can be identified in regions where the $dE/dx$ bands are not overlapping. In regions where bands are partially merged, simultaneous Gaussian fits to the bands in momentum slices are used to extract the probability that a track is of a given particle species. With the tracking field at 0.25 T, the STAR TPC has a one sigma $dE/dx$ resolution of about 8.2\% of the $dE/dx$ value. 
The knowledge of both a predicted $dE/dx$ value and the resolution of $dE/dx$ is a tool for performing particle identification. Using the predicted $dE/dx$, a measured $dE/dx$ and a sigma, a known fraction of a certain particle band can be sacrificed in order to eliminate other particles. This is accomplished by requiring the {\dedx} of all particles to fall within a certain number of sigma of a predicted band ({\dedx} deviant).

\section{Details of the Simulation Chain}
\label{sec:SimuChain}
We used the events generated with a Monte Carlo generator, described in the Chapter \ref{ch:MonteCarlo}. Next we used the GEANT detector description package\cite{Geant} to simulate the passage of electrons and positrons from the collision point through the detector material and ionization of the TPC active gas volume by the charged tracks. GEANT also simulates the creation of secondary particles in the detector material interactions. 

The secondary particles consisted mostly of knock-out electrons and a small number of \ee   pairs produced by hard Bremsstrahlung photons emitted by the original electrons or positrons. These particles are added to the table of Monte Carlo generated tracks with unique ID's, separating them from the original event generator tracks (primary particles).

The ionization produced by charged tracks traversing the TPC gas was passed through a TPC Response Simulator (TRS). TRS simulates the TPC response from the drift of the ionization in the TPC gas to the output of the front-end electronics. It includes the drift, diffusion, amplification and response of the electronics for the electrons created by ionization in the TPC gas. Simulated pad signals can then be then passed to the event reconstruction routine. 

\subsubsection{Additional Procedures Applied to the Monte Carlo}
An additional step in the event reconstruction routine for the simulated events is the association of the reconstructed tracks with the Monte Carlo tracks in the STAR TPC. The association routine takes the TPC hits used for a fit of a given track and checks if those hits might come from any single Monte Carlo track. If more than 5 TPC track hits are associated to the Monte Carlo track, the reconstructed track is associated to the Monte Carlo track. 
In the conditions of the low-multiplicity environment association has an efficiency of $\sim 99\% \pm 1\%$, nearly all  reconstructed tracks (with number of hits per track $>5$) were matched to the Monte Carlo generated tracks. We use only the reconstructed Monte Carlo tracks that are associated to the primary particles in Monte Carlo efficiency studies.

\begin{figure}
\centering
\includegraphics[width=200pt]{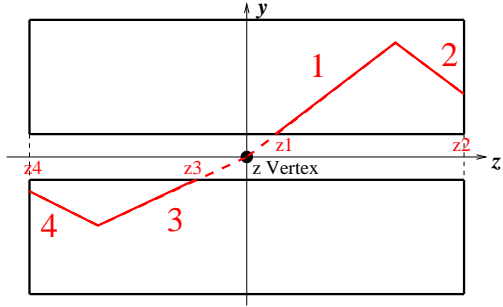}
\caption[Spiraling tracks in the STAR TPC]{Side-on view of the STAR TPC. Shown are two particles going through the TPC, each particle will be reconstructed as two separate tracks.}
\label{fig:spiraling}
\end{figure}

Also, we wish to exclude from Monte Carlo efficiency studies the spiraling tracks. Figure \ref{fig:spiraling} shows two particles in the STAR TPC, one in a positive $z$ direction, one in the negative. Each one of the tracks is reconstructed as two separate tracks. To exclude the second parts of the spirals (tracks 2 and 4) we apply the following requirement:

\begin{equation}
\label{eqn:spiraling}
\eta _{track} \left( z_{track} - z_{vertex} \right) > 0 
\end{equation}
where $z_{track}$ is the $z$ position of the first innermost hit on the track (shown as $z1$ , $z2$ , $z3$ and $z4$ in the Figure \ref{fig:spiraling}), $z_{vertex}$ is the $z$ position of the GEANT generated vertex, and $\eta _{track}$ is the reconstructed pseudorapidity of the track. Monte Carlo studies show that the requirement (\ref{eqn:spiraling}) eliminates 99\% of the spiraling tracks. Aditionally, we determined that for simulated \ee pairs with a reconstructed primary vertex, spiraling tracks make up only a 2\% contamination.

\section{Event Pre-Processing and Data Storage}
After the event reconstruction, all of the data is written to a Data Summary Tape (DST). The DSTs are split into chunks of approximately 130 events as each file is large ($\sim 500$ megabytes) by this stage. Each event is a separate entity on the DST, with a list of all primary and global tracks in the event, a map between the primary and global tracks, a vertex information, and general characteristics of the event, such as the time the event was taken, the trigger type for this event, the conditions of the STAR detector and so on. Additionally, if the event comes from the simulation, the full information about the simulated event kinematics and the association map between the simulated tracks and their reconstructed counterparts is included. 

This specific analysis utilized one additional stage of data processing, unique to the ultra-peripheral events. Such events can typically be easily separated from the hadronic events by either the low multiplicity of tracks in the event, or if an event was triggered with a specific ultra-peripheral type of trigger (such as the topology trigger, Section \ref{sec:trigger}, for instance). All DST files were filtered for the ultra-peripheral events based on the above conditions, and all the selected events were analyzed by a specific ultra-peripheral analysis routine. The routine takes all tracks and pairs them into all possible pair combinations. For each pair several characteristic variables are computed, such as the total charge of the pair, the total transverse and longitudinal momentum, the invariant mass of the pair assuming both tracks are electrons, pions or kaons, and a few more. This routine was run separately on the primary and global tracks in each event. The pairs comprised of primary tracks were called 'primary pairs', and the pairs comprised of the global tracks were called 'secondary pairs'.

\chapter{Analysis}
\label{ch:Analysis}

In order to compute the cross-section of a reaction, we need to count the number of events which are believed to come from a studied reaction ($N_{\text{observed events}}$), know the total integrated luminosity of the data set considered ($L_{tot}$) and the efficiency of triggering on the events of this type ($\text{Eff}^{\text{trig}}$), and to determine the efficiency for reconstructing the events of this type ($\text{Eff}^{\text{detector}}$):

\begin{equation}
\centering
\label{eqn:Sigma}
\sigma = \frac{N_{\text{observed events}}}{L_{tot}\times \text{Eff}^{\text{trig}} \times \text{Eff}^{\text{detector}}}
\end{equation}

This chapter describes the selection cuts we used for finding events of the type $AuAu \rightarrow Au^*Au^*+e^{+}e^{-}$ in the pool of reconstructed events in the STAR detector, and the computation of total luminosity times the trigger efficiency in the analysis dataset. The subsequent chapters (\ref{ch:Simu}, \ref{ch:Results}) will discuss the reconstruction efficiency and the measured cross-section.

\section{Data Set}
\label{sub:luminosity}
We desire to observe the ultra-peripheral electromagnetic production of $e^{+}e^{-}$ pairs with simultaneous nuclear excitation in gold-gold collisions: $AuAu \rightarrow Au^*Au^*+e^{+}e^{-}$. Events of this type constitute a sub-set of the events collected with Minimum Bias Trigger at 0.25 T magnetic field (half-field). During the 2001 data taking the largest half-field dataset was collected with a Minimum Bias Vertex trigger, this is the data set we used for the analysis. The dataset has about 800,000 events, about 60,000 of these events were found to have less than 10 global tracks per events. The presented analysis will consider these events.

\subsubsection{Luminosity}
The Equation \ref{eqn:Sigma} can be applied equally to ultra-peripheral reactions and to hadronic reactions. In particular, this equation can be applied to a total hadronic cross-section for gold-on-gold reactions, which has been calculated to be 7.2 b $\pm 10\%$\cite{BaltzWhite}. Simultaneous photonuclear excitations of both Au ions contribute additional 3.2 b cross-section in the Minimum Bias dataset\cite{ZDCpaper}. Analysis of the hadronic reactions in STAR detectors shows that events with more than 7 negatively charged hadrons in the STAR TPC in the pseudorapidity range $|\eta |< 0.5$ and with transverse momenta $p_{\perp} > 100$ MeV/c constitute 80\% of the total hadronic cross-section\cite{Zhangbu}. This allows us to measure the total luminosity times the trigger efficiency in the Minimum Bias Vertex data sample ('normalization to hadronic cross-section'):

\begin{equation}
\label{eqn:lumi}
L_{tot} \cdot \text{Eff}^{\text{trig}}  \! = \! \frac{ N_{\text{in MinBiasVertex dataset}} \left( \text{{\small 8+ positive hadrons in \ }}  | \eta | < 0.5 \text{{\small \ and }} p_{\perp} > 100 \text{{\small \ MeV/c}} \right) }{0.8 \times 7.2\text{{\small \ b}}}
\end{equation}

In the Minimum Bias Vertex data set, $L_{tot} \cdot \text{Eff}^{\text{trig}}$ for ultra-peripheral events can be determined by normalization to the hadronic cross-section only for events which have a longitudinal vertex position within $|z_{vert}|<25$ cm. This is due to the fact that to use normalization to hadronic cross-section,
we need to make sure that the trigger efficiency is the same for the hadronic events (which are usually high-multiplicity and produce a lot of neutrons in the ZDCs) and ultra-peripheral collisions (which are low multiplicity and typically produce just a few neutrons in the ZDCs). As we mention in Section \ref{sub:zdc}, the ZDCs have close to $100\%$ efficiency for detecting both single and multiple neutrons; therefore Minimum Bias trigger efficiency is the same for hadronic and ultra-peripheral events. However, the Minimum Bias Vertex trigger uses ZDC timing information for rejecting a portion of events that are accepted by the Minimum Bias trigger, and the efficiency of this rejection may be different for the events with a lot and a few neutrons. Selecting only the events within $|z_{vert}|<25$ cm, which is 5 cm tighter than the ZDC timing vertex cut,\footnote{Vertex position resolution in ZDC timing cut is on the order of $3.2$ cm (Section \ref{sub:zdc}).} allows us to avoid the problem of unequal trigger efficiencies for hadronic and ultra-peripheral events. 

The luminosity times trigger efficiency for the data set available to us with a cut $|z_{vert}|<25$ cm was found to be $62 \ \text{mb}^{-1}$, with systematic uncertainty of $10\%$ due to uncertainties in total hadronic cross-section.

\section{Event Selection}
We will be working with the sub-set of the Minimum Bias Vertex triggers, a total of about 60,000 events (with less than 10 global tracks per event). We will begin by identifying criteria for selecting ultra-peripheral events among the other low-multiplicity events.  We then present a Monte Carlo study of the STAR detector acceptance for low-momentum tracks, and  restrict our attention to the tracks which are in the region where STAR detector has good acceptance. Selecting $e^{+}e^{-}$ pairs among other ultra-peripheral events requires particle identification with $dE/dx$. Finally, we present a study of vertex finding efficiency and restrict the position of the primary vertex for selected events.

\subsection{Identifying Ultra-Peripheral Pairs}
\label{sub:UPC}
We require that the total pair charge be zero. If this requirement is not met the pair is guaranteed to be a background, and we will take advantage of this property later (Section \ref{sec:backgrounds}). We also require that there be exactly one primary pair in an event, and at most two more non-primary tracks - this allows for additional tracks due to occasional contamination by beam-gas interactions, spiraling tracks and secondary particles.  

A crucial way of distinguishing the pairs produced in ultra-peripheral coherent interactions of two gold ions is to cut on the maximal value of the total transverse momentum of the pair. The physics motivation for this cut was explained in Sections \ref{sec:EvgenResults} and \ref{sub:TransverseMomentum}. Figure \ref{fig:pperp} shows that the total transverse momentum of the coherent $e^{+}e^{-}$ pairs peaks at about 5 MeV/c and drops off significantly at 70 MeV/c. We set the $(p_{\perp}^{tot})^{\max}$ cut at 100 MeV/c for the coherent pairs.

\subsection{Track Acceptance Cuts}
\label{sub:AcceptanceCuts}
The STAR TPC has a limited acceptance for charged tracks. First of all, in transverse plane tracks must have a diameter  greater than the radius of the Inner Field Cage ($R_{IFC}=50$ cm), which translates into a requirement on the tracks' transverse momenta $p_{\bot} \! > \! p_{\min} = 0.3qB R_{IFC}$ $ = 37.5$ MeV/c. Secondly, since the TPC has a finite length ($L_{TPC}=4$ m), tracks must have pseudorapidity $ |\eta |<
 - \ln \left( {\tan \left( {0.5\tan ^{ - 1} \left[ {R_{IFC} /0.5L_{TPC} } \right]} \right)} \right)=2$ to make it inside the TPC. 

\begin{figure}
\centering
\includegraphics[width=200pt]{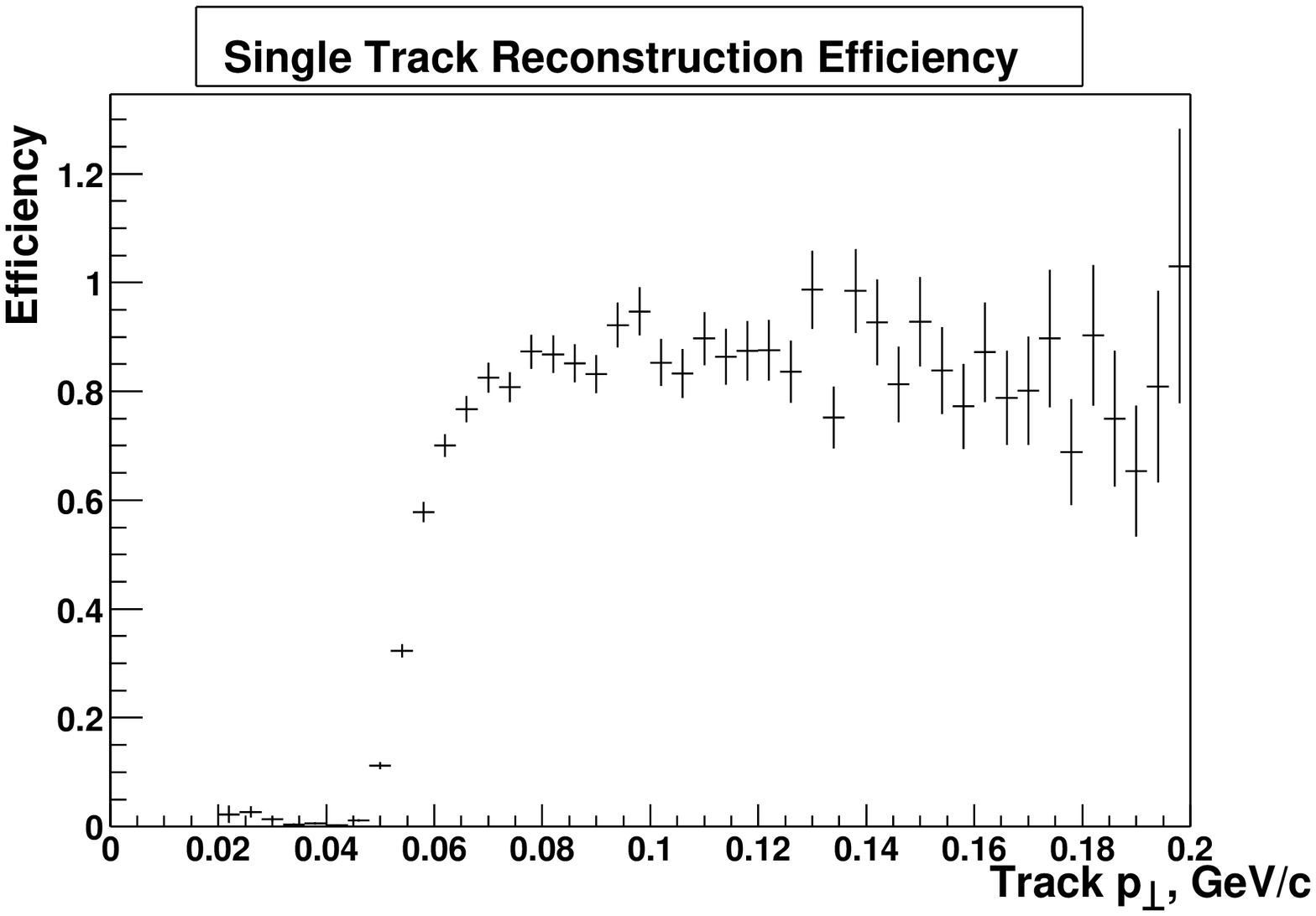}
\includegraphics[width=200pt]{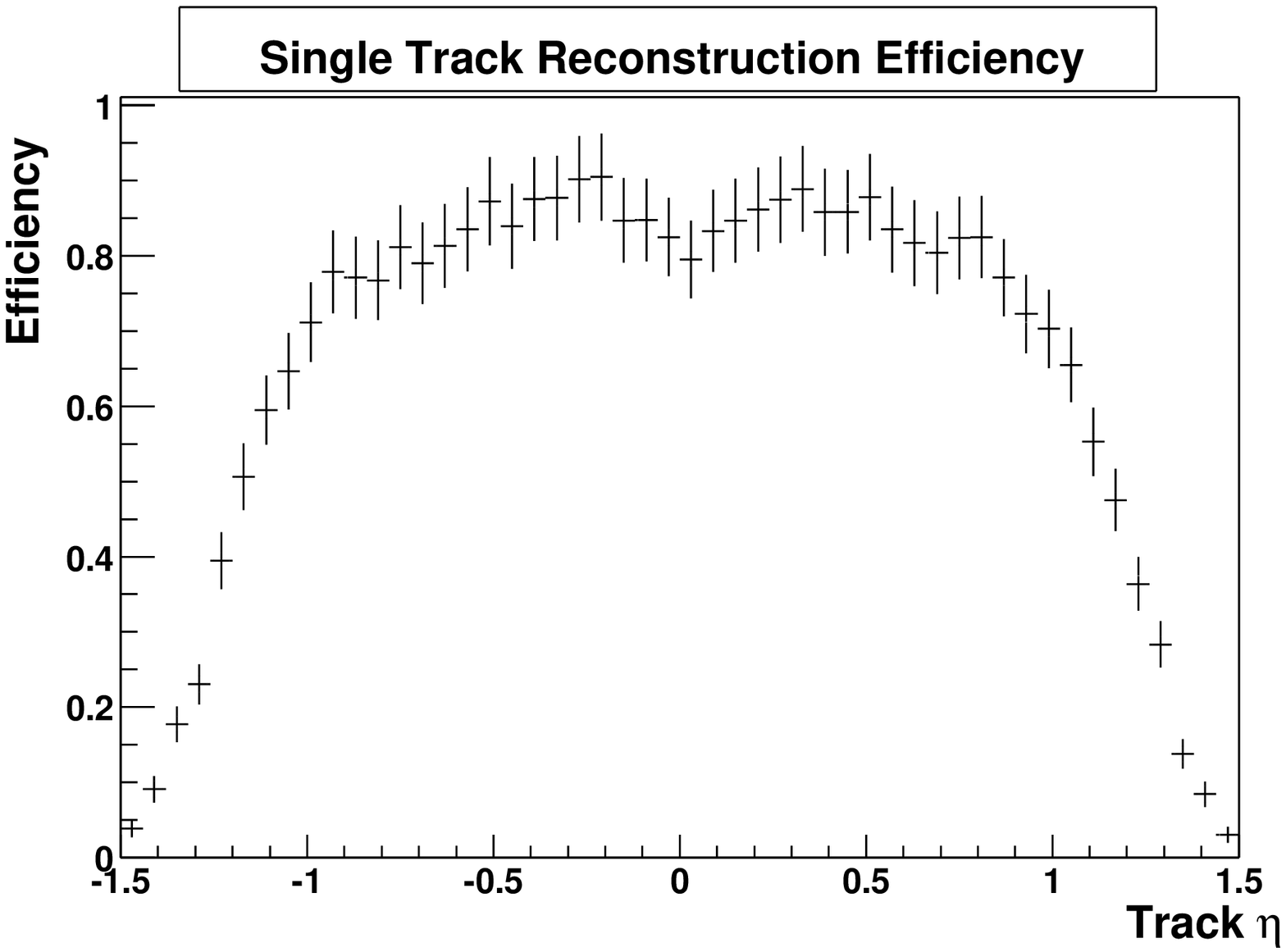}
\caption[TPC acceptance in $p_{\perp}$ and $\eta$]{Left: TPC tracking efficiency as a function of track $p_{\perp }$ for tracks within the $|\eta |<1.15$ cut. Right: acceptance as a function of track $\eta$, for tracks with $p_{\perp}>0.65$ GeV/c.}
\label{fig:acceptance}
\end{figure}

To find the regions of transverse momenta and pseudorapidity where the TPC tracking efficiency is relatively flat we used the Monte Carlo event sample described in Chapter \ref{ch:MonteCarlo} (individual track transverse momenta in the range $50 \text{ MeV/c} <p_{\perp} < 200 \text{ MeV/c}$ and pseudorapidity $ |\eta |<4 $ ). The tracking efficiency can be defined as:

\begin{equation}
\label{eqn:trackingEfficiency}
E_{\text{tracking}}  = \frac{{\# \text{single tracks reconstructed, associated to primary particles}}}{{\# \text{single primary particle tracks generated}}}
\end{equation}
The resulting distributions in single track $p_{\perp}$ and $\eta $ are shown in Figure \ref{fig:acceptance}. 

Figure \ref{fig:acceptance} demonstrates that STAR TPC has close to 1 acceptance for tracks with high transverse momentum at mid-rapidity. A slight dip in reconstruction efficiency as a function of $\eta $ at zero is due to the fact that tracks with $|\eta|<0.3$ are more likely to have low $p_{\perp}$ and not be reconstructed than tracks with $0.3<|\eta|<0.7$. A check procedure applying a cut $p_{\perp}>0.95$ GeV/c to Monte Carlo tracks, which ensures that reconstruction efficiency in $p_{\perp}$ is flat, returns an efficiency in $\eta$ which is flat for $|\eta|<0.7$. We chose to place acceptance cuts on the primary track $p_\perp$ and $\eta$ at 

\begin{equation}
\label{eqn:detAcceptance}
\begin{array}{*{20}l}
   { | \eta _1 | <   1.15} \hfill & {| \eta _2 | <   1.15} \hfill  \\
   {p_{1 \bot }  > 65{\rm  MeV/c}} \hfill & {p_{2 \bot }  > 65{\rm  MeV/c}} 
 \hfill  \\
 \end{array}
\end{equation}

Figure \ref{fig:acceptance} shows that the TPC has some residual acceptance outside of the acceptance region (\ref{eqn:detAcceptance}), however we chose to cut away the regions of low acceptance, where the simulation might not adequately describe the detector effect on the data.

\begin{figure}
\centering
\includegraphics[width=200pt]{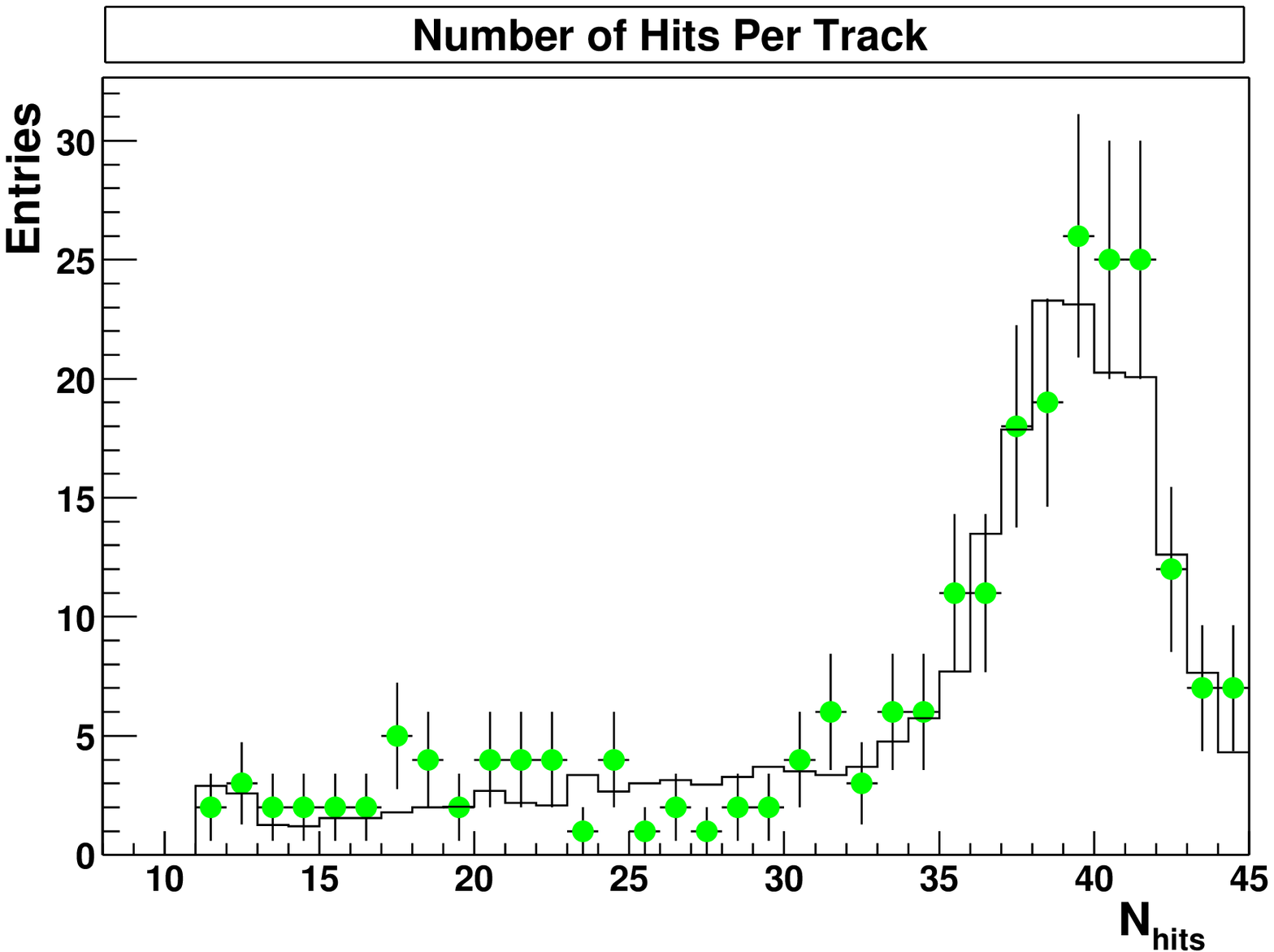}
\includegraphics[width=200pt]{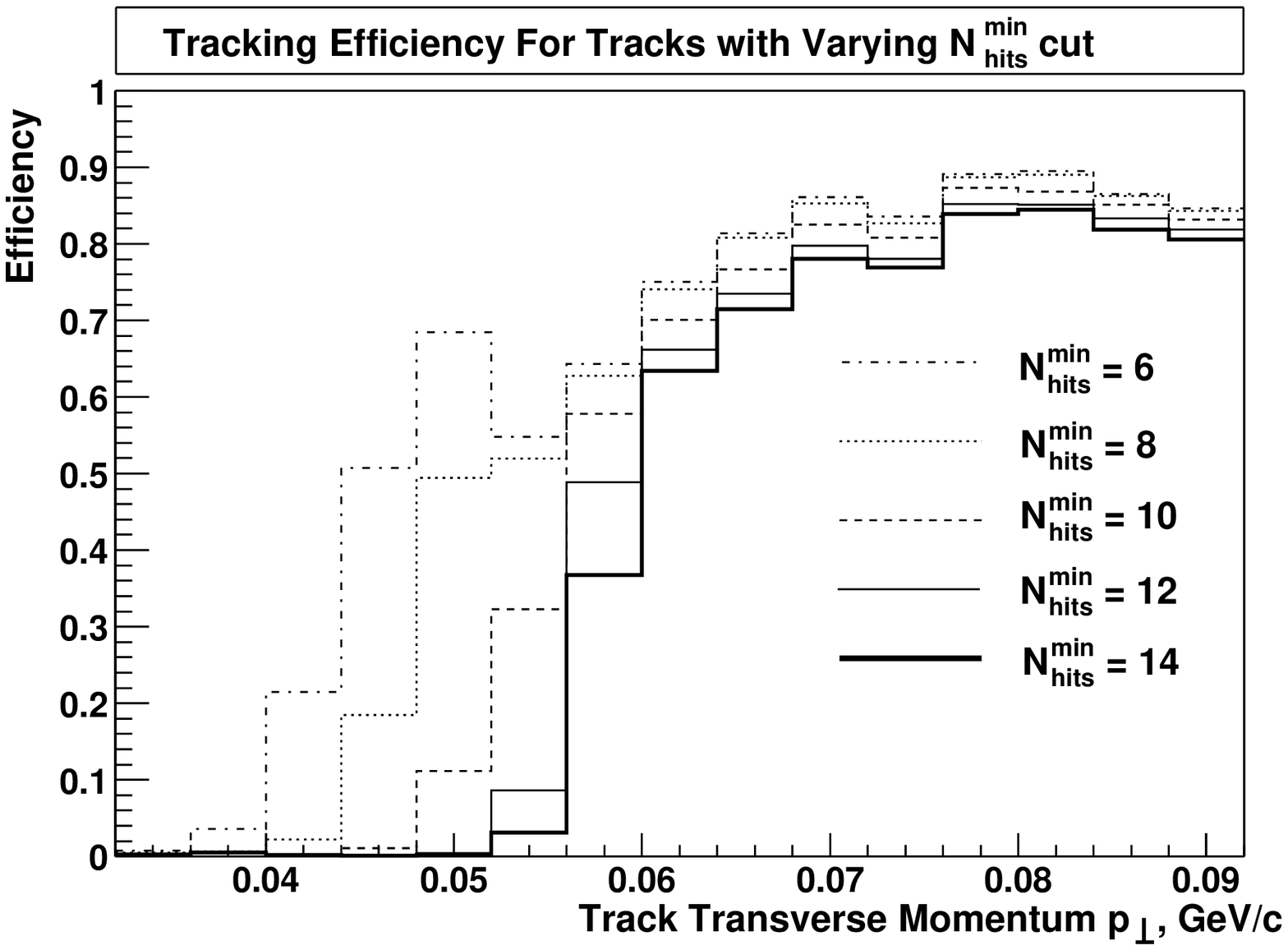}
\caption[Number of hits per track]{Left: distribution of $\text{N}_{hits}$ in the data (dots) and Monte Carlo (solid black). Right: TPC tracking efficiency for tracks with $\text{N}_{hits}^{\min}=6$, $\text{N}_{hits}^{\min}=8$, $\text{N}_{hits}^{\min}=10$, $\text{N}_{hits}^{\min}=12$, $\text{N}_{hits}^{\min}=14$. }
\label{fig:hits}
\end{figure}

Figure \ref{fig:hits} (left) shows the comparison of the number of hits per track ($\text{N}_{hits}$) distributions in the data with the cuts in Section \ref{sub:UPC} and Monte Carlo, and the agreement is quite good. We use Monte Carlo to study how the tracking efficiency depends on $\text{N}_{hits}$. Figure \ref{fig:hits} (right) shows the tracking efficiency vs. transverse momentum computed for tracks with more than 6,8,10,12 or 14 hits per track. For tracks with $p_{\perp}>65$ MeV/c, dependence of the tracking efficiency on the number of hits is negligible. STAR event reconstruction software writes on tape only TPC tracks with more than 10 hits per track, therefore we set $\text{N}_{hits}^{\min} = 10$ cut in the analysis. 

\subsection{Particle Identification via Specific Energy Loss}
\begin{figure}[t]
\centering
\includegraphics[width=200pt]{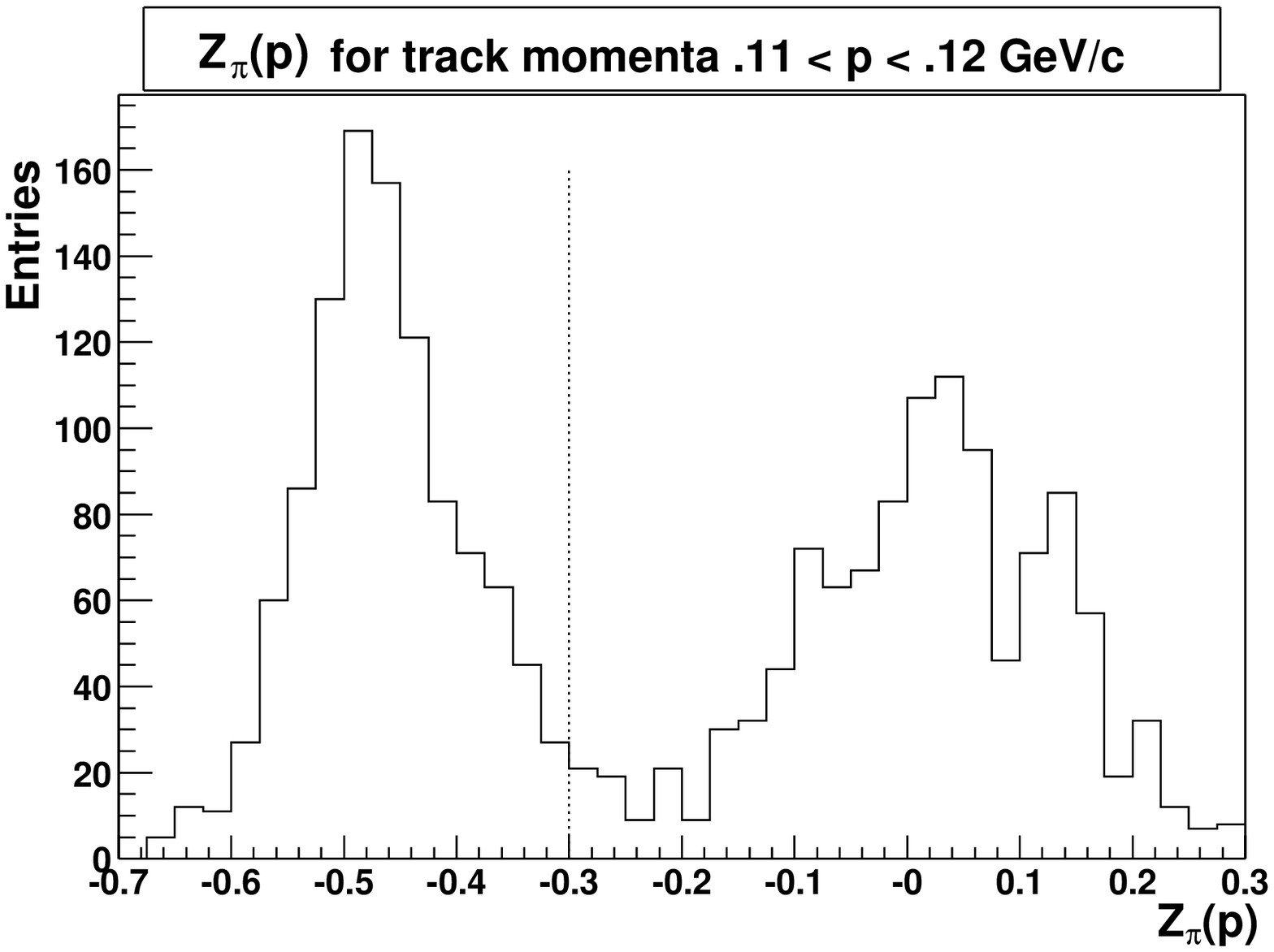}
\includegraphics[width=200pt]{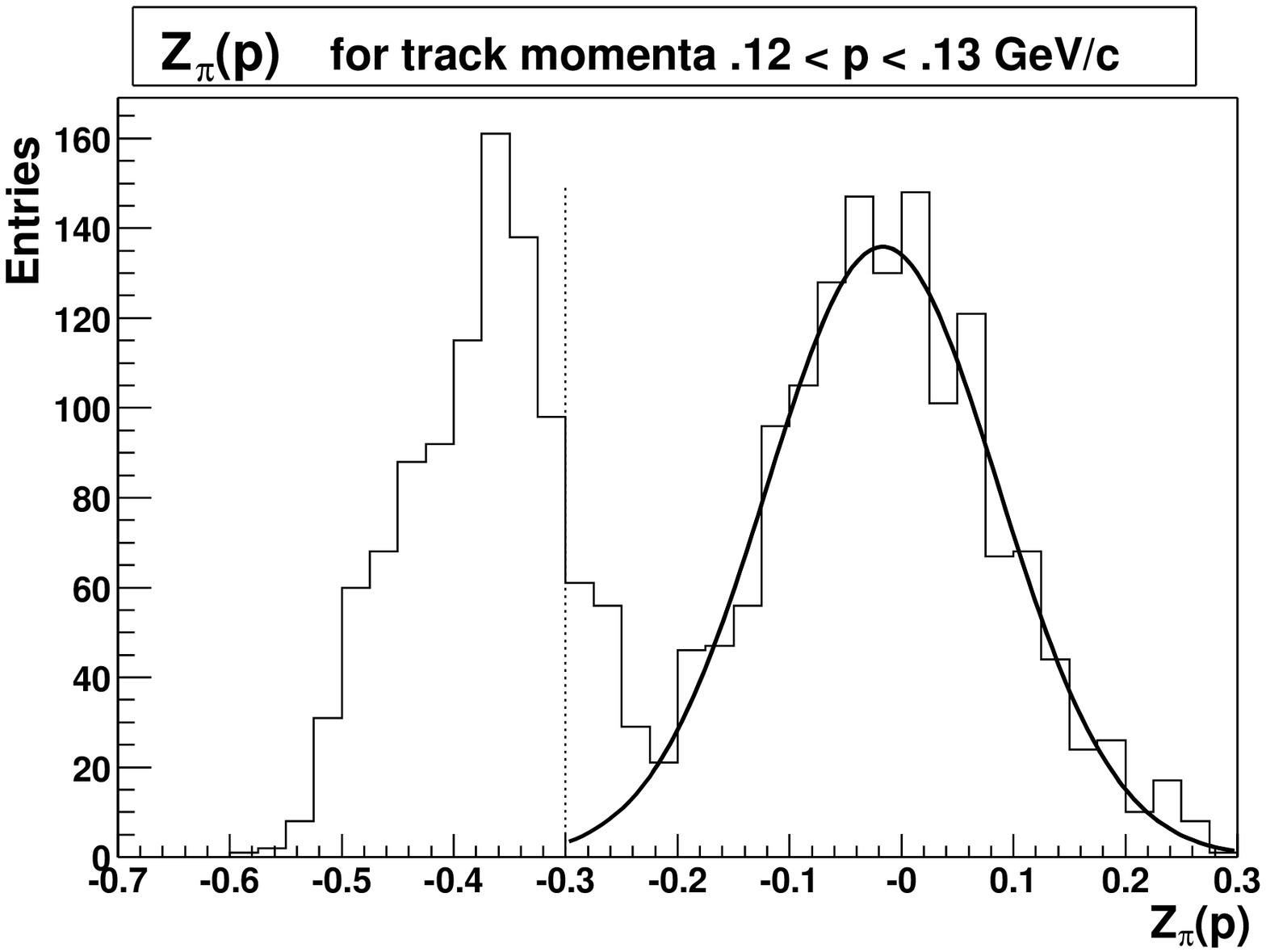}
\includegraphics[width=200pt]{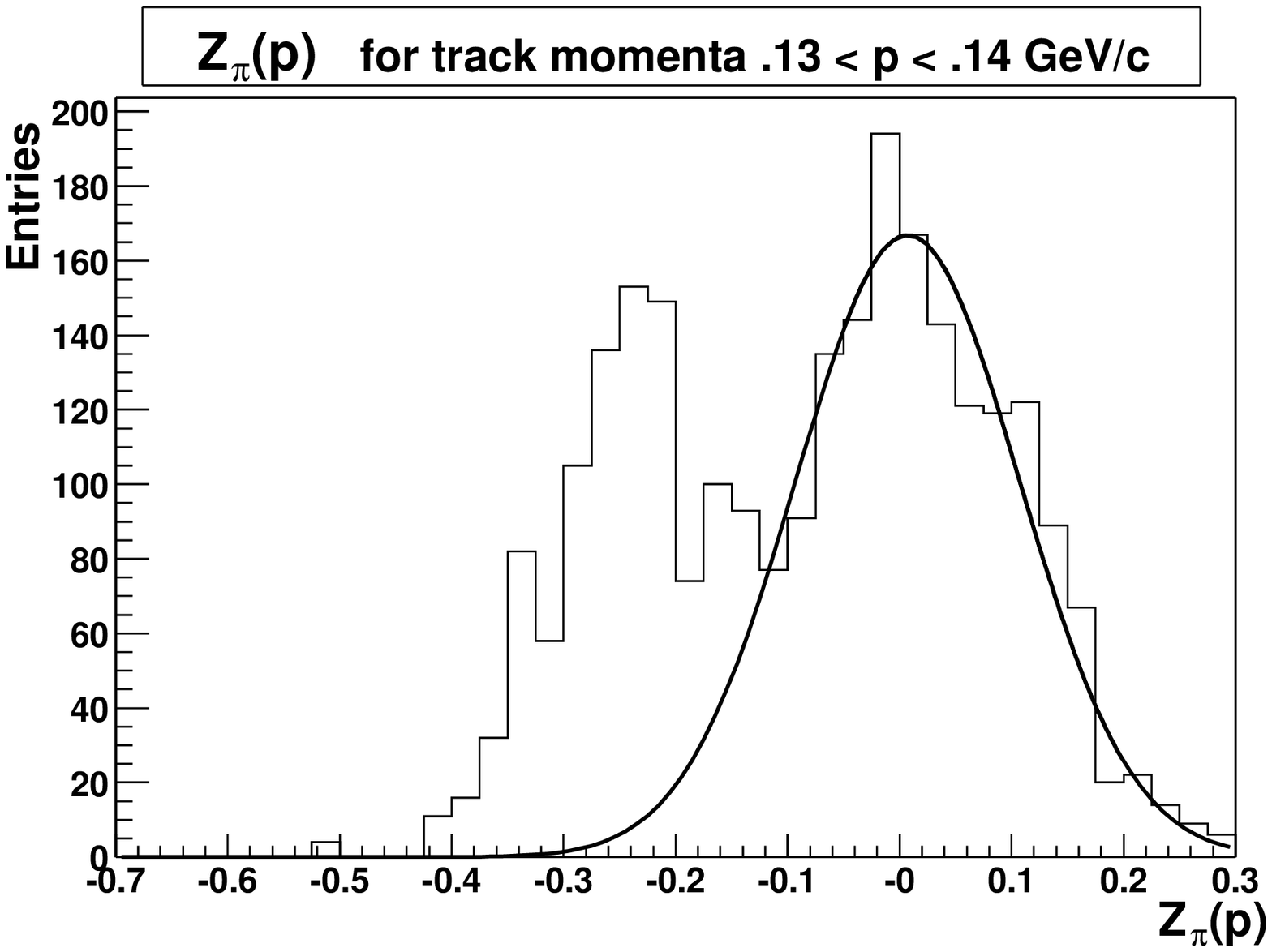}
\includegraphics[width=200pt]{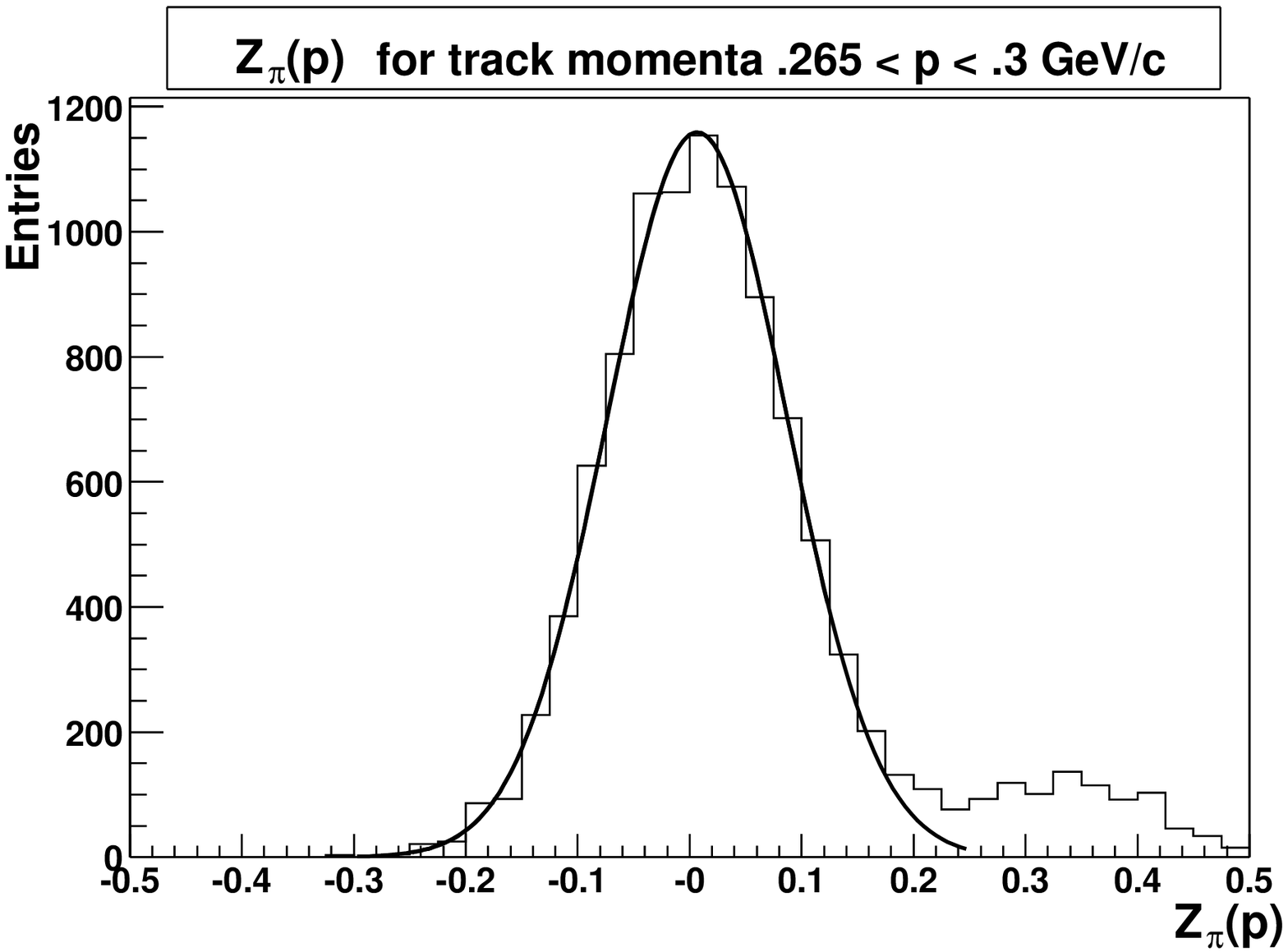}
\caption[$Z_\pi$ histograms, in $p$ slices]{$Z_\pi$ histograms, in $p$ slices. Solid curves - Gaussian fits to the pion peaks. Tracks to the right of the dotted lines (upper histograms) are rejected as pions.}
\label{fig:dedx_slices}
\end{figure}

Single track $dE/dx$ depends on the track's momentum, particle mass and charge and in a lesser degree, on the number of hits on the track. To eliminate the dependence of track $dE/dx$ on these variables, a new variable $Z$ is introduced. The definition depends on a species of particles this variable is defined for. For pions 
$Z_\pi =\log \left( \frac{dE/dx}{\left\langle dE/dx\right\rangle _\pi \left( p\right) }\right)$.\footnote{For a discussion of the expected value of $dE/dx$ for electrons and pions see Section \ref{sec:dedx}}
If this formula is applied to a pion track $dE/dx$, the variable Z is independent of the track momentum and length.
A set of data plots of track $Z_\pi$ for various slices of track momentum is shown in the Figure \ref{fig:dedx_slices}.

In the Figure \ref{fig:dedx_slices} pions are on the right, electrons/positrons are on the left, except for the bottom right histogram, where pion peak is located on the left. Shape and the mean of the pion band stay constant in all momentum slices. For a single track with momentum below 130 MeV/c, pions can be rejected at $95\%$ confidence level  by selecting tracks with $Z_\pi < -0.3$. For the tracks with momenta between 130 MeV/c and 140 MeV/c rejection of pions at $95\%$ confidence level cuts away a large fraction of electron/positron tracks, therefore we chose not to do particle identification for this $p$-slice tracks. For track momenta between 140 MeV/c and 265 MeV/c electron and pion bands overlap completely. Thus, for tracks with momenta between 130 MeV/c and 265 MeV/c particle identification based just on the information from a single track $dE/dx$ is impossible. However, doing a simultaneous $e^{+}e^{-}$ $dE/dx$ identification of $e^{+}$ and $e^{-}$ tracks allows us to resolve this problem partially. For tracks with momenta above 265 MeV/c electron/positron tracks can again be separated from pion tracks by requiring $Z_\pi > 0.25$.

The basic idea of the simultaneous pair identification is that if one track is identified as an electron (positron) with a high degree of likelihood, and the pair is identified as ultra-peripheral, the other track is most likely a positron (electron), since there are no processes that could produce a pair consisting of an electron and a charge-1 particle other than a positron in the ultra-peripheral AuAu collision.\footnote{Contribution from ultra-peripheral $\mu ^{+} \mu ^{-}$ pairs, where one muon decays via the channel $\mu  \to e + \nu \overline \nu $ is negligible due to the small $\mu ^{+} \mu ^{-}$ cross-section.}  Table \ref{tab:twotrackdedx} explains the details of the simultaneous $e^{+}e^{-}$ $dE/dx$ identification for a pair of tracks labeled track 1 (always a lower momentum track) and track 2 with momenta $p_1$ and $p_2$ respectively.

\small
\begin{table}[htbl]
\centering
\begin{tabular}{|l|l|l|}
\hline \multicolumn{2}{|l|}{\bf track momentum configuration} & {\bf accept as $e^{+}e^{-}$ if} \\
\hline
{\small \! 1} & $p_1 \! < \! 130$ {\small \ MeV/c, }$p_2 \! < \! 130${\small \ MeV/c}&  $Z_\pi^1<-0.3${\small and }$Z_\pi^2<-0.3$ \\
\hline
{\small \! 2} & $p_1 \! < \! 130${\small \ MeV/c, 130}$\! < \! p_2 \! < \! 265${\small \ MeV/c}& $Z_\pi^1 \! < \! -0.3${\small \ and }$Z_\pi^2${\small \ within \ }$2\sigma${\small \ of \ }$\left\langle dE/dx\right\rangle _e ^{mean}$ \\
\hline
{\small \! 3} & {\small 130}$\! < \! p_1 \! < \! 265${\small \ MeV/c, 130 }$\! < \! p_2 \! < \! 265${\small \ MeV/c}&{\small identification via \dedx \ impossible}\\
\hline
{\small \! 4} & $p_1 \! < \! 130${\small \ MeV/c, }$p_2>265${\small \ MeV/c}& $Z_\pi^1 \! < \! -0.3${\small \ and }$Z_\pi^2 \! > \! 0.25$ \\
\hline
{\small \! 5} & {\small 130}$ \! <p_1 \! <265${\small \ MeV/c, }$p_2 \! > \! 265${\small \ MeV/c}& $Z_\pi^2 \! > \! 0.25${\small \ and }$Z_\pi^1${\small \ within \ }$2\sigma${\small \ of \ }$\left\langle dE/dx\right\rangle _e ^{mean}$ \\
\hline
{\small \! 6} & $p_1 \! > \! 265${\small \ MeV/c, \ }$p_2 \! > \! 265${\small  \ MeV/c}& $Z_\pi^1>0.25${\small \ and }$Z_\pi^2>0.25$ \\
\hline
\end{tabular}
\caption[Simultaneous  $e^+e^-$ pair identification with \dedx]{Conditions that track 1 $Z_\pi^1$ and track 2 $Z_\pi^2$ must satisfy for a simultaneous \dedx \ $e^+e^-$ pair identification.}
\label{tab:twotrackdedx}
\end{table}
\normalsize

According to the table \ref{tab:twotrackdedx}, simultaneous $e^{+}e^{-}$ pair identification is possible when either one of the tracks has momentum below 130 MeV/c or above 265 MeV/c. As we discussed in Chapter \ref{ch:MonteCarlo}, the number of produced $e^{+}e^{-}$ pairs drops very rapidly with increasing invariant mass of the pair (equivalently, with increasing track momenta). For this reason, the number of pairs with a track having a total momentum above 265 MeV/c is expected to be extremely small, and in the sample of 60,000 events preliminary analysis did not find any such \ee pairs. Therefore, to identify a pair as an $e^{+}e^{-}$ we require that the pair tracks' $dE/dx$ satisfy either the requirement 1 or 2 in the table \ref{tab:twotrackdedx}.

The requirement that at least one of the tracks should have a momentum less than $130$ MeV/c and the other track should have the momentum less than $250$ MeV/c effectively limits the maximal invariant mass of the identified $e^+e^-$ pairs. We need to use Monte Carlo simulations to study the efficiency of simultaneous \ee   identification as a function of pair mass. 

\subsubsection{Pair Identification Efficiency from Monte Carlo}
\begin{figure}[!ht]
\centering
\includegraphics[width=200pt]{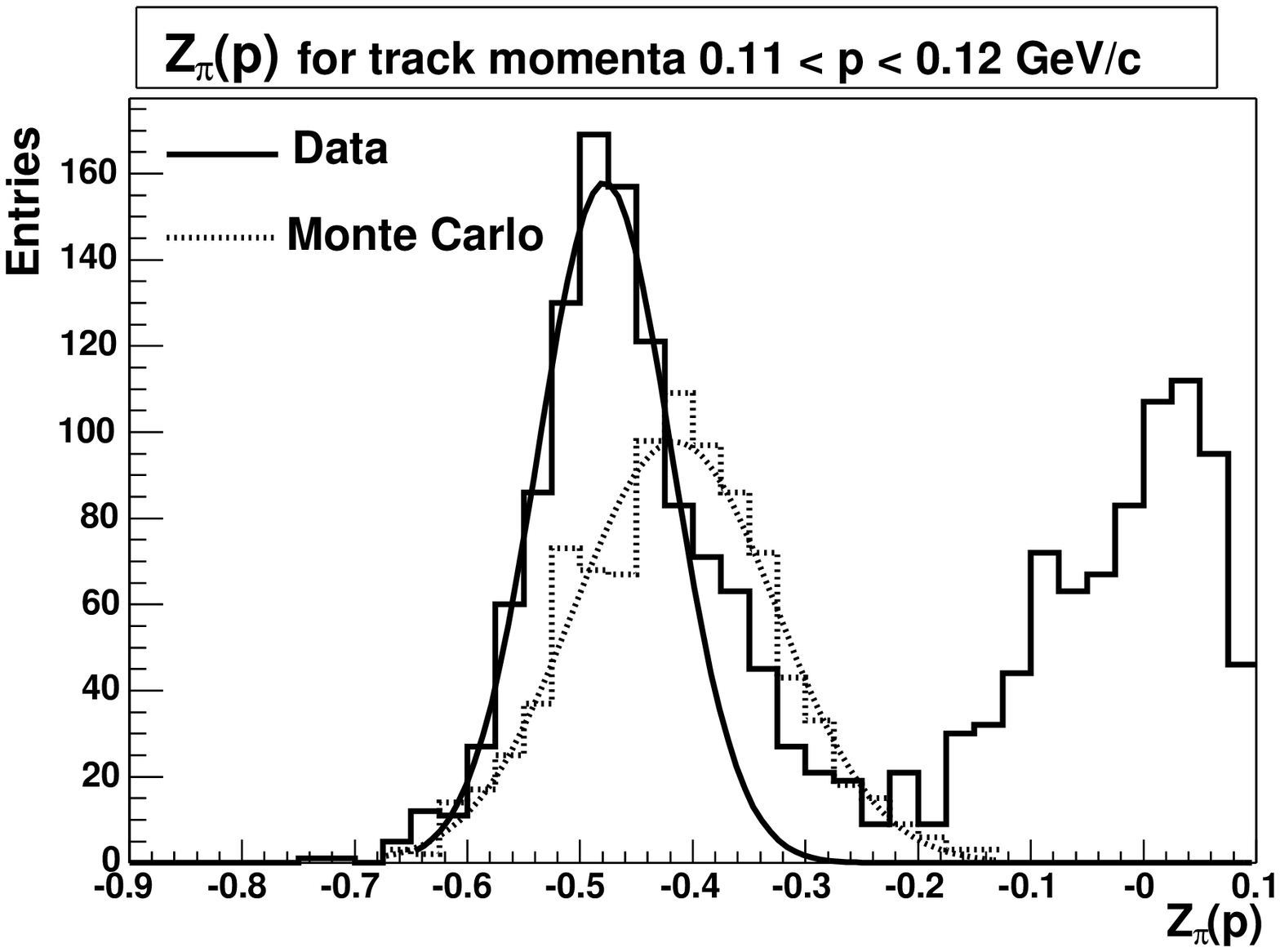}
\includegraphics[width=200pt]{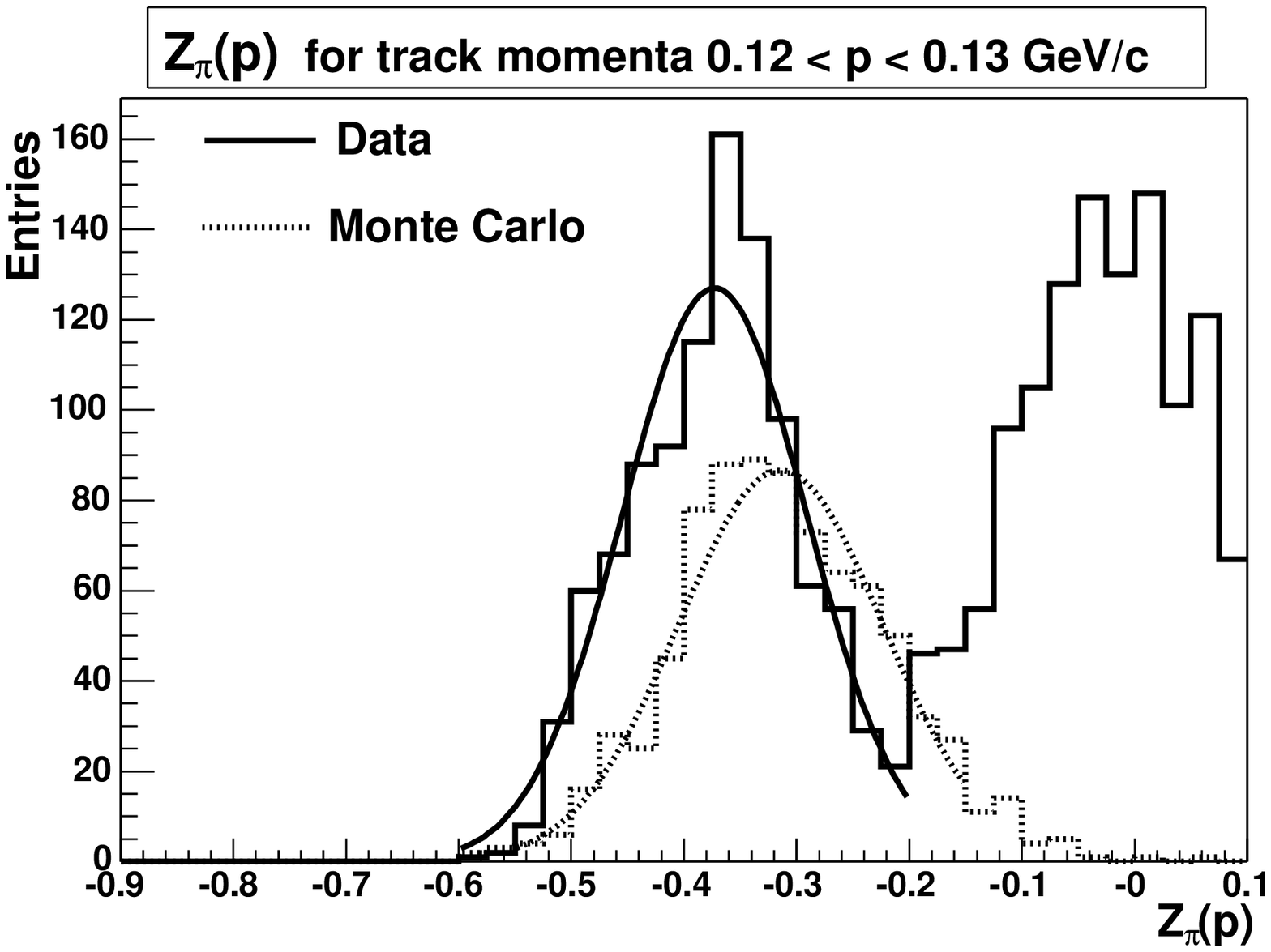}
\caption[Comparison of data and GEANT $Z_{\pi}(p)$]{Comparison of data and GEANT $Z_{\pi}(p)$. Solid histogram - data, dashed histogram - simulations. Data histograms show electron peaks (left) and pion peaks.  Electron peaks in the data and simulations were fit with Gaussian profiles to find mean and width of the peaks.}
\label{fig:DeDxSimuVsReco}
\end{figure}

We begin the pair identification study by comparing the electron track $Z_{\pi}$ distributions in STAR TPC for the data and simulations. Figure \ref{fig:DeDxSimuVsReco} shows the distribution of track $Z_{\pi}$  for 3 track momentum slices. There is a significant difference in the mean and the width of the real electron peak (solid black histogram, left peak is electrons) and the simulated electrons (dashed histogram). This is due to the fact that GEANT package has an imprecise model for the energy loss by electrons. We applied mean shift and the width scaling to the simulations to make sure that the corrected simulated electron $Z_{\pi}$ peak matched the data:
\begin{equation}
\label{eqn:correctZ}
Z_\pi ^{corrected \ simu} \left( p \right) = \left( {Z_\pi ^{simu} \left( p \right) - \left\langle {Z_\pi ^{simu} \left( p \right)} \right\rangle ^{mean} } \right) \cdot \frac{{\sigma _Z^{data} \left( p \right)}}{{\sigma _Z^{simu} \left( p \right)}} + \left\langle {Z_\pi ^{data} \left( p \right)} \right\rangle ^{mean} 
\end{equation}
where ${\left\langle {Z_\pi ^{data} \left( p \right)} \right\rangle ^{mean} }$, ${\left\langle {Z_\pi ^{simu} \left( p \right)} \right\rangle ^{mean} }$, ${\sigma _Z^{data} \left( p \right)}$ and ${\sigma _Z^{simu} \left( p \right)}$ are the means and widths of Gaussian fits to the $Z_{\pi}$ data and simulations distributions in momentum slices 0.11 GeV/c $<p<$ 0.12 GeV/c and 0.12 GeV/c $<p<$ 0.13 GeV/c in Figure \ref{fig:DeDxSimuVsReco}.

The pair identification efficiency can be computed as: 

\begin{equation}
\label{eqn:PIDeff}
E_{\text{pair identification}}  = \frac{{\# \text{generated pairs passing cuts 1 or 2}}}{{\# \text{generated pairs}}}
\end{equation}

\begin{figure}
\centering
\includegraphics[width=200pt]{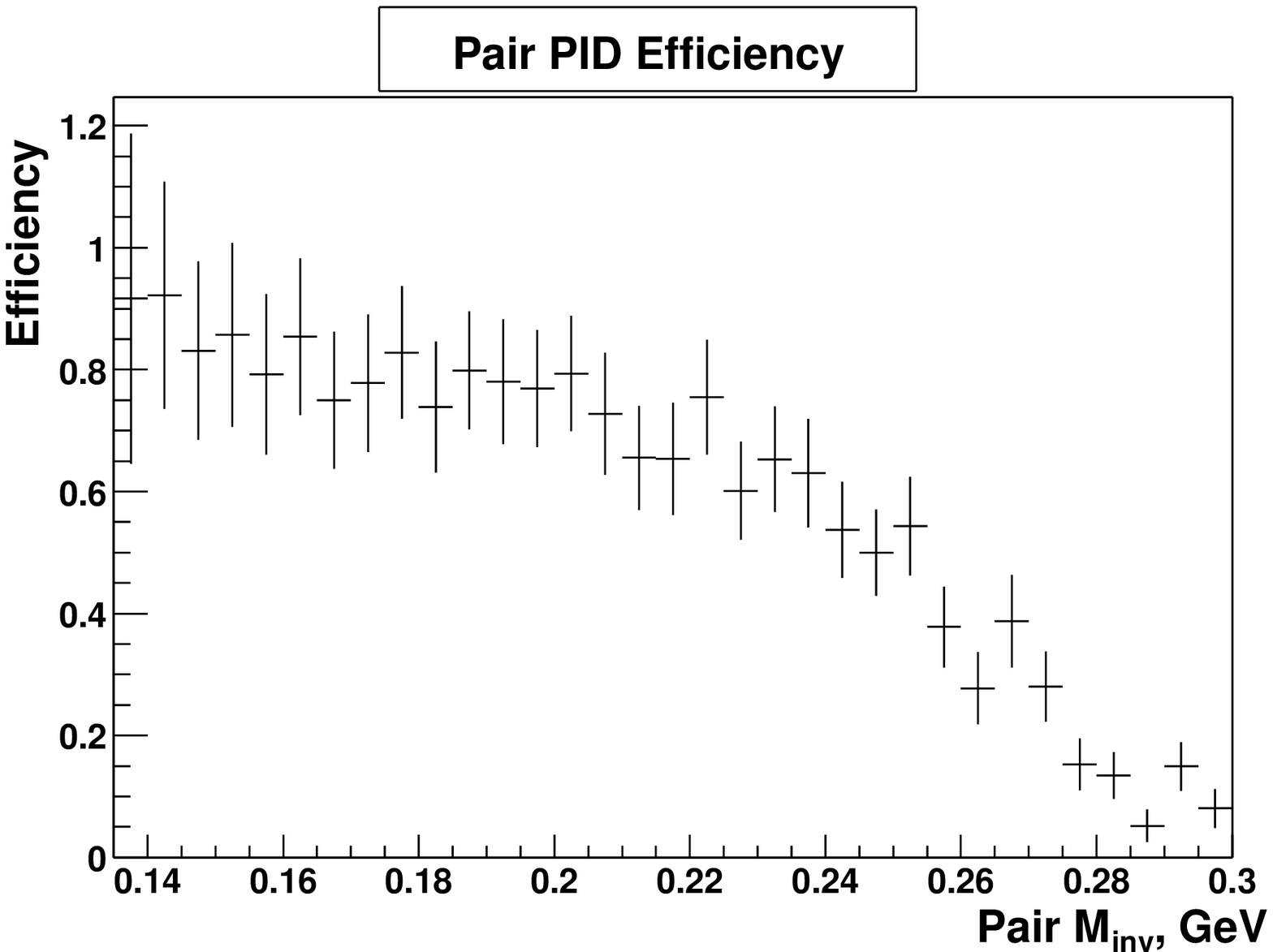}
\includegraphics[width=200pt]{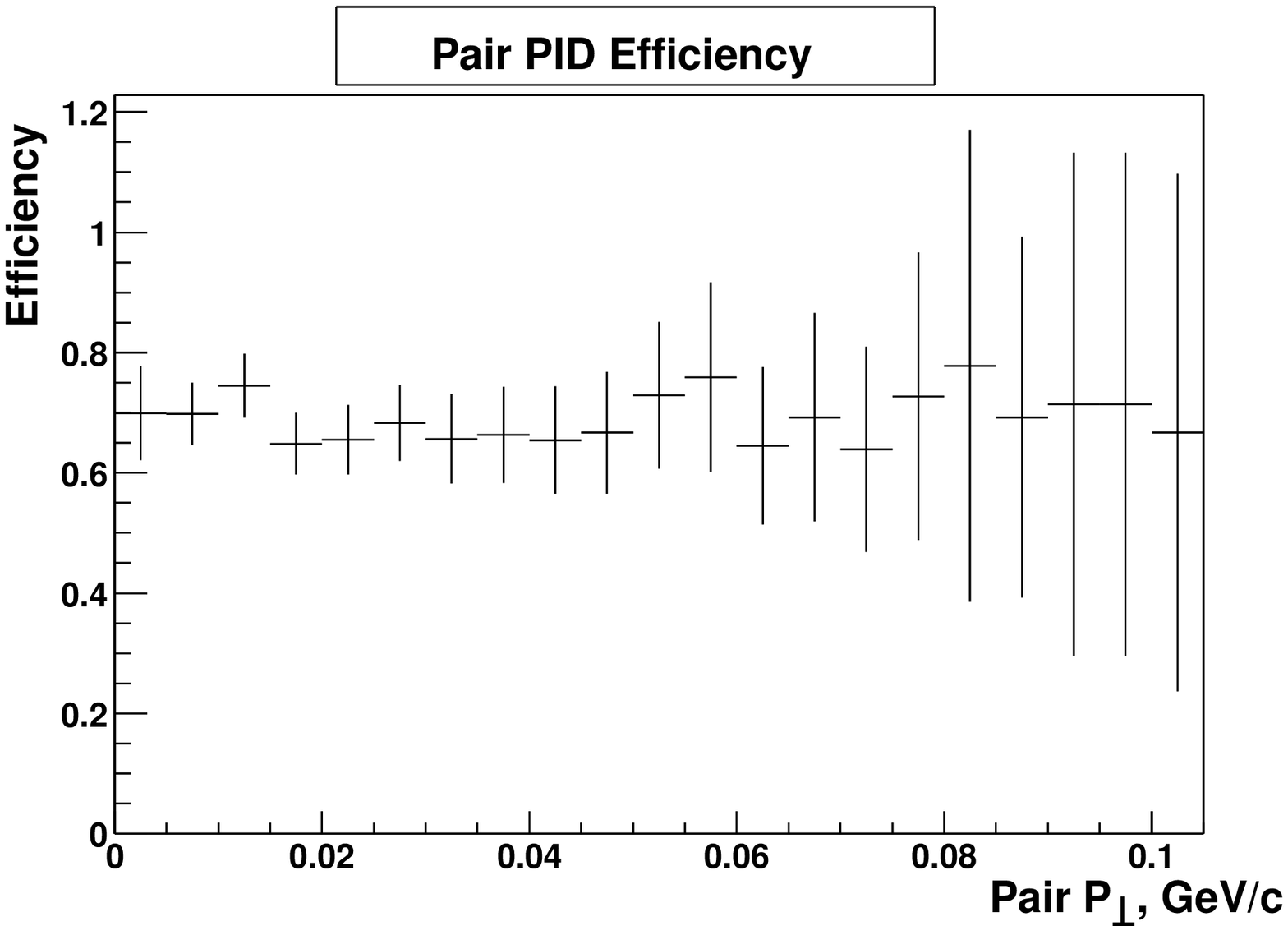}
\caption[Efficiency of pair identification]{Efficiency of pair identification vs. pair mass and total transverse momentum.}
\label{fig:PIDeffVSminv}
\end{figure}

The identification efficiency distribution as the function of pair mass and total pair transverse momentum is shown in Figure \ref{fig:PIDeffVSminv}. The pair identification efficiency is close to $1.0$ for low invariant masses (both tracks are in the region $p<130$ MeV/c, where electrons/positrons are distinguishable from pions), and drops to zero for  invariant masses on the order of $\sim 300$ MeV. We chose to place an upper cut-off on the pair invariant mass: $M_{inv}<265$ MeV to avoid the region where the efficiency is very low and the Monte Carlo does not provide an adequate description of the data.

\subsection{Vertex Finding}
\label{sub:VertexingEfficiency}
Reconstruction of an \ee   pair consists of track reconstruction of both tracks and finding a common vertex for these tracks. We need to study the vertex finding efficiency as a function of pair kinematical variables to ensure that we are not trying to detect pairs in the region where the vertex finding efficiency is low.

We define vertex finding efficiency as:

\begin{equation}
\label{eqn:Vertexing}
E_{\text{vert}}  = \frac{{\# \text{reconstructed $e^+e^-$ pairs with a found vertex}}}{{\# \text{reconstructed $e^+e^-$ pairs within acceptance}}}
\end{equation}

\begin{figure}
\centering 
\includegraphics[width=200pt]{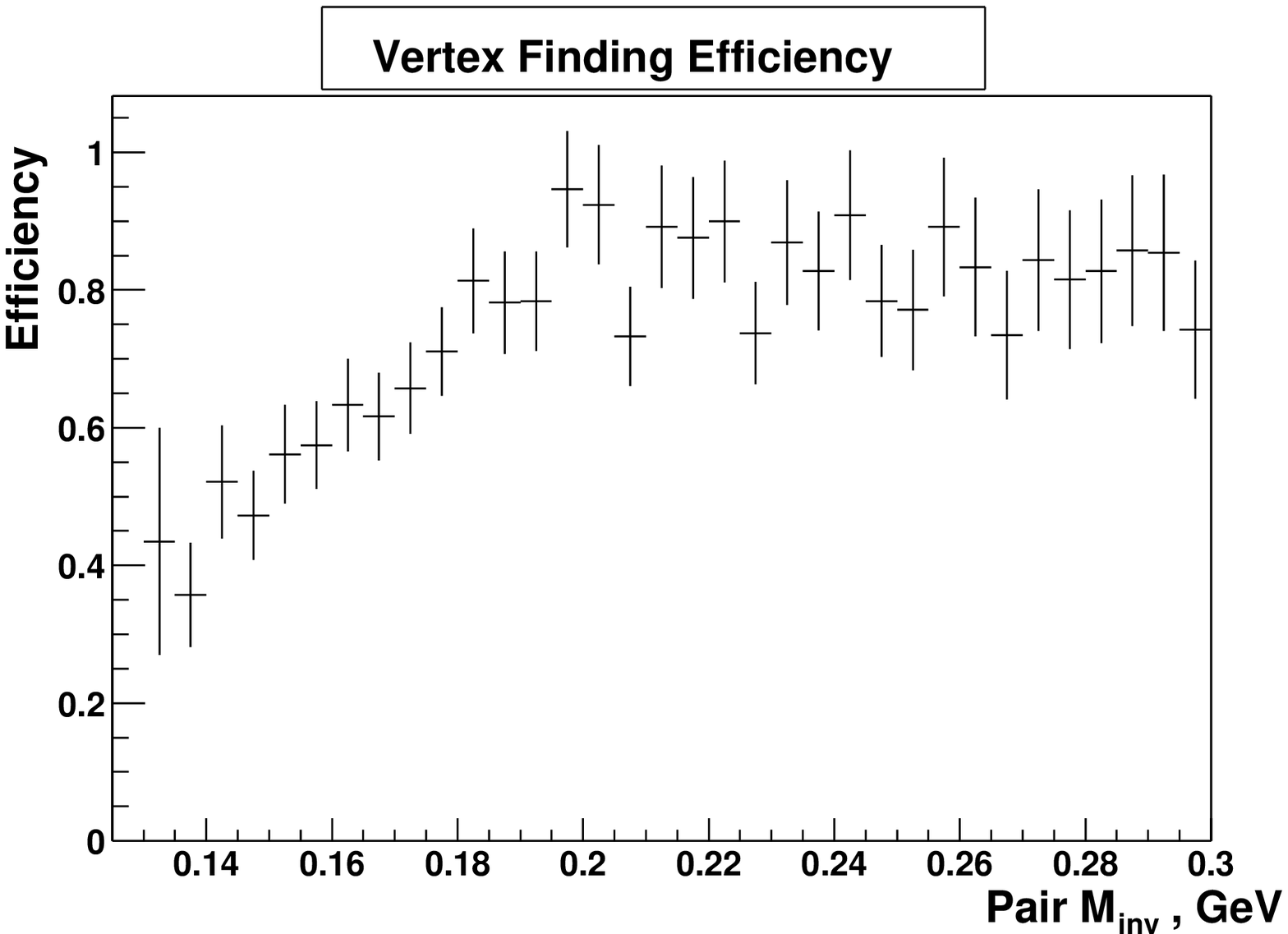}
\includegraphics[width=200pt]{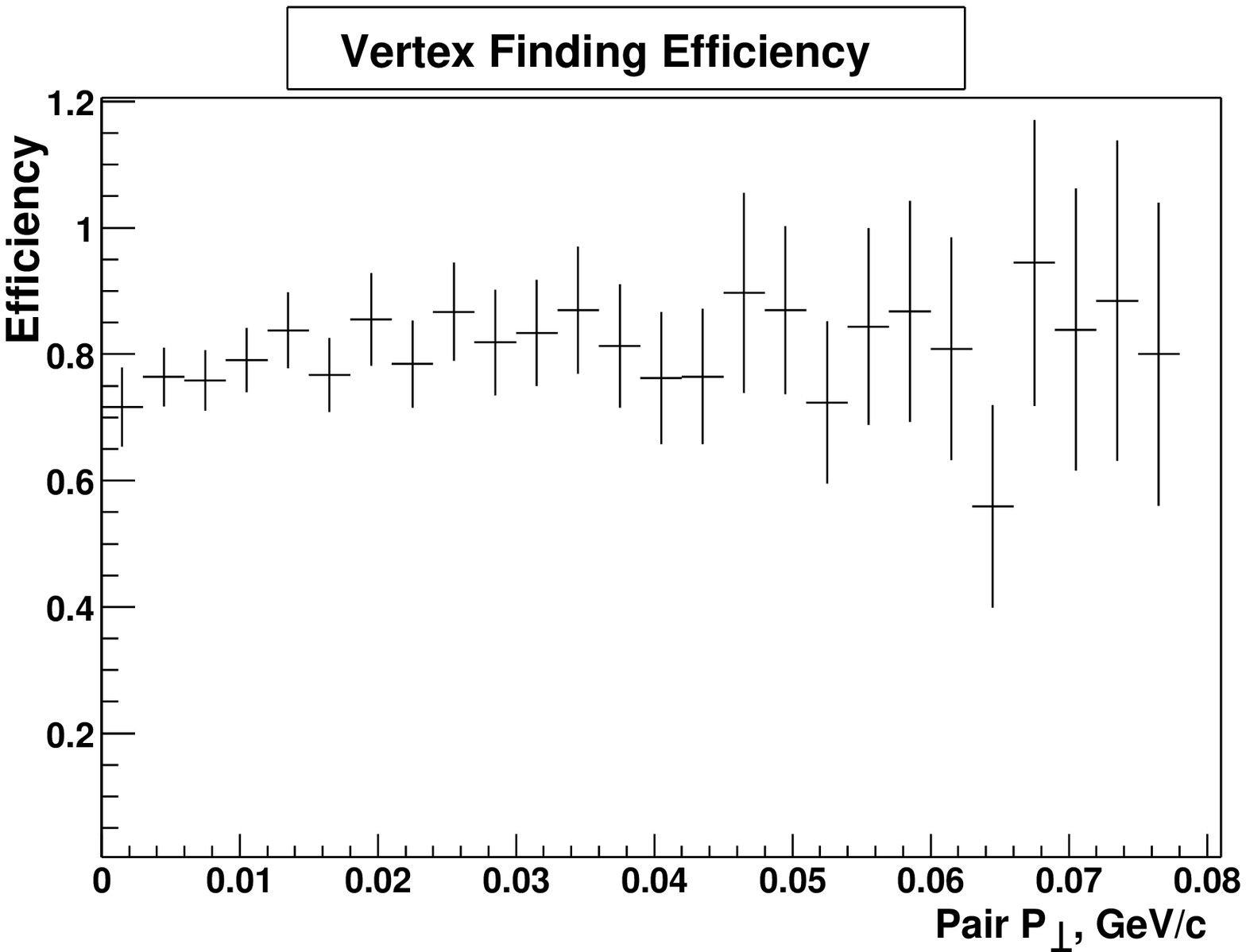}
\caption[Vertex finding efficiency]{Vertex finding efficiency as function of pair invariant mass and transverse momentum.}
\label{fig:Vertexing}
\end{figure}

Figure \ref{eqn:Vertexing} shows the vertex finding efficiency as a function of pair invariant mass and total transverse momenta. The efficiency is flat as a function of $p_{\perp}^{tot}$ and $M_{inv}$ for $M_{inv}>200$ MeV, but drops off for invariant masses below 200 MeV. This is due to the fact that the low invariant mass pairs are composed of low momentum tracks, and the low momentum tracks undergo a lot of multiple scattering in the beam pipe or the inner field cage. As the result, the momentum of these tracks is often mis-reconstructed and projecting the tracks back to the common origin yields a very large distance of closest approach between the tracks, which causes the vertex finding routine to fail. We set a cut on the minimal invariant mass of the reconstructed pair at $M_{inv}>140$ MeV.

\begin{figure}
\centering
\includegraphics[width=200pt]{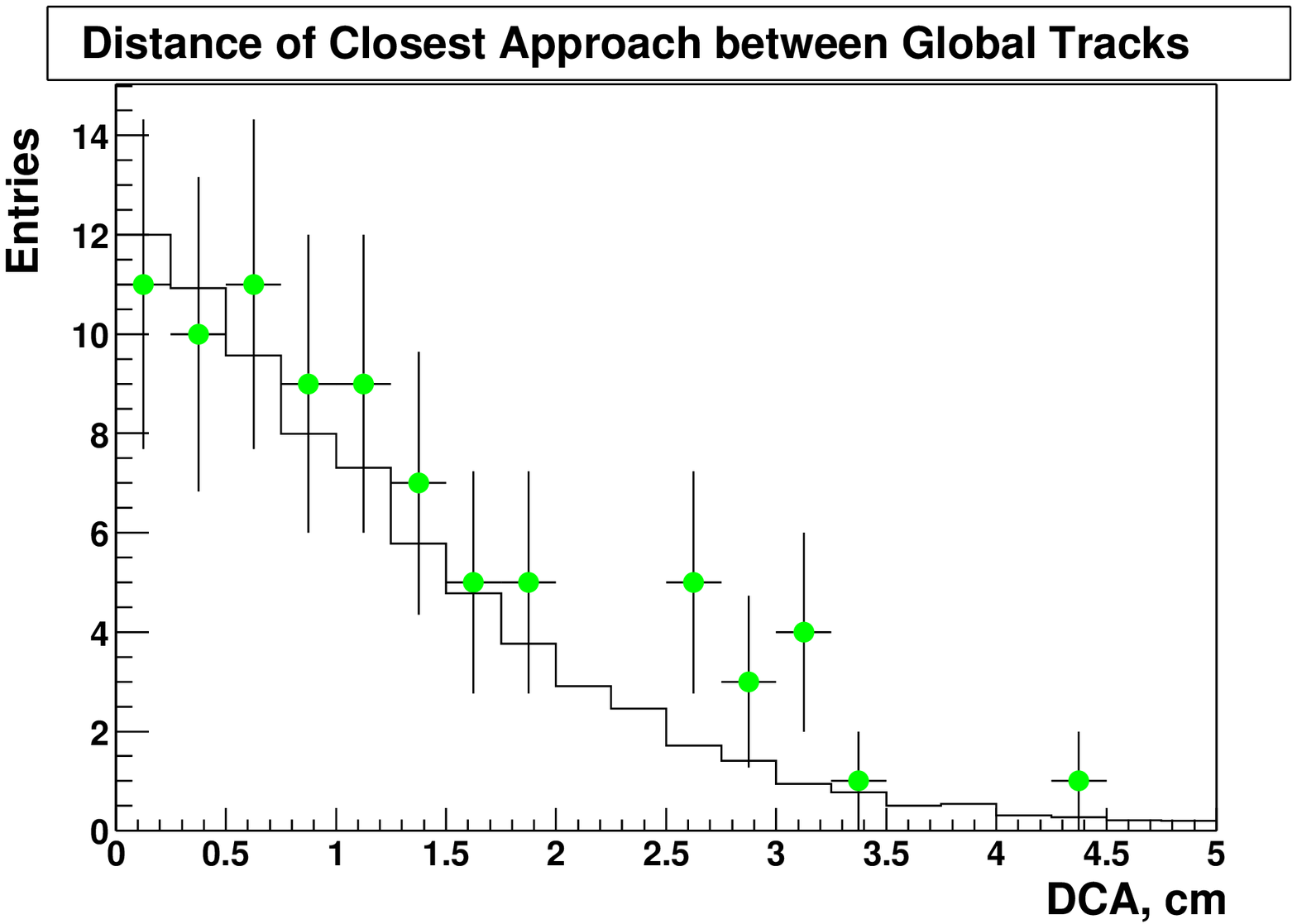}
\includegraphics[width=200pt]{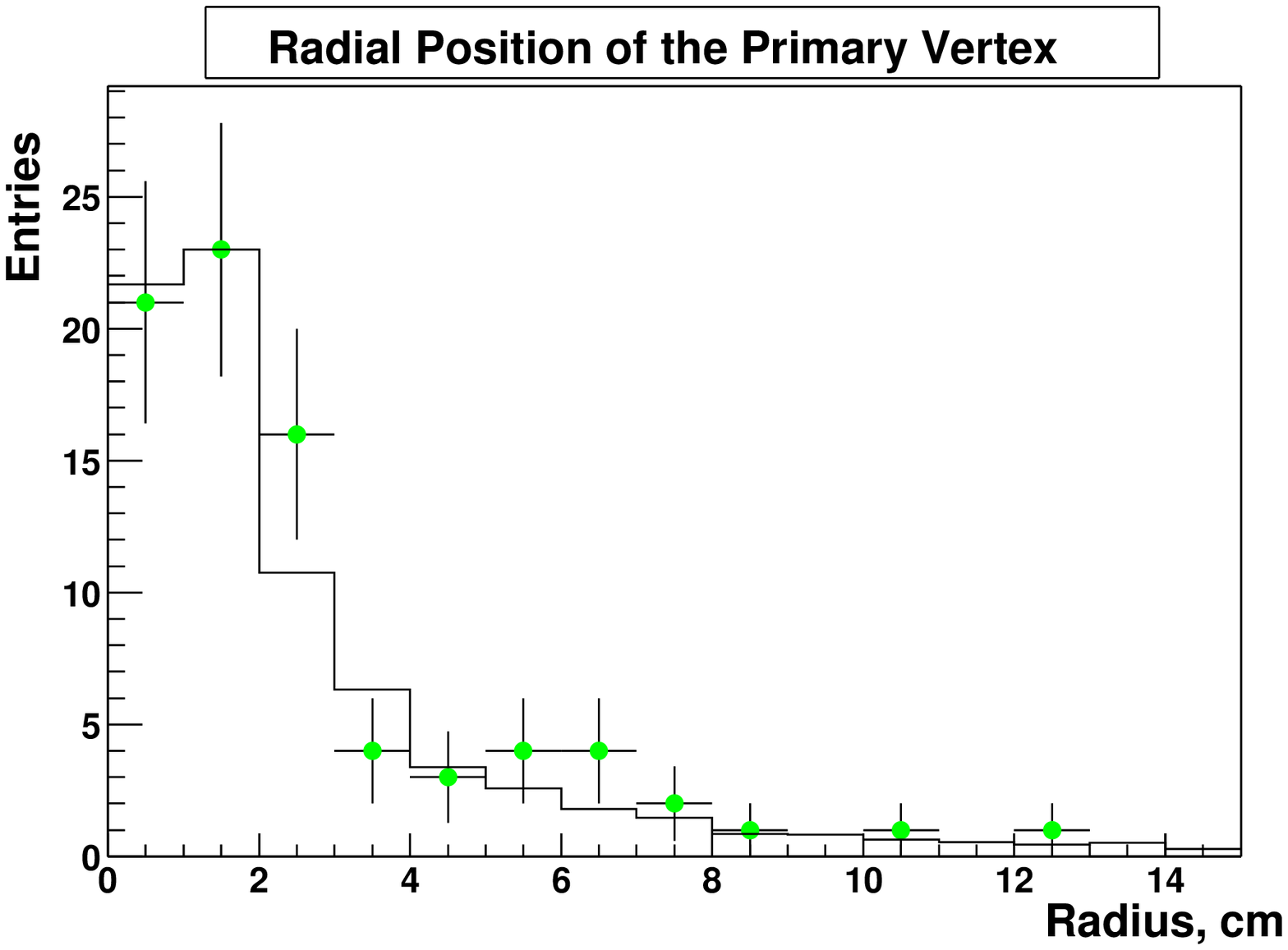}
\caption[Distance of closest approach between the tracks and the vertex radial position]{Distribution of the distance of closest approach between the global tracks and the vertex radial position.}
\label{fig:DCAandRVert}
\end{figure}

We wish to restrict our attention to the events with the good quality of vertex finding. 
For the events with only two tracks used for the determination of the vertex, the quality of the vertex fit depends on the distance of closest approach between the global tracks which are used for the vertex finding ($DCA$). We limit this variable to less than 4 cm. As a secondary check, we make sure that the found vertex is not too far (in the transverse plane) to the most likely position $(0,0)$. This is the second cut: $R_{vert} < 9.0$ cm. The distribution of $DCA$ and the vertex radius $R_{vert}$ are shown in Figure \ref{fig:DCAandRVert} for pairs which satisfy the definition of ultra-peripheral events (Section \ref{sub:UPC}). To increase event sample, particle identification was not applied, but we selected events with both tracks' momenta below 200 MeV/c. This ensures that $R_{vert}$ and $DCA$ distributions are close the distributions for a pure \ee sample. The agreement between the Monte Carlo and the data is excellent, which means that the simulation of the multiple scattering in the detector and its effect on tracking and vertex finding is very accurate.

\begin{figure}[t]
\centering
\begin{minipage}[t]{.45\textwidth}
\begin{center}
\includegraphics[width=200pt,clip=true]{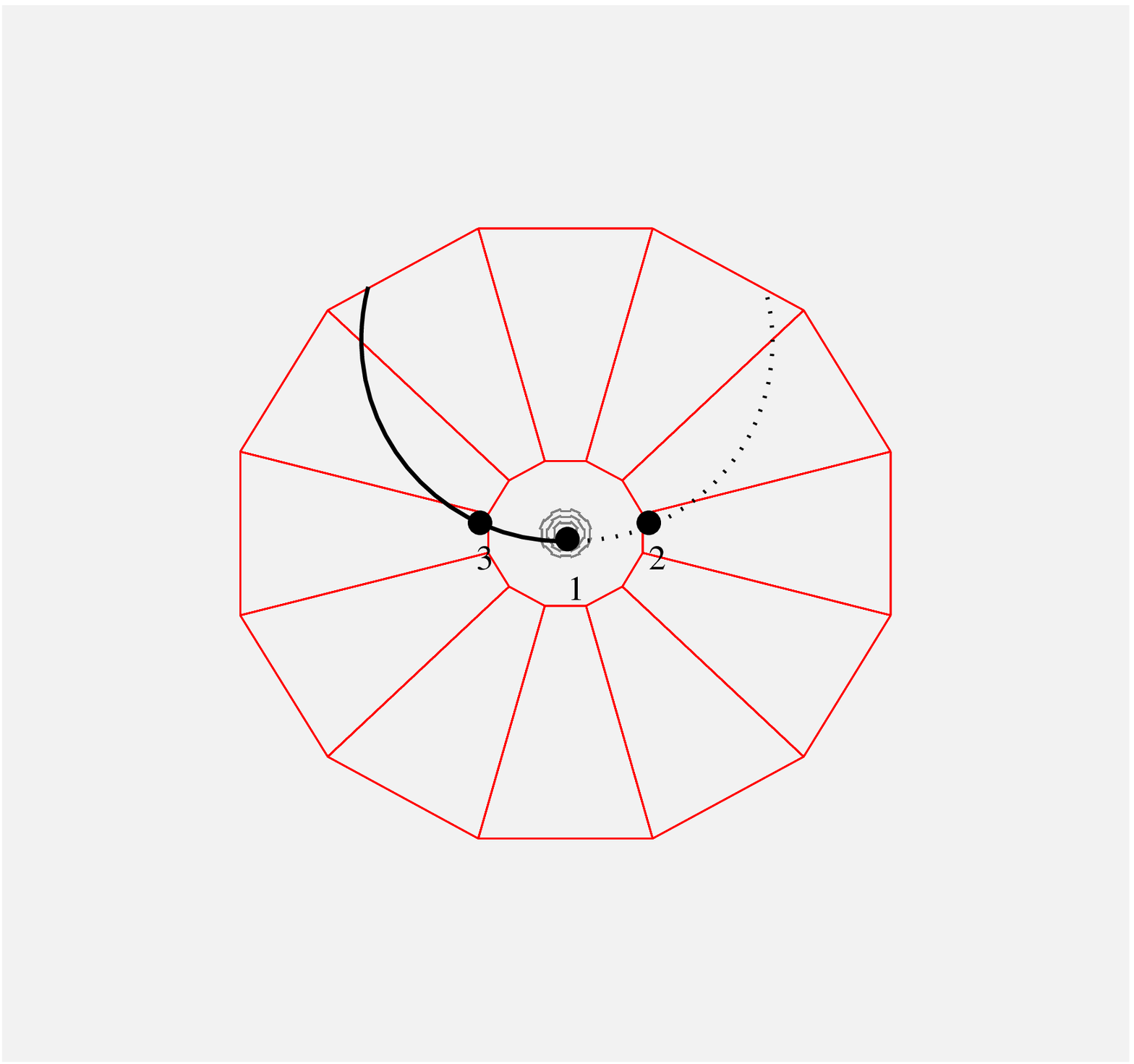}
\caption[Transverse vertex position in STAR TPC]{$xy$ projection of the STAR TPC. Solid  and dashed line arches represent electron/positron tracks. Points 1, 2 and 3 represent possible reconstructed transverse vertex positions.}
\label{fig:helices}
\end{center}
\end{minipage}
\hfill
\begin{minipage}[t]{.45\textwidth}
\vfil\null
\begin{center}
\includegraphics[width=200pt]{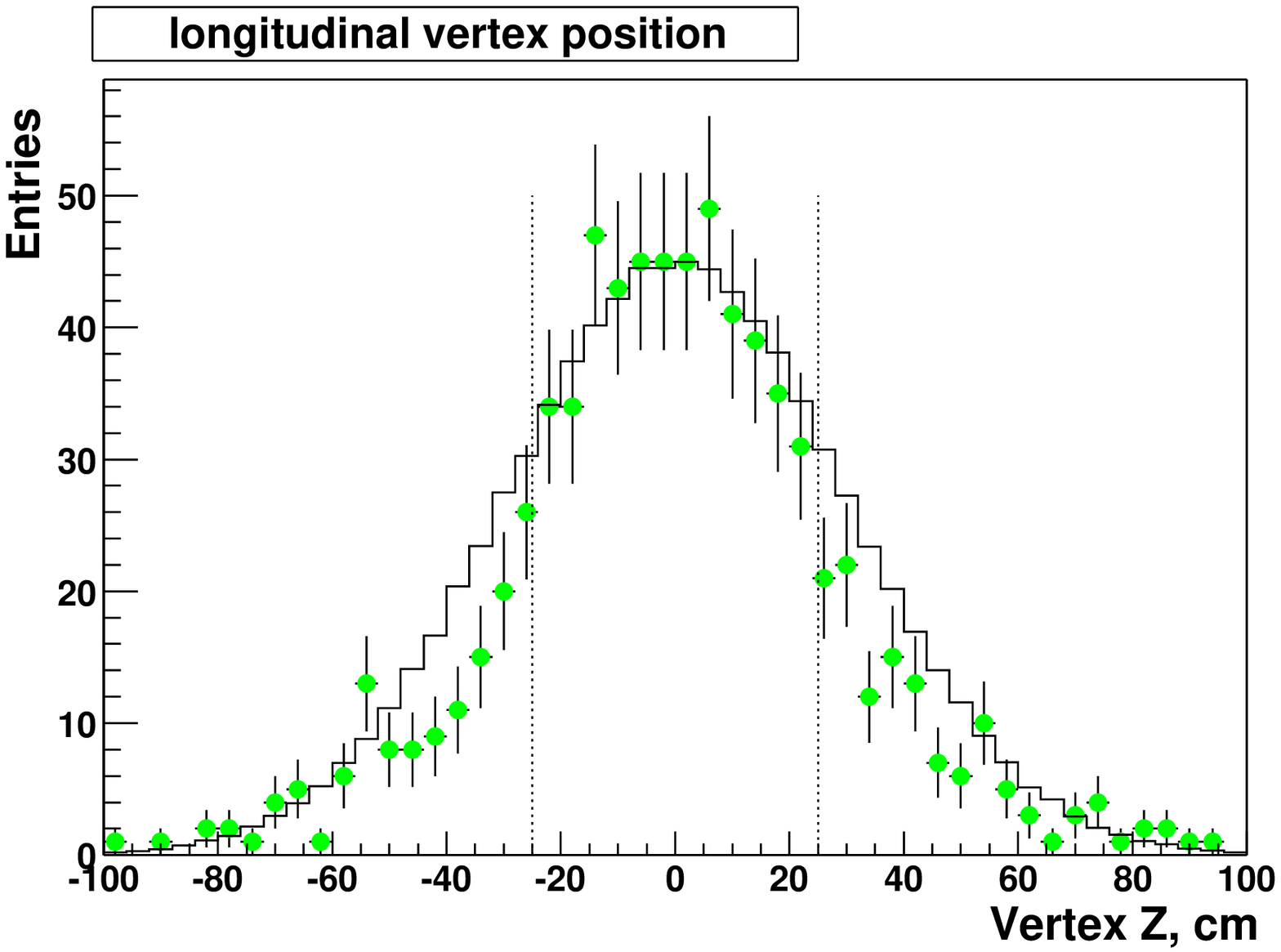}
\caption[Longitudinal vertex position distribution in Monte Carlo and data]{$z$ vertex position distribution in Monte Carlo (solid, scaled) and data (dots). Events with $|z_{vert}|<25$ cm are used for luminosity normalization to hadronic cross-section.}
\label{fig:zvertex}
\end{center}
\end{minipage}
\end{figure}

The distributions of $R_{vert}$ in the data and Monte Carlo have long tails, with $\sim 3\%$ of the evens having a reconstructed $R_{vert}>9$ cm. This is due to the fact that determination of the point of the closest approach between 2 helical tracks of similar curvatures and opposite transverse momenta is unstable. For instance, in the limiting case of electron and positron tracks having the total transverse momentum of zero, the position of the point of closest approach between the two tracks in the transverse plane is completely undefined (Figure \ref{fig:helices}). However, the uncertainty in the $R_{vert}$ determination has little effect on the determination of tracks' absolute momenta and the opening angle of the pair. 

\subsubsection{Longitudinal Vertex Position Distribution}
Figure \ref{fig:zvertex} compares the $z_{vert}$ vertex distribution for ultra-peripheral events in the data 
and \ee Monte Carlo simulation.  The Monte Carlo simulation (solid) has a Gaussian shape with half-width of 30 cm. The distribution of the vertices in the ultra-peripheral events in the data (dots) is close to Gaussian in the region $|z_{vert}|<25$ cm, but drops of steeper than the Gaussian distribution outside of this cut. This is the effect of the vertex position cut in the Minimum Bias Vertex Trigger. In the data, the ratio of the events with a vertex inside $|z_{vert}|<100$ cm to the events inside $|z_{vert}|<25$ cm is $1.52 \pm 0.07$. In the Monte Carlo, 99\% of the events are within $|z_{vert}|<100$ cm. We have performed a check procedure on Monte Carlo simulation, grouping events in different slices of the $z_{vert}$ vertex position, and the distribution of the kinematic variables of the reconstructed events was found to be identical for all $z_{vert}$ slices.

\subsubsection{Total Available Luminosity}

To increase the statistics, we would like to use ultra-peripheral events with a $z$ vertex inside a 100 cm range, instead of 25 cm (see Section \ref{sub:luminosity}). For such events, the luminosity times trigger efficiency can be estimated from the luminosity for ultra-peripheral events satisfying a $|z_{vert}|<25$ cm cut as:
\begin{equation}
\label{eqn:totlumi}
\centering
L_{tot}\cdot \text{Eff}^{\text{trig}}=\left( L_{tot}\cdot \text{Eff}^{\text{trig}}\right) _{|z_{vert}|<25}\times \frac{N_{{|z|<100}}}{N_{|z|<25}}
\end{equation}
The total available luminosity was found to be $93.8\text{ mb}^{-1} $ with a systematic error of 10\%.

\subsection{Summary of All Cuts}

Table (\ref{tab:allCuts}) summarizes the analysis cuts. The first part of the table \ref{tab:allCuts} contains cuts that select events with one primary $e^+e^-$ pair, produced in an ultra-peripheral electromagnetic interaction. The second part of the table selects only those of the identified $e^+e^-$ primary pairs which are produces within the defined acceptance region. The third part selects the pairs which have high quality of the vertex finding. 
 
\begin{table}[!h]
\centering
\begin{tabular}{|l|c|}
\hline \multicolumn{2}{|c|}{\bf Identification Cuts} \\
\hline \multicolumn{2}{|l|}{Event triggered with Minimum Bias Vertex trigger} \\
\hline \multicolumn{2}{|l|}{Number of global tracks in event $<$ 5} \\
\hline \multicolumn{2}{|l|}{Number of primary pairs in event $ = $ 1} \\
\hline \multicolumn{2}{|l|}{Primary pair total charge $ = $ 0} \\
\hline \multicolumn{2}{|l|}{Total transverse momentum of the primary pair $ < 100$ MeV/c} \\
\hline \multicolumn{2}{|l|}{The $z$ position of the primary vertex is within 100 cm from the zero} \\
\hline
EITHER $p_1 \! < \! 130$ MeV/c, $p_2 \! < \! 130$ MeV/c & and $Z_\pi^1<-0.3$ , $Z_\pi^2<-0.3$ \\
\hline
OR $p_1 \! < \! 130$ MeV/c, $130 \! < \! p_2 \! < \! 265$ MeV/c & and $Z_\pi^1 \! < \! -0.3$ , $Z_\pi^2$ within $2\sigma$ of $\left\langle dE/dx\right\rangle _e ^{mean}$ \\
\hline \multicolumn{2}{|l|}{ \ } \\
\hline \multicolumn{2}{|c|}{\bf Acceptance Cuts} \\
\hline \multicolumn{2}{|l|}{ $ \left| \eta_1 \right| < 1.15 $ , $ p_{\bot 1} >65$ MeV/c ,  $ \left| \eta_2 \right| < 1.15 $ , $ p_{\bot 2}  >65$ MeV/c }\\ 
\hline \multicolumn{2}{|l|}{140 MeV  $< M_{inv}  < $ 265 MeV} \\
\hline \multicolumn{2}{|l|}{ \ } \\
\hline \multicolumn{2}{|c|}{\bf Data Quality Cuts} \\
\hline \multicolumn{2}{|l|}{ Number of hits per track greater than 9 } \\
\hline \multicolumn{2}{|l|}{Radial position of the vertex $R_{vert}$ within 9.0 cm from $(0,0)$} \\
\hline \multicolumn{2}{|l|}{Global counterparts of the primary tracks have $DCA<4$ cm} \\
\hline 
\end{tabular}
\caption{Analysis cuts to select events of the type $AuAu \rightarrow Au^*Au^* + e^+e^-$.}
\label{tab:allCuts}
\end{table}
\section{Resulting Raw Distributions}
\begin{figure}
\centering
\includegraphics[width=450pt]{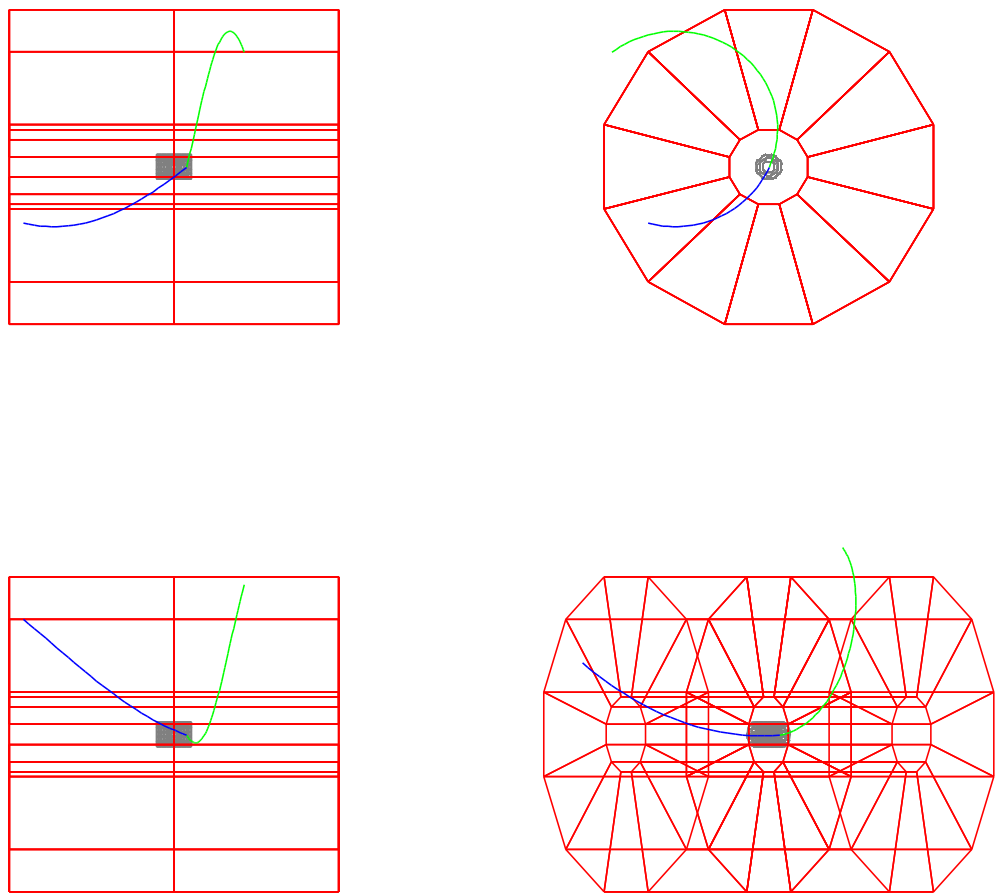}
\caption[Four views of an $e^+e^-$ pair in the STAR TPC]{Four views of an $e^+e^-$ pair in the STAR TPC. Top right: $zy$ projection, top left: $xy$ projection (transverse), bottom left: $zx$ projection, bottom right: 3-dimensional view.}
\label{fig:4views}
\end{figure}

After applying the cuts in table \ref{tab:allCuts} to the Minimum Bias Vertex dataset, we get 52 identified events of the type $AuAu \rightarrow Au^*Au^* + e^+e^-$. Figure \ref{fig:4views} presents an event display with $x$, $y$ and $z$ projections together with a side-on view of one of the identified events, showing the reconstructed tracks inside a STAR TPC. The event shows the typical characteristics of the reactions of this type: the tracks have a rather low transverse momenta (as evidenced by the strong curvature of the tracks) and are forward-peaked, with one track exiting the STAR TPC through the endcap. In the transverse projection (upper right figure) the tracks appear to be back-to-back and have very similar absolute values of the transverse momenta, this leads to a very low total transverse momentum of the pair, which is one of the distinctive characteristics of the $e^+e^-$ pairs produced in the coherent ultra-peripheral electromagnetic interactions of the heavy ions.

\begin{figure}[p]
\includegraphics[width=200pt]{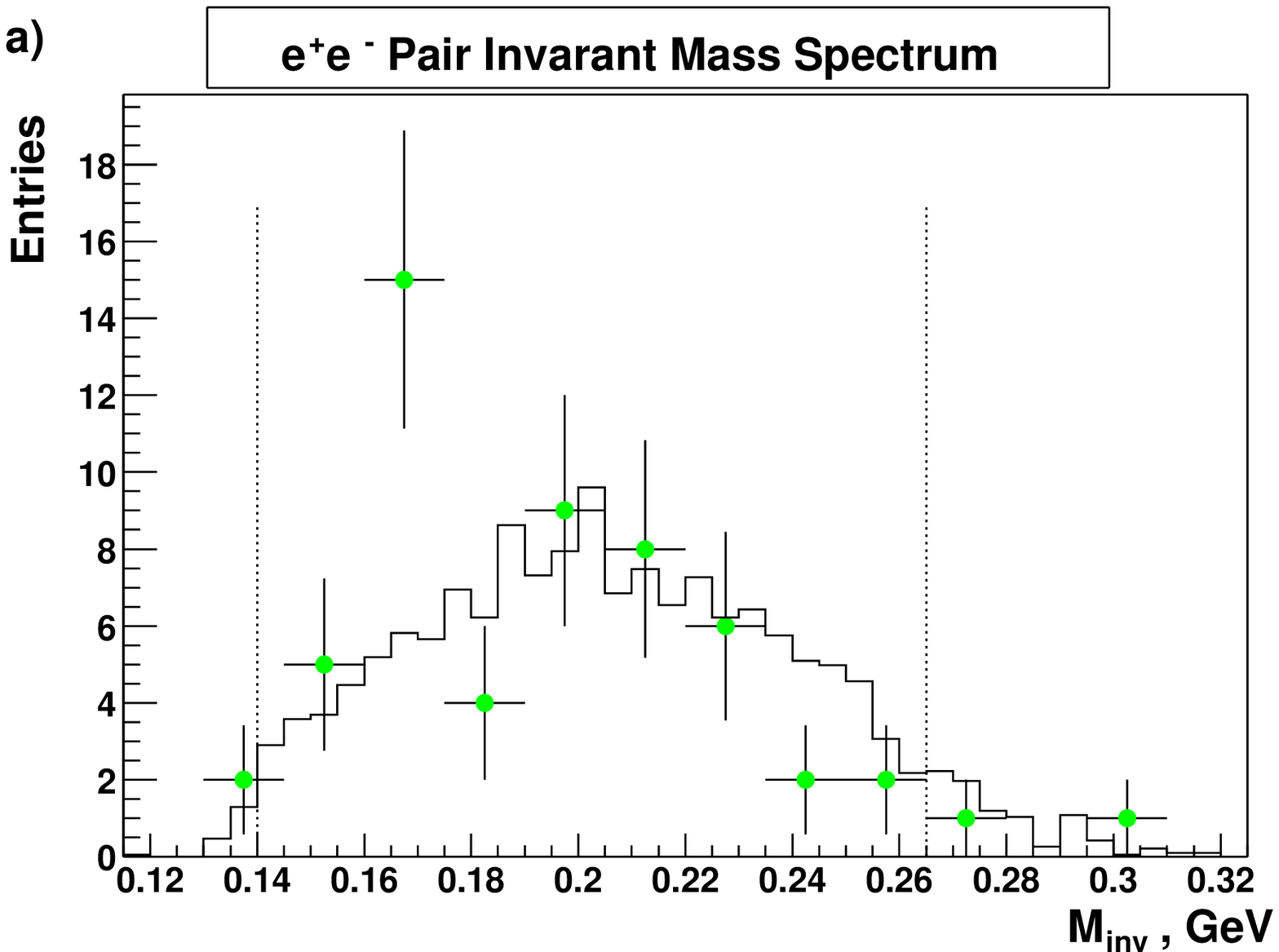}
\includegraphics[width=200pt]{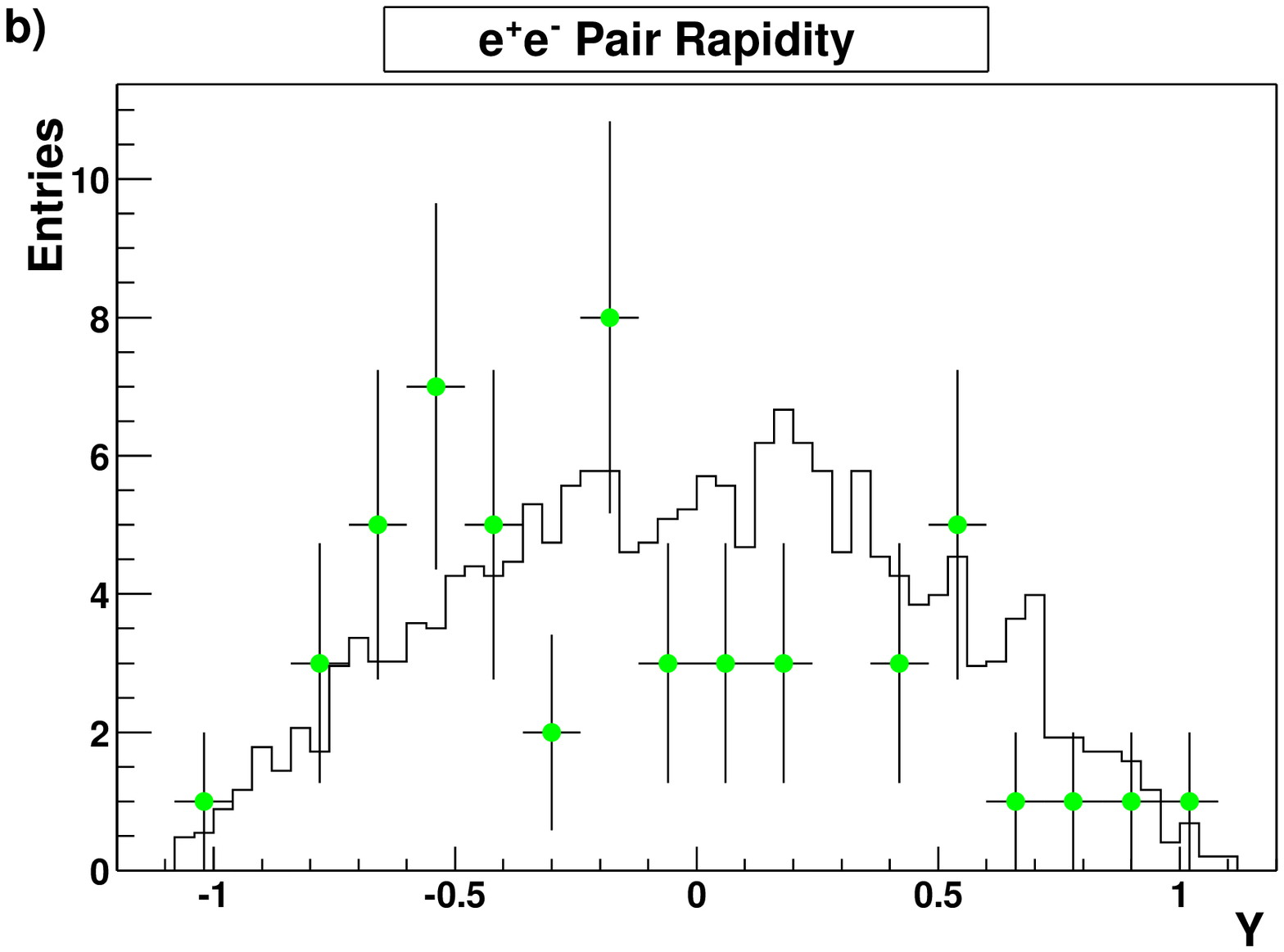}
\includegraphics[width=200pt]{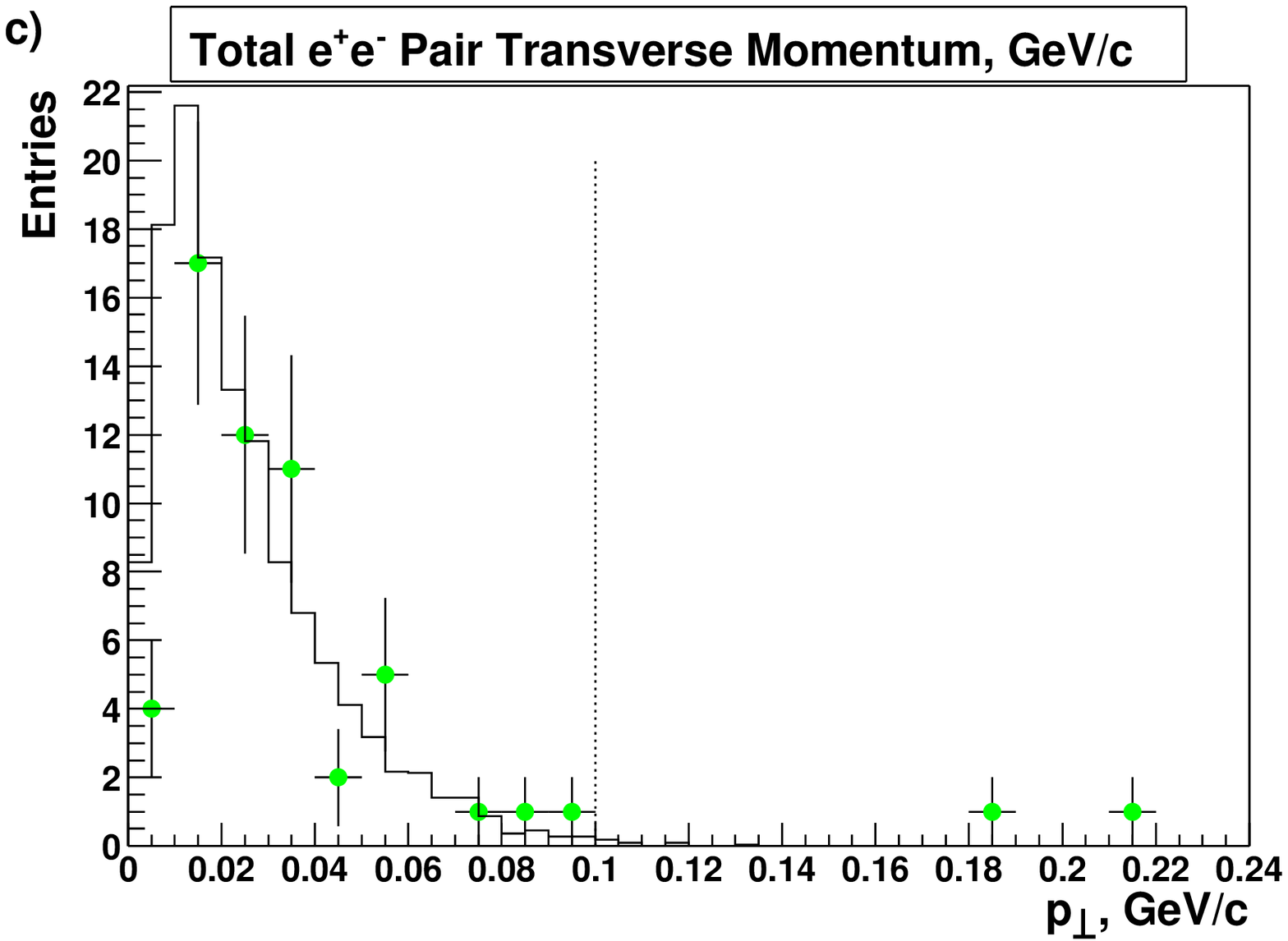}
\includegraphics[width=200pt]{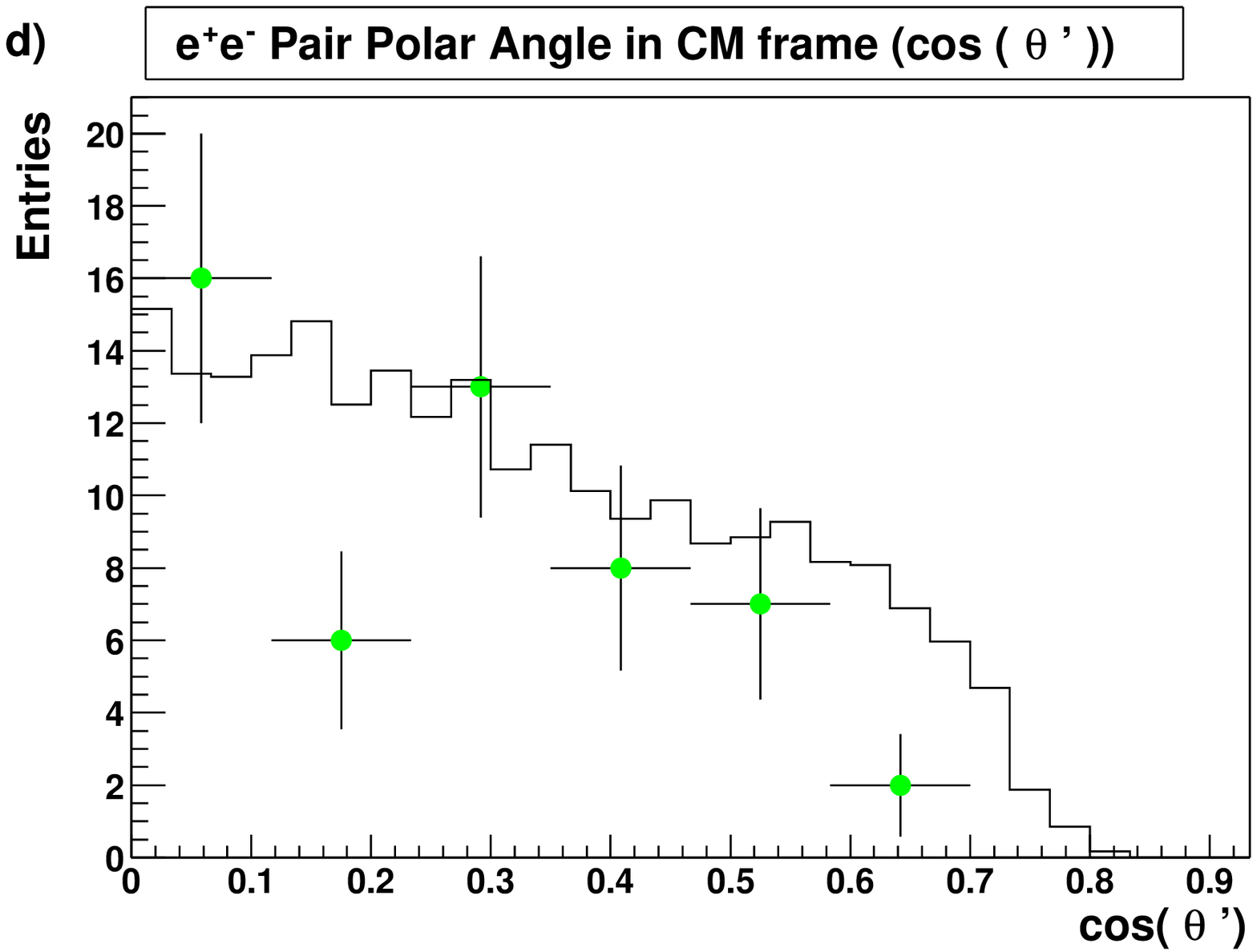}
\includegraphics[width=200pt]{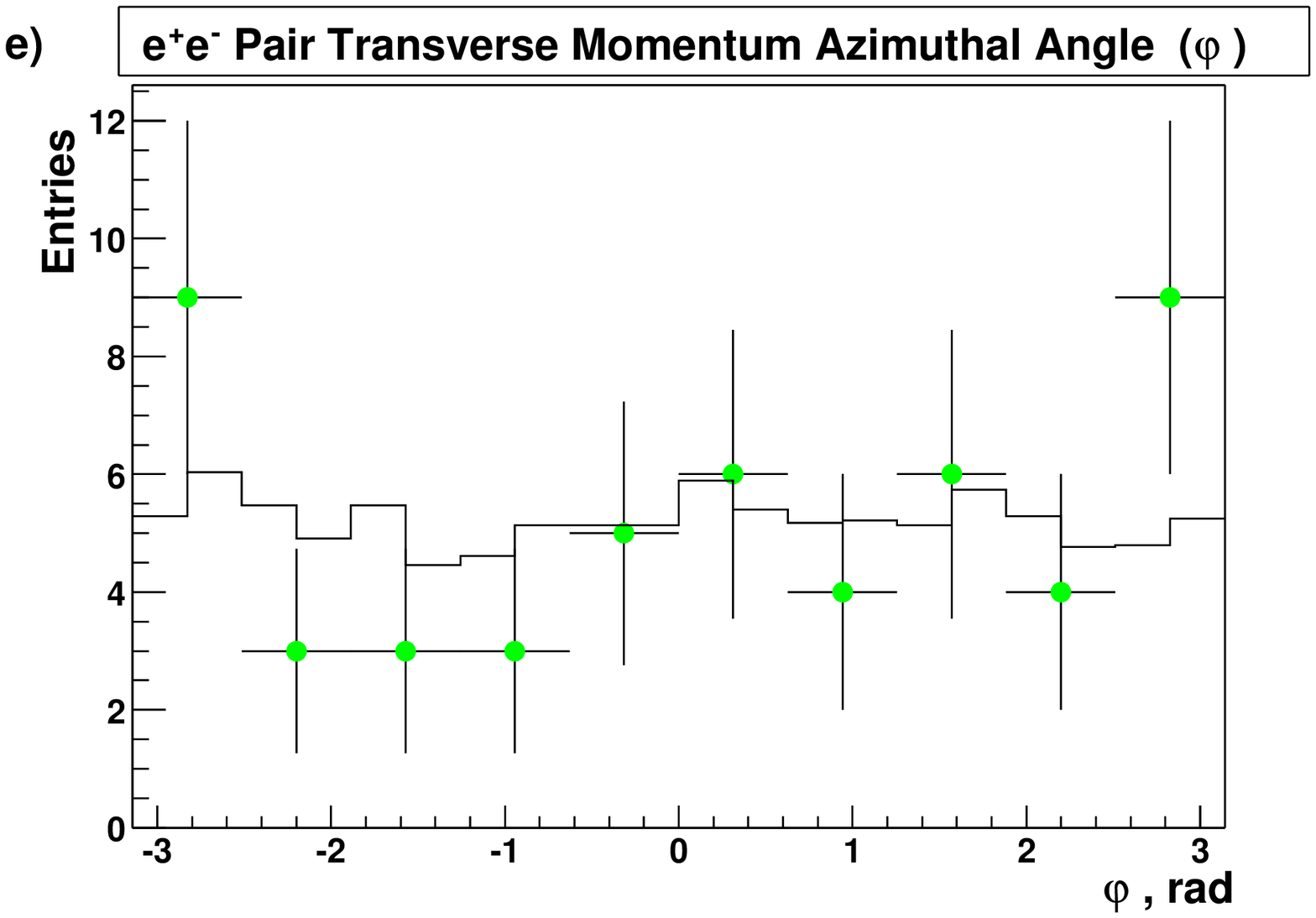}
\includegraphics[width=200pt]{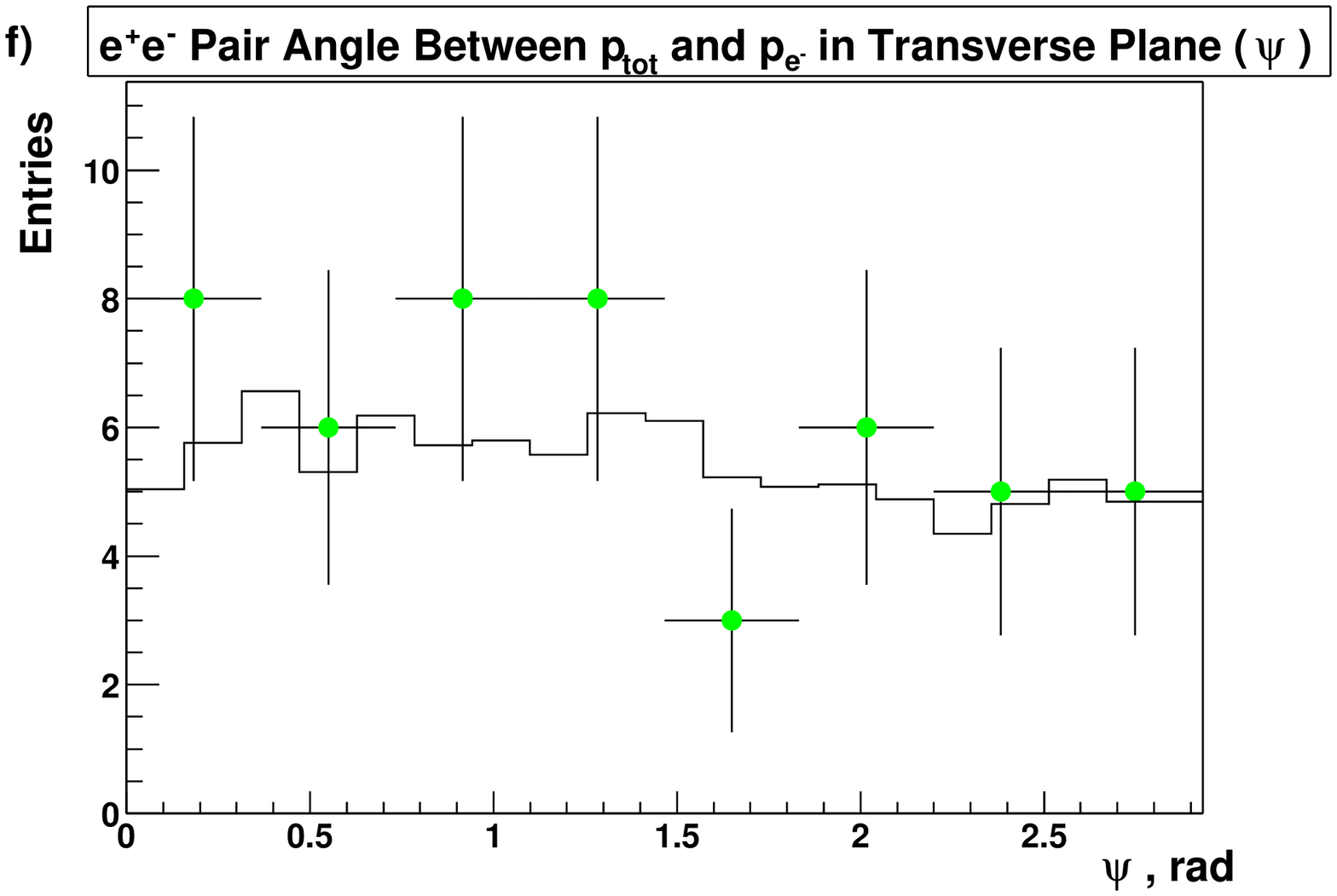}
\caption[Raw kinematic distributions of identified $e^+e^-$ pairs in STAR]{Raw kinematic distributions of identified $e^+e^-$ pairs in STAR. a)  $M_{inv}$, b) $Y$, c) $p_{\perp }^{tot}$, d) $\cos \left( \theta ' \right)$, e) $\varphi$, f) $\psi $. Dots - data, solid lines - Monte Carlo. Dashed lines in items a) and c) represent acceptance region cuts.}
\label{fig:rawDistributions}
\end{figure}

Figure \ref{fig:rawDistributions} presents the raw (not corrected for detector efficiency) distributions of the pair kinematic parameters in the selected 52 events of the type \AuAuee. The $p_{\perp}^{tot}$ distribution shows a clear peak at the low transverse momenta, which is a signature of ultra-peripheral \ee   pairs. The distributions of $\varphi$ and  $\psi $ are flat, as expected for $\gamma \gamma \rightarrow e^+e^-$. The agreement between the Monte Carlo results and the data is good, which will allow us to use the Monte Carlo simulations for the efficiency corrections. 

\begin{figure}
\centering
\includegraphics[width=200pt]{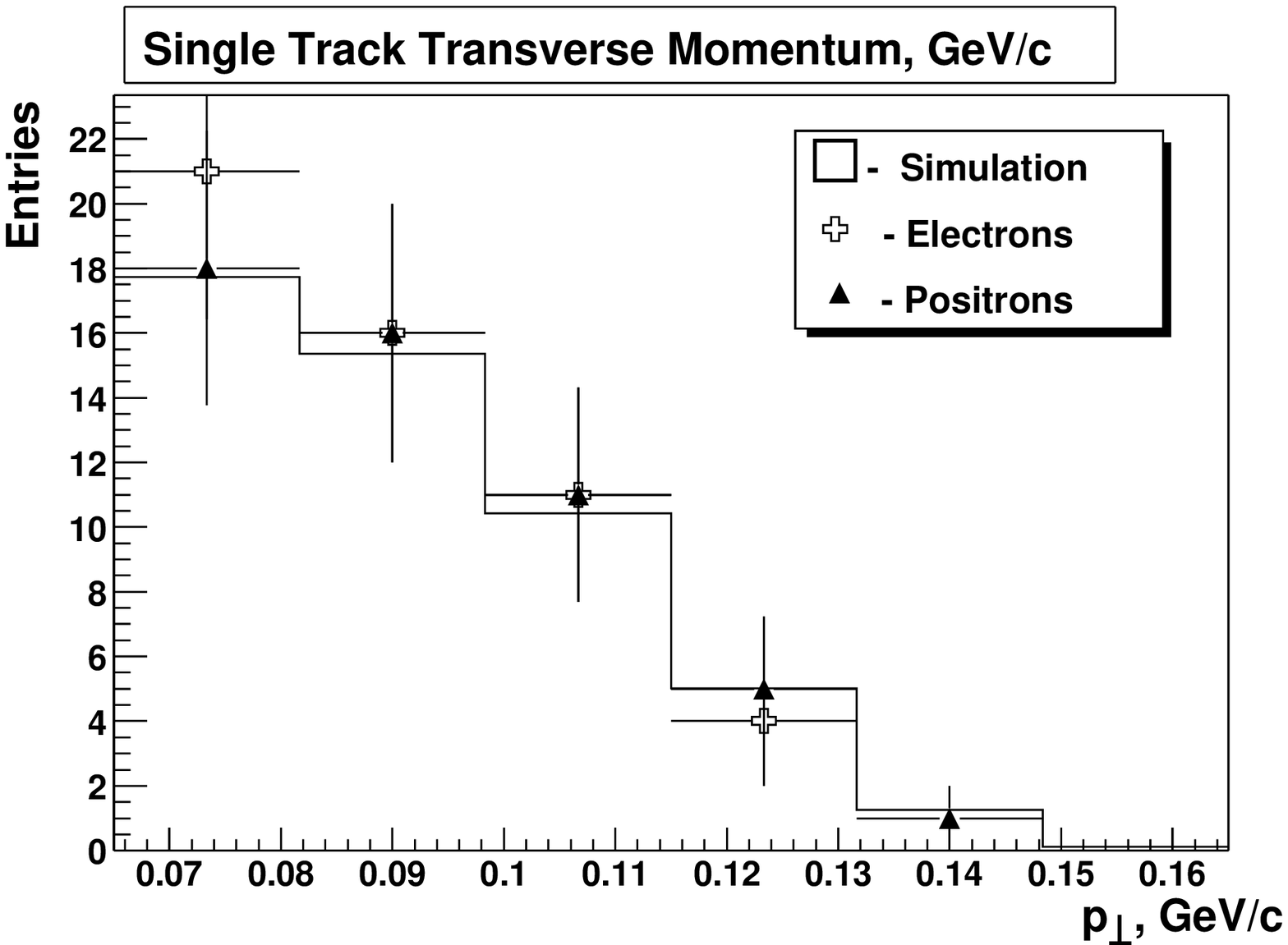}
\includegraphics[width=200pt]{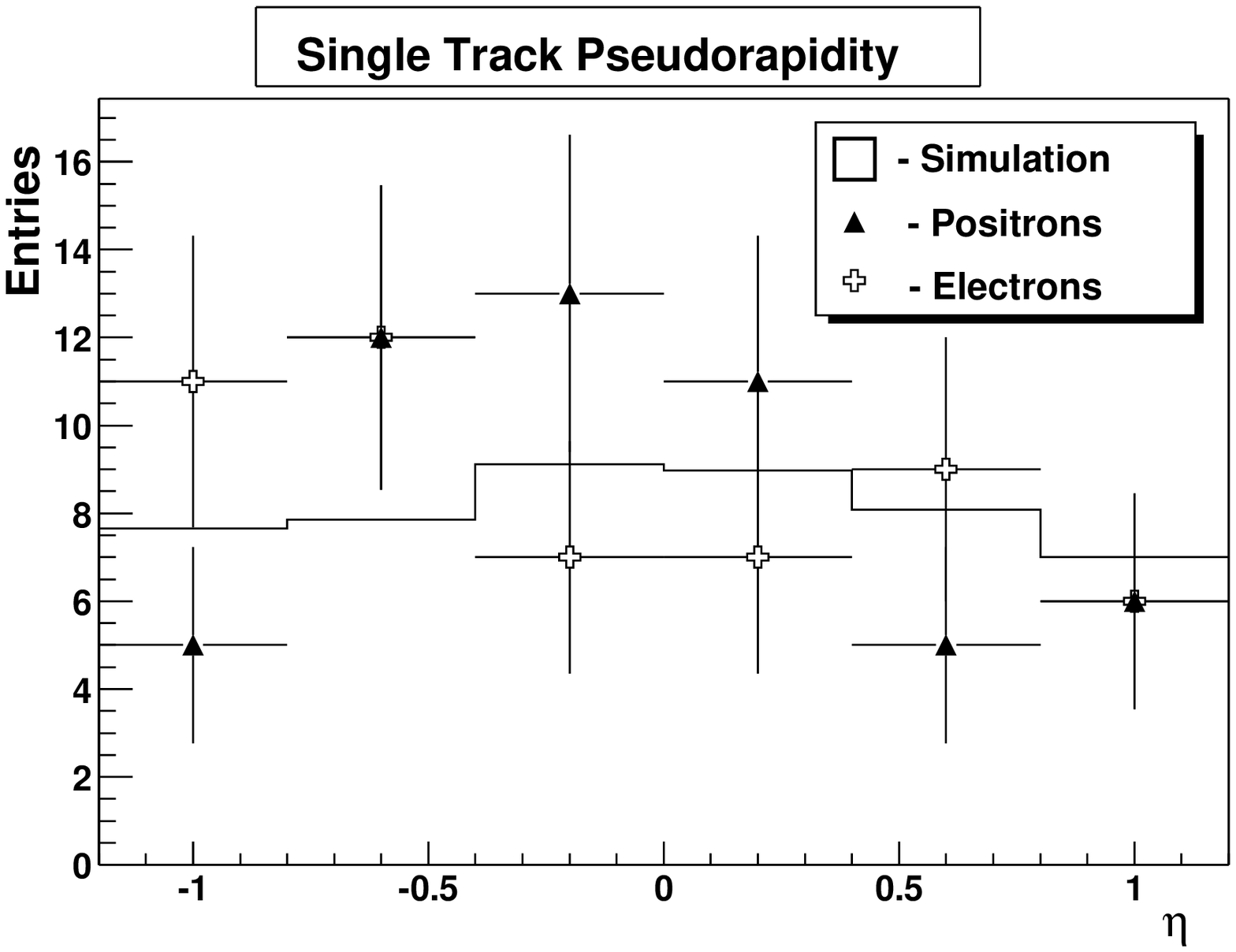}
\caption[Electron/positron track kinematical distributions]{Individual track $p_{\perp}$ spectrum (left) and $\eta$ spectrum (right). Triangles -- positrons, crosses -- electrons, solid line -- lowest-order QED simulation.}
\label{fig:ptSpectra}
\end{figure}

Figure \ref{fig:ptSpectra} shows the distribution of transverse momenta and pseudorapidities of electron tracks  and positron tracks. The distributions agree well with Monte Carlo and within available statistics no difference can be observed between electron/positron kinematical distributions, suggesting that Coulomb corrections are not observed.

\subsubsection{ZDC Spectra}
\begin{figure}
\includegraphics[width=200pt]{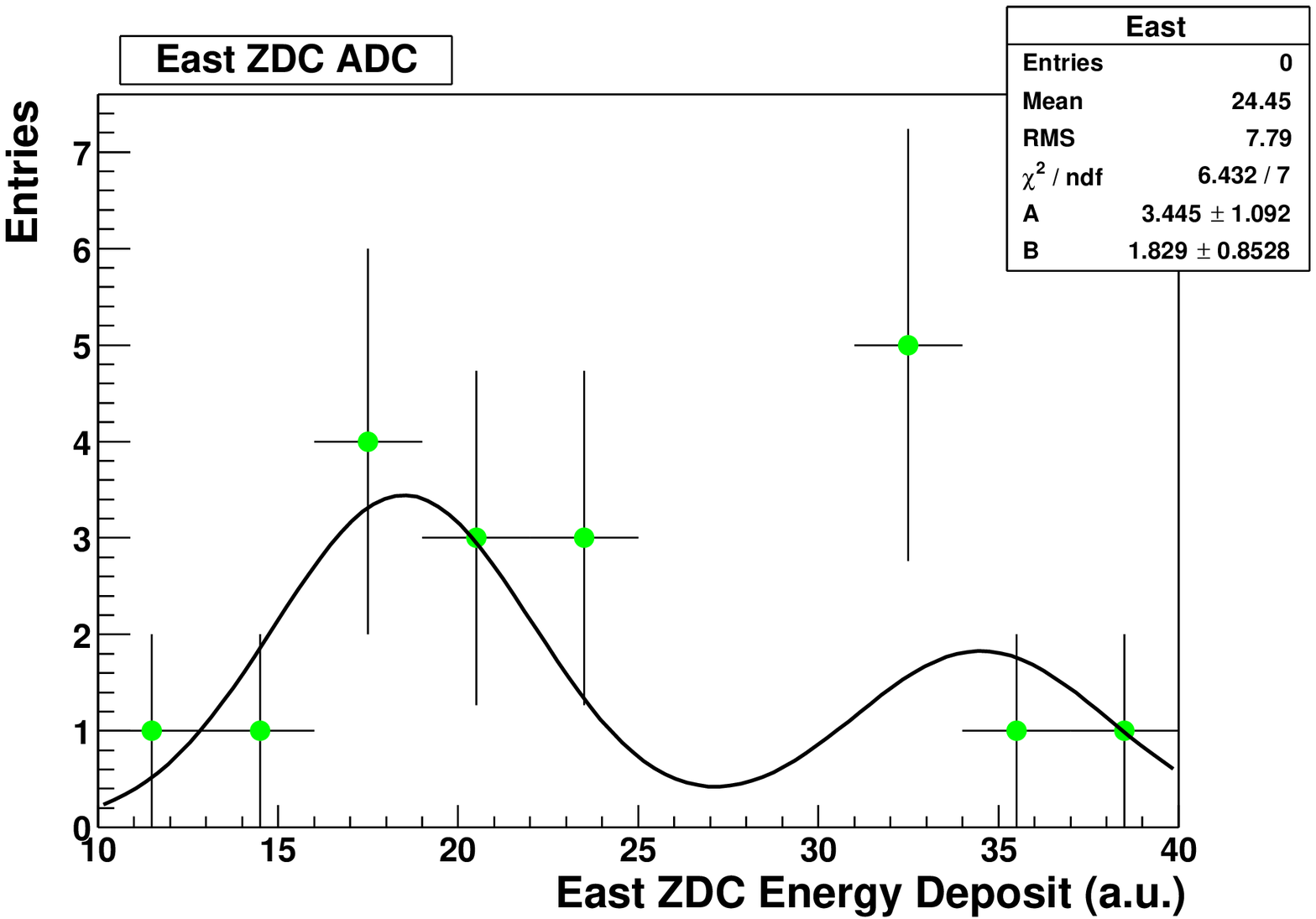}
\includegraphics[width=200pt]{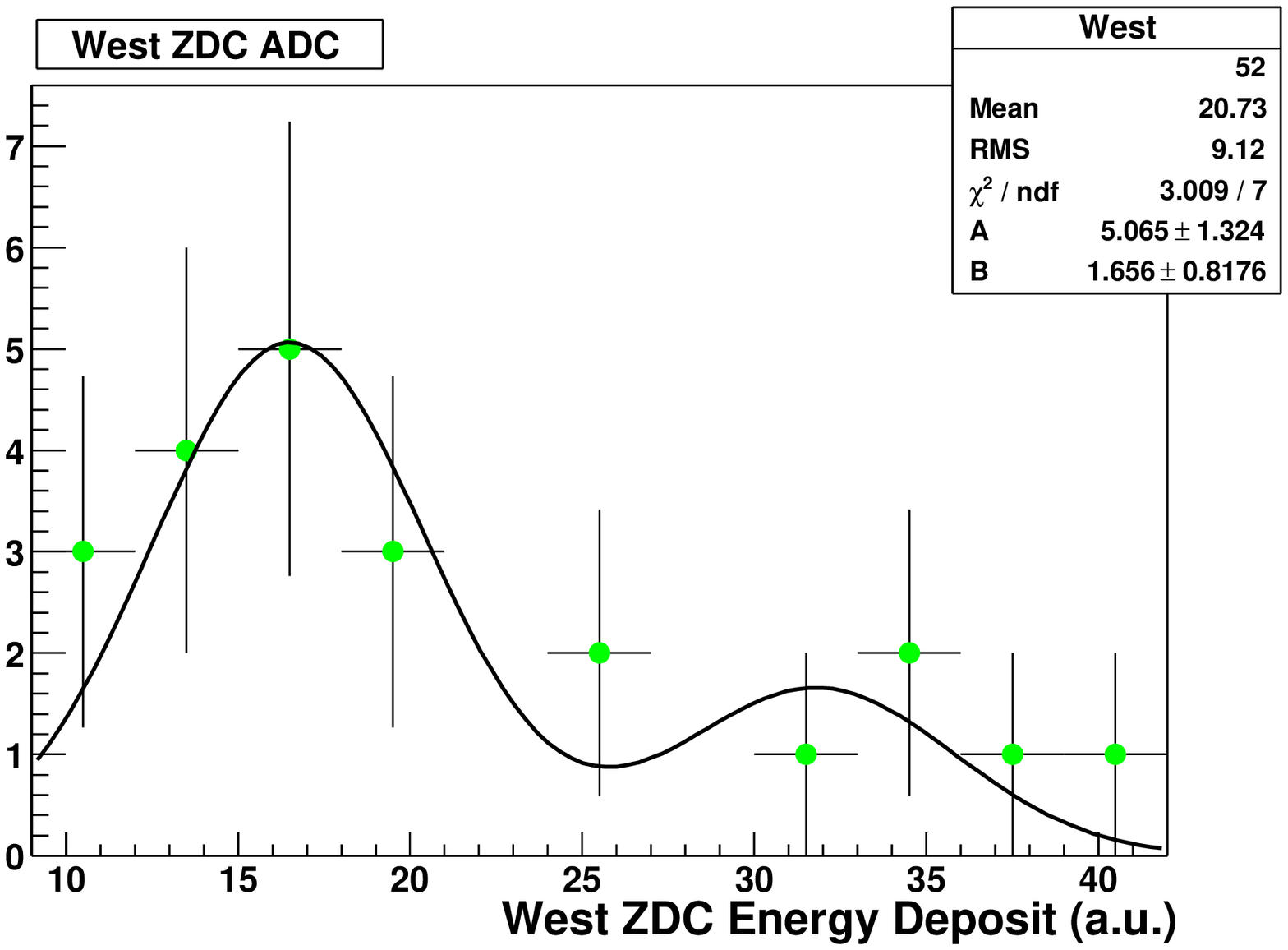}
\caption[ADC signal spectra in the ZDCs for events of the type \AuAuee]{\label{fig:ZDCspectra}
ADC signal spectra in the ZDCs for events of the type \AuAuee. Dots - data, solid lines - Gaussian fit.}
\end{figure}

We do a check of the ADC signal spectra in the East and West ZDCs for the selected events. The signals in the ZDCs are from mutual electromagnetic excitations of Au ions emitting one, two or more neutrons. Figure \ref{fig:ZDCspectra} shows the distribution of ADC in the East and West ZDCs and the fits of the data to the sum of two Gaussians, representing one neutron peak and two neutrons peak:
\begin{equation}
f(x) = A\exp \left( { - \frac{{\left( {x - x_0  - \delta } \right)^2 }}{{\sigma ^2 }}} \right) + B\exp \left( { - \frac{{\left( {x - 2x_0  - \delta } \right)^2 }}{{2\sigma ^2 }}} \right) 
\label{eqn:ZDC}
\end{equation}
To reduce the number of free parameters in the fit, only parameters $A$ and  $B$  were found with a fit to the \AuAuee data. The values of pedestal offset ($\delta$), first Gaussian mean ($x_0$) and width ($\sigma$) were taken from the analysis of ultra-peripheral $\rho ^0$ production with mutual nuclear excitation\cite{Falk200GeV}. 

\begin{figure}
\centering
\includegraphics[width=200pt]{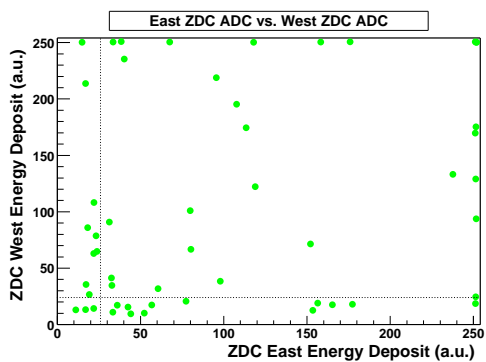}
\caption[Correlation of East ZDC ADC and West ZDC ADC for events of the type \AuAuee]{\label{fig:ZDCEastWest}Correlation of East ZDC ADC and West ZDC ADC for events of the type \AuAuee. Dashed lines separate 1 neutron peaks from multiple neutrons peaks.}
\end{figure}

Both histograms and their Gaussian fits show that 1 neutron emission events can be clearly separated from multiple neutron events by selecting events with ADC count below 27 in the East ZDC and below 25 in the West ZDC. Figure \ref{fig:ZDCEastWest} shows the correlation of the East vs. West ZDC signals in the selected 52 events. 3 events satisfy the cut $\left [ \text{ ZDC East}<27, \text{ZDC West}<25\right ]$ - these are $(1n, 1n)$ events. The ratio of $(1n,1n)$ events to $(Xn,Xn)$ events in the sample ($0.06 \pm 0.04$) agrees within available statistics with the ratio of $(1n,1n)$ events to $(Xn,Xn)$ in the ultra-peripheral $\rho ^0$ study ($0.095 \pm 0.05$)\cite{Falk200GeV}. Since \ee production with nuclear excitation and $\rho^{0}$ production with excitation have the same mechanism of electromagnetic nuclear excitation, this confirms that the observed signals in the ZDCs are from mutual nuclear excitation decays.

\chapter{Efficiency Corrections and Background Subtraction}
\label{ch:Simu}
We use Monte Carlo simulations to do the efficiency correction for the observed spectra and to estimate the backgrounds. This chapter also describes the extrapolation technique we utilized to extrapolate the experimentally accessible cross-section to the $4\pi$ cross-section within the kinematical range of $140 \text{ MeV } <M_{inv}<265 \text{ MeV}$ and $ |Y|<1.15$.

\section{Kinematical Parameter Resolution Study}
\label{sec:smearing}
\begin{figure}
\centering
\includegraphics[width=200pt]{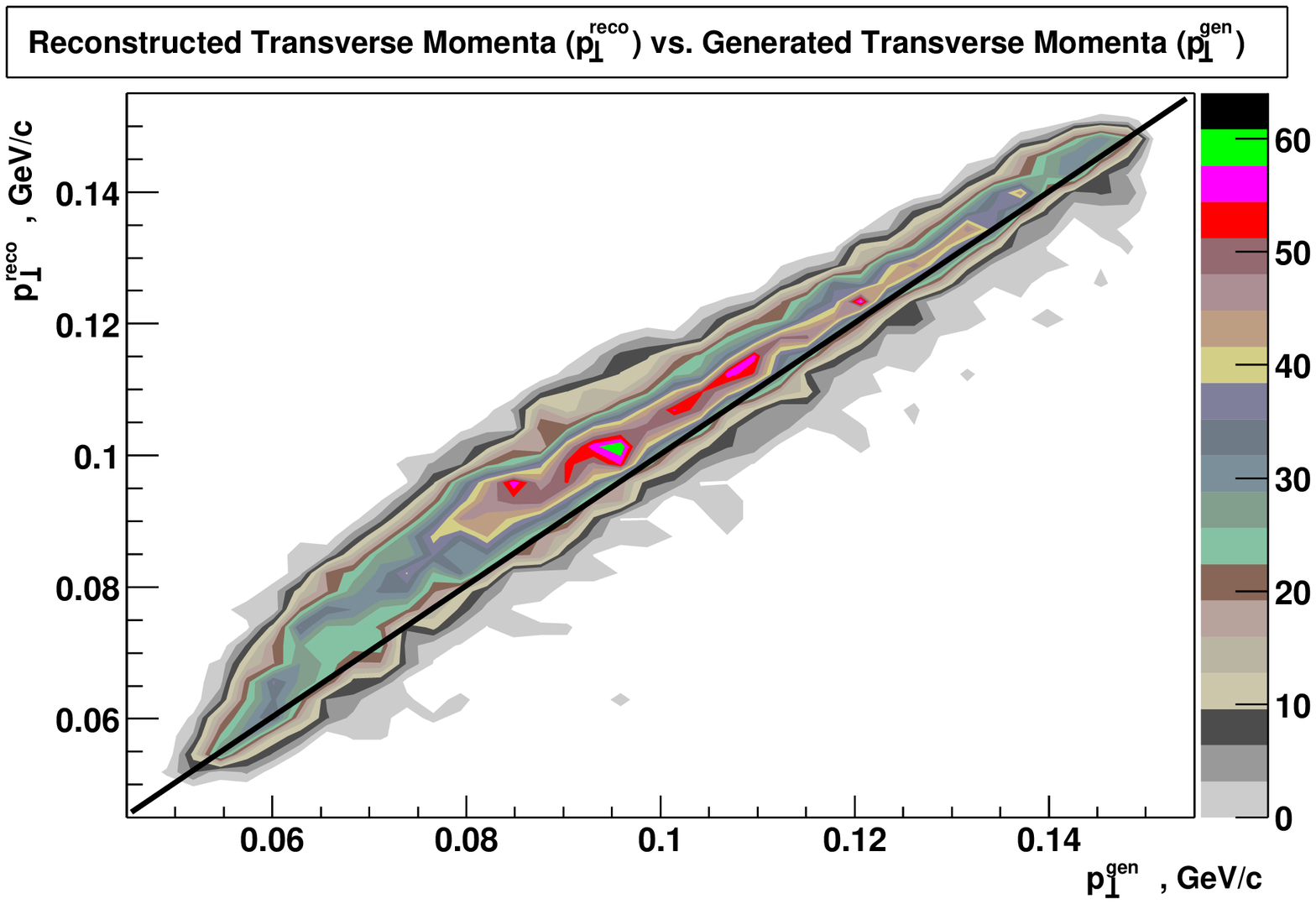}
\includegraphics[width=200pt]{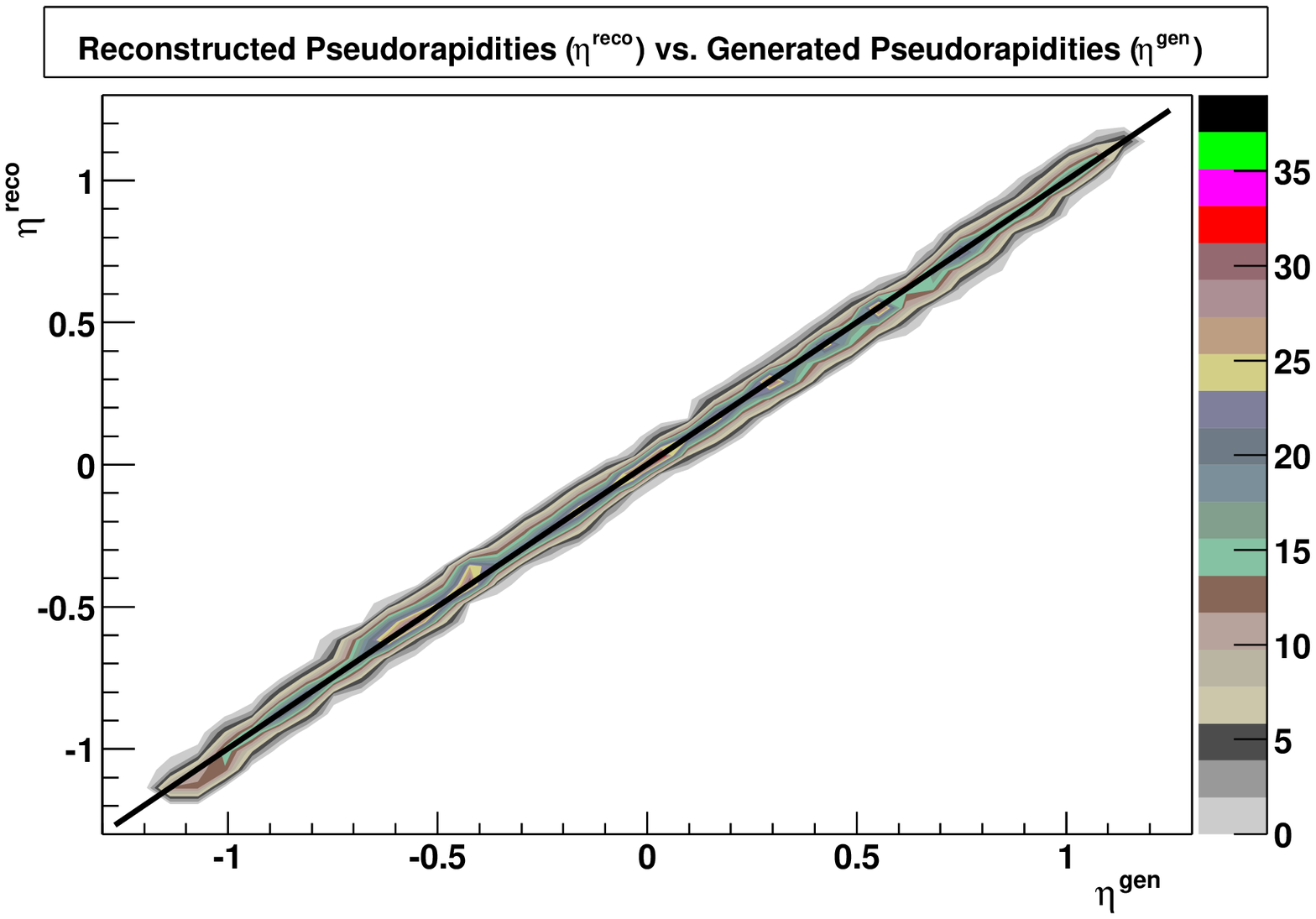}
\caption[Transverse momentum and pseudorapidity for the generated tracks vs. reconstructed]{Contour plots of transverse momentum and pseudorapidity for the reconstructed tracks vs. generated values. Solid lines represent equal generated and reconstructed values.}
\label{fig:CompareKinematics}
\end{figure}

The GEANT detector simulation together with the event reconstruction routine allows us to study the  effects of secondary scattering in the detector material, energy loss and particle mis-identification on the determination of the tracks' kinematics. Figure \ref{fig:CompareKinematics} compares the track transverse momenta and pseudorapidity for the simulated generated and the reconstructed primary tracks (electrons and positrons), which were successfully associated to the generated tracks.

From Figure \ref{fig:CompareKinematics} we see that the reconstructed tracks' $p_{\perp}^{reco}$  are distorted compared to the true generated values $p_{\perp}^{gen}$. The distortion can be separated into the smearing of the reconstructed momentum value for the fixed value of the generated $p_{\perp}^{gen}$ and the shift of the mean $\left <p_{\perp}^{reco}\right >^{mean}$ compared to the original value. The smearing can be attributed to the secondary scattering of the electron/positron tracks in the material of STAR. The shift of the mean value is due to the use of the pion mass hypothesis in the fitting of the primary tracks and in the corrections for the energy loss applied to the primary track (Section \ref{sub:Kalman}). The shift and smearing are most significant for $p_{\perp}^{gen} \sim 95$ MeV/c, where the mean of the distribution $p_{\perp }^{reco} - p_{\perp }^{gen}$ is equal to $8.5$ MeV/c ($8.9\%$ of the value of $p_{\perp}^{gen}$) and the standard deviation of the distribution is $10.3$ MeV/c ($10.8\%$ of $p_{\perp}^{gen}$).

The distortion in $\eta$ is much less significant. The mean of the distribution $\eta ^{reco} - \eta ^{gen}$ is zero for all values of $\eta ^{gen}$, and the standard deviation of the distribution is $0.017$. For a typical track with pseudorapidity of $\eta = 0.6$, this represents only a $2.8\%$ distortion.

\begin{figure}[!h]
\label{fig:PtCorrection}
\centering
\includegraphics[width=250pt]{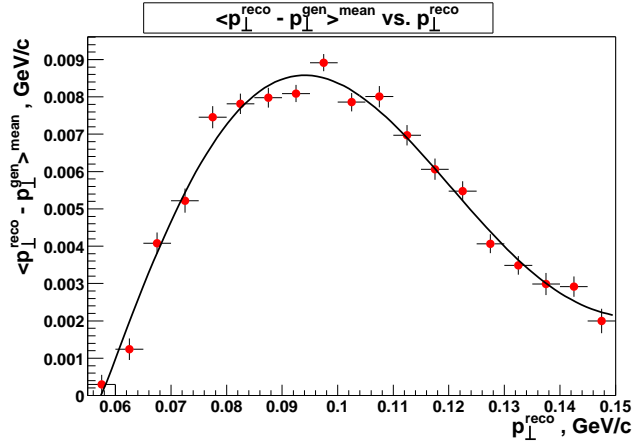}
\caption[Transverse momentum correction]{Value of $\delta p_{\perp , i}$ vs. $p_{\perp , i}^{reco}$ (dots) with statistical uncertainties. Solid line - 4-th order polynomial fit $f(p_{\perp})$.}
\end{figure}

We can correct the data for the shift in the mean value of reconstructed transverse momenta. We slice the simulated reconstructed transverse momentum spectrum between 50 and 150 MeV/c into 20 slices of $p_{\perp , i}^{reco}$ and find the mean value $\delta p_{\perp , i} = \left < p_{\perp }^{reco} - p_{\perp }^{gen} \right> _{i}$ for each slice. Figure \ref{fig:PtCorrection} shows all 20 values of $\delta p_{\perp , i}$ vs. $p_{\perp , i}^{reco}$ and also a fit to the 4-th order polynomial function $f(p_{\perp})$. Each value of the data $p_{\perp}$ is then corrected by the value of $f(p_{\perp})$. For a typical value of $p_{\perp} \sim 75$ MeV/c, the uncertainty of the correction $f(p_{\perp})$ is on the order of $10\%$ of the value of the correction.

\section{Efficiency Corrections}
\label{sec:efficiency}

\subsection{Reconstruction Efficiency}

We would like to find the reconstruction efficiency for \ee  pairs in the defined acceptance region. From the analysis chapter we established the following definition of the acceptance region:

\begin{equation}
\label{eqn:Acceptance}
\begin{array}{c}
 \left| {\eta _{e^ -  } } \right| < 1.15 \ , \ p_{ \bot e^ -  }  > 65 \text{ MeV/c } \ , \ \left| {\eta _{e^ +  } } \right| < 1.15 \ , \ p_{ \bot e^ +  }  > 65 \text{ MeV/c} \\ 
140 \text{ MeV }  < M_{inv}  < 265 \text{ MeV } \\ 
 \end{array}
\end{equation}

For this acceptance region, we will compute the efficiency correction as:
\begin{equation}
\label{eqn:FullEfficiency}
\frac{1}{\text{Reconstruction Efficiency}}=\frac{\# \text{generated events in acceptance region }}{\# \text{events after detector simulation and all analysis cuts}}
\end{equation}

This efficiency correction represents the convolution of the efficiencies at several stages of analysis, such as tracking, vertex finding and pair identification for the tracks in the defined acceptance region. These efficiencies are not independent from each other, therefore the reconstruction efficiency cannot be computed as a simple product of separate efficiencies. Equation \ref{eqn:FullEfficiency} also includes the smearing effects. 

\begin{figure}[!t]
\includegraphics[width=200pt]{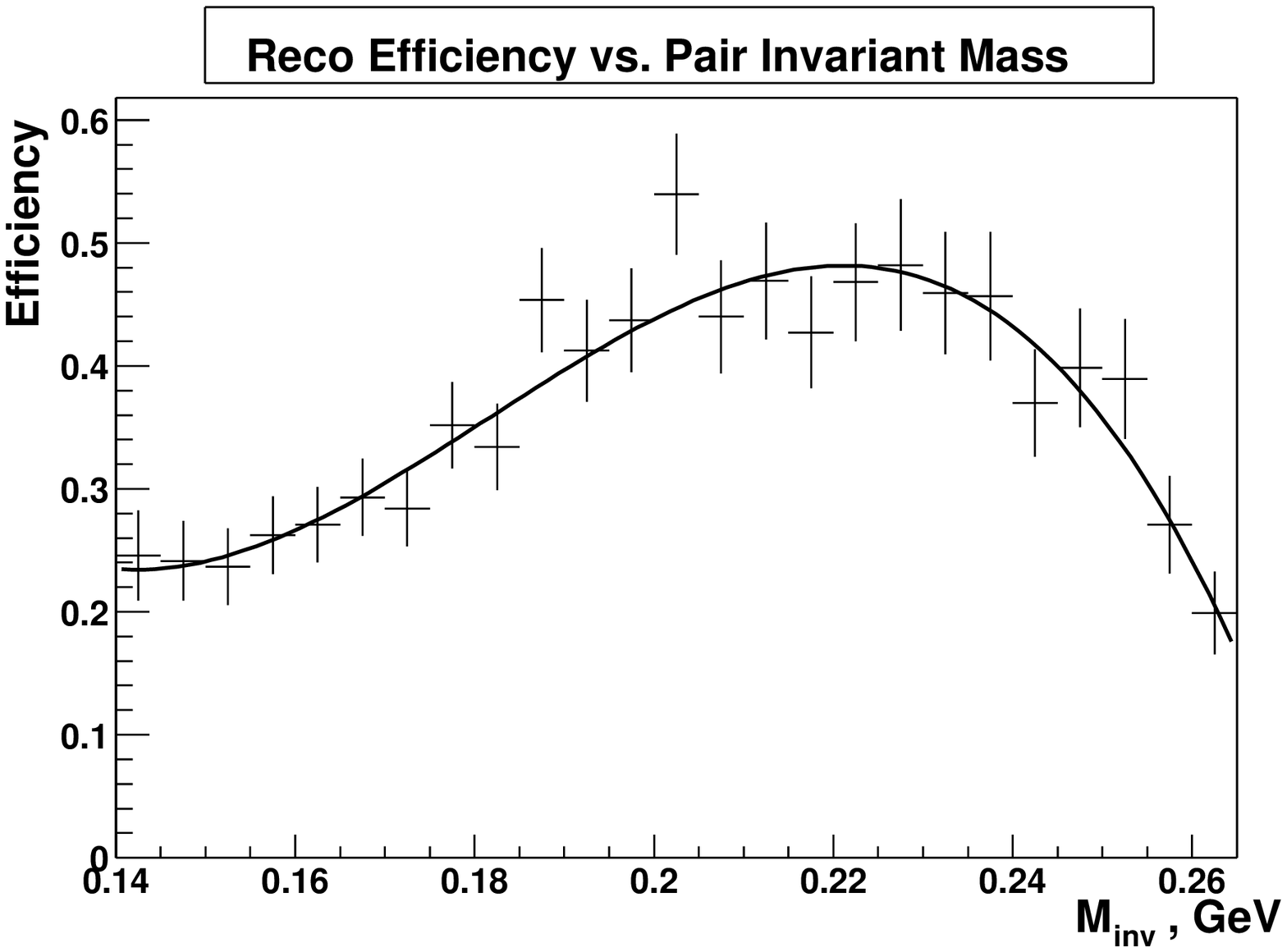}
\includegraphics[width=200pt]{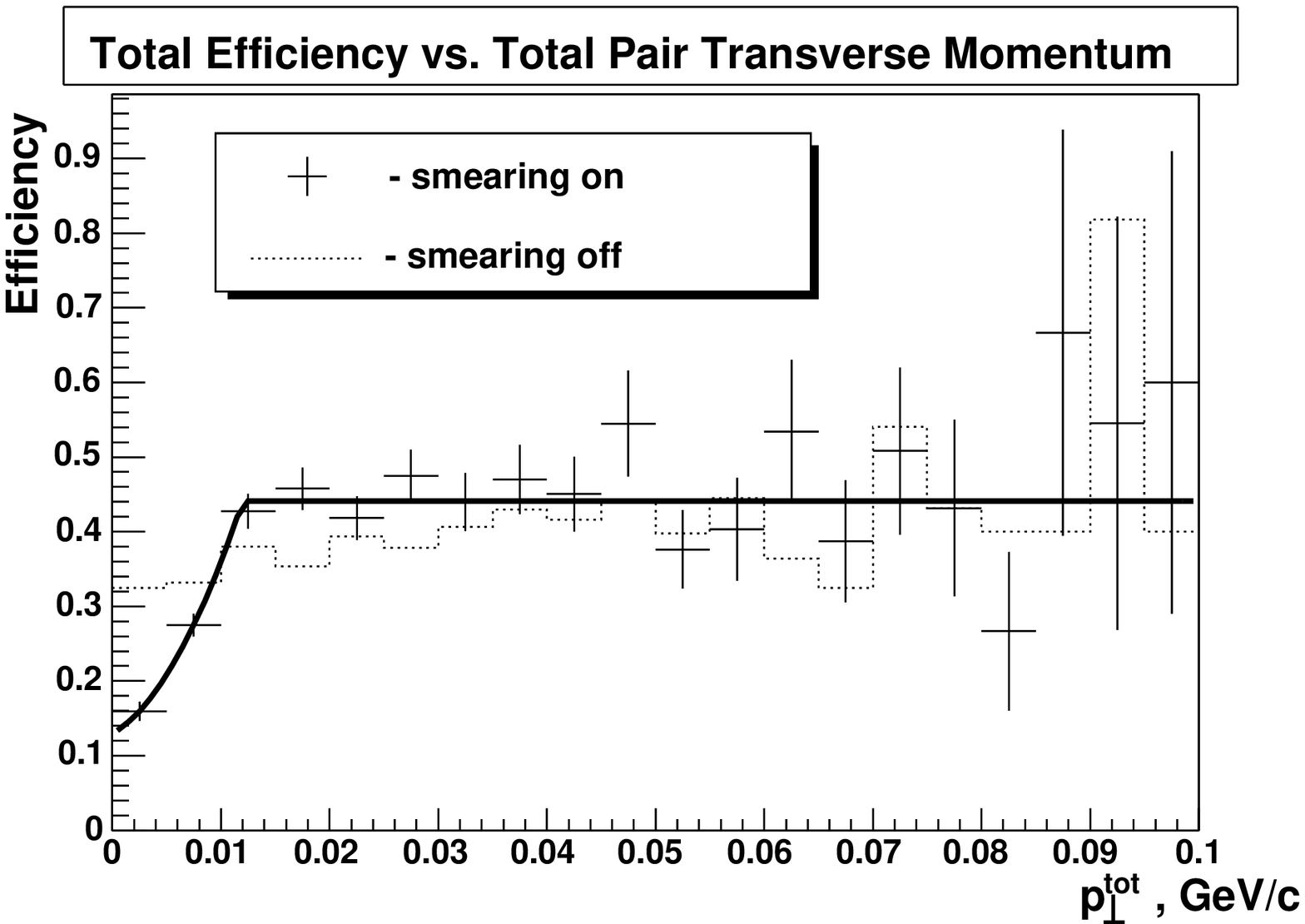}
\includegraphics[width=200pt]{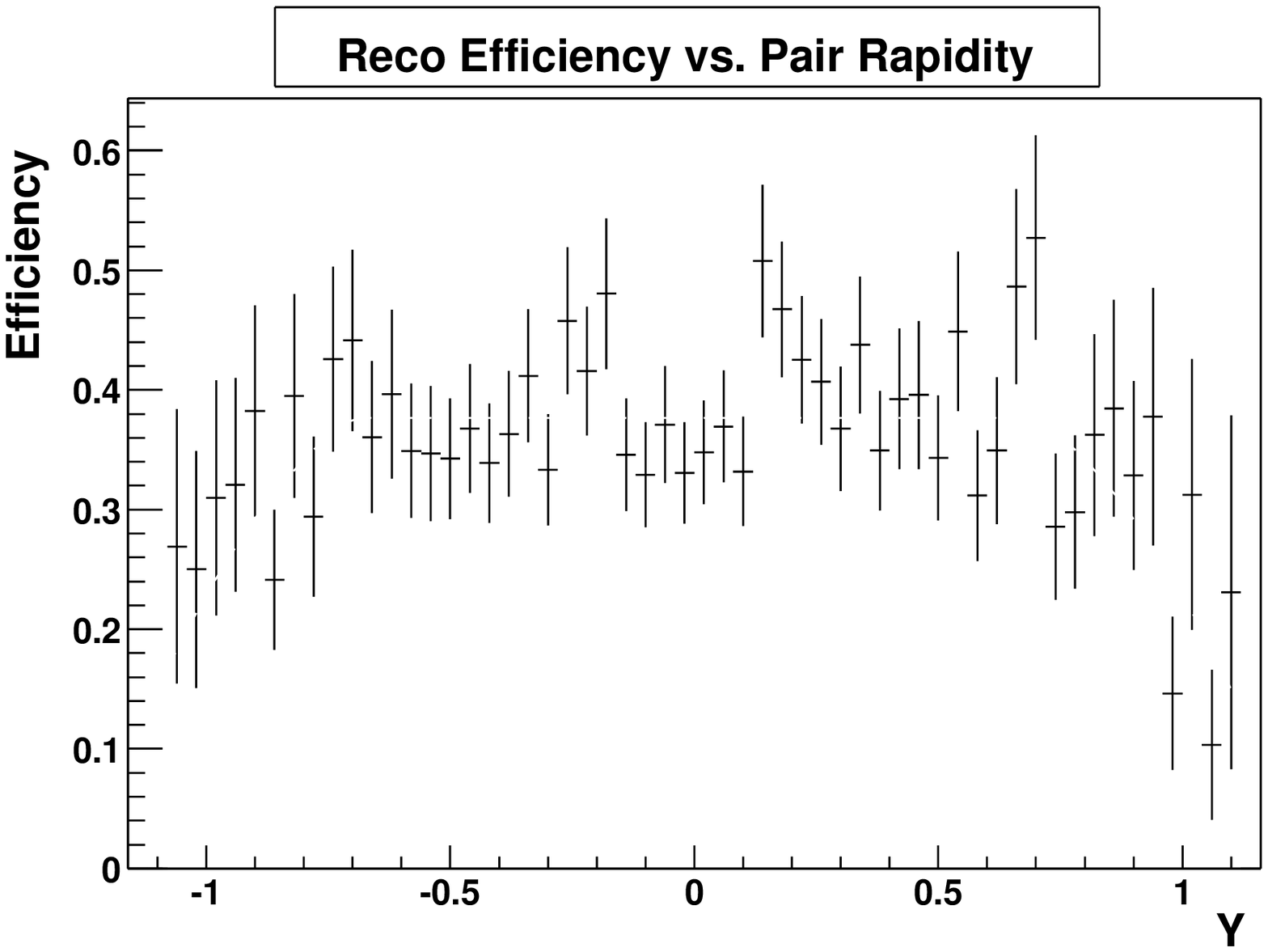}
\includegraphics[width=200pt]{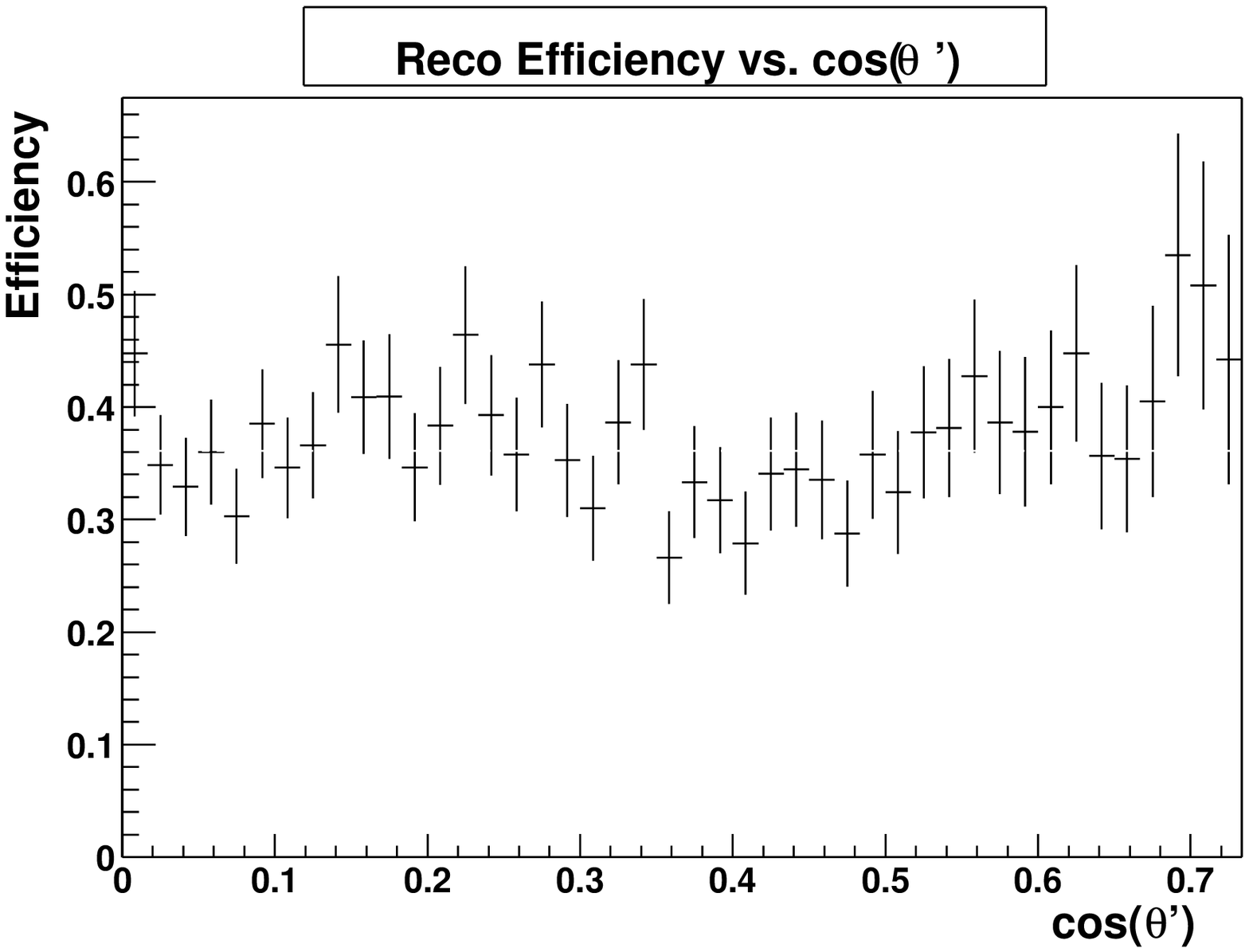}
\caption[Reconstruction efficiency vs. $p_{\perp}^{tot}$, $M_{inv}$, $Y$ and $\cos(\theta ')$]{Reconstruction efficiency vs. $p_{\perp}^{tot}$, $M_{inv}$, $Y_{Pair}$ and $\cos(\theta ')$. Efficiencies in $M_{inv}$ and $p_{\perp}^{tot}$ are fit with smooth functions $e(M_{inv})$ and $e(p_{\perp}^{tot})$ (solid lines). $p_{\perp}^{tot}$ efficiency is also presented for the simulations without $p_{\perp}$ smearing (dashed histogram).}
\label{fig:totalEfficiency}
\end{figure}

Figure \ref{fig:totalEfficiency} shows the dependence of the reconstruction efficiency on the total transverse momentum ($p_{\perp }^{tot}$), invariant mass ($M_{inv}$), polar angle in the center of mass frame ($\cos(\theta ')$) and pair rapidity ($Y$). Efficiencies if $M_{inv}$ and $p_{\perp}^{tot}$ can be fit with smooth functions $e(M_{inv})$ and $e(p_{\perp}^{tot})$, which we will use as efficiency correction functions. The peaked shape of the reconstruction efficiency  $e(M_{inv})$ is a product of raising tracking and vertex finding efficiency and dropping pair identification efficiency as a function of $M_{inv}$. The reconstruction efficiency in $p_{\perp}^{tot}$ is flat, except for momenta under 12 MeV/c, where the efficiency drops, mostly as a result of the $p_{\perp}$ smearing. As a check procedure, we turned off $p_{\perp}$ smearing in the simulations and the resulting reconstruction efficiency in $p_{\perp}^{tot}$ (dashed histogram in Figure \ref{fig:totalEfficiency}) shows a much smaller dip for pairs with $p_{\perp}^{tot}<12$ MeV/c.

\subsection{Systematic Errors}

Systematic errors in the reconstruction efficiency determination occur because the simulations do not perfectly describe the data. The following subsections discuss the systematic errors in the reconstruction efficiency from various sources.

\subsubsection{Tracking}

\begin{figure}
\centering
\includegraphics[width=200pt]{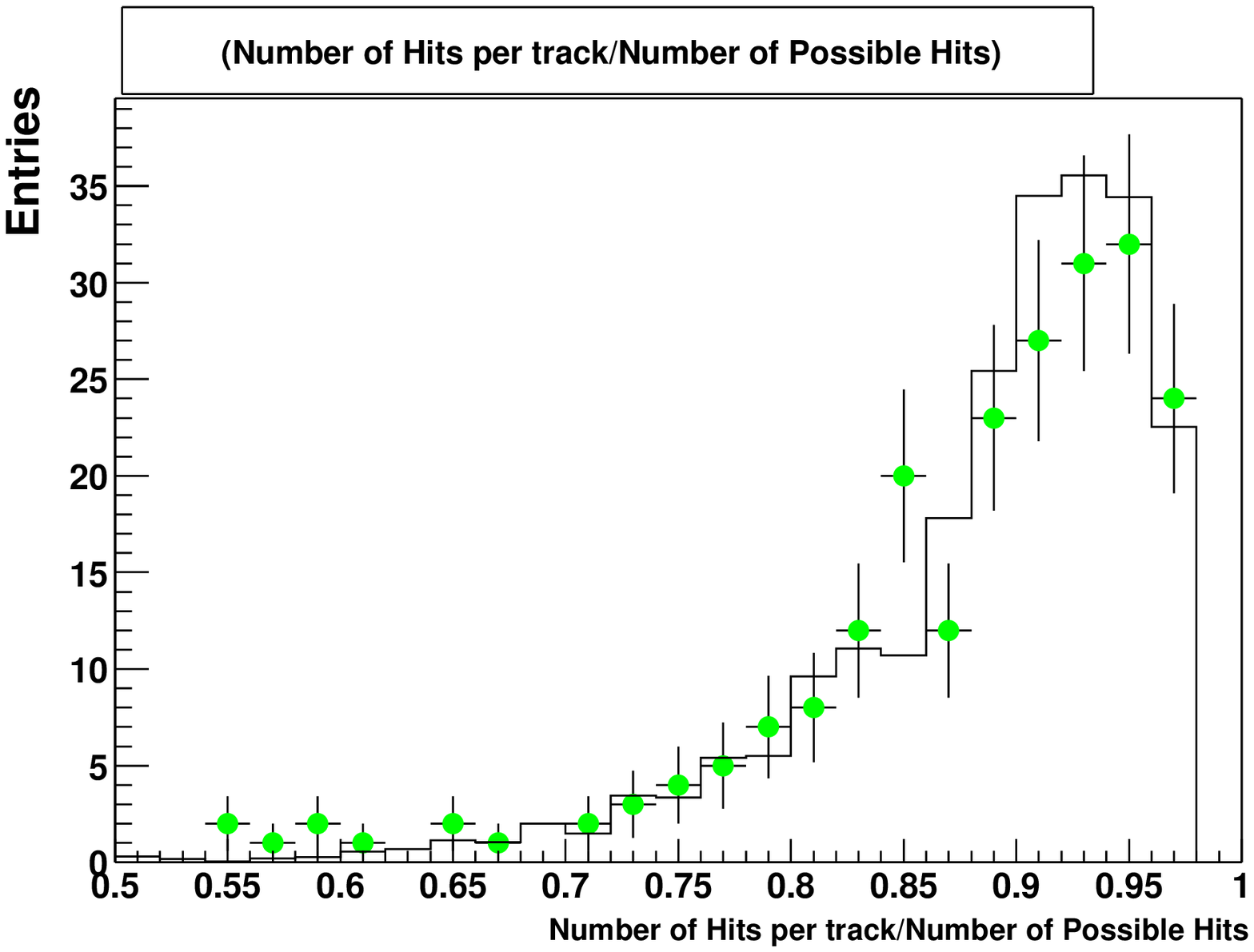}
\includegraphics[width=200pt]{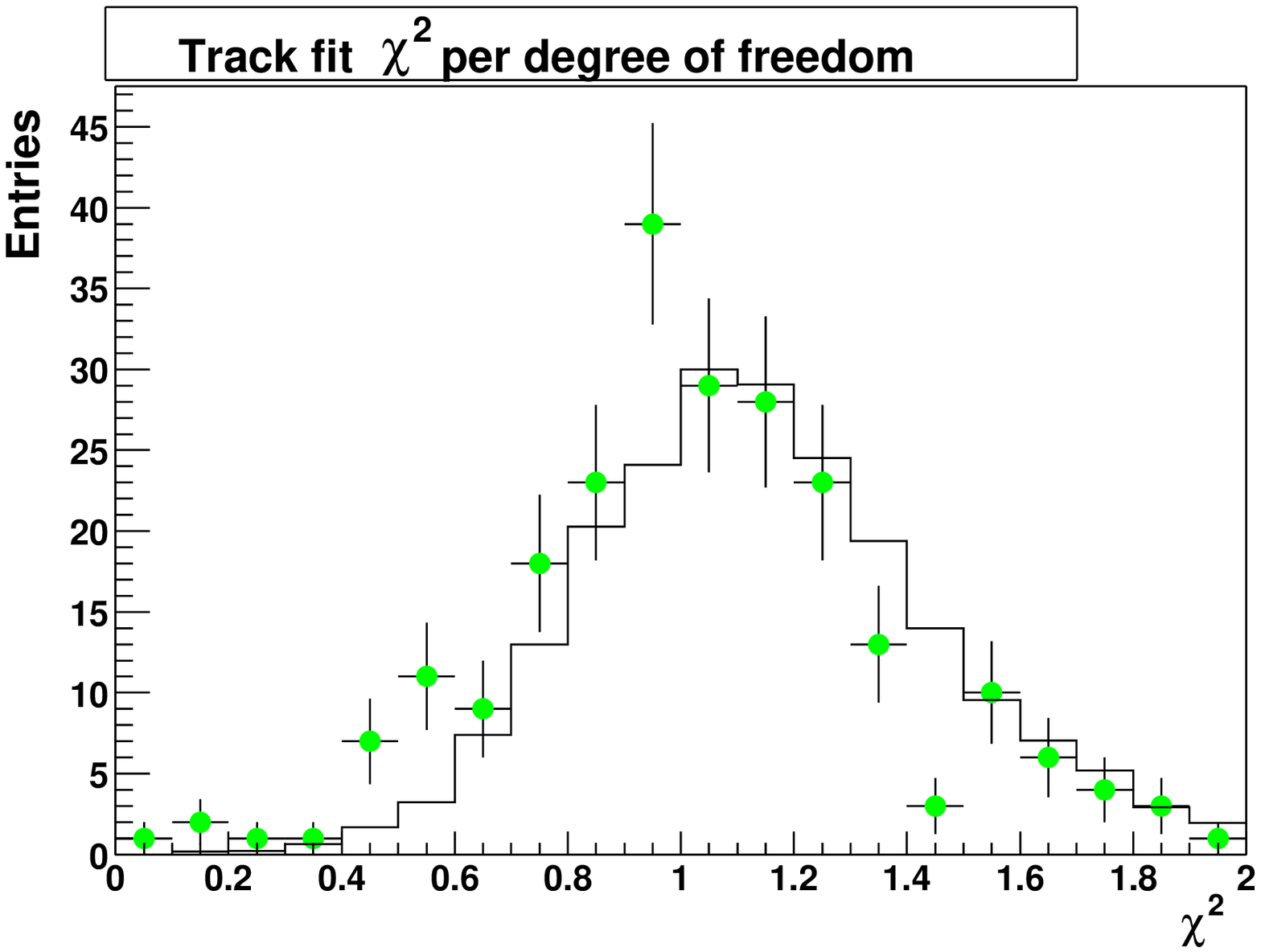}
\caption[Comparing track parameters for data and Monte Carlo]{Distribution of the $(\text{N}_{hits}/\text{N}_{possible \ hits})$ and the track fit $\chi _{\text{per D.O.F}}$. Dots - data, solid - Monte Carlo (scaled).}
\label{fig:ChiSq}
\end{figure}

To check how well the simulations describe the tracking performance of the STAR TPC, we compare the ratio of $\text{N}_{hits}$ to the number of possible hits on the track $\text{N}_{possible \ hits}$ in data and Monte Carlo.  The number of possible hits was calculated by extrapolating a track's geometry over the pad plane. The calculation incorporated both the detector geometry and features of the track finding routine. This calculation also correctly handled daughters from photon conversions that occurred within the tracking volume of the TPC\cite{IanThesis}. The distribution of the ratio ($
hit{\rm \ }ratio = {{{\rm N}_{hits} } \mathord{\left/
 {\vphantom {{{\rm N}_{hits} } {{\rm N}_{possible{\rm  \ }hits} }}} \right.
 \kern-\nulldelimiterspace} {{\rm N}_{possible{\rm \ }hits} }} $) is shown in Figure \ref{fig:ChiSq} (left). 

The quality of the track fit in data and Monte Carlo is characterized by the  $\chi _{\text{per D.O.F}}$ of the track fit. Figure \ref{fig:ChiSq} (right) shows the distribution of $\chi _{\text{per D.O.F}}$ in data and Monte Carlo. To increase the statistics, events were selected with all analysis cuts in the table \ref{tab:allCuts} except the particle identification cut. The individual track momenta were limited to $p<200$ MeV/c. The agreement between Monte Carlo and data is good for both distributions. 

In a related analysis (\cite{Manuel}) the authors studied the variation in the tracking efficiency-corrected $p_{\perp}$ and $\eta$ spectra due to small variations in track quality cuts. The systematic error was found to be 6.4\% for each reconstructed track, which we accept as a tracking efficiency systematic error for the present analysis.

\subsubsection{Vertex Finding}

We use data to produce an estimate of vertex finding efficiency for \ee pairs and compare the result to the efficiency obtained from Monte Carlo (see Section \ref{sub:VertexingEfficiency} and Figure \ref{fig:Vertexing}). In the data, we plot invariant mass spectra for identified secondary \ee pairs comprised of any two global tracks (diamonds) and for pairs comprised of two global tracks that also were primary tracks (triangles). The first sub-set of pairs is larger than the second sub-set, since it includes events with no primary vertex. The ratio of the \ee pairs with the primary tracks to the total number of pairs gives an estimate of the vertex finding efficiency in the data. Due to the low statistics, coarse binning had to be used. 

\begin{figure}
\centering
\includegraphics[width=200pt]{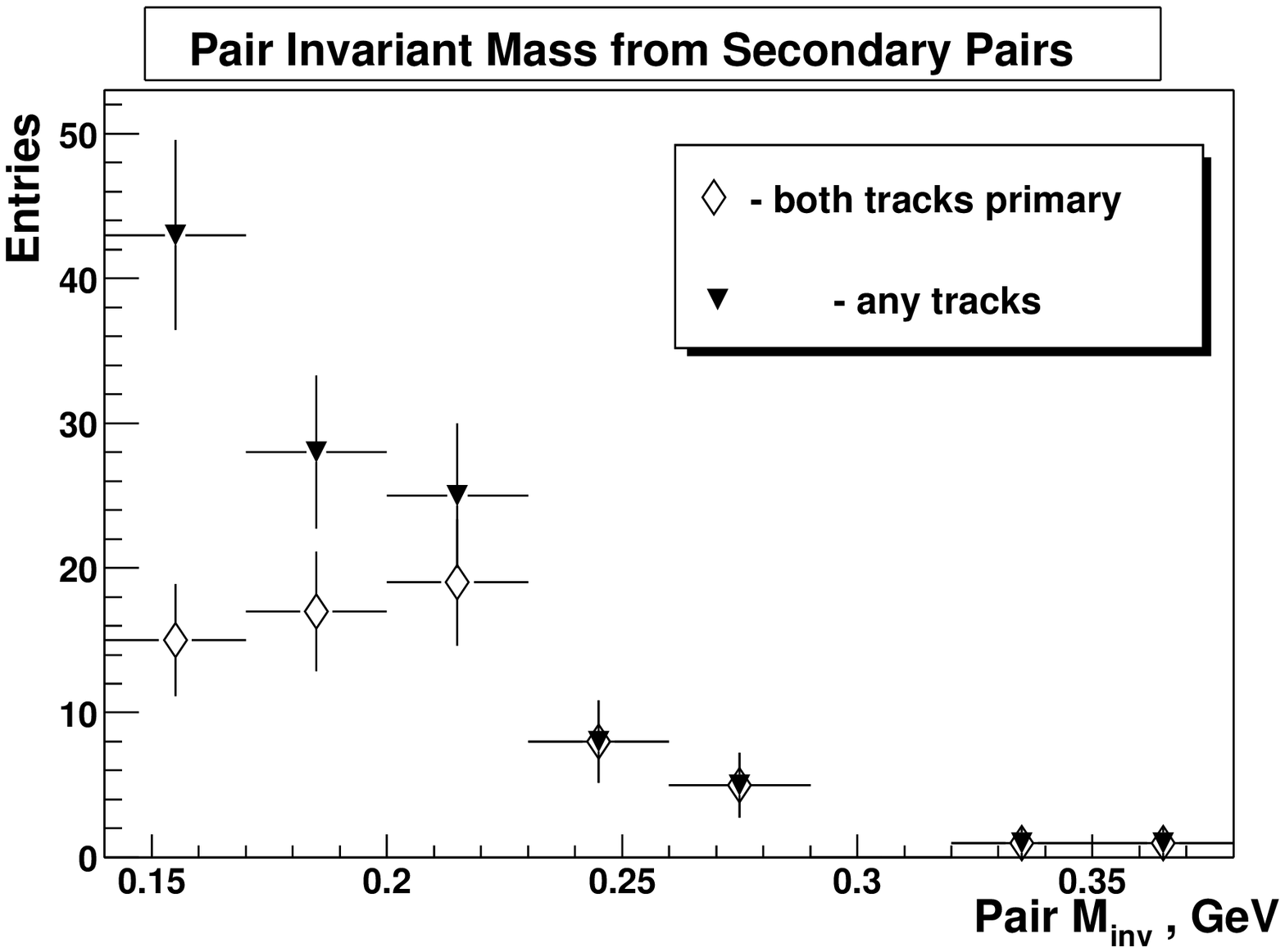}
\includegraphics[width=200pt]{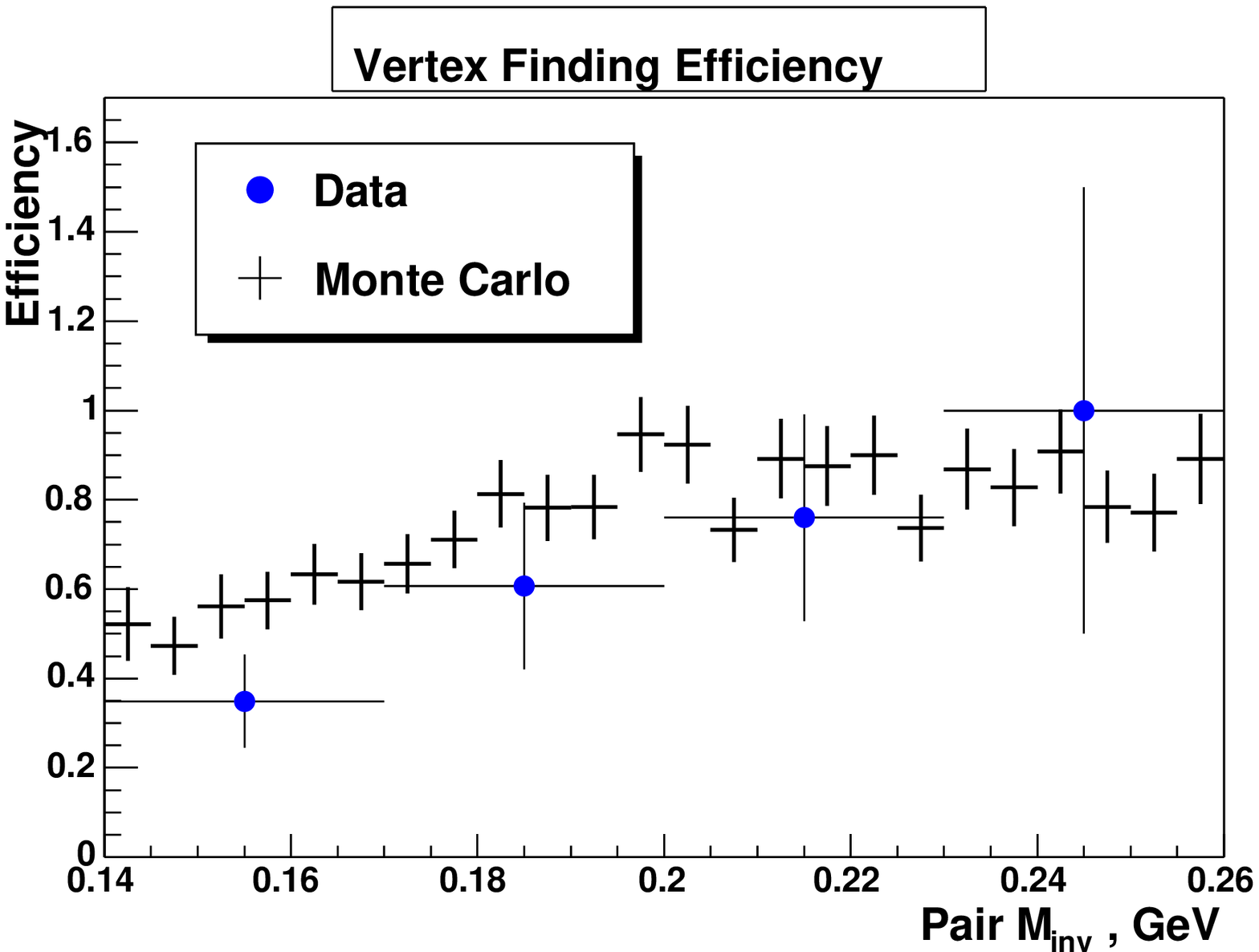}
\caption[Vertex finding efficiency systematic error]{Left: invariant mass spectrum from secondary pairs (diamonds - pairs comprised of two primary tracks, triangles - pairs comprised of any tracks). Right: comparison of vertex finding efficiency from the simulations (crosses) and data (dots).}
\label{fig:VtxEffSyst}
\end{figure}

From Figure \ref{fig:VtxEffSyst} the general shapes of vertex finding efficiency distributions vs. pair invariant mass agree for data and Monte Carlo. The most sizeable difference between Monte Carlo efficiency and efficiency from data appears to be between 0.14 GeV $<M_{inv}<$ 0.2 GeV. Analysis of the $p_{\perp}^{tot}$, $cos(\theta ')$ and $Y$ distributions in the identified secondary \ee pairs shows significant deviations from Monte Carlo spectra, which are not present in the distributions of primary \ee pairs. We believe that this is due to the fact that without a requirement of a primary vertex, the selected secondary \ee pair sample contains a large fraction of events which consist of electron or positron tracks that suffered extreme energy loss (e.g. hard Bremsstrahlung) in the detector material, or the spiraling tracks (Section \ref{sec:SimuChain}). The reconstructed momentum of such tracks differs from the true momentum by more than $50\%$, and therefore these events are unusable for the \ee pair spectrum determination, and are a background to the 'good' \ee events. These events contaminate the secondary \ee pair sample, causing an under-estimation of the vertex finding efficiency from data. The primary vertex requirement is essential in rejecting these backgrounds, and must be enforced in the event selection.

Formally, the ratio of the integrated vertex finding efficiency from Monte Carlo to the vertex finding efficiency from the data is $1.17 \pm 0.12$. This suggests a $17\%$ systematic error in vertex finding efficiency from Monte Carlo. To correct for the unaccounted backgrounds in the secondary \ee pairs sample, we take $50\%$ of $17\%$ ($8.5\%$) as an estimate of the systematic error in vertex finding efficiency. 

\subsubsection{Pair Identification}

To estimate systematic uncertainty in pair identification efficiency via specific energy loss we compute the systematic error on the value of shifted $Z_{\pi} ^{corrected \ simu}$ from the correction in Equation (\ref{eqn:correctZ}). According the Gaussian error propagation law: 
\[
\left( {\delta Z_\pi ^{corrected \ simu} } \right)^2  = 
\]
\begin{equation}
\label{eqn:PIDSyst}
\left( {\frac{{\sigma _Z^{data} }}{{\sigma _Z^{simu} }}} \right)^2 \! \left( {\delta \left\langle {Z_\pi ^{simu} } \right\rangle  } \right)^2 \! +  \! \left( {\delta \left\langle {Z_\pi ^{data} } \right\rangle  } \right)^2 \! + \! \left( {Z_\pi ^{simu} \! - \! \left\langle {Z_\pi ^{simu} } \right\rangle  } \right)^2 \! \left( {\left( {\frac{{\delta \sigma _Z^{data} }}{{\sigma _Z^{simu} }}} \right)^2 \! + \! \left( {\frac{{\sigma _Z^{data} }}{{\sigma ^{2 \ \ simu} _Z }} \cdot \delta \sigma _Z^{simu} } \right)^2 } \right)
\end{equation}
where errors ${\delta \left\langle {Z_\pi ^{simu} } \right\rangle }$, ${\delta \left\langle {Z_\pi ^{data} } \right\rangle }$, ${\delta \sigma _Z^{simu} }$ and ${\delta \sigma _Z^{data} }$
 are the fit errors on the determination of the mean and width of the Gaussian fits to the data and simulations peaks in Figure \ref{fig:DeDxSimuVsReco}.\footnote{We scale up each error by the $\chi ^2 _{\text{per D.O.F}}$ of the fit to account for the general disagreement of the distribution shape in the data and the fit function. For full explanation of the method see standard reference\cite{pdg}.}
  
Given a single track systematic error $\delta Z_{\pi} ^{corrected \ simu}$, we systematically shift each track's $Z_{\pi} ^{corrected \ simu}$ in the simulation at first by $+ \delta Z_{\pi}$ and then by $- \delta Z_{\pi}$ and compare the resulting particle identification efficiencies. The difference between the two curves was found to be at most at the level of 2\%; therefore the systematic error in pair identification is negligible compared to the statistical error.
 
\subsubsection{Transverse Momenta Mis-Reconstruction}

\begin{figure}
\centering
\includegraphics[width=200pt]{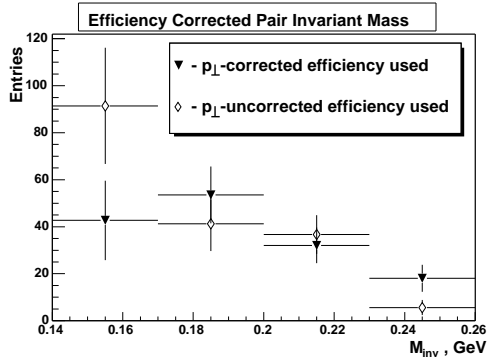}
\caption[Systematic error due to transverse momenta mis-reconstruction]{Efficiency-corrected $M_{inv}$ data spectrum. Diamonds - $p_{\perp}$ shift correction was not used in efficiency computation. Triangles - efficiency includes $p_{\perp}$ shift correction.}
\label{fig:SmearingSyst}
\end{figure}

To study the systematic error due to shift in reconstructed $\left\langle {p_ \bot ^{reco} } \right\rangle$ compared to the true value of transverse momenta, we re-compute the reconstruction efficiency without applying the shift correction to the simulations. The raw $M_{inv}$ distribution (which is computed with $p_{\perp}$ correction) is then efficiency-corrected with two different efficiency functions -- one with account of $p_{\perp}$ shift, and the other without. Figure \ref{fig:SmearingSyst} compares the two resulting efficiency-corrected $M_{inv}$ spectra.

Figure \ref{fig:SmearingSyst} shows that the effect of $p_{\perp}$ distortion is most significant  in the last and first bins. From Section \ref{sec:smearing} the error of the $p_{\perp}$ correction is $\sim 10\%$ of the value of the correction. Therefore, the systematic error on the number of events in the first bin is $
10\% \cdot \left( {{{\left( {91 - 43} \right)} \mathord{\left/
 {\vphantom {{\left( {91 - 43} \right)} {43}}} \right.
 \kern-\nulldelimiterspace} {43}}} \right) = 11\% 
$ and in the last bin it is $
10\% \cdot \left( {{{\left( {20 - 5} \right)} \mathord{\left/
 {\vphantom {{\left( {20 - 5} \right)} {20}}} \right.
 \kern-\nulldelimiterspace} {20}}} \right) = 7.5\% $. However, the errors in the first and last bin are strongly anti-correlated. Therefore, for the measure of the total cross-section systematic error due to $p_{\perp}$ distortion we use the total number of entries in the two histograms:
 \begin{equation}
 \label{eqn:SmearingSyst}
 \delta \sigma _{syst}^{p_{\bot} \ distortion}  \approx 10\% \left( {{{\left( {177 - 146} \right)} \mathord{\left/
  {\vphantom {{\left( {177 - 146} \right)} {146}}} \right.
  \kern-\nulldelimiterspace} {146}}} \right) = 2\% 
 \end{equation}

\subsubsection{Asymmetry}

Asymmetry in the number of reconstructed tracks in the East side of the TPC vs. the West side of the TPC has been observed in the high-statistics analyses of the hadronic events in STAR\cite{Manuel}.
Since the simulations assume a perfect symmetry between East/West sides of the TPC, a systematic error could be introduced in the efficiency determination. To test if there might be any asymmetry in observed \ee events we compute the ratio:
\begin{equation}
\label{eqn:AsymmetrySyst}
A = \frac{{N_{ +  + }  - N_{ -  - } }}{{N_{ +  + }  + N_{ -  - } }} \cdot \frac{{N_{ +  + }  + N_{ -  - } }}{{N_{ +  + }  + N_{ -  - }  + N_{ +  - } }} = \frac{{14 - 22}}{{14 + 22 + 52}} = 0.15
\end{equation}
where $N_{ +  + }$, $N_{ -  - }$ and $N_{ +  - }$ are the numbers of events in the identified pool of 52 \ee events with two tracks with positive longitudinal momenta ($p_{z}$), two tracks with negative $p_{z}$ and two tracks with $p_{z}$ of opposite charges.

The ratio $A$ is within $2\sigma \sim 2/\sqrt{52}$ from zero, therefore systematic error due to asymmetry is not appreciable in comparison with the statistical uncertainty.

\subsubsection{Sensitivity to Generated Spectra}

The effect that $p_{\perp}$ smearing and particle identification have on the the reconstruction efficiency depends in part on the spectra distributions generated by event generator (e.g. $(p_{\perp}^{tot})^{gen}$ and $M_{inv}^{gen}$). If the physics model of the event generator spectra is wrong, there may be a systematic error introduced in the reconstruction efficiency computation. 

To test the sensitivity of reconstruction efficiency to variations in generated spectra, we can generate an event sample with, for instance, flat $p_{\perp}^{tot}$ distribution, pass this event sample through the GEANT simulation, event reconstruction and analysis, and determine reconstruction efficiency in that way. Alternatively, we can use the same event generator event sample that we have been using so far (with pair transverse momentum distributed according to the probability distribution $f(p_{\perp}^{tot})$), but compute the efficiency by the formula:
\[
\text{Reconstruction Efficiency From Flat } p_{\perp}^{tot}\text{ Spectrum} =
\]
\begin{equation}
\label{eqn:Weighting}
=\frac{\# \text{weighted events after detector simulation and all analysis cuts}}{\# \text{weighted generated events in acceptance}}
\end{equation}
where the weight for each event is inversely proportional to $f(p_{\perp}^{tot})$. Incidentally, the effect of weighting ensures that the denominator in Equation (\ref{eqn:Weighting}) is a flat distribution. 

\begin{figure}
\centering
\includegraphics[width=200pt]{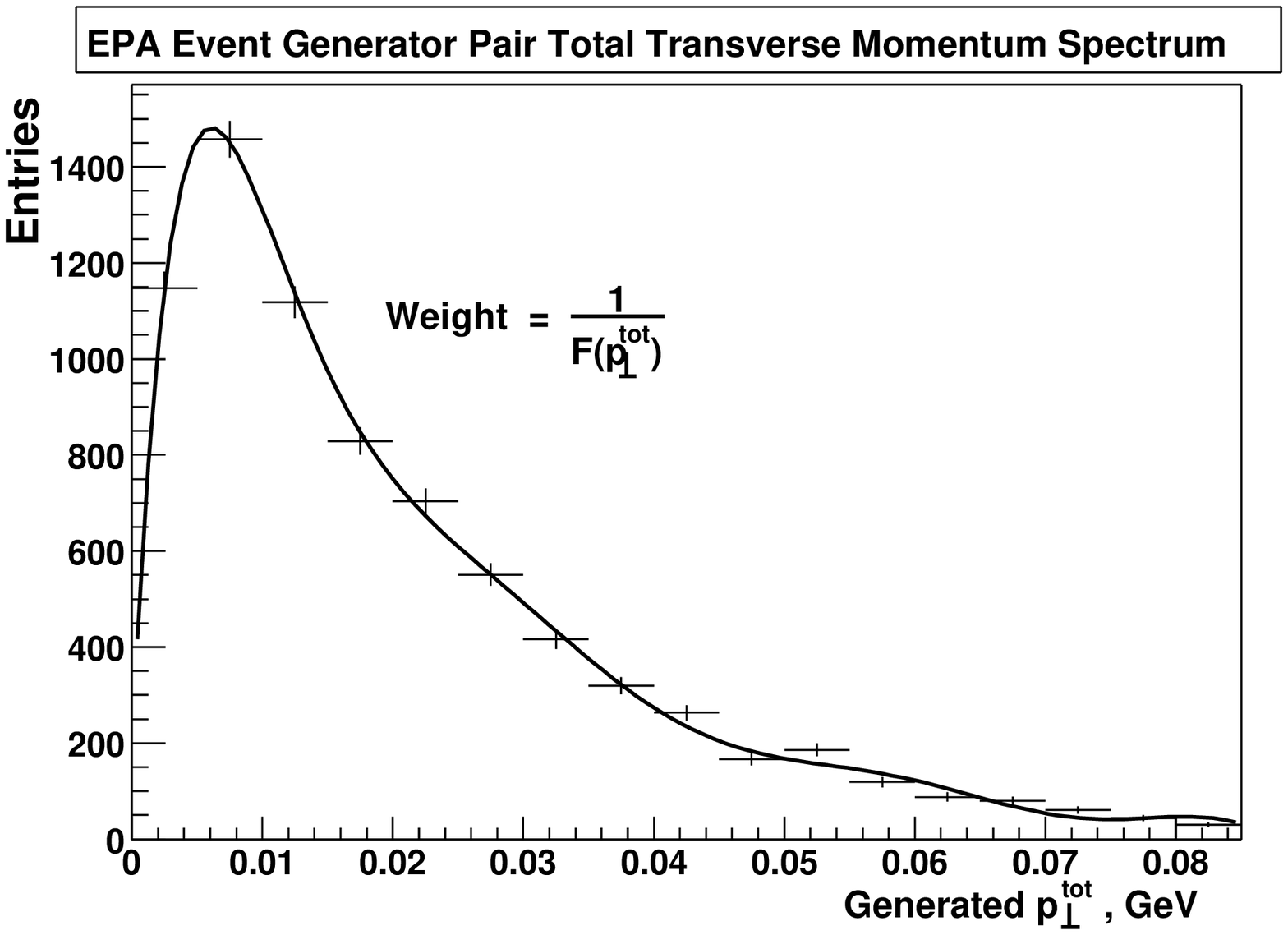}
\includegraphics[width=200pt]{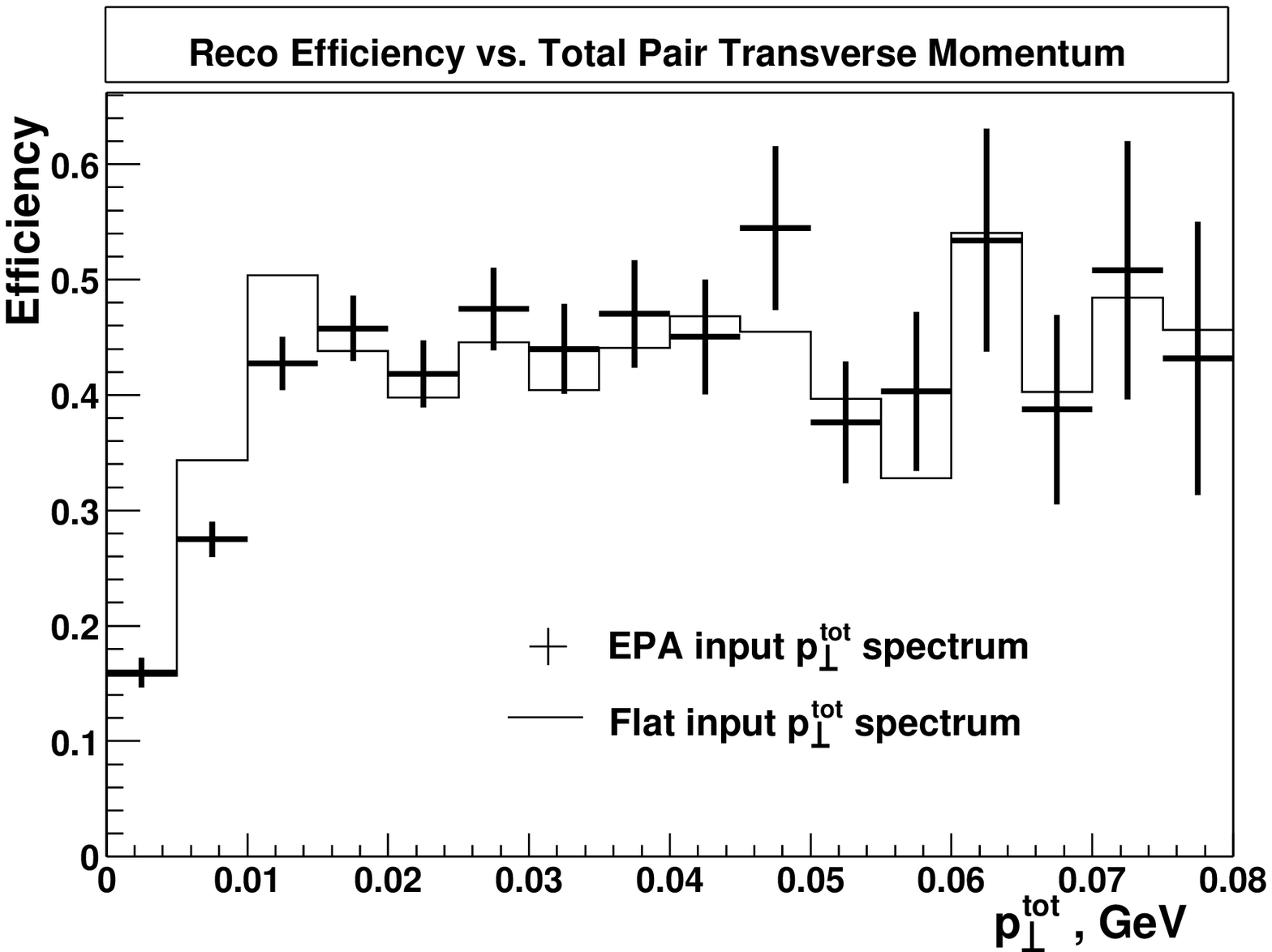}
\caption[Sensitivity of reconstruction efficiency to generated spectra]{Left: EPA generated $p_{\perp}^{tot}$ distribution is fit with a smooth function $F(p_{\perp}^{tot})$ and each event in the sample is weighted by $1/F(p_{\perp}^{tot})$. Right: reconstruction efficiency vs. $p_{\perp}^{tot}$ for flat generated spectrum (solid line) and EPA spectrum (crosses).}
\label{fig:Weighting}
\end{figure}

Figure \ref{fig:Weighting} shows how we get weights for each events (left). Using weighting, we compute the reconstruction efficiency from a flat $p_{\perp}^{tot}$ distribution, and compare it to the previously obtained efficiency from EPA Monte Carlo (Figure \ref{fig:Weighting}, right).  The difference between the two distributions appears only for low transverse momenta ($p_{\perp}^{tot} \sim 10$ MeV/c).  The EPA transverse momentum distribution has a peak at $p_{\perp}^{tot} \sim 5 \div 10$ MeV/c and therefore smearing the transverse momentum value upwards in the reconstructed events affects EPA distribution more strongly than a flat distribution. This explains why the reconstruction efficiency is smaller for the EPA distribution than for the flat distribution. For most values of the transverse momenta ($p_{\perp}^{tot} >20 $ MeV/c) the reconstruction efficiency is not affected by the shape of the generated distribution.

\subsubsection{Results of Varying all Analysis Cuts}

We compare the sensitivity of data and simulations to small variations in cuts in table \ref{tab:allCuts} by systematically varying each analysis cut and observing the change in the number of events which pass the new cuts relative to the number of evens that pass the original cuts. All cuts were varied by $\pm$2.5\%, $\pm$5.0\% and $\pm$10\%.\footnote{The cut on the total primary pair charge was not varied.}
The relative event number variations were found to be very close (differ on the average by $  \sim 4.5\%$) in data and simulations, therefore the simulations describe the data response to the analysis cuts with excellent accuracy. 

\section{Backgrounds}
\label{sec:backgrounds}

\subsubsection{Combinatorial Backgrounds}
\begin{figure}
\centering
\includegraphics[width=200pt]{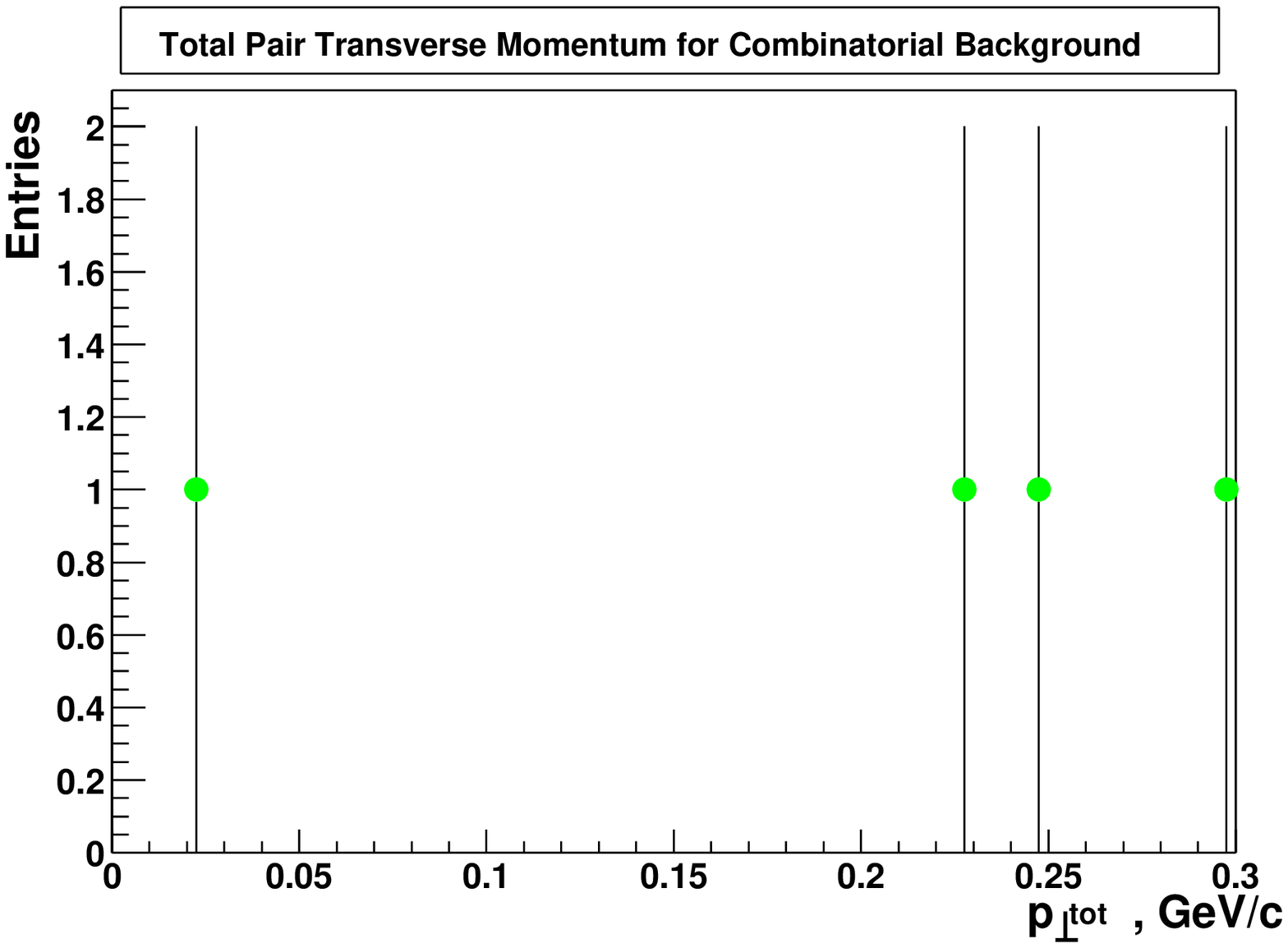}
\includegraphics[width=200pt]{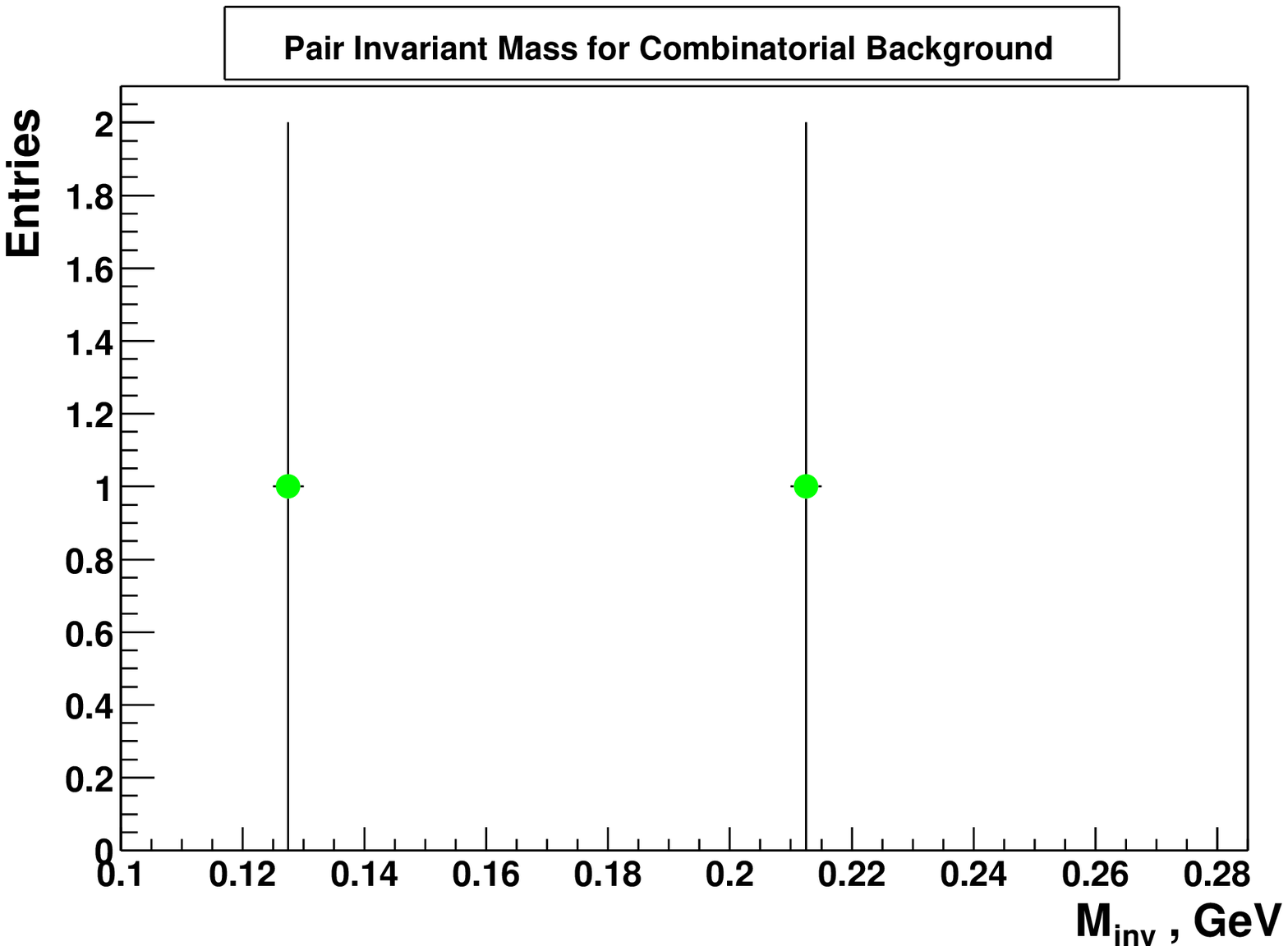}
\caption[Total pair transverse momentum and invariant mass spectrum of combinatorial background pairs]{Total pair transverse momentum and invariant mass of like-sign pairs. Left: events selected with 0.14 GeV$<M_{inv}<$0.265 GeV. Right: events selected with $p_{\perp}^{tot}< 0.1 $ GeV/c.}
\label{fig:incoherentPt}
\end{figure}

Random combinations of tracks in low-multiplicity events (e.g. grazing nuclear collisions) can occasionally yield an event which satisfies all analysis cuts. We use the like-sign pairs (events with primary pair total charge $Q_{tot} \ne 0$) as a model of combinatorial backgrounds. Figure \ref{fig:incoherentPt} shows a total pair transverse momentum spectrum of such pairs. The majority of like-sign pairs are above $p_{\perp}^{tot} > 0.15 $ GeV/c, and there are very few like-sign events below $p_{\perp}^{tot} < 0.15 $ GeV/c, where the coherent \ee events are concentrated. After applying cuts 0.14 GeV$<M_{inv}<$0.265 and $p_{\perp}^{tot}< 0.1 $ GeV/c to the data, there is one background event left. Therefore we neglect the effect of combinatorial backgrounds in the data.

\subsubsection{Coherent Pion Background}
A background to ultra-peripheral \ee events arises from ultra-peripheral  $\pi \pi$  pairs, which come from coherent photo-nuclear $\rho ^0$ production or direct $\pi \pi$ production (Section \ref{sub:rho0}) and are mis-identified as \ee pairs. 
To estimate the number of $\pi ^+ \pi ^-$ pairs which might have passed the analysis cuts, we used a Monte Carlo for this process. 12,000 events were generated with 0.12 GeV $<M_{inv}^{e^+e^-}<$ 0.3 GeV and $|Y^{e^+e^-}|<$1.2. From this sample 5 events were found to pass all analysis cuts in the table \ref{tab:allCuts}; therefore reconstruction efficiency for coherent $\pi \pi$ pairs is E$_{\pi \pi} = 0.0042\%$. Total expected  $\pi \pi$ pair contribution in the selected pool of 52 events is then $\sigma _{AuAu \rightarrow Au^*Au^* + \rho ^0 + \text{ direct } \pi \pi } \cdot \text{L}_{tot} \text{Eff}^{trig} \cdot \text{E}_{\pi \pi} = 0.026$ events.  The value of the cross-section $\sigma _{AuAu \rightarrow Au^*Au^* + \rho ^0}$ was taken from the theory (5.3 mb)\cite{SpencerWithBaltz} and the value of $\text{L}_{tot} \text{Eff}^{trig}$ was set equal to the total luminosity times trigger efficiency in our data sample. 

The $\pi \pi$ contribution is extremely low because of a very small fraction of the pairs with invariant masses (in $e^+e^-$ hypothesis) in the range of 140 to 265 MeV and a strong suppression of these pairs by particle identification requirement.

\subsubsection{Other sources of backgrounds}
Coherent ultra-peripheral $\rho ^0$ mesons can decay into \ee   pairs. Due to a combination of the small $\rho ^0$ production cross-section and a small $\rho ^0 \rightarrow e^+e^-$ branching rate ($\sim 5\cdot 10^{-5}$) this reaction does not produce any appreciable backgrounds.

Coherent $\mu^+ \mu^-$ are not expected to contribute any backgrounds due to the small cross-section of this reaction ($\sigma _{\mu \mu } / \sigma _{ee} \sim \left( m_e / m_\mu \right)^4 $).

Individual protons inside the heavy ions can emit photons independently of the rest of the protons in the ion. These photons and their interactions are called 'incoherent', to distinguish them from the photons emitted by the ions a a whole. The cross-section for incoherent two-photon processes is suppressed by a factor of $1/Z^2 \sim 1.6 \cdot 10^{-4}$, therefore their contribution is negligible.

Cosmic muons can be reconstructed as a pair of tracks in STAR TPC with the total charge zero. These spurious pairs are strongly suppressed by the requirement of the coincident neutron signals in the ZDCs. Additional suppression of this background is provided by the vertex cut $R_{vert}<9$ cm and $|z_{vert}|<100$ cm. The remaining cosmics rate is negligible. The vertex requirement also rejects the pairs consisting of the tracks which come from the beam-gas interactions.

Photons produced in hadronic AuAu collisions can pair-convert into an \ee pair in the material of SVT (located at $ \sim 6 \text{ cm } \div 15$ cm and in the STAR TPC gas volume). However, the opening angle of the \ee from photon conversions is very small ($\sim 2m_e /E_{photon} \approx 0.01$ radians for a photon of energy $E_{photon} = 100$ MeV) and the invariant mass of such pairs is extremely small, well outside of our acceptance range\cite{Perkins}.

\section{Efficiency Corrected Spectra}

\begin{figure}[!h]
\includegraphics[width=200pt]{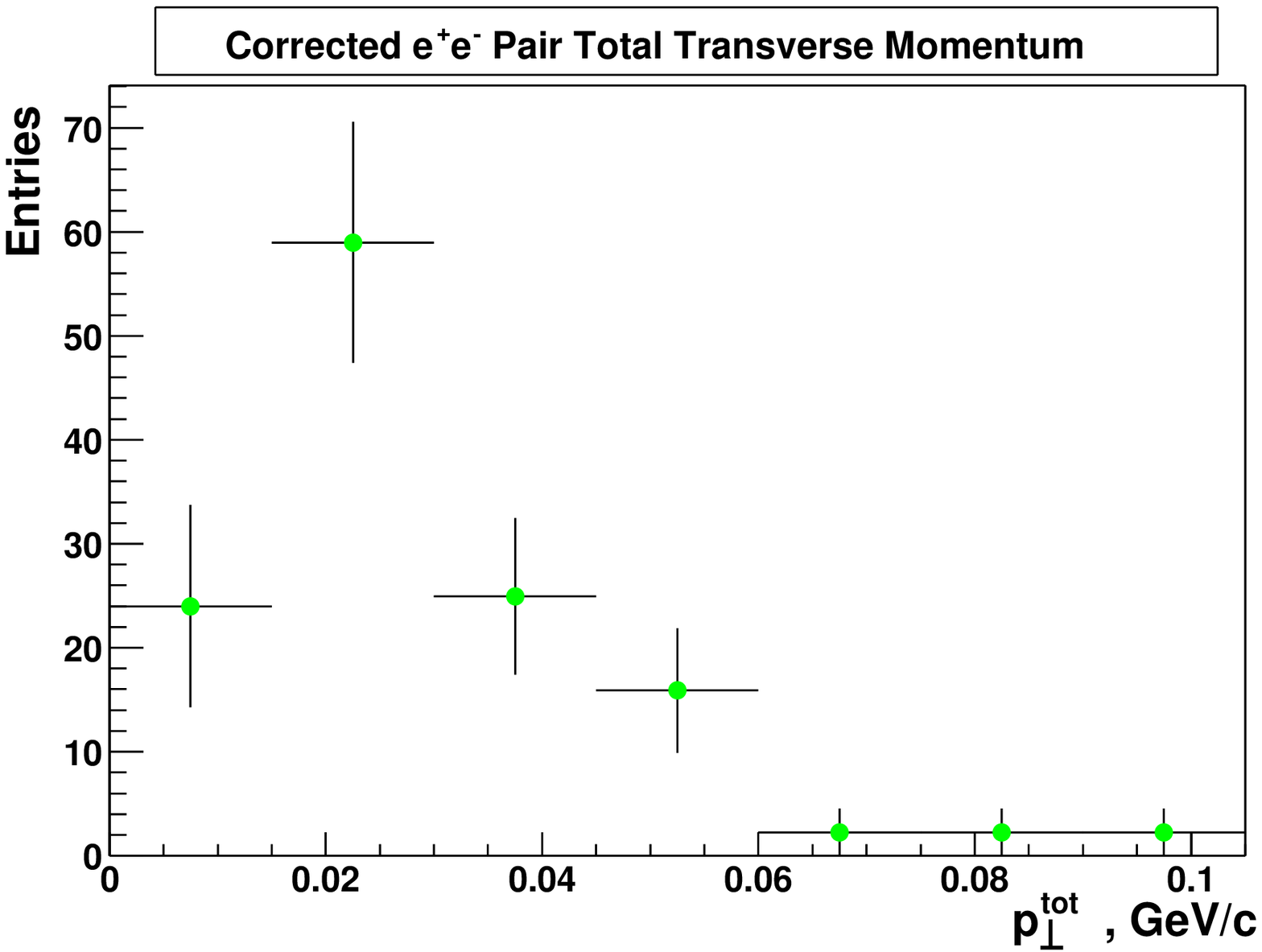}
\includegraphics[width=200pt]{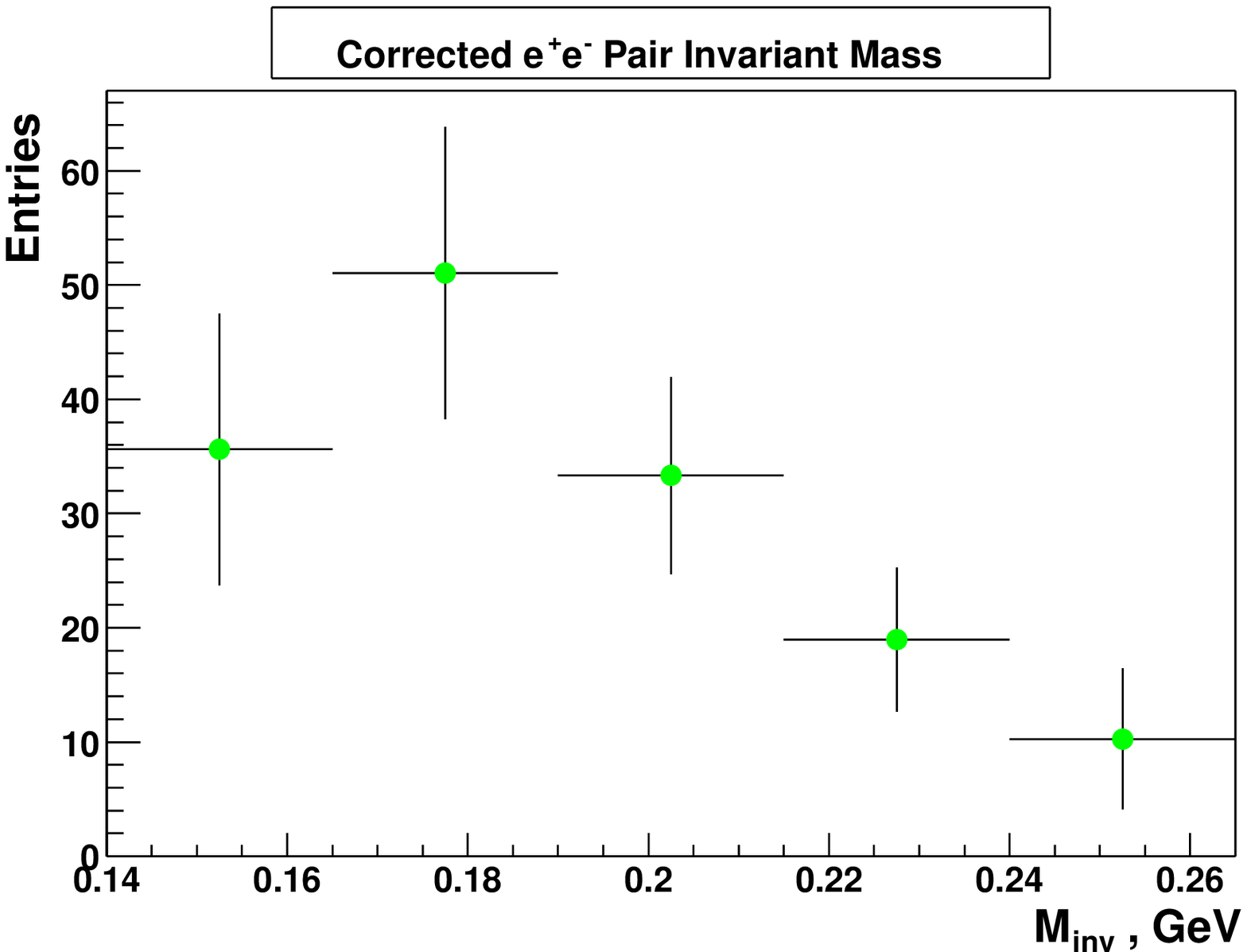}
\includegraphics[width=200pt]{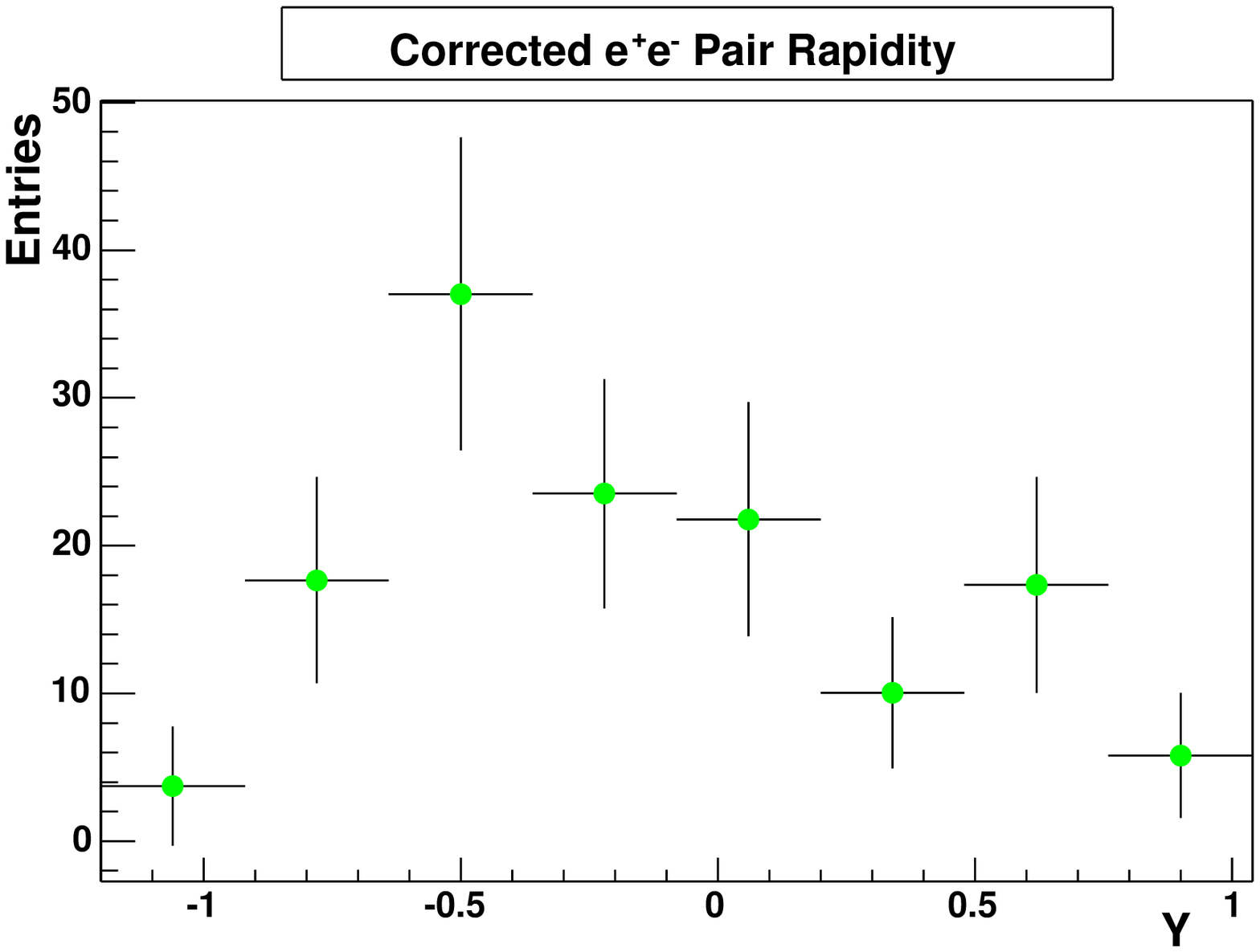}
\includegraphics[width=200pt]{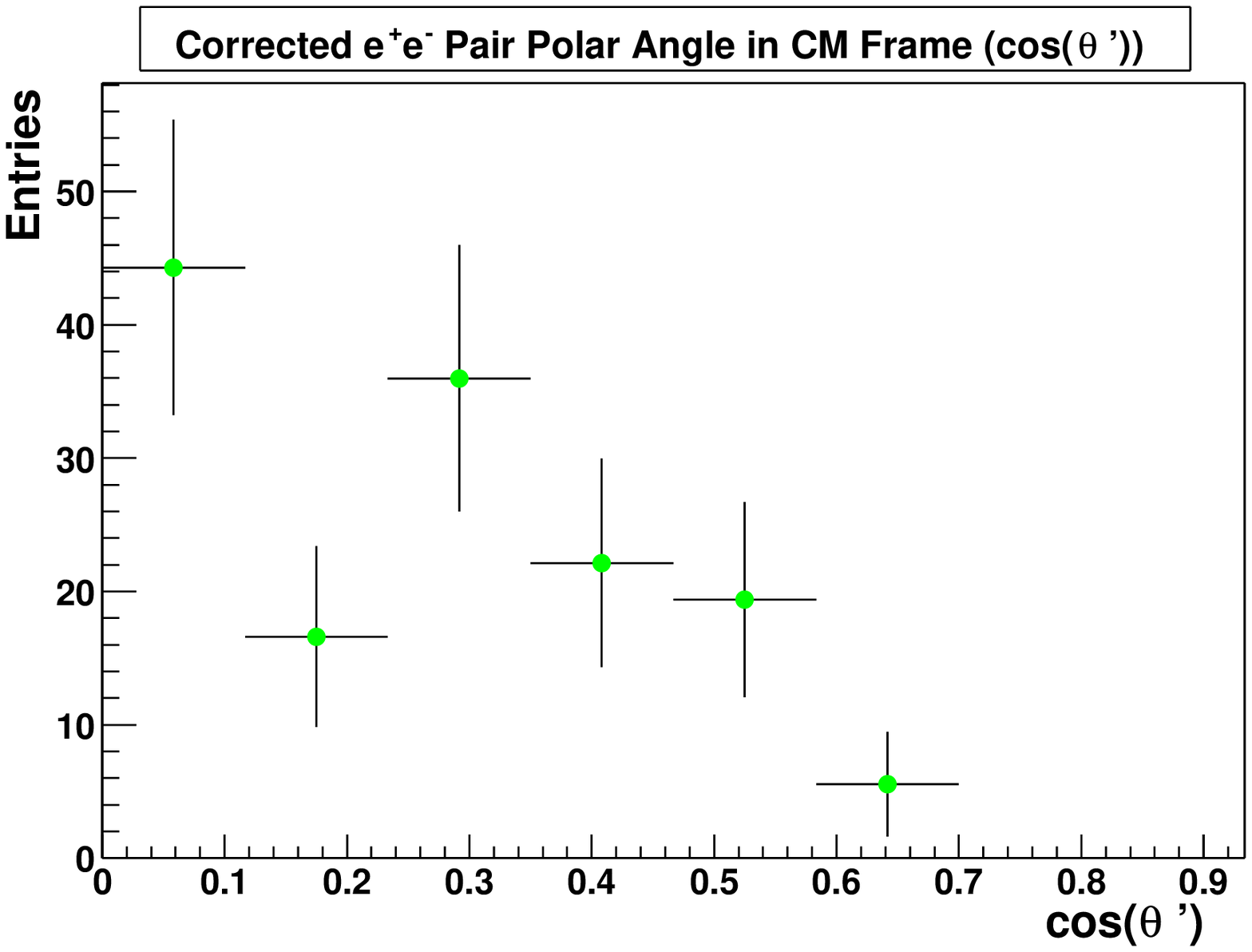}
\caption[Resulting efficiency corrected spectra]{Resulting efficiency corrected spectra.}
\label{fig:CorrectedSpectra}
\end{figure}

We correct the raw event spectra by the reconstruction efficiency. Figure \ref{fig:CorrectedSpectra} shows the resulting $M_{inv}$ , $Y$, $cos(\theta ')$ and $p_{\perp}^{tot}$ distributions.

\section{Extrapolation to $4\pi$}
The acceptance cuts $p_{\perp}^{track}>65$ MeV/c and $ |\eta ^{track}|<1.15$ are specific to the STAR detector. We would like to extrapolate the observed cross-section to the full $4\pi$ range. The kinematic range of the \ee   pair (i.e. $M_{inv}$ and $Y$) will need to be limited, since we do not want to extrapolate to the regions where we have no detecting power. 

First of all, the electron/positron tracks are required have transverse momenta of more than $p_{\perp}^{min} = 65$ MeV/c and the tracks are nearly back-to-back in the transverse plane. This means that the invariant mass of the pair should be greater than $2p_{\perp}^{min}=130$ GeV. The analysis cut $M_{inv} > 140$ MeV already ensures that the pair invariant mass is above this cut-off. The invariant mass of the pair is also limited from above: $M_{inv}<265$ MeV/c due to dropping pair identification efficiency at high masses. We have to carry this cut on to the pair kinematics cuts. The second cut we would like to translate into the pair kinematics is the cut $|\eta ^{track}|<1.15$. As we have shown in Chapter \ref{ch:MonteCarlo} this limits the total pair rapidity to $|Y|<1.15$. This is a final pair kinematics cut.

Below is a summary of the previously used cuts ('detector acceptance cuts') and the limits of the pair kinematic range we would like to extrapolate to ('$4\pi$ kinematic range').

\begin{table}[!h]
\centering
\begin{tabular}{|c|c|}
\hline STAR Detector Acceptance & $4\pi$ Kinematic Range \\
\hline $\left| \eta_{e^-} \right| \! < \! 1.15 $ , $p_{\bot e^-}\! > \! 65 ${\small \ MeV/c  , } $\left| \eta_{e^+} \right| \! < \! 1.15 $ ,    $p_{\bot e^+} \! > \! 65${\small \ MeV/c} & $ |Y| \! < \! 1.15 $ \\ 
{\small 140  MeV}$\! < \! M_{inv} \! < \!${\small 265 MeV} & {\small 140  MeV}$\! < \! M_{inv} \! < \!${\small 265 MeV} \\
\hline 
\end{tabular}
\caption{Comparison of the detector acceptance cuts and pair kinematics cuts.}
\label{tab:DetCutsVsAcceptanceCuts}
\end{table}

Monte Carlo distributions show that all events passing the detector acceptance group of cuts are within the $4\pi$ kinematic range. The reverse is not true, however. The $4\pi$ extrapolation factor is the inverse ratio of the number of events passing the detector acceptance cuts to the number of events within the $4\pi$ kinematic range.

\begin{figure}[!h]
\centering
\includegraphics[width=200pt]{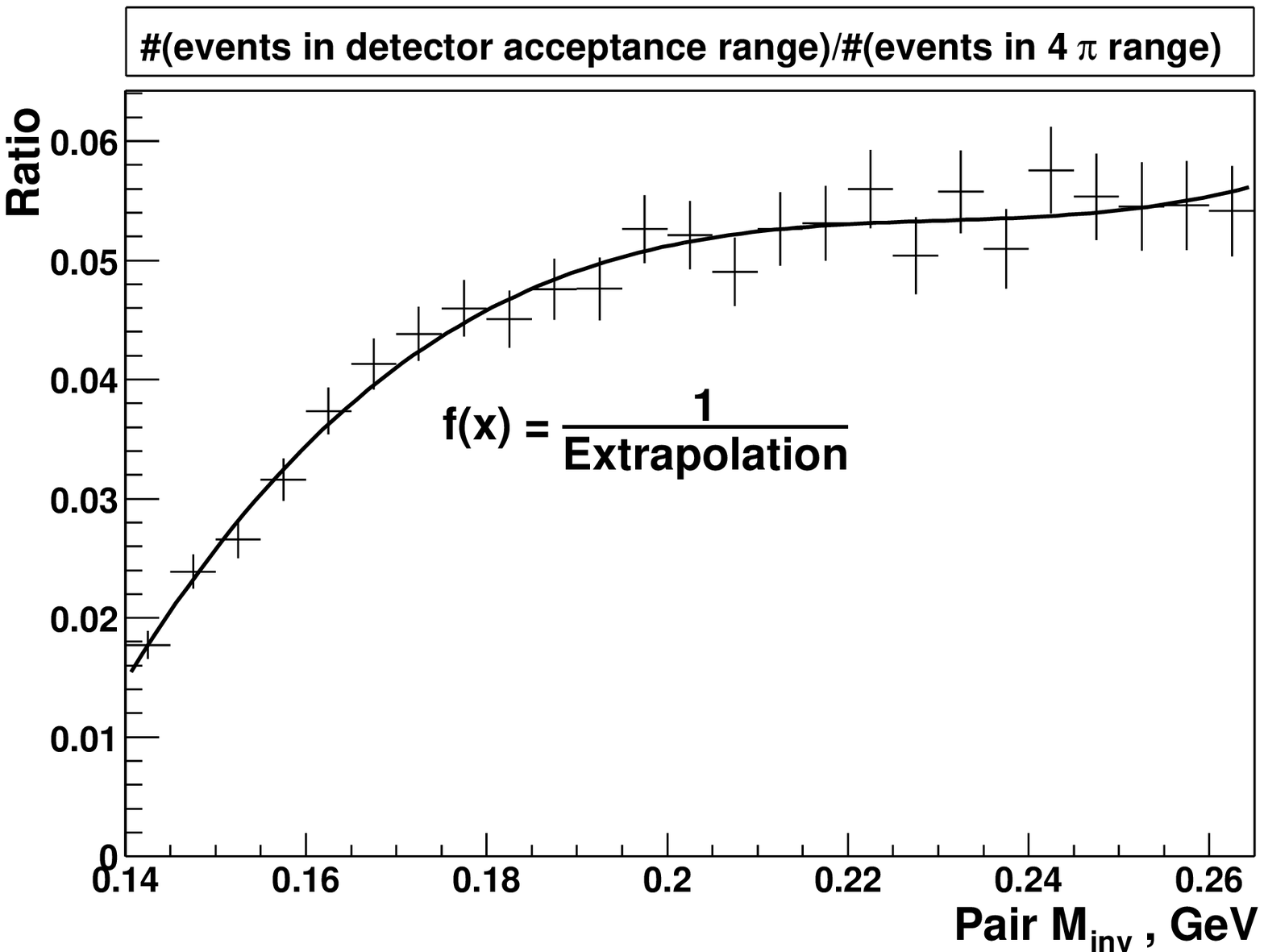}
\includegraphics[width=200pt]{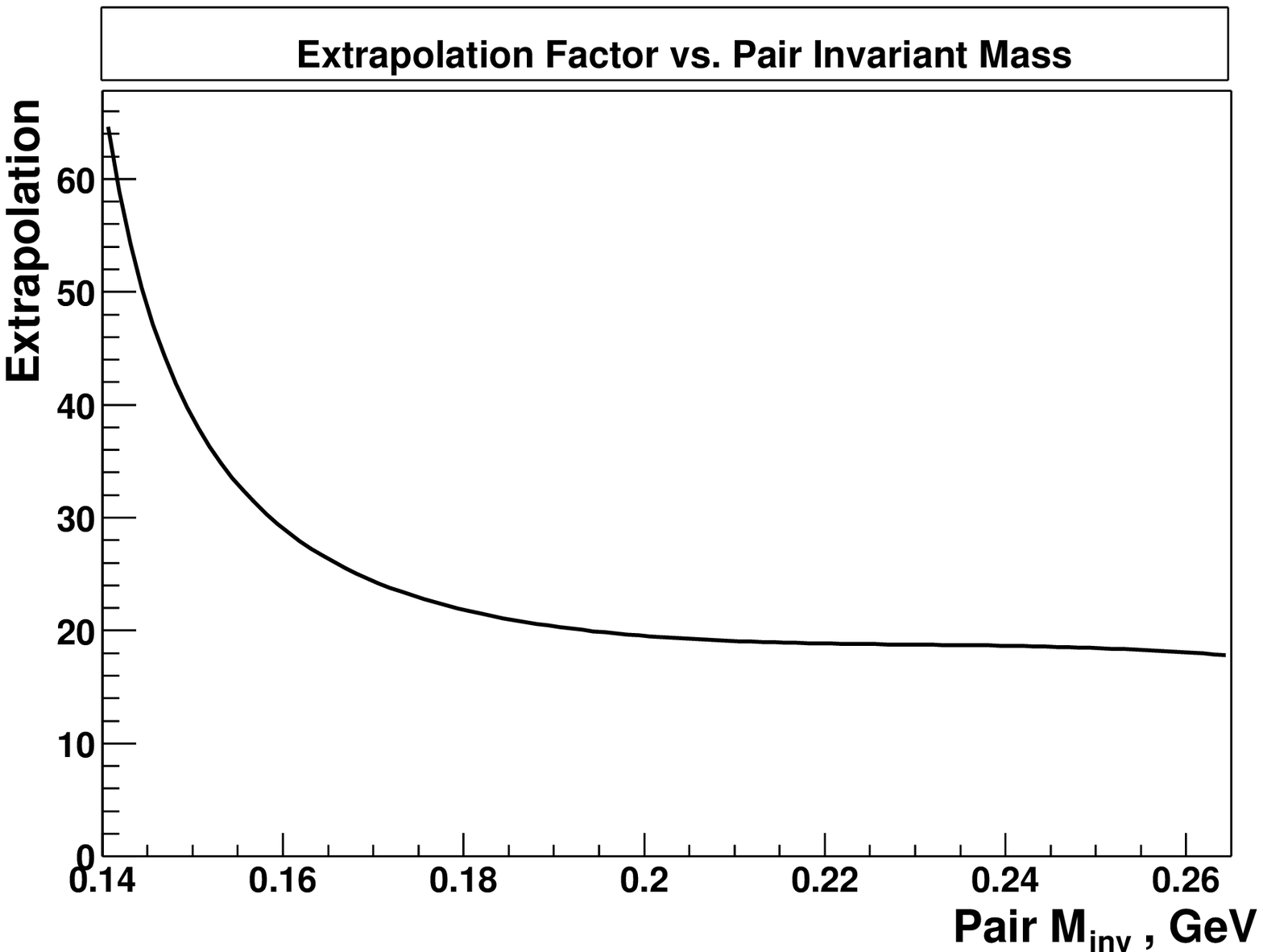}
\caption[$4\pi$ extrapolation factor vs. pair invariant mass]{$4\pi$ extrapolation factor vs. pair invariant mass.}
\label{fig:Extrapolation}
\end{figure}

Figure \ref{fig:Extrapolation} shows the ratio of the number of events in the detector acceptance to the number of events in the $4\pi$ kinematic range and the $4\pi$ extrapolation factor as functions of pair invariant mass.
The extrapolation factor is rather large. This is due to the fact that the electron and positron tracks are very forward peaked (see Figure \ref{fig:Eta1VsEta2}). As a result, for the pairs in the central rapidity range (within the $4\pi$ kinematic range of $ |Y|<1.15$) a large fraction of electron or positron tracks has high values of pseudorapidity ($\eta \sim 4$) outside of the detector acceptance range. 

As a function of pair invariant mass, the extrapolation factor is larger for the lower invariant masses. By restricting the individual track transverse momentum to $p_{\perp}>65$ MeV we suppress pairs with invariant masses of the order of $M_{inv} \sim 130 \div 180$ MeV, but this doesn't affect the pairs with higher invariant masses $M_{inv} \sim 180 \div 265$ MeV, which are typically comprised of higher transverse momentum tracks ($p_{\perp} \sim 100 $ MeV/c).

\chapter{Results}

\label{ch:Results}
We present the differential cross-section distributions of the \ee pairs, which we obtain normalizing the \ee pair spectra by the total luminosity, according to the Equation (\ref{eqn:Sigma}). We compare the experimental results to the event generator results, and to independent theoretical predictions.

\section{Cross-section in Detector Acceptance}

\begin{figure}
\centering
\includegraphics[width=200pt]{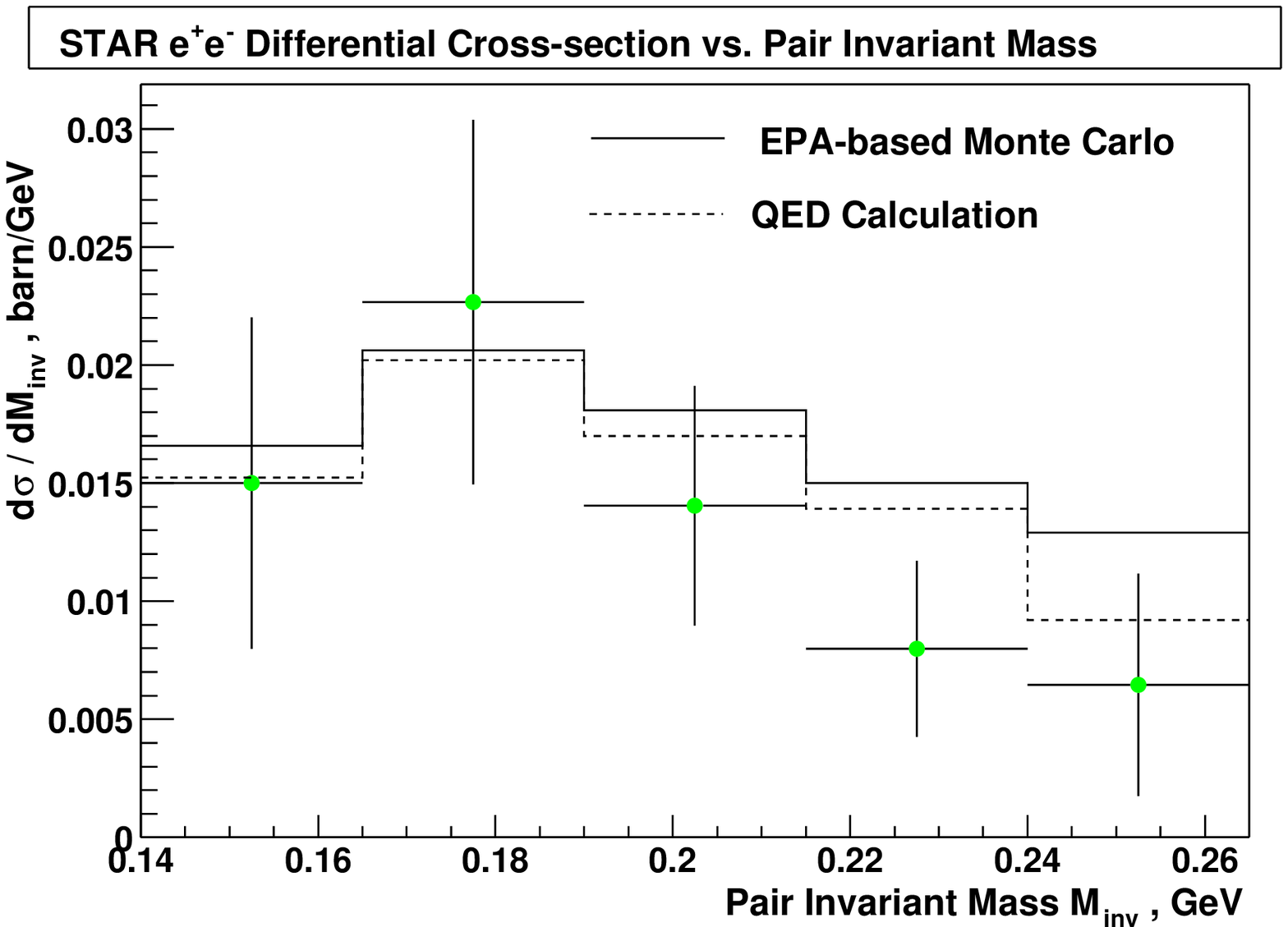}
\includegraphics[width=200pt]{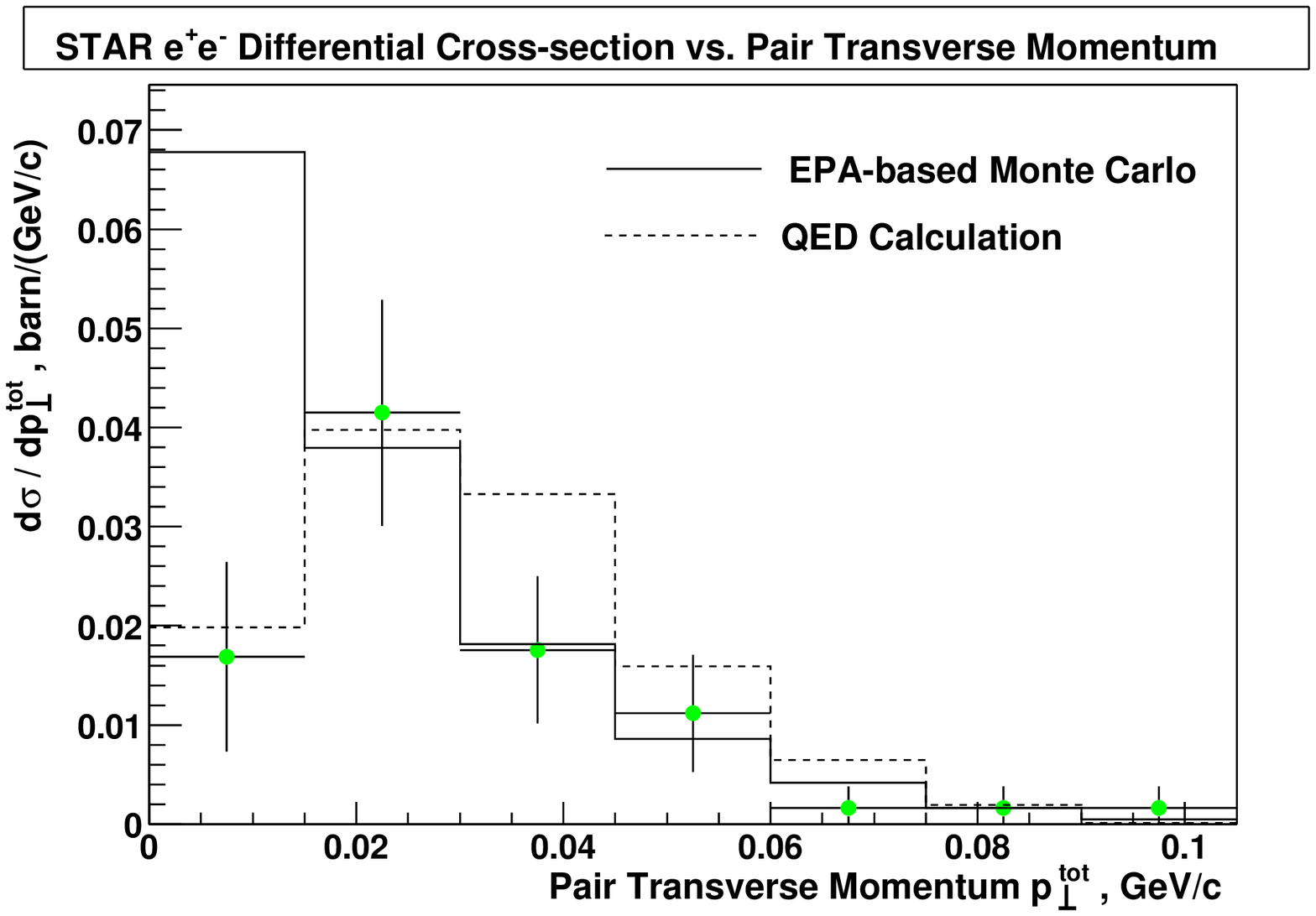}
\includegraphics[width=200pt]{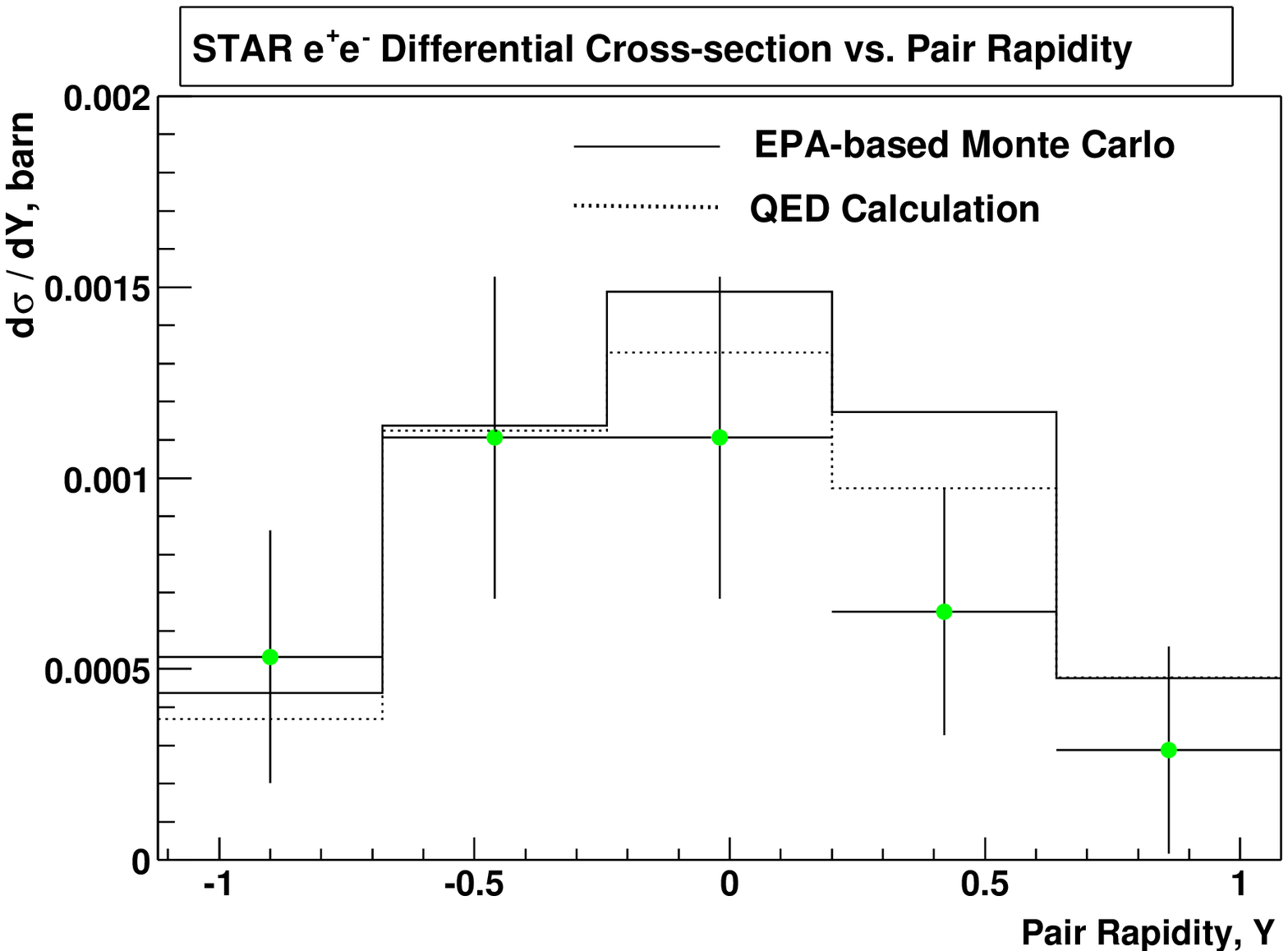}
\includegraphics[width=200pt]{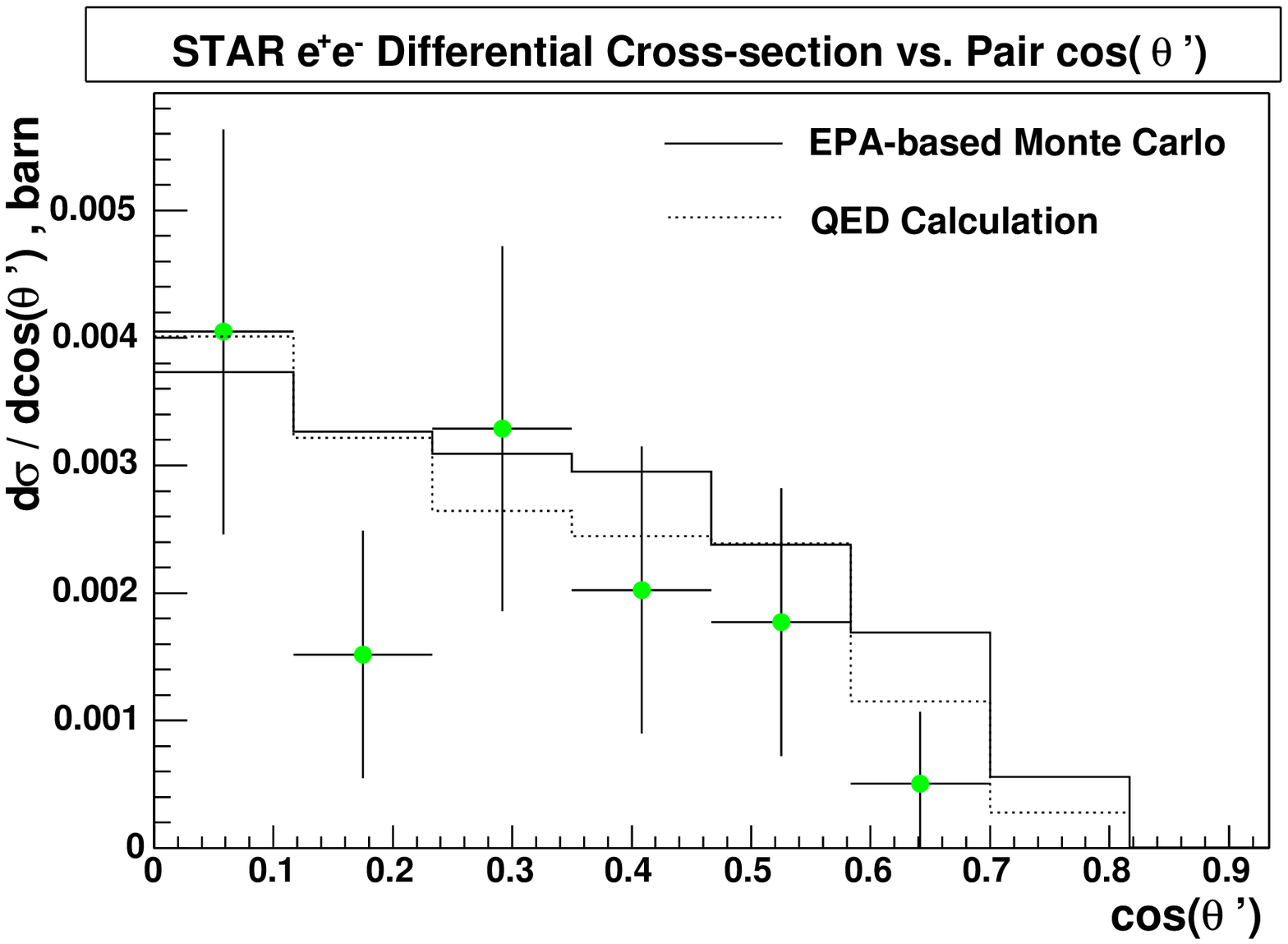}
\caption[Electron-positron production cross-section in detector acceptance range]{Comparison of measured electron-positron production cross-sections (dots) with event generator output (solid histogram), and lowest-order QED calculation (dashed). Error bars incorporate statistical and systematic uncertainties.}
\label{fig:sigma}
\end{figure}

Figure \ref{fig:sigma} shows the measured differential cross-sections $d\sigma_{e^+e^-}/dM_{inv}$, $d\sigma_{e^+e^-}/dp_{\perp}^{tot}$, $d\sigma_{e^+e^-} /dY$ and $d\sigma_{e^+e^-} /dcos(\theta ')$ and compares them to the cross-sections from the event generator ('EPA Monte Carlo') and the QED calcuation.

The shape of the measured distributions $d\sigma /dM_{inv}$, $d\sigma /dY$ and $d\sigma /dcos(\theta ')$ agrees well with the EPA Monte Carlo distribution, with all data points lying no more than $2\sigma$ away from the Monte Carlo. The distribution of $d\sigma_{e^+e^-}/dp_{\perp}^{tot}$ also agrees well with the EPA Monte Carlo, except in the lowest bin. In that bin, for \ee pairs with $p_{\perp}^{tot}<$15 MeV/c, the measured cross-section is $4.5\sigma$ away from the Monte Carlo prediction. However, due to very low statistics (number of raw events in the first bin is 6), the Gaussian '$2\sigma$ deviation' criterion does not apply, and we cannot conclusively reject the possibility that EPA Monte Carlo describes the data accurately.

In the STAR detector acceptance 
\begin{equation}
\label{eqn:resultsAcceptance}
\begin{array}{c}
 \left| {\eta _{e^ -  } } \right| < 1.15 \ , \ p_{ \bot e^ -  }  > 65 \text{ MeV/c } \ , \ \left| {\eta _{e^ +  } } \right| < 1.15 \ , \ p_{ \bot e^ +  }  > 65 \text{ MeV/c} \\ 
140 \text{ MeV }  < M_{inv}  < 265 \text{ MeV } \\ 
 \end{array}
\end{equation}
we find the cross-section of ultra-peripheral \ee production with mutual nuclear excitation at RHIC to be:
\begin{equation}
\label{eqn:result}
\sigma_{AuAu \to Au^*Au^* + e^+e^-} = 1.65 \pm 0.23 \text{(stat.)} \pm 0.30 \text{(syst.)} \text{ mb}
\end{equation}
which is $1.2\sigma$ less than a EPA Monte Carlo prediction of 2.08 mb. The systematic uncertainty quoted consists of a 10\% contribution from luminosity normalization, 6.4\% contribution per track from tracking (total $13\%$ tracking systematic uncertainty) and $8.5\%$ systematic uncertainty from vertex finding. These are added in quadratures for a total systematic error of $18.5\%$.

Additionally, we have recently been provided with a  theoretical prediction of the \AuAuee cross-section at RHIC in the STAR detector acceptance by K. Hencken\cite{HenckenRecent}. This computation is based on the full QED lowest-order term for the process $\gamma \gamma  \to e^ +  e^ -  $, taking into account photon virtualities and treating the external Coulomb fields of the Au ions in the external field approximation\cite{Vidovic}. The computed differential cross-sections are presented in Figure \ref{fig:sigma} as dashed histograms ('QED calculation').

All four differential cross-section distributions in the data  show excellent agreement with the QED calculation. The \ee production cross-section prediction from the QED calculation is 1.88 mb, which is only $0.6\sigma$ higher than the experimental measurement. The agreement of the data vs. QED calculation for  $d\sigma_{e^+e^-}/dp_{\perp}^{tot}$ distribution is much better than the agreement of data vs. EPA Monte Carlo. We believe that this is due to the fact that the QED calculation takes photon virtuality and transverse momenta into the account in computing the \ee production cross-section. The Equivalent Photon Approximation, on the other hand, assumes   that the photons have zero virtuality and the transverse momenta of the photons are ignored in the cross-section computation.

\section{$4 \pi$-extrapolated Cross-section}
Using the extrapolation factor in Figure \ref{fig:Extrapolation}, we extrapolated the measured differential cross-section $d\sigma /dM_{inv}$ in the STAR detector acceptance to the differential cross-section in the full $4\pi$ solid angle for the limited kinematic range:
\begin{equation}
\label{eqn:resultsAcceptance4pi}
140 \text{ MeV } < M_{inv} < 265 \text{ MeV }  \ , \ |Y| < 1.15
\end{equation}

\begin{figure}[t]
\centering
\includegraphics[width=300pt]{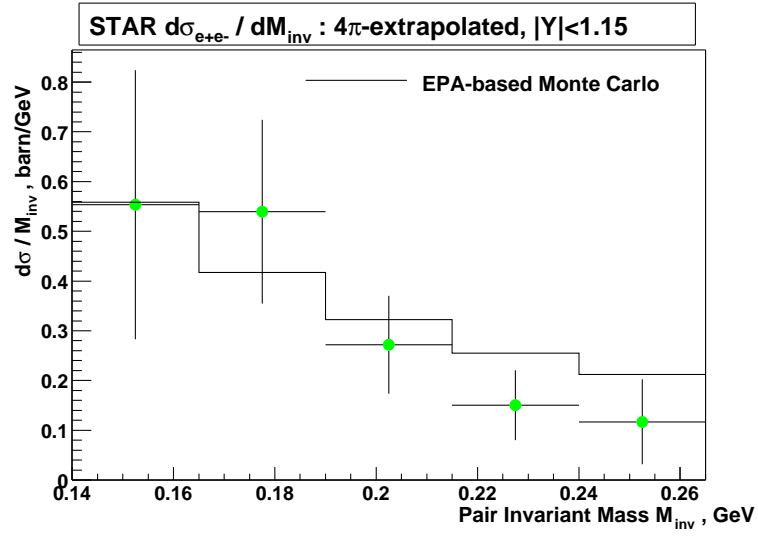}
\caption[Electron-positron production cross-section extrapolated to $4\pi$ range]{$4\pi$ extrapolated cross-section for reaction \AuAuee at STAR.  Solid histogram - EPA Monte Carlo.}
\label{fig:sigma4pi}
\end{figure}

The cross-section was found to be:
\begin{equation}
\label{eqn:result4pi}
\sigma_{AuAu \to Au^*Au^* + e^+e^-} \left| _{4\pi} \right.= 40.8 \pm 5.7 \text{(stat.)} \pm 7.5 \text{(syst.)} \text{ mb}
\end{equation}
and the EPA Monte Carlo prediction of the \ee production cross-section in the $4\pi$ solid angle is 46.4 mb. The extrapolated cross-section and the EPA Monte Carlo prediction are shown in Figure \ref{fig:sigma4pi}.

\addvspace{-20pt}
\chapter{Conclusions}
\vspace{-0.3in}
We observe coherent ultra-peripheral $e^{+}e^{-}$ pairs with mutual nuclear excitation in STAR. The cross-section of \ee production agrees with the EPA Monte Carlo simulation result and an independent theoretical prediction based on the full QED calculation of the \ee production. With the available statistics the data is consistent with the assumption that \ee production is independent from the mutual nuclear excitations at RHIC. The data is also consistent with the lowest-order two-photon QED approximation of the electromagnetic \ee production.

The differential cross-section distributions $d\sigma /dM_{inv}$, $d\sigma /dp_{\perp}^{tot}$, $d\sigma /dY$ and $d\sigma /dcos(\theta ')$ agree with both the EPA Monte Carlo and QED calculation, except for the low $p_{\perp}^{tot}$ events. For events with $p_{\perp}^{tot} \! < \! 15$ MeV/c the data favors the QED calculation over the EPA Monte Carlo. 

Due to low statistics in the identified \ee dataset, we are not able to conclusively rule out the validity of EPA approach for \ee pair production. A tenfold increase in statistics would allow us to answer the question of the EPA applicability at RHIC for \ee production. Increasing statistics by a factor of 100 would most likely allow us to observe the higher-order QED contributions to the \ee production cross-section.

Many improvements can be made in the experimental setup of STAR to enhance the collected \ee statistics. 
Lowering the magnetic field to 0.125 T could extend the TPC tracking to tracks with low $p_{\perp}$, thus greatly increasing the statistics. Forward Time Projection Chambers can also provide tracking of low $p_{\perp}$ or high $|\eta|$  tracks. Using the Multi-wire Proportional Chambers for triggering on the high $|\eta|$ tracks would provide triggering on the \ee pairs produced without the mutual nuclear excitations. The cross-section of the reaction $AuAu \! \to \! AuAu + e^ +  e^ -$ is much larger than the cross-section of the \ee production with mutual nuclear excitation, therefore the statistics for such events will be much higher. With the growing interest in the ultra-peripheral physics results within the RHIC community, the perspectives of \ee production studies in STAR are promising.
\vspace{-0.1cm}
\newpage

\bibliographystyle{plain}
\bibliography{refs}

\appendix 

\chapter{Formulae Reference}
\label{app:variables}

\subsubsection{Determination of the track momenta from the helix parameters}

\begin{table}[h]
\begin{tabular}{lll}
\hline
Symbols  & Name (units) & Definition \\
\hline
$\left(x_0,y_0,z_0\right)^\dagger$  &coordinate of first point (cm) \\
$\Psi^\dagger$        &{\xypl} direction at first point \\
$\lambda^\dagger$     &dip angle ($^ \circ$)     & $\tan^{-1}(p_z / p_{\bot})$ \\ 
$(x_c,y_c)^\dagger$     &coordinate of helix center (cm) \\
R$^\dagger$           &helix radius (cm) \\
                  &rigidity                    & $\frac{0.3RB}{\cos\left(\lambda\right)}$, units $R$(m) and $B$(T)\\
$q^\ast$          &charge (relative to the charge of one $e^+$) \\ 
$p_{\bot}^\ast$        &transverse momentum, p$_{xy}$, (GeV/c) & $0.3RqB$, units $R$(m) and $B$(T)\\
$p_z$               &momentum along the beam axis (GeV/c)   & $p_{\bot}\cdot\tan\left(\lambda\right)$ \\
p                 &total momentum (GeV/c)                 & $\sqrt{p_{\bot}^2+p_z^2}$ \\
E                 &Energy assuming particle of mass(m) (GeV) & $\sqrt{p^2+m^2}$ \\
$y$               &rapidity     
& $\frac{1}{2}\ln\left(\frac{E+p_z}{E-p_z}\right)$ \\
$\eta$            &pseudorapidity     & $-\ln \left[ \tan \left( \frac{180^{\circ }-\lambda }2\right) \right] $ \\
 
\hline

\end{tabular}
\caption[Determination of the track momenta from the helix parameters]
{Formulae for determination of the track momenta from the helix parameters. The symbols $\dagger$ and $\ddagger$ denote variables that are obtained directly from the helix parameters of tracks and the photon finding algorithm, respectively. The $\ast$ symbol is used to denote that in the tracking algorithm, the magnitude of a particle's charge is assumed to be the charge of a positron. Later, a more accurate assumption of a particle's charge can be made through the combination of {\dedx} and rigidity.}
\end{table}

\newpage

\subsubsection{Lorenz Transform}
We work in the $(1,-1,-1,-1)$ metric system. If covariant 4-vector $p_{\mu}$ is equal $(E, p_x, p_y, p_z)$ then contravariant 4-vector $p^\mu$ is equal $(E, -p_x, -p_y, -p_z)$.
The Lorentz transform tensor as a function of $\overrightarrow \beta = (\beta _x , \beta _y , \beta _z ) $ is given by: 
\[
\Lambda (\overrightarrow \beta )  = 
\left( {\begin{array}{*{20}c}
   \gamma  & { - \gamma \beta _x } & { - \gamma \beta _y } & { - \gamma \beta _z }  \\
   { - \gamma \beta _x } & {1 + \frac{{(\gamma  - 1)\beta _x^2 }}{{\beta ^2 }}} & {\frac{{(\gamma  - 1)\beta _x \beta _y }}{{\beta ^2 }}} & {\frac{{(\gamma  - 1)\beta _x \beta _z }}{{\beta ^2 }}}  \\
   { - \gamma \beta _y } & {\frac{{(\gamma  - 1)\beta _x \beta _y }}{{\beta ^2 }}} & {1 + \frac{{(\gamma  - 1)\beta _y^2 }}{{\beta ^2 }}} & {\frac{{(\gamma  - 1)\beta _y \beta _z }}{{\beta ^2 }}}  \\
   { - \gamma \beta _z } & {\frac{{(\gamma  - 1)\beta _x \beta _z }}{{\beta ^2 }}} & {\frac{{(\gamma  - 1)\beta _y \beta _z }}{{\beta ^2 }}} & {1 + \frac{{(\gamma  - 1)\beta _z^2 }}{{\beta ^2 }}}  \\
\end{array}} \right)
\]
\[
\text{where } 
\beta  = \sqrt {\beta _x^2  + \beta _y^2  + \beta _z^2 } 
\text { and }
\gamma  = \frac{1}{{\sqrt {1 - \beta ^2 } }}
\]

The Lorentz transform of vector $p^\mu$ is:
\[
\left( {p^{boost} } \right)^\mu   = \Lambda^\mu _{\ \nu} p^\nu  
\]

\subsubsection{Electron-Positron Pair Kinematics}
Let 
\[
p_1^\mu =\left( \left| \overrightarrow{p_1}\right| ,\overrightarrow{p_1}\right) =\left( p_1,\overrightarrow{p}_{1\bot },p_{1z}\right) =\left( p_1,p_{1x},p_{1y},p_{1z}\right) \text{ and}
\]
\[
p_2^\mu =\left( \left| \overrightarrow{p_2}\right| ,\overrightarrow{p_2}\right) =\left( p_2,\overrightarrow{p}_{2\bot },p_{2z}\right) =\left( p_2,p_{2x},p_{2y},p_{2z}\right) \text{\ \ \ \ }
\]
be electron and positron 4-momenta (correspondingly) in the lab frame\footnote{Electron/positron mass is ignored compared to the momentum.}. 
Lorentz transform to the center of mass frame is specified by the boost factor
\[
\overrightarrow{ \beta }^{CM}  = - \frac{{\left( {\overrightarrow p _1  + \overrightarrow p _2 } \right)}}{ p _1  + p _2 }
\]
In the center of mass frame electron/positron momenta are given by 
\[
\left( p_1^\mu \right) ^{CM}=\Lambda^\mu _{\ \nu} \left( \overrightarrow{\beta }^{CM}\right) \cdot p_1^\nu =\left( p_1^{CM},p_{1x}^{CM},p_{1y}^{CM},p_{1z}^{CM}\right) \text{ and}
\]
\[
\left( p_2^\mu \right) ^{CM}=\Lambda^\mu _{\ \nu} \left( \overrightarrow{\beta }^{CM}\right) \cdot p_2^\nu =\left( p_2^{CM},p_{2x}^{CM},p_{2y}^{CM},p_{2z}^{CM}\right)  \text{\ \ \ \ }
\]

Pair kinematical variables can be then expressed as:
\[
p_{\bot }^{tot}=\left| \overrightarrow{p}_{1\bot }+\overrightarrow{p}_{2\bot }\right| 
\]
\[
Y^{pair}=-\ln \left( \frac{(p_1+p_2)-(p_{1z}+p_{2z})}{(p_1+p_2)+(p_{1z}+p_{2z})}\right)  
\]
\[
M_{inv}=\sqrt{p_1^\mu \cdot p_{2\mu }^{}}
\]
\[
\varphi =\cos ^{-1}\left( \frac{\left( \overrightarrow{p}_{1\bot }+\overrightarrow{p}_{2\bot }\right) _x}{p_{\bot }^{tot}}\right) 
\]
\[
\psi =\cos ^{-1}\left( \frac{\left( \overrightarrow{p}_{1\bot }+\overrightarrow{p}_{2\bot }\right) \cdot \overrightarrow{p}_{2\bot }}{p_{\bot }^{tot}\cdot \left| \overrightarrow{p}_{2\bot }\right| }\right) 
\]
\[
\theta ' = \left\{ {\begin{array}{*{20}c}
   {\cos ^{ - 1} \left( {{{p_{1z}^{CM} }}/{{p_1^{CM} }}} \right),\text{ if }p_{1z}^{CM}  > 0}  \\
   {180^ \circ   - \cos ^{ - 1} \left( {{{p_{1z}^{CM} }}/{{p_1^{CM} }}} \right),\text{ if }p_{1z}^{CM}  < 0}  \\
\end{array}} \right.{\rm  }
\]

\subsubsection{Euclidean Rotation}
A Euclidean Rotation can be specified by an axis of rotation $\overrightarrow \nu = (\nu _x ,\nu _y, \nu _z )$ and angle of rotation $\phi$ around the axis $\overrightarrow \nu$. Then a transformation matrix $R$ is  
\[
R(\overrightarrow \nu  ,\phi ) = \left( {\begin{array}{*{20}c}
   {\nu _x^2  + (1 - \nu _x^2 )\cos \phi } & {\nu _x \nu _y (1 - \cos \phi ) - \nu _z \sin \phi } & {\nu _x \nu _z (1 - \cos \phi ) + \nu _y \sin \phi }  \\
   {\nu _x \nu _y (1 - \cos \phi ) + \nu _z \sin \phi } & {\nu _y^2  + (1 - \nu _y^2 )\cos \phi } & {\nu _y \nu _z (1 - \cos \phi ) - \nu _x \sin \phi }  \\
   {\nu _x \nu _z (1 - \cos \phi ) - \nu _y \sin \phi } & {\nu _y \nu _z (1 - \cos \phi ) + \nu _x \sin \phi } & {\nu _z^2  + (1 - \nu _z^2 )\cos \phi }  \\
\end{array}} \right)
\]

A rotation of vector $\overrightarrow a = (a_x, a_y, a_z)$ can be specified as  
\[
\overrightarrow a^{{\rm rotated}}  = R \cdot \overrightarrow a
\]

\subsubsection{Effects of Photon Polarization}
Cross-section formulae (\ref{eqn:AngularCrossSection})  and (\ref{eqn:CrossSectionGammaGammaEE}) assume that the photons in the center of mass frame are circularly polarized. If this is not the case, we need to distinguish two cross--sections: scalar cross--sections $\sigma ^{s}$ (two photon  polarizations are parallel)  and pseudoscalar cross--section $\sigma^{ps}$ (perpendicular polarizations). For two photons in the lab frame with momenta $\overrightarrow {k_1} = (0,0,\omega_1)$, $\overrightarrow {k_2} = (0,0,\omega_2)$ and polarizations $\overrightarrow {\epsilon _1 }  = (\overrightarrow {\epsilon _ {1\bot}  } ,0) $ , $\overrightarrow {\epsilon _2 }  = (\overrightarrow {\epsilon _ {2\bot}  } ,0) $ the total cross--section is then\cite{Vidovic}:
\begin{equation}
\label{eqn:ScalarVsPseudoscalar}
\sigma \left( {\omega _1 ,\omega _2 ,\overrightarrow {\epsilon _1 } ,\overrightarrow {\epsilon _1 } } \right) = \left( {\overrightarrow {\epsilon _1 }  \cdot \overrightarrow {\epsilon _1 } } \right)^2 \sigma ^s \left( {\omega _1 ,\omega _2 } \right) + \left( {\overrightarrow {\epsilon _1 }  \times \overrightarrow {\epsilon _1 } } \right)^2 \sigma ^{ps} \left( {\omega _1 ,\omega _2 } \right)
\end{equation}
where $\sigma ^{s}$ and $\sigma ^{ps}$ are given by\cite{Vidovic}:
\[
\sigma _{\gamma \gamma \to e^{+}e^{-}}^s\left( \!s\right) \!=\!8\pi \alpha ^2\frac 1s\left( 1\!+\!\frac{4m^2}s\!-\!\frac{12m^4}{s^2}\right) \ln \left( \frac{\sqrt{s}}{2m}\!+\!\sqrt{\frac s{4m^2}\!-1}\right) \!+\!4\pi \alpha ^2\left( \frac 1s\!+\!\frac{6m^2}{s^2}\right) \sqrt{1\!-\!\frac{4m^2}s} \text{ and }
\]
\[
\sigma _{\gamma \gamma \to e^{+}e^{-}}^{ps}\left( \!s\right) \!=\!8\pi \alpha ^2\frac 1s\left( 1\!+\!\frac{4m^2}s\!-\!\frac{4m^4}{s^2}\right) \ln \left( \frac{\sqrt{s}}{2m}\!+\!\sqrt{\frac s{4m^2}\!-1}\right) \!+\!4\pi \alpha ^2\left( \frac 1s\!+\!\frac{2m^2}{s^2}\right) \sqrt{1\!-\!\frac{4m^2}s} \text{ \ \ }
\]

In the range of center of mass energies where STAR can observe \ee \ pairs, the two cross--sections are approximately equal. In this case, the expression (\ref{eqn:ScalarVsPseudoscalar}) evaluates to 
\[
\sigma \left( {\omega _1 ,\omega _2 ,\overrightarrow {\epsilon _1 } ,\overrightarrow {\epsilon _1 } } \right) = \left[ {\left( {\overrightarrow {\epsilon _1 }  \cdot \overrightarrow {\epsilon _1 } } \right)^2  + \left( {\overrightarrow {\epsilon _1 }  \times \overrightarrow {\epsilon _1 } } \right)^2 } \right] \cdot \sigma \left( s \right) = \sigma \left( s \right)
\]
since polarization vectors have unit length ($|\overrightarrow {\epsilon }| = 1$). Therefore we do not need to consider two cases of polarization separately.

\subsubsection{Two-Photon Luminosity}

We start with an expression defining the two-photon luminosity
\begin{equation}
\label{eqn:sigma}
\sigma _{AuAu\to AuAu+e^{+}e^{-}}=\iint \frac{dL_{\gamma \gamma }}{dM_{inv}dY}\cdot \sigma _{\gamma \gamma \to e^{+}e^{-}}\left( M_{inv},Y\right) dM_{inv}dY
\end{equation}
on the other hand, $\sigma _{AuAu\to AuAu+e^{+}e^{-}}$ can be computed directly from photon densities (\ref{eqn:WeizsackerWilliams}) :
\[
\sigma \! = \! \iiiint \sigma _{\gamma \gamma }\left( M_{inv}\left( \omega _1,\omega _2\right) ,Y\left( \omega _1,\omega _2\right) \right) \frac{d^2n_1}{d^2 \overrightarrow b_1}\left(\overrightarrow  b_1,\omega _1\right) \frac{d^2n_2}{d^2 \overrightarrow b_2}\left( \overrightarrow b_2,\omega _2\right) \frac{d\omega _1}{\omega _1}\frac{d\omega _2}{\omega _2}d^2\overrightarrow{b_1}d^2\overrightarrow{b_2}
\]
\begin{equation}
\label{eqn:integral}
=\iint \sigma_{\gamma \gamma } \left( \omega _1,\omega _2\right) \cdot \frac{F_{\gamma \gamma }\left( \omega _1,\omega _2\right) }{\omega _1\omega _2}\cdot d\omega _1d\omega _2
\end{equation}

Using the change of variable theorem, we can convert integral in Equation (\ref{eqn:integral}) into the integral over variables $( M_{inv},Y)$ using the conversion $( \omega _1,\omega _2) = ( M_{inv} \exp (Y) /2 , M_{inv} \exp (-Y) /2 ) $  :
\[
\sigma \! = \! \iint \sigma_{\gamma \gamma } \left( \omega _1\left( M_{inv},Y\right) ,\omega _2\left( M_{inv},Y\right) \right) \cdot \frac{F_{\gamma \gamma }\left( \omega _1\left( M_{inv},Y\right) ,\omega _2\left( M_{inv},Y\right) \right) }{\omega _1\left( M_{inv},Y\right) \cdot \omega _2\left( M_{inv},Y\right) }\cdot \left| \frac{\partial \left( \omega _1,\omega _2\right) }{\partial \left( M_{inv},Y\right) }\right| dM_{inv}dY
\]
where the Jacobian $\left| \frac{\partial \left( \omega _1,\omega _2\right) }{\partial \left( M_{inv},Y\right) }\right| $ is equal to:
\[
\left| {\frac{{\partial \left( {\omega _1 ,\omega _2 } \right)}}{{\partial \left( {M_{inv} ,Y} \right)}}} \right| = \left| {\begin{array}{*{20}c}
   {\frac{{\partial \omega _1 }}{{\partial M_{inv} }}} & {\frac{{\partial \omega _1 }}{{\partial Y}}}  \\
   {\frac{{\partial \omega _2 }}{{\partial M_{inv} }}} & {\frac{{\partial \omega _2 }}{{\partial Y}}}  \\
\end{array}} \right| = \left| {\begin{array}{*{20}c}
   {\frac{{\exp \left( Y \right)}}{2}} & {\frac{{M_{inv} \exp \left( Y \right)}}{2}}  \\
   {\frac{{\exp \left( { - Y} \right)}}{2}} & {\frac{{ - M_{inv} \exp \left( { - Y} \right)}}{2}}  \\
\end{array}} \right| = \frac{{M_{inv} }}{2}
\]
Thus, Equation (\ref{eqn:integral}) can be re-written as:
\[
\sigma _{AuAu\to AuAu+e^{+}e^{-}}  =  \iint \frac{F_{\gamma \gamma }\left( \omega _1\left( M_{inv},Y\right) ,\omega _2\left( M_{inv},Y\right) \right) }{M_{inv}^2/4} \cdot \frac{M_{inv}}2 \cdot \sigma _{\gamma \gamma \to e^{+}e^{-}}\left( M_{inv},Y\right) \cdot dM_{inv}dY
\]
Comparing this to the Equation (\ref{eqn:sigma}) we can conclude
\[
\frac{dL_{\gamma \gamma }}{dM_{inv}dY}=\frac{2F_{\gamma \gamma }\left( \omega _1\left( M_{inv},Y\right) ,\omega _2\left( M_{inv},Y\right) \right) }{M_{inv}}
\]

\end{document}